%% file: master.tex
\documentclass[a4paper,fleqn,10pt]{article}
\pdfoutput=1
\input{text/global}

\hypersetup{
  pdfauthor={Wan-Li Ju, Marek Schoenherr},
  pdftitle={The qT and dphi spectra in W and Z production at the LHC at N3LL'+N2LO}
}
\preprint{IPPP/20/117\\MCnet-21-12}
\author{Wan-Li Ju, Marek Sch{\"o}nherr}
\title{The \texorpdfstring{\qT}{qT} and \texorpdfstring{\dphi}{dphi} spectra in \texorpdfstring{$W$}{W} and \texorpdfstring{$Z$}{Z} production at the LHC at \texorpdfstring{\NNNLLNNLO}{N3LL'+N2LO}\hspace*{-10mm}}
\institute{Institute for Particle Physics Phenomenology, Durham University, Durham DH1 3LE, United Kingdom}
\begin{document}
\vspace*{10mm}
\maketitle
\vspace*{20mm}
\begin{abstract} 
  The production of weak gauge bosons, $W^\pm$ and $Z$, 
  are at the core of the LHC precision measurement program. 
  Their transverse momentum spectra as well as 
  their pairwise ratios are key 
  theoretical inputs to many high-precision analyses,
  ranging from the $W$ mass measurement to the 
  determination of parton distribution functions. 
  Owing to the different properties of the $W$ and $Z$ boson 
  and the different accessible fiducial regions for their 
  measurement, a simple one-dimensional correlation is insufficient 
  to capture the differing vector and axial-vector dynamics of the 
  produced lepton pair. 
  We propose to correlate them in two observables, the transverse 
  momentum \qT\ of the lepton pair and its azimuthal separation \dphi. 
  Both quantities are purely transverse and therefore accessible 
  in all three processes, either directly or by utilising the 
  missing transverse momentum of the event. 
  We calculate all the single-differential \qT\ and \dphi\ 
  as well as the double-differential $(\qT,\dphi)$ spectra 
  for all three processes at \NNNLLNNLO\ accuracy, resumming small 
  transverse momentum logarithms in the soft-collinear effective 
  theory approach and including all singlet and non-singlet contributions.
  Using the double-differential cross sections 
  we build the pairwise ratios \RWpZ, \RWmZ, and \RWpWm\ 
  and determine their uncertainties assuming fully correlated, 
  partially correlated, and uncorrelated uncertainties in the 
  respective numerators and denominators. 
  In the preferred partially correlated case we find uncertainties 
  of less than 1\% in most phase space regions and up to 3\% in the 
  lowest \qT\ region.
\end{abstract}
\newpage
\tableofcontents
\input{text/introduction}

\input{text/methods}

\input{text/results}
\input{text/conclusions}
\appendix
\input{text/Appendixes}

\bibliographystyle{amsunsrt_mod}
\bibliography{refs}

  \end{document}

%% file: text/global.tex
\usepackage{amsmath}
\usepackage{amssymb}
\usepackage{array}
\usepackage{calc}
\usepackage{longtable}
\usepackage{multirow}
\usepackage{slashed}
\usepackage{pstricks}
\usepackage{graphicx}
\usepackage{xspace}
\usepackage{units}
\usepackage{tikz}
\usepackage{textcomp}

\numberwithin{equation}{section}
\usepackage{cite}
\usepackage[pdfborder={0 0 0}]{hyperref}
\usepackage[format=hang,labelfont=bf,hypcap=true]{caption}
\usepackage{subcaption}
\usepackage{sectsty}
\usepackage{enumitem}
\allsectionsfont{\sffamily}
\subsubsectionfont{\mdseries\itshape\large}
\makeatletter
\DeclareRobustCommand*{\bfseries}{%
  \not@math@alphabet\bfseries\mathbf
  \fontseries\bfdefault\selectfont
  \boldmath
}
\makeatother
\let\spreprint\empty
\newcommand{\preprint}[1]{\def\spreprint{\protect#1}}
\let\sinstitute\empty
\newcommand{\institute}[1]{\def\sinstitute{\protect#1}}
\makeatletter
\renewcommand{\maketitle}{\begingroup
  \null\thispagestyle{empty}%
    \ifx\spreprint\empty
      \vskip 5ex
    \else
      \flushright\large\spreprint\vskip 10ex
    \fi
    \vskip 5ex
    \flushleft
      {\sffamily\bfseries\huge\@title}\vskip 6ex
      \@author\vskip 2ex
      \ifx\sinstitute\empty
      \else
        {\small\sinstitute}
      \fi
    \vskip 5ex
  \endgroup
}
\makeatother
\renewenvironment{abstract}{\begin{center}
  {\large\sffamily\bfseries Abstract: }
  \begin{minipage}[t]{0.75\textwidth}
}{\end{minipage}\end{center}\vskip 10ex}

\numberwithin{equation}{section}
\allowdisplaybreaks[2]

\newcommand{\LHAPDF}{L\protect\scalebox{0.8}{HAPDF}\xspace}

\newcommand{\Rivet}{R\protect\scalebox{0.8}{IVET}\xspace}

\newcommand{\CutTools}{C\protect\scalebox{0.8}{UT}T\protect\scalebox{0.8}{OOLS}\xspace}
\newcommand{\OneLoop}{O\protect\scalebox{0.8}{NE}L\protect\scalebox{0.8}{OOP}\xspace}

\newcommand{\OpenLoops}{O\protect\scalebox{0.8}{PEN}L\protect\scalebox{0.8}{OOPS}\xspace}
\newcommand{\Collier}{C\protect\scalebox{0.8}{OLLIER}\xspace}

\newcommand{\Cuba}{C\protect\scalebox{0.8}{UBA}\xspace}
\newcommand{\Sherpa}{S\protect\scalebox{0.8}{HERPA}\xspace}

\newcommand{\Amegic}{A\protect\scalebox{0.8}{MEGIC}\xspace}

\long\def\symbolfootnote[#1]#2{\begingroup%
\def\thefootnote{\fnsymbol{footnote}}\footnote[#1]{#2}\endgroup}

\newcommand{\done}{{\rm d}}
\newcommand{\order}{\mathcal{O}}

\newcommand{\mr}[1]{\mathrm{#1}}

\newcommand{\bea}{\begin{eqnarray}}
\newcommand{\eea}{\end{eqnarray}}
\newcommand{\bi}{\begin{itemize}}
\newcommand{\ei}{\end{itemize}}
\newcommand{\hl}{\vphantom{$\int_A^B$}}

\newcommand{\mhhl}{\vphantom{\frac{\pi^2}{6}}}

\newcommand{\hfs}{\hspace*{0.02\textwidth}}

\newcommand{\shortequal}{\!\!=\!\!}

\newcommand{\Gmu}{\ensuremath{G_\mu}}
\newcommand{\alphaGmu}{\ensuremath{\alpha_{\Gmu}}}
\newcommand{\alphamZ}{\ensuremath{\alpha_{m_Z}}}
\newcommand{\sssS}{{\scriptscriptstyle \Sigma}}
\newcommand{\ff}{f\!\!f}

\newcommand{\pT}{\ensuremath{p_\mathrm{T}}}

\newcommand{\pTmis}{\ensuremath{\displaystyle{\not}\pT}}

\newcommand{\qT}{\ensuremath{q_\mathrm{T}}}
\newcommand{\qTvec}{\ensuremath{\vec{q}_\mathrm{T}}}
\newcommand{\qTcut}{\ensuremath{\mu_Q}}
\newcommand{\qTref}{\ensuremath{q_\mathrm{T}^\text{ref}}}
\newcommand{\dqT}{\ensuremath{\Delta q_\mathrm{T}}}
\newcommand{\mT}{\ensuremath{m_\mathrm{T}}}
\newcommand{\dphi}{\ensuremath{\Delta\phi}}
\newcommand{\bT}{\ensuremath{b_\mathrm{T}}}
\newcommand{\bTvec}{\ensuremath{\vec{b}_\mathrm{T}}}
\newcommand{\RWpZ}{\ensuremath{\mathcal{R}_{W^+/Z}}}
\newcommand{\RWmZ}{\ensuremath{\mathcal{R}_{W^-/Z}}}
\newcommand{\RWpWm}{\ensuremath{\mathcal{R}_{W^+/W^-}}}

\newcommand{\NLO}{\ensuremath{\text{NLO}}}\xspace
\newcommand{\NNLO}{\ensuremath{\text{N$^2$LO}}}\xspace
\newcommand{\NNNLO}{\ensuremath{\text{N$^3$LO}}}\xspace
\newcommand{\NLOs}{\ensuremath{\text{NLO$_\text{s}$}}}\xspace
\newcommand{\NNLOs}{\ensuremath{\text{N$^2$LO$_\text{s}$}}}\xspace
\xspace
\newcommand{\NLL}{\ensuremath{\text{NLL$'$}}}\xspace
\newcommand{\NNLL}{\ensuremath{\text{N$^2$LL$'$}}}\xspace
\newcommand{\NNNLL}{\ensuremath{\text{N$^3$LL$'$}}}\xspace
\newcommand{\NLLNLO}{\ensuremath{\text{NLL$'$+NLO}}}\xspace
\newcommand{\NNLLNNLO}{\ensuremath{\text{N$^2$LL$'$+N$^2$LO}}}\xspace
\newcommand{\NNNLLNNLO}{\ensuremath{\text{N$^3$LL$'$+N$^2$LO}}}\xspace
\xspace
\xspace

\newlist{myitemize}{itemize}{3}
\setlist[myitemize]{leftmargin=14em}

\newcolumntype{C}{>{\centering\arraybackslash}p{0.14\textwidth}}

\newlength{\unitcharwidth}


%% file: text/introduction.tex
\section{Introduction}
\label{sec:intro}
 
One of the most important observables at the LHC is the 
differential spectrum of the electroweak gauge bosons in 
their leptonic decay channels \cite{Aad:2011fp,Aad:2012wfa,
Aad:2014xaa,Aad:2015auj,Aad:2019wmn,Chatrchyan:2011wt,
Khachatryan:2015oaa,Khachatryan:2016nbe,Sirunyan:2017igm,
Sirunyan:2019bzr,Aaij:2015zlq}. 
The extraordinary precision reached by the ATLAS and CMS 
collaborations in their measurements enables the precision 
extraction of the parameters of the Standard Model (SM), 
such as the $W$ boson mass \cite{Aaboud:2017svj} 
and parton densities \cite{Ball:2017nwa,Boughezal:2017nla,Bacchetta:2017gcc,Bertone:2019nxa}.
In order to exploit this precision data to its fullest, 
however, theory calculations of equal precision are 
indispensable.
Of particular interest here are angular observables of 
the final state leptons, such as $\phi^*$ in $Z$ production 
\cite{Vesterinen:2008hx,Banfi:2010cf},
as the angular resolution of charged objects is much 
more precise than their energy resolution, which is needed 
for \pT-type observables. 
The purely transverse azimuthal decorrelation \dphi\ of 
the lepton pair carries the same experimental advantages 
as $\phi^*$, while being less favoured theoretically as it 
is not weighted by the scattering angle and, thus, less 
sensitive to the vector boson transverse momentum \qT.

Unfortunately, the measurement of $W$ production always 
involves the determination of the event's missing momentum 
as a proxy for the inaccessible neutrino momentum. 
Furthermore, only the transverse part of the missing momentum 
can be determined at hadron colliders due to the composite 
nature of the incident protons and the incomplete detector 
geometry, and thus all $W$ observables have to be constructed 
solely in the transverse plane. 
As such, observables such as $\phi^*$ can not be used, 
in contrast to its simpler version \dphi, the azimuthal 
decorrelation of the lepton and the missing transverse momentum. 
For its lepton ingredient it has the same experimental 
advantages as $\phi^*$, depending only on the lepton direction 
but not its momentum.
Conversely, however, the missing transverse momentum's 
transverse direction resolution, being determined by the sum of 
all other measurable particles' momentum vectors, is not 
significantly improved as compared to its magnitude 
\cite{Khachatryan:2014gga,Aad:2016nrq,Aaboud:2018tkc,Sirunyan:2019kia}. 
Still, in combination, a better resolution for \dphi\ should 
be achievable as compared to the transverse momentum \qT\ of the 
reconstructed $W$ boson.

Similarly, due to the purely transverse nature of measurable 
missing momentum vector as well as the rapidity limitations 
of the physical lepton detectors (electromagnetic calorimeter and 
muon chambers), the fiducial regions for 
lepton-neutrino and lepton-pair final states differ by 
definition. 
Thus, any correlation of the two production processes that 
aims for the precise extrapolation from one to the other, 
as is paramount in the $W$ mass measurements for example,
should take into account detector-acceptance-induced difference 
of the internal dynamics of both systems.
Hence, also multidifferential precision predictions 
of the $W^+$, $W^-$ and $Z$ production cross sections and their 
ratios are needed.

On the theory side, the Drell-Yan processes have drawn extensive 
attention for decades.
The total cross section at LO \cite{Drell:1970wh} was one of the earliest processes 
calculated in the Standard Model, and 
its NLO QCD corrections have been derived soon after 
\cite{KubarAndre:1978uy,Altarelli:1978id,Altarelli:1979ub,Harada:1979bj,Aurenche:1980tp}. 
The NNLO QCD corrections \cite{Hamberg:1990np,vanNeerven:1991gh,Anastasiou:2003yy,Anastasiou:2003ds,Melnikov:2006di,Melnikov:2006kv,Catani:2009sm,Catani:2010en} and, very recently, 
the third-order QCD results \cite{Duhr:2020sdp,Duhr:2020seh} 
are known as well. 
Similarly, the transverse momentum spectrum of the gauge boson 
at finite \qT\  
is known also up to $\order(\alpha_s^3)$ \cite{Arnold:1988dp,Gonsalves:1989ar,Giele:1993dj,Boughezal:2015ded,Boughezal:2016isb,Boughezal:2015dva,Boughezal:2016dtm,Ridder:2015dxa,Ridder:2016nkl,Gehrmann-DeRidder:2016jns,Gauld:2017tww,Gehrmann-DeRidder:2017mvr}.
In addition to the QCD corrections higher-order electro-weak (EW) 
effects \cite{Baur:1997wa,Baur:2001ze,Dittmaier:2001ay,Arbuzov:2005dd,CarloniCalame:2006zq,Zykunov:2005tc,Arbuzov:2007db,Denner:2009gj,Denner:2011vu,Hollik:2015pja,Kallweit:2014xda,Kallweit:2015dum} 
and QCD-EW mixed corrections emerging at two-loop order 
\cite{Dittmaier:2014qza,Dittmaier:2015rxo,deFlorian:2018wcj,Bonciani:2019nuy,Delto:2019ewv,Buccioni:2020cfi,Bonciani:2020tvf,Cieri:2020ikq,Dittmaier:2020vra,Behring:2020cqi,Buonocore:2021rxx}
are of importance at this level of accuracy as well.

In addition, the small transverse momentum region of the gauge boson 
is of particular interest as it contains the bulk of the production 
cross section. 
In light of its sensitivity to the soft and collinear radiations, 
fixed-order calculations are dominated by the powers of large 
logarithms of the form $\ln \lambda_{\mathrm T}$, with 
$\lambda_{\mathrm{T}}\equiv \qT/m$ and $m$ being the mass of the 
produced (off-shell) gauge boson, spoiling the convergence 
of the perturbative expansion.
It is thus imperative to resum these logarithms to all perturbative 
orders.

Based on the infrared-collinear (IRC) properties of QCD, the 
exploration of the small-\qT\ exponentiation has been a topic 
of investigation since the formulation of the theory 
\cite{Dokshitzer:1978hw,Parisi:1979se,Curci:1979bg,Bassetto:1979nt,Collins:1981uk,Collins:1981va,Kodaira:1981nh,Kodaira:1982az,Catani:1988vd,Davies:1984hs}, and the first all-order proof 
was achieved by Collins, Soper and Sterman~\cite{Collins:1984kg}.
After that, a formalism of recombining the occurring ingredients 
has been proposed in Refs.~\cite{Catani:2000vq,Bozzi:2005wk,Bozzi:2007pn},
such as to arrive at a process-independent Sudakov form factor.
In the recent decades, many efforts have been devoted to this 
theme and alternative schemes have been proposed and implemented, 
such as the distributional space resummation \cite{Ebert:2016gcn}, 
the direct momentum-space resummation 
\cite{Monni:2016ktx,Bizon:2017rah,Bizon:2019zgf,Bizon:2018foh} 
and a number of variants within the soft-collinear effective 
theory (SCET) \cite{Bauer:2000yr,Bauer:2001yt,Beneke:2002ph}. 
As one of the more popular factorisation techniques, SCET enables 
a formal but flexible way to explore the factorisation properties 
in the small-\qT\ domain \cite{Becher:2010tm,GarciaEchevarria:2011rb,Becher:2011dz,Chiu:2011qc,Chiu:2012ir,Li:2016axz,Li:2016ctv}.

With the progresses made in the framework development, 
the resummation accuracy has increased steadily.  
Throughout this work, we take the following counting 
rule for the resummation results, 
\begin{equation}\label{eq:intro:def_res_order}
  \begin{split}
    \frac{\done \sigma}{\done^2 \qTvec}
    \sim&\;
      \sigma_\text{Born}\cdot
      \exp\left[\;
             \underbrace{\ln\lambda_{\mathrm{T}} f_0(\alpha_s\ln \lambda_{\mathrm{T}})}_{(\text{LL})}
            +\underbrace{f_1(\alpha_s\ln\lambda_{\mathrm{T}})}_{(\text{NLL,NLL$'$})}
            +\underbrace{\alpha_s f_2(\alpha_s\ln\lambda_{\mathrm{T}})}_{(\text{N$^2$LL,N$^2$LL$'$})}
            +\underbrace{\alpha^2_s f_3(\alpha_s\ln\lambda_{\mathrm{T}})}_{(\text{N$^3$LL,N$^3$LL$'$})}
            +\ldots\;
          \right]
    \\
    &\;\phantom{\sigma_\text{Born}}\cdot
      \Big\{
        1          (\text{LL,NLL});
        \alpha_s   (\text{NLL$'$,N$^2$LL}); 
        \alpha_s^2 (\text{N$^2$LL$'$,N$^3$LL});
        \alpha_s^3 (\text{N$^3$LL$'$,N$^4$LL});
        \ldots
      \Big\} \, ,
  \end{split}
\end{equation}
where the exponent indicates the anomalous dimension level 
required by the resummation accuracy, the desired perturbative 
level for the fixed-order functions is presented in the curly brackets.   
We have taken $\alpha_s \ln \lambda_{\mathrm{T}}\sim \mathcal{O}(1)$ here.
In the previous investigations, the N$^2$LL calculations were 
carried out in Refs.~\cite{Becher:2010tm,Bozzi:2010xn,Becher:2011xn,Banfi:2011dx,Banfi:2011dm,Banfi:2012du,Catani:2015vma,Scimemi:2017etj}, 
whilst the authors in Refs.~\cite{Bizon:2018foh,Bacchetta:2019sam,Bizon:2019zgf,Becher:2020ugp,Ebert:2020dfc} have accomplished N$^3$LL 
very recently. 
Thanks to the developments in fixed-order perturbative calculations, 
the cusp anomalous dimension has achieved four-loop 
accuracy~\cite{Moch:2018wjh} and the fixed-order ingredients 
(including the non-singlet quark form factor~\cite{Gehrmann:2010ue}, 
soft~\cite{Li:2014afw,Li:2016ctv} and beam 
functions~\cite{Luo:2019szz,Luo:2020epw,Ebert:2020yqt}) at N$^3$LO are 
available now. 
The non-cusp anomalous dimensions can be extracted from the latter sectors.

Consequently, we present in this work N$^3$LL$'$ accurate calculations 
where the fixed-order functions are improved by one 
power of $\alpha_s$ order with respect to the unprimed accuracy.\footnote{ 
  During the preparation of this work, two papers \cite{Re:2021con,Camarda:2021ict} at N$^3$LL$'$ 
  accuracy have appeared very recently.
}
During our calculation, not only are the ingredients mentioned 
above assembled, the singlet contributions in the neutral Drell-Yan 
process are also addressed for completeness. 
In spite of its expected smallness, the singlet contribution in 
fact acts as the essential constituent starting from N$^2$LL$'$ or N$^3$LL. 
To include it, the low energy effective field theory (LEEFT) 
\cite{Chetyrkin:1997un,Appelquist:1974tg,Chetyrkin:1993ug,
  Chetyrkin:1993jm,Larin:1993tq,Chetyrkin:1993hk} 
resulting from integrating out the top quark has been 
employed in this work.
With the help of the availability of the complete four-momenta 
of the final state leptons and neutrinos, we compute the 
\qT\ spectrum in an experimentally accessible fiducial region. 
Additionally, this work will also project the resummation of 
small-\qT\ logarithms on the azimuthal decorrelation $\dphi$, and 
compute single- and double- differential cross sections and their 
ratios.

The paper is organised as follows: In Sec.\ \ref{sec:methods} we detail 
the ingredients of our calculation, emphasising on the specifics of the 
resummation. Sec.\ \ref{sec:results} then presents the results for off-shell 
$Z$, $W^+$ and $W^-$ production and the respective $W^\pm/Z$ ratios, 
double differential in $(\qT,\dphi)$. 
Sec.\ \ref{sec:conclusions} summarises our findings. 
Finally, the appendices collect the details on the specific size and impact 
of the singlet contributions and leptonic power corrects. 
They also detail the process-specific hard functions.

%% file: text/methods.tex
\section{Details of the calculation}\label{sec:methods}

In this section we detail the construction of our resummed results 
using the SCET formalism for both the \qT\ and \dphi\ observables. 
These expressions, fully differential in the lepton momenta, are then 
matched to the respective fixed-order calculation.

\subsection{Factorisation and fixed-order functions}
\label{sec:methods:fact}
 
From the QCD factorization theorem~\cite{Collins:1989gx},  
the differential cross section for the Drell-Yan (DY) process 
can be expressed as
\begin{equation}\label{eq:methods:fact:qcd_fac}
  \begin{split}
    \frac{\done^5\sigma}{\done^2\qTvec\,\done Y_L\,\done M_L^2\,\done\Omega_{L}}
    =&\;
      \sum_{i,j}\int^1_{\tau_{\text{min}}} d\tau  \, \ff_{ij}(\tau,\mu_F) \, \frac{\done^5\hat{\sigma}_{ij}(\tau,\mu_R,\mu_F)}
           {\done^2\qTvec\,\done Y_L\,\done M_L^2\,\done\Omega_{L}}\;,
  \end{split}
\end{equation}
where  $Y_L$ and $M_L$ stand for the rapidity and the invariant mass 
of the final state lepton pair, respectively. 
$\Omega_L$ represents the solid angle of one of the final leptons in 
the lepton-pair rest system.   
$\hat{\sigma}_{ij}$ denotes the partonic cross section which depends 
on the renormalisation scale $\mu_R$ as well as the factorisation 
scale $\mu_F$. 
$\ff_{ij}$ is  the effective parton luminosity function.  
It is defined as
\begin{equation}
  \begin{split}
    \ff_{ij}(\tau,\mu_F)
    =&\; 
      \int^1_{\tau} \frac{d\xi}{\xi}\,
      f_{i/N_+}(\xi,\mu_F) \, f_{j/N_-}(\tau/\xi,\mu_F) \; ,
  \end{split}
\end{equation}
where $f_{i/N}$ is the parton distribution function (PDF) for the 
parton $i$ out of the nucleon $N_\pm$ traveling in the $\pm z$ direction.
In addition, eq.~\eqref{eq:methods:fact:qcd_fac} also involves the 
parameter $\tau\equiv\hat{s}/s$, where $s$ and $\hat{s}$ are the 
square of the hadronic and partonic colliding energies, respectively.
Its minimal value is a function of the invariant mass and transverse 
momentum to be produced, 
\begin{equation}
  \begin{split}
    \tau_{\text{min}}
    =&\;
      \frac{1}{s} 
      \left[
        \cosh(Y_L)\overline{M}_{\mathrm{T}}
        +\sqrt{\qT^2+\sinh^2(Y_L)\overline{M}^2_{\mathrm{T}}}
      \right]^2\,.
  \end{split}
\end{equation}
Therein, $\overline{M}_{{\mathrm T}}=\sqrt{M_L^2+\qT^2}$. 
In the small \qT\ regime particularly concerned, the partonic cross 
section $\hat{\sigma}_{ij}$ can be factorised further.
In the context of SCET, one can in principle utilise 
the decoupling transformation~\cite{Bauer:2001yt} to express 
$\hat{\sigma}_{ij}$ as a convolution of hard, collinear and soft sectors.
 However, neither the collinear nor soft sector at this stage is 
well-defined due to the appearance of the rapidity singularity \cite{Manohar:2006nz}.
Various different regulators have been proposed in the 
recent years, such as the analytic  regulator 
\cite{Becher:2010tm,Becher:2011dz}, 
the $\Delta$ regulator \cite{Chiu:2011qc,Chiu:2012ir} and 
exponential regulators \cite{Li:2016axz,Li:2016ctv}.  
In light of its particular performance in the fixed-order calculations, 
the framework with the  exponential regulator will be employed in this work. 
In this approach, the factorisation formula of 
eq.~\eqref{eq:methods:fact:qcd_fac} can be rewritten 
as~\cite{Li:2016axz,Li:2016ctv}, 
\begin{equation}\label{eq:methods:TMD_factorisation}
  \begin{split}
    \frac{\done^5\sigma}{\done^2\qTvec\,\done Y_L\,\done M_L^2\,\done\Omega_{L}}
    =&\;
      \frac{1}{16s(2\pi)^4M_{L}^2}\;
      \sum_{i,j}\Big[ \int\done^2 \bTvec\;
      e^{i \bTvec \cdot \qTvec}\;
      \mathcal{H}^{V}_{ij}(M_L,\Omega_L,\mu_R)\;
      \mathcal{S}( {\bTvec}, \mu_R,\nu)\\
    &\hspace*{40mm}{}\;
      \mathcal{B}_+^i(\eta_+,\bTvec, \mu_R,\nu)\;
      \mathcal{B}^j_-(\eta_-,\bTvec, \mu_R,\nu)\; \Big]+\mathcal{O}(\lambda_{\mathrm{T}})\;.
      \hspace*{-10mm}
  \end{split}
\end{equation}
Therein the light-cone decomposition has been carried out upon 
the impact parameter $b$, i.e., 
\begin{equation}
  \begin{split}
    b^{\mu}
    =&\;
      \frac{b\cdot n_+}{2}\,n_-^\mu + \frac{b\cdot n_-}{2}\,n_+^\mu + \bTvec
    \equiv
      b_+ n_-^\mu + b_- n_+^\mu + \bTvec\;.
  \end{split}
\end{equation}
Here $n_\pm$ are two reference vectors satisfying $n_\pm^2 = 0$ and 
$n_+ \cdot n_- = 2$.  
For later reference, we also introduce the light-cone coordinate 
$b^\mu\equiv(b_+,b_-,\bTvec)$. 

The integrand of eq.~\eqref{eq:methods:TMD_factorisation} comprises 
three kinds of ingredients. 
$\mathcal{S}(\bTvec,\mu_R,\nu)$ is the soft function encoding all 
the soft quantum fluctuations surrounding the beam.  
As a result of the ultraviolet (UV) and rapidity divergences, 
it possesses explicit dependences on the virtuality scale $\mu_R$ 
and the rapidity scale $\nu$. 
The definition of $ \mathcal{S}$  reads~\cite{Li:2016axz}
\begin{equation}\label{eq:methods:fact:def_softfunc}
  \begin{split} 
    \mathcal{S}(\bTvec,\mu_R,\nu)
    =&\;
     \mathcal{Z}^{-1}_{\mathcal{S}}\,
     \lim_{\tau\to0^+} \frac{1}{N_c}
     \langle 0|\mathrm{Tr}
     \left[
       \overline{Y}^{\dagger}_+\overline{Y}_-
       (-{i b_0\tau} ,-{i b_0\tau},\bTvec)\,
       \overline{Y}^{\dagger}_-\overline{Y}_+(0)
     \right]
     |0\rangle \bigg|_{\tau=\frac{1}{2\nu}}\;,
  \end{split}
\end{equation} 
where $ \mathcal{Z}^{-1}_{\mathcal{S}} $ is the soft renormalisation 
constant, $\overline{Y}_\pm$ stands for the incoming Wilson 
line along the $n_\pm$ direction, and $\tau$ here denotes the rapidity 
regulator and $b_0=2e^{\gamma_E}$.  
Currently, the soft function $\mathcal{S}(\bTvec,\mu_R,\nu)$ is 
known at N$^3$LO accuracy \cite{Li:2016ctv}. 

The beam functions $\mathcal{B}_{\pm}$ contain the collimated 
contributions along the $\pm z$ direction. 
Introducing the momentum fractions $\eta_\pm=M_L\exp(\pm Y_L)/\sqrt{s}$, 
the field-operator definition for $\mathcal{B}_{+}$ is~\cite{Li:2016axz,Luo:2019hmp} 
 \begin{equation}\label{eq:methods:fact:def_beamfunc}
  \begin{split} 
    \mathcal{B}_{+}^i(\eta_+,\bTvec,\mu_R,\nu)
    =&\;
      \mathcal{Z}^{-1}_{\mathcal{B}_+}\mathcal{Z}^{-1}_{0}\,
      \lim_{\tau\to0^+}
      \int\frac{\done b_+}{2\pi}\, e^{-2i \eta_+ b_{+} P_{-}}
       \langle P| \bar{\chi}_{+,i}
       (-{i b_0\tau},-{i b_0\tau}+b_+,\bTvec)\frac{\slashed{n}_-}{2}
       {\chi}_{+,i}(0) |P \rangle\bigg|_{\tau=\frac{1}{2\nu}},
  \end{split}
\end{equation} 
where $\mathcal{Z}^{-1}_{\mathcal{B}_+} $ is the collinear 
renormalisation constant, $\mathcal{Z}^{-1}_{0}$ is the 
zero-bin subtractor to remove the soft-collinear overlapping 
contribution, ${\chi}_{+,i}$ is the gauge-invariant building 
block for the collinear quark $q_{i}$~\cite{Beneke:2002ni,Hill:2002vw} 
in $+z$ direction.  
The expression for $\mathcal{B}_{-}$ or the anti-quark case 
can be obtained through changing the light-cone components 
or field operators in eq.~\eqref{eq:methods:fact:def_beamfunc}, 
respectively.  
Working in the hierarchy $\qT\gg\Lambda_{\mathrm{QCD}}$, 
the beam function can be further factorised into the following 
form~\cite{Li:2016axz,Luo:2019hmp},
\begin{equation}
  \begin{split}\label{eq:methods:fact:Beam_fac}
    \mathcal{B}_\pm^i(\eta,\bTvec,\mu_R,\nu)
    =&\;
      \sum_{j} \int^{1}_{x_\pm}\frac{d\xi}{\xi}\;
      \mathcal{I}_{ij}(\xi,\bTvec,\mu_R,\mu_F,\nu)\;
      f_{j/N_\pm}\left(\frac{\eta}{\xi},\mu_F\right)\;,
  \end{split}
\end{equation} 
where the factorisation scale $\mu_F$ emerges in the right-hand 
side in the usual way.  
Considering the $\mu_F$ dependence of the PDFs will cancel 
against that in $\mathcal{I}_{ij}$ order by order, we suppress 
$\mu_F$ in the $\mathcal{B}_\pm$ arguments. 
Currently, the hard-collinear function  $ \mathcal{I}_{ij}$ is known 
at the N$^3$LO accuracy \cite{Luo:2019hmp,Luo:2019bmw,Luo:2019szz}.

In addition to the soft and beam functions, the factorisation formula in 
eq.~\eqref{eq:methods:TMD_factorisation} also involves the hard sector 
$\mathcal{H}^{V}_{ij}$ ($V=\gamma/Z,W^\pm$).  
$\mathcal{H}^{V}_{ij}$ can be calculated from the UV renormalised partonic 
amplitudes, 
\begin{equation}\label{eq:methods:fact:def_hardfunc}
  \begin{split}
    \mathcal{H}^{V}_{ij}
    =&\;
      \overline{\sum_{\mathrm {col,pol}}}\,
      \mathcal{Z}_{\mathcal{B}_+}\mathcal{Z}_{\mathcal{B}_-}\mathcal{Z}_{\mathcal{S}}
     \Big|\,\mathcal{M}(q_{i}\bar{q}_{j}\to V^*\to\ell \bar{\ell}) \Big|^2\,,
  \end{split}
\end{equation}
where the sum runs over all the colours and polarisations of the 
initial partons and includes the appropriate averaging factors. 
As the result of renormalisation, $|\mathcal{M}|^2$ is free of UV 
divergences but presents manifest IRC singularities. 
From the strategy of asymptotic expansion 
\cite{Beneke:1997zp,Smirnov:1997ct}, these IRC behaviours should be 
exactly removed by the product of $\mathcal{Z}_{\mathcal{B}_+}$, 
$\mathcal{Z}_{\mathcal{B}_-}$ and $\mathcal{Z}_{\mathcal{S}}$ and 
thus $\mathcal{H}^{V}_{ij}$ is left as a finite quantity. 
In contrast to $\mathcal{B}_\pm^i$ and $\mathcal{S}$, which are 
universal within the three processes under consideration in this paper, 
the hard function depends on the process and encodes the specifics 
of the hard partonic interaction. 
In Sec.~\ref{sec:methods:HijV} their structure will be detailed.

\subsection{Hard function: non-singlet and singlet contributions}
\label{sec:methods:HijV}

\begin{figure} 
\centering
\begin{subfigure}{.28\textwidth}
  \centering
  \includegraphics[width=.45\linewidth, height=0.65\linewidth]{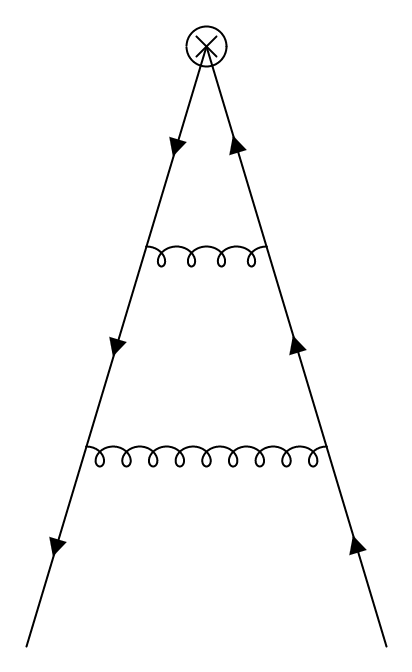}  
  \caption{Non-singlet amplitude.\\\phantom{H}}
  \label{fig:methods:feyndia_hard:nonsing}
\end{subfigure}
\hspace*{0.05\textwidth}
\begin{subfigure}{.28\textwidth}
  \centering
  \includegraphics[width=.45\linewidth, height=0.65\linewidth]{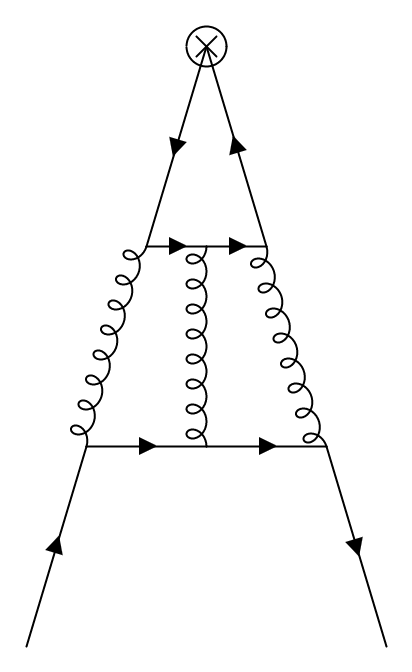}  
  \caption{Singlet amplitude induced by a vector current.}
  \label{fig:methods:feyndia_hard:sing_vector}
\end{subfigure}
\hspace*{0.05\textwidth}
\begin{subfigure}{.28\textwidth}
  \centering
  \includegraphics[width=.45\linewidth, height=0.65\linewidth]{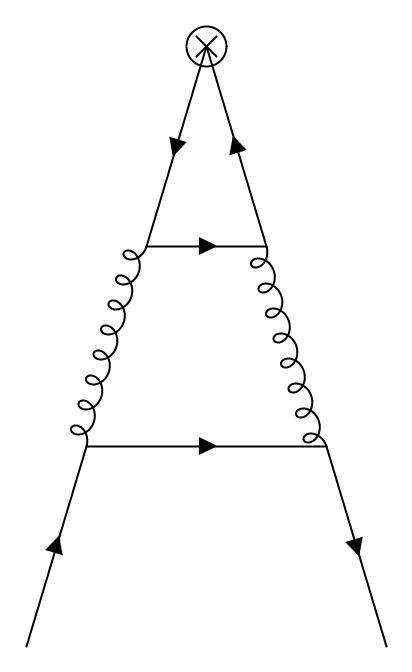}  
  \caption{Singlet amplitude induced by an axial-vector current.}
  \label{fig:methods:feyndia_hard:sing_axial}
\end{subfigure}
\caption{
  Representative Feynman diagrams which contribute to the hard functions. 
  The curly and straight lines denote the gluon and quark propagators, respectively. 
  The crossed circle represents the electroweak current operator.
}
\label{fig:methods:feyndia_hard}
\end{figure}

The Feynman diagrams contributing to the three hard processes under 
consideration in this paper, off-shell $Z$ or $W^\pm$ production, can 
be grouped into two categories according to whether or not the 
incident quark lines are connected to the EW vertex or 
not: the singlet and non-singlet contributions (see Fig.~\ref{fig:methods:feyndia_hard}).
Here, the non-singlet contribution of Fig.\ \ref{fig:methods:feyndia_hard:nonsing} 
collects all configurations that connect the external quark lines to 
the EW vertex while all other configurations are part of the singlet 
contribution. 
The latter can be subdivided according to whether they couple to 
the vector or axial-vector part of the EW vertex,
as depicted in Fig.\ \ref{fig:methods:feyndia_hard:sing_vector} and 
Fig.\ \ref{fig:methods:feyndia_hard:sing_axial}, respectively.

\paragraph*{$Z$ production.}
Here, the quarks coupling to the EW vertex are always of the same flavour. 
Hence, both singlet and non-singlet amplitudes contribute. 
In the following, their construction and embedding in the 
resummation framework is detailed. 
Their numerical impact is fully examined in App.\ \ref{app:numeric:singlet}. 
In essence, while at \NNLO\ the impact of the singlet contributions 
is numerically small,  it reaches the same size at \NNNLO\  as the 
standard non-singlet contribution and it is essential to include it.

All needed amplitudes can in principle be calculated in the full SM and evaluated loop by loop. 
However, the presence of multiple mass scales complicates the loop 
integrations considerably, rendering this method somewhat involved. 
Alternatively, in this work we will employ LEEFT 
resulting from the SM by integrating out the top quark. 
For the vector current, due to its conservative nature, the matching 
from SM onto LEEFT amounts to the re-definitions of the strong 
coupling $\alpha_s$ as well as the field operators 
\cite{Chetyrkin:1997un,Appelquist:1974tg}.\footnote{
  This work takes $m_{u,d,c,s,b}=0$ throughout and 
  hence the redefinition of quark masses is not essential here.
} As a results, one can collect the effective vector currents as 
\begin{equation}\label{eq:methods:HijV:JZv_EFT}
  \begin{split}
    {V}_{\gamma}^{\mu}
      =&\;
        \sum_{q_i=u,d,c,s,b} g_{\gamma}^{q_i}\, \bar{q}_i\gamma^{\mu}  q_i\,,\\
    {V}_{Z}^{\mu}
      =&\;
        \sum_{q_i=u,d,c,s,b}g_{V}^{q_i}\, \bar{q}_i\gamma^{\mu}  q_i\,.
  \end{split}
\end{equation}
It is apparent that the expressions for SM light quark vector 
currents are formally retained  in ${V}_{\gamma/Z}^\mu$.
For later convenience the currents induced by $\gamma^* q\bar{q}$ 
and $Z^* q\bar{q}$ vertices are listed separately.  
In absence of the EW corrections, one can always distinguish them.  
The coupling factors $g_{\gamma,V}^{q_i}$ collect the EW coupling 
constants,
\begin{equation}\label{eq:methods:HijV:AZqq_cpl}
  \begin{split}
    g_{\gamma}^q
    =&\;
      e\,Q_q\;,\hspace*{10mm}{}
    g_V^q
    =
      \frac{e}{s_w\,c_w}\left(\frac{T^3_q}{2}\,-Q_q\,s_w^2 \right)\,.
    \end{split}
\end{equation}
Here we utilise $e$ as the electromagnetic coupling and $s_w$ and $c_w$ 
as the sine and cosine of the weak-mixing angle $\theta_w$.
$T^3_q$ and $Q_q$ are the third component of the 
weak isospin and charge for the quark $q$, respectively. 

The axial-vector current on the other hand, which exists only for 
$Z$ boson exchange, requires a more 
delicate matching procedure. 
As the renormalisability of SM relies on the anomaly 
cancellation within each quark generation, 
the removal of the top quark by brute force will break the 
renormalisation group invariance (RGI) explicitly and thus it 
necessitates an additional renormalisation constant in the LEEFT for 
its restoration. 
To this end, the following operator basis is proposed for the 
axial-vector effective current
\cite{Chetyrkin:1993ug,Chetyrkin:1993jm,Larin:1993tq,Chetyrkin:1993hk},
\begin{equation}\label{eq:methods:HijV:Ja_EFT}
  \begin{split}
    A_{Z}^{\mu}
    =&\;
      g_A\left[ \sum_{i=1,2,3} \Delta_i^{\mathrm{ns}}+C_t \mathcal{O}_s\right]\;,
  \end{split}
\end{equation}
where the coupling factor $g_A$ again collects the EW coupling constant, 
i.e.
\begin{equation}
  \begin{split}
    g_A=&\;-\frac{e}{4\,s_w\,c_w}\;.
  \end{split}
\end{equation}
It is independent of the quark flavour.
$\Delta_i^{\mathrm{ns}}$ stands for the non-singlet operators of the 
$i$th quark generation,
\begin{equation}\label{eq:methods:def:ns_operator}
  \begin{split}
    \Delta^{\mathrm{ns}}_1
    =\bar{u} \gamma^{\mu}\gamma_5u-\bar{d} \gamma^{\mu}\gamma_5d  ,\,~~\, 
    \Delta^{\mathrm{ns}}_2
    =\bar{c} \gamma^{\mu}\gamma_5c-\bar{s} \gamma^{\mu}\gamma_5s  ,\,~~\,
    \Delta^{\mathrm{ns}}_3
    =\frac{\mathcal{O}_s}{N_F}-\bar{b}\gamma^{\mu}\gamma_5b \;.
  \end{split}
\end{equation}
For the first two quark generations, the operators 
$\Delta^{\mathrm{ns}}_{1,2}$ are formally identical to those in SM, 
whilst for the third one, $\Delta^{\mathrm{ns}}_{3}$ is artificially 
constructed to facilitate the anomalous cancellation. 
Here $N_F$ represents the number of massless quarks, taking the value 
$N_F=5$ throughout this work. 
$\mathcal{O}_s$ is the singlet operator, which is defined as 
\begin{equation}
  \begin{split}
    \mathcal{O}_s=\sum_{q_i=u,d,c,s,b}\bar{q}_i\gamma^{\mu}\gamma_5 q_i.
  \end{split}
\end{equation} 
In addition, $A_{Z }^{\mu}$ also comprises a novel structure  
$C_t \mathcal{O}_s$ beyond what is present in the SM. 
In presence of the axial-anomaly, the renormalised singlet 
$\mathcal{O}_s$ operator exhibits the explicit dependence 
on the renormalisation group transformations (RGT).  
However, this RGT-dependence will be eliminated order by order 
by the Wilson coefficient $C_t$ such that the RGI is still 
maintained in LEEFT. 
To calculate $C_t$, one needs to compare the SM amplitudes 
induced by $\bar{t}\gamma^{\mu}\gamma_5t-\bar{b}\gamma^{\mu}\gamma_5b$ 
with those from $\mathcal{O}_s$ in the limit $M_{L}\ll m_t$. 
Here $m_t$ denotes the pole mass of the top quark. 
In this work we follow the conventions of \cite{Larin:1993tq}, 
\begin{equation}\label{eq:methods:def:ct}
  \begin{split}
    C_t
    =&\;
      -\frac{1}{N_F}
      +\left(\frac{\alpha_s}{4\pi}\right)^2
       \left(-8L_t+4\mhhl\right)
      +\left(\frac{\alpha_s}{4\pi}\right)^3
       \left(-\frac{184}{3}L_t^2-\frac{784}{9}L_t+208\zeta_3-\frac{6722}{27}\right)\;.
  \end{split} 
\end{equation}
where $L_t=\ln(\mu^2/m_t^2)$.  
Here the tree-level result $(-1/N_F)$ balances 
the $\mathcal{O}_s/N_F$ term in $\Delta^{\mathrm{ns}}_3$. 
The singlet contribution starts from the two-loop level and 
the $\mathcal{O}(\alpha^2_s)$ results can be either 
straightforwardly read from the axial-anomaly form factor 
in Ref.~\cite{Bernreuther:2005rw}, or extracted from 
\cite{Collins:1978wz,Chetyrkin:1993jm}. 
The third order expressions can be found in Ref.\ \cite{Chetyrkin:1993ug}.
Particular attention needs to be noted that in 
Refs.~\cite{Chetyrkin:1993ug,Chetyrkin:1993jm} the $\mathcal{O}_s$ 
operator is renormalised in a different prescription from those 
in Refs.~\cite{Larin:1993tq,Bernreuther:2005rw}.
Their conversion is essential and we present 
the corresponding details in App.\ \ref{app:Ct}.   
  
Now that we have all ingredients at hand we can finally 
calculate the hard function. 
From eq.~\eqref{eq:methods:fact:def_hardfunc}, the hard 
sector can be addressed from the square of IRC-subtracted 
on-shell amplitudes.  
In absence of EW corrections, the amplitudes involved 
can be naturally decomposed into two parts: the hadronic 
and the leptonic currents. 
The hadronic current can be expressed using the 
effective vector and axial-vector currents, 
$V_{Z/\gamma }^{\mu}$ and $A_{Z}^{\mu}$, respectively. 
More explicitly, we have
\begin{equation}\label{eq:method:def:hadcur}
  \begin{split}
    H^{\mu,ij}_{\gamma}
    =&\;
      \sqrt{\mathcal{Z}_{\mathcal{B}_+}\mathcal{Z}_{\mathcal{B}_-}\mathcal{Z}_{\mathcal{S}}}\,
      \langle 0|V_{\gamma}^{\mu}|q_i\bar{q}_j\rangle 
    =
      \left(g_{\gamma}^{q_i}C_{\mathrm{ns}}+g_{\gamma}^\sssS C_{\mathrm{s}}^V \right)
      \mathcal{V}^{\mu}_{ij}\,,\\
    H^{\mu,ij}_{Z,V}
    =&\;
      \sqrt{\mathcal{Z}_{\mathcal{B}_+}\mathcal{Z}_{\mathcal{B}_-}\mathcal{Z}_{\mathcal{S}}}\,
      \langle 0|V_{Z}^{\mu}|q_i\bar{q}_j\rangle
    =
      \left(g_{V}^{q_i} C_{\mathrm{ns}}+ g_{V}^\sssS C_{\mathrm{s}}^V \right)
      \mathcal{V}^{\mu}_{ij}\,,\\
    H^{\mu,ij}_{Z,A}
    =&\;
      \sqrt{\mathcal{Z}_{\mathcal{B}_+}\mathcal{Z}_{\mathcal{B}_-}\mathcal{Z}_{\mathcal{S}}}\,
      \langle 0|A_{Z}^{\mu}|q_i\bar{q}_j\rangle
    =
      g_{A}\left[\left(2T_{q_i}+\frac{1}{N_F}\right)C_{\mathrm{ns}}+C_tC_{\mathrm{s}}^A\right]
      \mathcal{A}^{\mu}_{ij}\,,
  \end{split}
\end{equation}
  where $g_{\gamma/V}^\sssS$ is introduced to collect the EW coupling, namely,
\begin{equation}
  \begin{split}
    g_{\gamma}^\sssS
    \equiv
      \sum_{q_i=u,d,s,c,b} g^{q_i}_{\gamma}\;,
    \hspace*{10mm}
    g_{V}^\sssS
    \equiv
      \sum_{q_i=u,d,s,c,b} g^{q_i}_{V}\;.
\end{split}
\end{equation}
$\mathcal{V}^{\mu}_{ij}$ and $ \mathcal{A}^{\mu}_{ij}$ are born 
level amplitudes induced by the vector and axial-vector vertices, 
respectively. 
Their expressions read, 
\begin{equation}
  \begin{split}
    \mathcal{V}^{\mu}_{ij}
    =
      \delta_{ij}\,
      \langle 0|\bar{q_i}\gamma^{\mu} q_i|q_i\bar{q}_i\rangle\Big|_{\mathrm{Born}}\,,
    \hspace*{10mm}
    \mathcal{A}^{\mu}_{ij}
    =
      \delta_{ij}\,
     \langle 0|\bar{q_i}\gamma^{\mu}\gamma_5 q_i|q_i\bar{q}_i\rangle \Big|_{\mathrm{Born}}\,.  
  \end{split}
\end{equation}
As the massless QCD interactions conserve chirality, 
one can always factor them out after the renormalisation 
and IRC-pole subtraction. 
In addition, we have also utilised a set of hard coefficients 
in eq.~\eqref{eq:method:def:hadcur} encoding the loop corrections. 
$C_\text{ns}$ contains all the non-singlet contributions (see 
Fig.~\ref{fig:methods:feyndia_hard:nonsing}) and can be extracted 
from the $\gamma^*q\bar{q}$ form factor. 
In the recent years, the $\gamma^*q\bar{q}$ form factor has been 
calculated up to three-loop level~\cite{Moch:2005id,Baikov:2009bg,
  Gehrmann:2010ue}. 
Further, $C_\text{s}^V$ stems from the singlet contribution to 
the vector current (see Fig.~\ref{fig:methods:feyndia_hard:sing_vector}). 
As Furry's theorem forbids contributions at two-loop order, 
the lowest order result enters at $\mathcal{O}(\alpha_s^3)$, see 
\cite{Baikov:2009bg,Gehrmann:2010ue}.
Similarly, $C_\text{s}^A$ represents the QCD corrections induced 
by $\mathcal{O}_s$. 
Its N$^2$LO expression can be extracted from Refs.~\cite{Moch:2005id,Baikov:2009bg,
  Gehrmann:2010ue,Bernreuther:2005rw}, while the logarithmic dependences 
at third-loop order accuracy can be obtained from the anomalous 
dimensions. The specific expressions for all three coefficient functions 
$C_\text{ns}$, $C_\text{s}^V$ and $C_\text{s}^A$ are listed 
in App.\ \ref{app:funcs}.

Besides the hadronic contributions, the hard function also 
comprises the leptonic ones, namely the leptonic currents, 
including the vector boson propagators,
\begin{equation}\label{eq:methods:HijV:HardAmp_PPZ_AZ}
  \begin{split}
    L^{\mu}_{\gamma}
    =&\;
      \frac{g_\gamma^\ell}{M^2_L}\;
      \langle \ell^{+}\ell^-|\bar{\ell}\gamma^{\mu}\ell|0\rangle\,,   \\
    L_{Z,V}^{\mu}
    =&\;
      \frac{g_V^\ell}{M_L^2-\mu_Z^2}\;
      \langle \ell^{+}\ell^-|\bar{\ell}\gamma^{\mu}\ell|0\rangle\,, \\
    L_{Z,A}^{\mu}
    =&\;
      \frac{g_A^\ell}{M_L^2-\mu_Z^2}\;
      \langle \ell^{+}\ell^-|\bar{\ell}\gamma^{\mu}\gamma_5\ell|0\rangle \;. 
  \end{split}
\end{equation}
Therein, $\mu_Z$ stands for the complex mass of $Z$ boson, 
which will be introduced properly in Sec.\ \ref{sec:results:setup}.
In absence of EW corrections, the couplings of the leptonic sector 
$g_{\gamma,V,A}^\ell$ can be inferred from the leading order vertices, 
\begin{equation}\label{eq:methods:HijV:AZll_cpl}
  \begin{split}
    g_\gamma^\ell=e\,Q_\ell\,,\qquad
    g_V^\ell=\frac{e}{s_w\,c_w}\left[\frac{T^3_\ell}{2}-Q_\ell\,s_w^2\right]\,,\qquad\text{and}\qquad
    g_A^\ell=-\frac{e}{\,s_w\,c_w}\cdot\frac{T^3_\ell}{2}\;.
  \end{split}
\end{equation} 

Putting all pieces together, the hard function can be expressed as 
\begin{equation}\label{eq:methods:HijV:HF_ppZ}
  \begin{split}
    \mathcal{H}^{\gamma/Z}_{ij}(m_t,M_L,\Omega_L,\mu)
    =&\; \overline{\sum_{\mathrm {col,pol}}}\,
      \left| H^{\mu,ij}_{\gamma}   L_{\gamma,\mu}
            + \left(H^{\mu,ij}_{Z,V}+ H^{\mu,ij}_{Z,A}\right)
              \left(L_{Z,V,\mu}+L_{Z,A,\mu}\vphantom{H^{j}_{Z}}\right)
      \right|^2\;.
  \end{split}
\end{equation}

\paragraph{$W^\pm$ production.}
In this process, as the quarks participating in the EW vertex are always 
of different flavours, only non-singlet diagrams will contribute.
In analogy to eqs.\ \eqref{eq:method:def:hadcur}-\eqref{eq:methods:HijV:HF_ppZ}, 
we write the hard functions as
\begin{equation}\label{eq:methods:HijV:HF_ppW}
  \begin{split}
    \mathcal{H}^{W}_{ij}(M_L,\Omega_L,\mu)
        =&\;  \overline{\sum_{\mathrm {col,pol}}}  \,\big |  H^{\mu,ij}_{W}   L_{W,\mu}\big|^2\;,
  \end{split}
\end{equation}
where
\begin{equation}\label{eq:methods:def_Hw_Lw}
\begin{split}
H^{\mu,ij}_{W} = &\frac{e\,V_{ji}}{2\sqrt{2}\,s_w}\,C_\text{ns} \, \langle 0|\bar{q_j}\gamma^{\mu} (1-\gamma_5)q_i|q_i\bar{q}_j\rangle\Big|_{\mathrm{Born}}\,,\\
 L_{W}^{\mu}=&\frac{e }{2\sqrt{2}\,s_w}  \frac{1}{M^2_L-\mu_W^2}\;
      \langle \ell^{\pm}\nu(\bar{\nu})|\bar{\ell}\gamma^{\mu}(1-\gamma_5)\ell|0\rangle \,.  \\
\end{split}
\end{equation}
Here $V_{ji}$ signals the Cabibbo-Kobayashi-Maskawa (CKM) matrix, and
$\mu_W$ represents the complex mass of the $W$ boson.

\subsection{Resummation}

With the help of eq.~\eqref{eq:methods:TMD_factorisation}, the scales 
relevant to the small \qT\ regime  can be successfully separated.  
However, in presence of the scale hierarchy, such as $\qT\ll M_L$, 
any fixed-order expansion of eq.~\eqref{eq:methods:TMD_factorisation} 
will suffer from the large logarithmic terms reducing the perturbativity. 
To address this issue, we employ the solutions of RGEs as well as 
the rapidity group equations (RaGEs) to carry out the scale evolution.
In this way, the fixed-order functions contribute only at their 
intrinsic scales and the large logarithmic terms arising from 
the scale hierarchy can collectively be resummed within the 
exponential functions of the evolution.
The procedure is detailed in the following.

We begin with the RGEs and RaGEs for each sector. 
The RGE for $C_{t}$ reads 
\begin{equation}\label{eq:methods:HijV:ct_RGE}
  \begin{split}
    \frac{\done}{\done\ln \mu^2}\ln C_t (m_t, \mu)=-\gamma_t\;,
  \end{split}
\end{equation}
where $\gamma_t$ is the 
anomalous dimension and its N$^3$LO result can be found in 
Ref.~\cite{Larin:1993tq}. The evolution equations for $C_\text{ns}$, $C_\text{s}^A$ and $C_\text{s}^V$ can be expressed as 
\begin{equation}
  \begin{split}\label{eq:methods:res_rge_rage_C}
    \frac{\done}{\done\ln \mu^2} \ln C_\text{ns} (M_L, \mu)
    =&\;
      \frac{\Gamma_{\mathrm{cusp}}}{2}\,
      \ln\bigg[ \frac{-M_L^2- i\epsilon}{\mu^2}\bigg]
      +\gamma_h\;,\\
       \frac{\done}{\done\ln\mu^2} \ln C_\text{s}^A (M_L, \mu)
    =&\;
      \frac{\Gamma_{\mathrm{cusp}}}{2}\,
      \ln\bigg[\frac{-M_L^2-i\epsilon}{\mu^2}\bigg]+\gamma_h+\gamma_t\;,\\
       \frac{\done}{\done\ln\mu^2} \ln C_\text{s}^V (M_L, \mu)
    =&\;
      \frac{\Gamma_{\mathrm{cusp}}}{2}\,
      \ln\bigg[\frac{-M_L^2-i\epsilon}{\mu^2}\bigg]+\gamma_h\;,
  \end{split}
\end{equation}
where $\Gamma_{\mathrm{cusp}}$ stands for the cusp anomalous dimension.
Its result is known at the three-loop accuracy~\cite{Moch:2004pa}, 
whilst the four-loop evaluation has been accomplished very recently \cite{Moch:2018wjh}. 
$\gamma_h$ represents the hard anomalous dimensions and can be extracted 
from \cite{Gehrmann:2010ue}. 
The remaining equations for the soft and beam functions, $\mathcal{S}$ 
and $\mathcal{B}$, read~\cite{Li:2016axz}
\begin{equation}
  \begin{split}\label{eq:methods:res_rge_SB}
    \frac{\partial}{\partial\ln \mu^2}\ln \mathcal{S}(\bTvec,\mu,\nu)
    =&\;
      \Gamma_{\mathrm{cusp}} \,\ln \bigg[ \frac{\mu^2}{\nu^2}\bigg]
      -\gamma_s\;,\\      
    \frac{\partial}{\partial\ln \mu^2}\ln \mathcal{B}(\eta, \bTvec,\mu,\nu)
    =&\;
      \Gamma_{\mathrm{cusp}}\, \ln \bigg[ \frac{\nu}{\eta\sqrt{s}}\bigg]
      +\gamma_b\;,
  \end{split}
\end{equation}
and
\begin{equation}
  \begin{split}\label{eq:methods:res_rage_SB}
    \frac{\partial}{\partial\ln \nu^2}\ln \mathcal{S}(\bTvec,\mu,\nu)
    =&\;
      -2\,\frac{\partial}{\partial\ln \nu^2}\ln \mathcal{B}(\eta,\bTvec,\mu,\nu)
    =
     \gamma_r\left[\alpha_s\left(\frac{b_0}{\bT}\right)\right]
     +\int_{\mu^2}^\frac{b_0^2}{\bT^2}
      \frac{\done\bar{\mu}^2}{\bar{\mu}^2}\;
      \Gamma_{\mathrm{cusp}}\left[\alpha_s(\bar{\mu}) \right]\;.
  \end{split}
\end{equation}
Therein, $\gamma_s$ is the soft anomalous dimension, on which the accuracy has already arrived at 
N$^3$LO~\cite{Li:2014afw}.  
$\gamma_b$ is the non-cusp anomalous dimension for the 
beam function, which can be derived from the consistency relationship 
$\gamma_b=(\gamma_s-2\gamma_h)/2$.  
In addition to the ingredients of the RGEs, the rapidity 
anomalous dimension, $\gamma_r$, is involved within the RaGEs. 
Its universality (or the rapidity regulator independence) has 
been discussed in Refs.~\cite{Vladimirov:2016dll,Li:2016axz}, 
and the N$^3$LO results are detailed in 
Refs.~\cite{Li:2016ctv,Vladimirov:2016dll}.

\begin{table}[h!]
  \centering
  \begin{tabular}{|c|c|c|c|c|c|} \hline
    Logarithmic accuracy\hl & $C_{\text{ns}}$, $C_{\text{s}}^{A}$, $C_{\text{s}}^{V}$, $C_t$, $\mathcal{S}$, $\mathcal{B}$ & $\Gamma_{\text{cusp}}$& $\gamma_{t,h,s,b}$ \\ \hline
    \NLL\hl & $\mathcal{O}(\alpha_s)$& $\mathcal{O}(\alpha^2_s)$& $\mathcal{O}(\alpha_s)$ \\ \hline
    \NNLL\hl & $\mathcal{O}(\alpha^2_s)$& $\mathcal{O}(\alpha^3_s)$& $\mathcal{O}(\alpha^2_s)$ \\ \hline
    \NNNLL\hl & $\mathcal{O}(\alpha^3_s)$& $\mathcal{O}(\alpha^4_s)$& $\mathcal{O}(\alpha^3_s)$ \\ \hline
  \end{tabular}
  \caption{
    Needed accuracy of the fixed-order inputs to achieve a given 
    logarithmic accuracy of the resummation, in accordance with eq.~\eqref{eq:intro:def_res_order}.
    \label{tab:methods:res_accuracy}
  }
\end{table}

With all ingredients collected, we can now carry out the resummation. 
In this work, the resummed distributions are calculated at \NLL, \NNLL, 
and \NNNLL\ level. 
For each level of the logarithmic accuracy, the perturbative order for each ingredient can be 
found out in eq.\ \eqref{eq:intro:def_res_order} and has been also 
summarised in Tab.\ \ref{tab:methods:res_accuracy}. 
Note that as the accuracy of the fixed-order functions 
is counted with respect to Born cross section, only the 
$\mathcal{O}(\alpha_s^3)$ terms for $C_{\text{s}}^V$ are needed 
in this paper. 
The resummed spectrum is then obtained by substituting the 
solutions of the RGEs as well as RaGEs into 
eq.~\eqref{eq:methods:TMD_factorisation}, giving
\begin{equation}\label{eq:methods:res:TMD_Resummation}
  \begin{split}
    \frac{\done^4\sigma_{\text{res}}}    
    {\done^2\qTvec\,\done Y_L\,\done M_L^2\,\done\Omega_{L}}
    =&\;
      \frac{1}{16s(2\pi)^4M_{L}^2}\;
      \sum_{i,j}\int\done^2 \bTvec\;
      e^{i \bTvec \cdot \qTvec}\;
      \mathcal{U}_{V}(\mu_h,\mu_{b_{\pm}},\mu_s,\nu_{b_\pm})\;
      \mathcal{U}_R(\mu_s,\nu_{b_{\pm}},\nu_s)\\
    &\hspace*{50mm}\times\vphantom{\sum_{i,j}}
      \widetilde{\mathcal{H}}^{V,\mathrm{res}}_{ij}(M_L,\Omega_L,\mu_h)\;
      \mathcal{S}( {\vec{b}_{T}}, \mu_s,\nu_s)\\
    &\hspace*{50mm}\times
      \mathcal{B}_+^i(\eta_+,\vec{b}_{T}, \mu_{b_+},\nu_{b_+})\;
      \mathcal{B}_-^j(\eta_-,\vec{b}_{T}, \mu_{b_-},\nu_{b_-})\;,\hspace*{-5mm}
  \end{split}
\end{equation}
where the kernels $\mathcal{U}_{V,R}$, respectively, effect 
the virtuality and rapidity evolutions. 
Their explicit expressions read
\begin{equation}
  \begin{split}
    \mathcal{U}_{V} 
    &=\;
      \exp\bigg\{
        \int_{\mu^2_h}^{\mu^2_{b_\pm}}\frac{\done\bar{\mu}^2}{\bar{\mu}^2}
        \Big[
          -\Gamma_{\mathrm{cusp}}\ln\left(\frac{\bar{\mu}^2}{M_L^2}\right)
          +2\gamma_h
        \Big]
      +\int_{\mu^2_s}^{\mu^2_{b_\pm}}\frac{\done\bar{\mu}^2}{\bar{\mu}^2} 
       \Big[
         \Gamma_{\mathrm{cusp}}\ln\left(\frac{\bar{\mu}^2}{\nu^2_{b_\pm}}\right)
         -\gamma_s
       \Big]
     \bigg\}\,,\\
    \mathcal{U}_{R}
    &=\;
      \exp\bigg\{
        \ln\left[\frac{\nu^2_{b_\pm}}{\nu^2_s}\right]
        \Bigg[
          \gamma_r\left(\alpha_s\left(\frac{b_0}{\bT}\right)\right)  
          +\int^\frac{b^2_0}{\bT^2}_{\mu^2_s}\frac{\done\bar{\mu}^2}{\bar{\mu}^2}\;
           \Gamma_{\mathrm{cusp}}\left[\alpha_s(\bar{\mu})
          \right]
        \Bigg]
      \bigg\}\,.
  \end{split}
\end{equation}
Finally, $\widetilde{\mathcal{H}}^{V,\mathrm{res}}_{ij}$ is the part 
of the hard function participating in the resummation. 
In $W^\pm$ production, as $M_L$ is the only physical scale in the 
hard sector, the natural choice of scale $\mu_h$ can be easily identified (assuming it is related to $M_L$). 
$\widetilde{\mathcal{H}}^{V,\mathrm{res}}_{ij}$ then coincides directly 
with $\mathcal{H}^{W}_{ij}$ from eq.~\eqref{eq:methods:HijV:HF_ppW}, 
giving
\begin{equation}
  \begin{split}
    \widetilde{\mathcal{H}}^{W,\mathrm{res}}_{ij}(M_L,\Omega_L,\mu_h)
    =
      \mathcal{H}^{W}_{ij}(M_L,\Omega_L,\mu_h)\;. 
  \end{split}
\end{equation}

Conversely, the hard function in $Z$ production, contains a second 
intrinsic scale in addition to $M_L$, $m_t$, originating in the 
singlet contributions.
In order to resum the logarithm associated with their scale hierarchy, 
$\ln[m^2_t/M^2_L]$, we substitute the Wilson coefficient $C_t$ in 
eq.~\eqref{eq:methods:HijV:AZqq_cpl} with its resummed version, 
\begin{equation}
  \begin{split}
    C_{t}^{\mathrm{res}}(m_t,\mu_h)
    =&\;
      C_t (m_t,\mu_t) \, \exp \left[ -\int_{\mu^2_t}^{\mu^2_h}   \frac{d\mu^2}{\mu^2}\,\gamma_t         \right]\;,
  \end{split}
\end{equation}
and then evaluate all other Wilson coefficients $C_\text{ns}$, 
$C_\text{s}^V$ and $C_\text{s}^A$, at the scale ${\mu_h}$. 
As a result, the part of the hard function participating in the 
resummation can be expressed as 
\begin{equation}
  \begin{split}
    \widetilde{\mathcal{H}}^{\gamma/Z,\mathrm{res}}_{ij}(m_t,M_L,\Omega_L,\mu_h)
    =&\;     
\mathcal{H}^{\gamma/Z}_{ij}
      \Bigg|_{\genfrac{}{}{0pt}{3}
                      {C_t\to C_{t}^{\mathrm{res}}(m_t,\mu_h),
                       C_\text{ns}\to C_\text{ns}(M_{L},\mu_h)}
                      {C_\text{s}^V\to C_\text{s}^V(M_{L},\mu_h),
                       C_\text{s}^A\to C_\text{s}^A(M_{L}, \mu_h)}}\;.
  \end{split}
\end{equation}

\subsection{Power corrections}
\label{sec:methods:leppow}

To extrapolate eq.~\eqref{eq:methods:res:TMD_Resummation} to the 
entire phase space, it is crucial to properly account for power 
corrections stemming from the truncation of the asymptotic 
series in $\qT/{M_L}$ (recalling that 
eq.~\eqref{eq:methods:TMD_factorisation} only contains the leading 
power contributions).
They can impact both the leptonic currents 
$L_{W,Z,\gamma}^{\mu}$ (in eq.~\eqref{eq:methods:HijV:HardAmp_PPZ_AZ} 
and eq.~\eqref{eq:methods:def_Hw_Lw}) as well as the hadronic ones. 
To improve the lepton currents, we substitute the  
$L_{W,Z,\gamma}^{\mu}$ by their pre-expanded results, i.e. 
\begin{equation}\label{eq:methods:extra:leppowcorr}
  \begin{split}
    L_{V}^{\mu}\longrightarrow \Lambda^{\mu}_{\ell,\nu}(\qTvec) L_V^\nu\;,
  \end{split}
\end{equation}
where $V=W,Z,\gamma$.  $\Lambda^{\mu}_{\ell,\nu}$ accounts for the 
Lorentz transformation from the leptonic centre of mass reference 
frame to the rest frame of the initial proton pair or lab frame.
Noting that in eq.~\eqref{eq:methods:HijV:HardAmp_PPZ_AZ} and 
eq.~\eqref{eq:methods:def_Hw_Lw}, as a result of the $\qT/M_{L}$ 
expansion in the hard sector, $L_{V}^{\mu}$ is defined in the lepton centre 
of mass frame.  
Thus, the replacement in eq.~\eqref{eq:methods:extra:leppowcorr} 
restores the leptonic currents to their form in the exact SM calculation. 
Their impact is assessed in App.\ \ref{sec:app_lpc}. 
It is worth noting that even though the substitution in 
eq.~\eqref{eq:methods:extra:leppowcorr} manifestly breaks 
energy-momentum conservation when coupling the hadronic and the corrected 
leptonic tensor, it induces no gauge violations in practice. 
This is because the longitudinal component of the $\gamma$, $Z$, or $W^\pm$ 
propagators can always be eliminated by the attached leptonic currents 
in the massless limit. 
Alternative methods with the analogous effects can also be found in 
Refs.~\cite{Balazs:1997xd,Balazs:1995nz,Ellis:1997sc,Catani:2015vma,Re:2021con}.  
Very recently, a systematic classification has been carried out in 
Ref.~\cite{Ebert:2020dfc} as to the leptonic power corrections.
 
On the other hand, it is also necessary to handle the power 
corrections to the hadronic currents.  
In principle, one can incorporate them power by power with 
the aid of the subleading SCET Lagrangians and Hamiltonians 
\cite{Beneke:2002ni,Beneke:2002ph,Pirjol:2002km}. 
Nevertheless, not only can more structures beyond 
eq.~\eqref{eq:methods:TMD_factorisation} be introduced by the subleading 
vertices, the inhomogeneity in asymptotic series, which arises from 
the rapidity regulation, may further complicate the investigations 
\cite{Ebert:2018gsn,Buonocore:2019puv,Cieri:2019tfv}. 	
So for simplicity, this work only considers the leading 
power factorisation and the according resummation in the hadronic sector, treating 
the power corrections through the matching 
to fixed-order.

\subsection{Matching to fixed-order QCD and observable calculation}
\label{sec:methods:extra}

Finally, we match the resummation to the exact QCD fixed-order calculation. 
We therefore introduce the following matching procedure, 
\begin{equation}\label{eq:methods:extra:TMD_matched}
  \begin{split}
  \frac{\done\sigma_{\text{mat}}}{\done\Phi}
  =&\;
    \left\{
      \frac{\done\sigma_{\text{res+lpc}}}{\done\Phi}
      -\frac{\done\sigma_{\text{exp+lpc}}}{\done\Phi}(\mu_R,\mu_F)
    \right\}
    f\left(\frac{\qT}{\qTcut}\right)
    +\frac{\done\sigma_{\text{f.o.}}}{\done\Phi}(\mu_R,\mu_F)\;,
  \end{split}
\end{equation}
where $\sigma_{\text{exp+lpc}}$ is the fixed-order expansion 
of $\sigma_{\text{res+lpc}}$ and the ``lpc'' addition in the 
subscript signals the inclusion of the leptonic power corrections.
$\sigma_{\text{f.o.}}$ denotes the exact fixed-order perturbative 
result at the conventional renormalisation and factorisation scales 
$\mu_R$ and $\mu_F$.  
Of course, both $\sigma_{\text{exp+lpc}}$ and $\sigma_{\text{f.o.}}$ 
must be expanded to the same order.
$\done\Phi$ stands for the differential phase space element 
$\done^2\qTvec\,\done Y_L\,\done M_L^2\, \done\Omega_{L}$. 
The transition function $f$ is introduced to assure that the resummation 
is only active in the asymptotic region, more explicitly,
\begin{equation}\label{eq:methods:extra:f}
  \begin{split}
    f\left(\frac{\qT}{\qTcut}\right)=\frac{1}{1+a^{\frac{\qT}{\qTcut}-1}}
    \hspace*{0.2\textwidth}
    \begin{minipage}{0.48\textwidth}
      \includegraphics{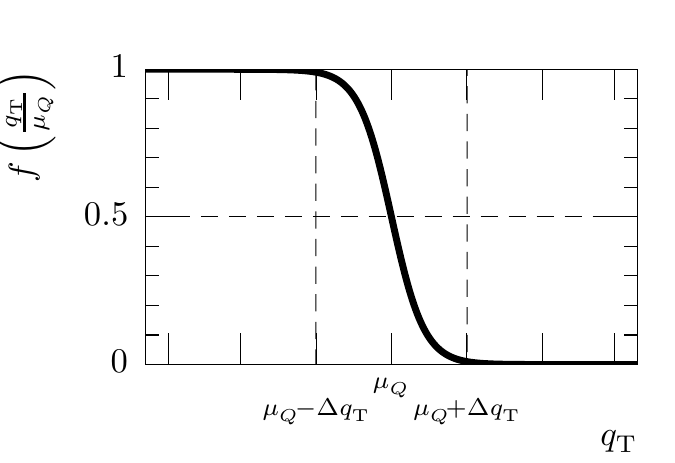}
    \end{minipage}
  \end{split}
\end{equation}
The numerical choice of the matching scale \qTcut\ will be discussed in 
Sec.~\ref{sec:results}. 
The base of the exponent $a$ can be thought of as a 
free parameter governing the shape of the transition 
function. 
We choose to link it to \qTcut\ and 
define it as $a=\exp(\qTcut/\qTref)$, 
\qTref\ is taken to be 1\,GeV.
It is important to note that although $f$ is smaller than 
unity for all physical \qT, the precise choice of \qTcut\ 
will ensure that it differs from unity by amounts much smaller 
than one permille in a wide region around the Sudakov peak 
where the resummation effects dominate. 
Likewise, $\qT=\qTcut$ does not mark the endpoint of the 
resummation region, but where the resummation effects are 
faded out to half their intrinsic size, $f(1)=\tfrac{1}{2}$. 
The functional form then induces a symmetric reduction 
of the influence of the resummation in the region 
$[\qTcut-\dqT,\qTcut+\dqT]$, wherein $\dqT=\qTref\ln((1-f)/f)$ 
and $f$ is the fraction of reduction.
Thus, with the above value for \qTref\ the resummation effects
are reduced from 99\% to 1\% of their actual size in the interval 
$\qTcut\mp4.6\,\text{GeV}$. 
The functional form of the base of the exponent $a$ ensures 
that the width of the transition region, \dqT, is directly 
proportional to \qTref, with \qTref\ taking the role of an easily 
interpretable shape parameter.

When \qTvec\ enters the hadronic regime 
$\Lambda_{\mathrm{QCD}}\sim \qT\ll M_L$, the non-perturbative 
contributions become non-negligible. 
Hence, a modification factor parametrising the non-perturbative 
dynamics of this regime, $S_{\mathrm{mod}}(\bT^2)$, 
encoding all the hadronic contributions within the low $\qT$ 
region~\cite{Korchemsky:1994is,Tafat:2001in}, should be introduced.
In the recent years, a number of parametrisations and 
input extractions of $S_{\mathrm{mod}}(\bT^2)$ have been 
discussed in the literatures~\cite{Aidala:2014hva,Echevarria:2014rua,DAlesio:2014mrz,Bertone:2019nxa,Bacchetta:2019tcu,Bacchetta:2017gcc,Scimemi:2016ffw}. 
However, as the main concern of this work is with the perturbative 
domain, we ignore $S_{\mathrm{mod}}(\bT^2)$ here and leave its 
incorporation to the future investigations.

Taking the integral of eq.~\eqref{eq:methods:extra:TMD_matched} 
together with the customised theta and delta functions, one is 
able to implement the fiducial cuts and carry out the observable 
calculations, more explicitly, 
\begin{equation}
  \begin{split}\label{eq:methods:extra:obs_calc}
    \frac{\done\sigma}{\done O}
    =&\;
      \int\done\Phi\;
      \Theta\Big[\mathcal{G}\big(\Phi\big)\Big]\;
      \delta\Big[O-\mathcal{F}_{O}\big(\Phi\big)\Big]\;
      \frac{\done\sigma_{\mathrm{mat}}}{\done\Phi},
  \end{split}
\end{equation}
where $\mathcal{G}$ is a function of the phase space encoding 
the fiducial cuts.  
$O$ stands for a generic observable and $\mathcal{F}_{O}$ is its 
corresponding functional definition.
This work focusses on the \qT\ and \dphi\ spectra, giving
\begin{equation}
  \begin{split}
    \mathcal{F}_{\qT}
    =&\;
     |\qTvec|,\\
    \mathcal{F}_{\dphi}
    =&\;
      \arccos\left[\frac{\vec{p}_{_{1,\mathrm{T}}}\cdot\vec{p}_{_{2,\mathrm{T}}}}{ |\vec{p}_{_{1,\mathrm{T}}}||\vec{p}_{_{2,\mathrm{T}}}|  }\right].
  \end{split}
\end{equation}
Here $ \vec{p}_{_{1(2),\mathrm{T}}}$ denotes the transverse momentum of 
the final (anti-)lepton in the lab frame. 
In particular, both variables are purely transverse and are 
thus well-defined for both $W$ and $Z$ production in a 
realistic detector environment where only the transverse 
part of the neutrino's momentum is observable through as 
missing transverse momentum.
Further, in calculating the phase-space integral of eq.~\eqref{eq:methods:extra:obs_calc}, we retain the full 
kinematic dependences  in  $\mathcal{G}(\Phi)$ and  $\mathcal{F}_{O}(\Phi)$   and do not carry out any multipole-expansions.

%% file: text/results.tex
\section{Numerical Results}
\label{sec:results}

In this section we are discussing numerical results obtained with 
the methods outlined in the previous section for a representative 
fiducial region.

\subsection{Setup and fiducial region}
\label{sec:results:setup}

Throughout this paper, all distributions for $W^\pm$ production 
are calculated in the $G_\mu$ scheme while those for $Z$ production 
are calculated in the $\alpha(m_Z)$ scheme.
We use the following input parameters \cite{Zyla:2020zbs}
\begin{center}
  \begin{tabular}{rclrcl}
    $G_\mu$ & \shortequal & $1.166378\times 10^{-5}\,\text{GeV}^{-2}$ &&& \\
    $m_W$ & \shortequal & 80.379\,\text{GeV}&
    $\Gamma_W$ & \shortequal & 2.085\,\text{GeV} \\
    $m_Z$ & \shortequal & 91.1876\,\text{GeV}&
    $\Gamma_Z$ & \shortequal & 2.4955\,\text{GeV} \\
    $m_t$ & \shortequal & 173.2\,\text{GeV}\;.&&&\\
  \end{tabular}
\end{center}
The width of the top quark as well as the mass and the width 
of the Higgs boson have no phenomenological impact and are neglected. 
All other particles are considered massless.
We work in the complex-mass scheme \cite{Denner:2005fg}, 
with the complex masses and mixing angles defined through 
\begin{equation}
  \mu_i^2=m_i^2-\mr{i}m_i\Gamma_i
  \qquad\text{and}\qquad
  \sin^2\theta_w=1-\frac{\mu_W^2}{\mu_Z^2}\;.
\end{equation}
The electromagnetic coupling constant is defined as
\begin{equation}
  \alphaGmu = \left|\frac{\sqrt{2}G_\mu \mu_W^2 \sin^2\theta_w}{\pi}\right|
  \qquad\text{and}\qquad
  \alphamZ=1/128.802
\end{equation}
in the $G_\mu$ and $\alpha(m_Z)$ schemes, respectively. 
Other input schemes, in particular for $Z$ boson production, 
may be chosen in order to improve the description of the correlations 
of the lepton pair system and the hadronic recoil, e.g.\ by using 
the weak mixing angle as an input parameter \cite{Chiesa:2019nqb}. 
Furthermore, we use the \texttt{NNPDF31\_nnlo\_as\_0118} \cite{Ball:2017nwa} 
parton distribution functions, interfaced through \LHAPDF 
\cite{Buckley:2014ana}, with the corresponding value 
of the strong coupling $\alpha_s(m_Z)=0.118$. 
In accordance with this choice we neglect all photon induced 
contributions. 
The CKM matrix is chosen as diagonal, i.e.\ $V_{ij}=\delta_{ij}$.

To carry out the resummation entering our matched results 
of eqs.\ \eqref{eq:methods:extra:TMD_matched} and \eqref{eq:methods:extra:obs_calc}, 
we employ the \Cuba library \cite{Hahn:2014fua,Hahn:2004fe} performing the respective 
integrations and make use of 
\LHAPDF evolving the factorisation scale. 
To tackle the harmonic poly-logarithms participating in the $\mathcal{I}_{ij}$ function (see eq.~\eqref{eq:methods:fact:Beam_fac}), the package 
HPOLY~\cite{Ablinger:2018sat} is used.
The LO hard functions in eqs.\ 
\eqref{eq:methods:HijV:HF_ppW} and \eqref{eq:methods:HijV:HF_ppZ} 
are computed with FeynCalc \cite{Mertig:1990an,Shtabovenko:2016sxi,Shtabovenko:2020gxv} 
and FeynArts~\cite{Hahn:2000kx,Kublbeck:1990xc}. 
All remaining parts of the hard function, in particular the non-singlet 
and singlet contributions $C_\text{ns}$, $C_\text{s}^V$, and $C_\text{s}^A$
as well as the top-quark contribution $C_t$, are implemented analytically.
The fixed-order contribution to eq.\ \eqref{eq:methods:extra:TMD_matched} 
and, hence, the matched result is computed using 
\Sherpa \cite{Bothmann:2019yzt,Gleisberg:2008ta} in combination with 
\OpenLoops \cite{Buccioni:2019sur,Cascioli:2011va}. 
In this framework, renormalised virtual amplitudes are provided 
by \OpenLoops, which uses \Collier tensor reduction library \cite{Denner:2016kdg} 
as well as \CutTools \cite{Ossola:2007ax} together with the \OneLoop library 
\cite{vanHameren:2010cp}.
All remaining tasks, i.e.\ the bookkeeping of partonic subprocesses, 
phase-space integration, and the subtraction of all infrared 
singularities, are provided by \Sherpa
using the matrix element generator \Amegic \cite{Krauss:2001iv,Schonherr:2017qcj,Gleisberg:2007md}. 

In estimating the theoretical uncertainties, we emphasise two aspects. 
The first one is the uncertainty arising from higher order corrections, 
which can be estimated by examining the sensitivity of the results to 
the scale variations. 
As presented in Sec.\ \ref{sec:methods:extra} the matched result 
depends on a set of auxiliary scales, originating both in the resummation 
and the fixed-order calculation, 
$\{\mu,\nu\}\equiv\{\mu_t,\mu_h,\mu_{b_+},\mu_{b_-},\mu_s,\nu_{b_+},\nu_{b_-},\mu_R,\mu_F\}$.  
During the numerical implementation, we take  $\mu_b =\mu_{b_{\pm}}$ 
as well as $\nu_b=\nu_{b_{\pm}}$ in accordance with identical initial 
state particles, and set $\mu_R=\mu_F$ throughout to simplify the 
matching procedure of the fixed-order full QCD calculation to the 
results obtained in the soft-collinear effective theory. 
We set their default values, well away from the non-perturbative 
regime, to
\begin{equation}
  \begin{split}\label{eq:results::setup:scales_I}
    \mu_R^\text{def}=\mu_F^\text{def}=M_L.
  \end{split}
\end{equation} 
Please note that while $M_L=m_{\ell\ell}$ in $Z$ production, it 
is equal to invariant chanrged-lepton-neutrino mass in $W$ production. 
For the sake of reducing the logarithmic contribution in the 
$\mathcal{H},\mathcal{B},\mathcal{S}$ functions, the default 
intrinsic scales are taken as
\begin{equation}
  \begin{split}\label{eq:results::setup:scales_II}
    \mu^{\mathrm{def}}_t
    =&\;
    m_t\,,
    \quad
    \mu^{\mathrm{def}}_h=\nu^{\mathrm{def}}_{b}
    =M_{L}\,,
    \quad
    \mu^{\mathrm{def}}_s=\nu^{\mathrm{def}}_s=\mu^{\mathrm{def}}_{b}
    =b_0/\bT\,.
  \end{split}
\end{equation}  
It is worth noting that with the choice of 
$\mu^{\mathrm{def}}_s=\mu^{\mathrm{def}}_{b} =b_0/\bT$, 
the impact-parameter space integration of 
eq.\ \eqref{eq:methods:res:TMD_Resummation} 
can not be carried out straightforwardly due to presence of 
the Landau singularity at small momentum transfers, or large 
impact parameters, of the strong coupling constant $\alpha_s$. 
To cope with this issue, the prescription in Ref.~\cite{Neill:2015roa} 
is adopted in this work to suppress the higher-$\bT$ influences.
For estimating the  uncertainties from the choices in 
eqs.\ \eqref{eq:results::setup:scales_I} and 
\eqref{eq:results::setup:scales_II}, we vary the scales 
$\mu_{t,h,b,s}$, $\mu_{R,F}$ and $\nu_{b,s}$ independently to 
twice and half their default values, and then combine the deviations 
from the results using their above defined default values in the quadrature.  
The thus constructed uncertainty estimate is denoted $\delta_\text{scale}$ 
hereafter.

This leaves the uncertainty originating in the matching parameter \qTcut. 
As our default we take $\qTcut=16\,\text{GeV}$. 
The effectiveness of this choice will be illustrated and 
discussed in Sec.~\ref{sec:results:val}.   
For the estimation of its error, we alter the value in 
the interval $[15,17]\,\text{GeV}$ and denote the uncertainty 
from this by $\delta_\text{mat}$. 
Together with $\delta_\text{scale}$, the total uncertainty can 
be evaluated as 
$\delta_\text{tot}=\sqrt{\delta_\text{scale}^2+\delta^2_\text{mat}}$.

\begin{table}[t!]
  \centering
  \begin{tabular}{c|c|c|c}
    \hl & $W^+$ & $W^-$ & $Z$ \\\hline
    $\pT(\ell^+)$\hl & $[20,\infty]\,\text{GeV}$ & -- & $[20,\infty]\,\text{GeV}$ \\
    $|\eta(\ell^+)|$\hl & $[-2.4,2.4]$ & -- & $[-2.4,2.4]$ \\
    $\pT(\ell^-)$\hl & -- & $[20,\infty]\,\text{GeV}$ & $[20,\infty]\,\text{GeV}$ \\
    $|\eta(\ell^-)|$\hl & -- & $[-2.4,2.4]$ & $[-2.4,2.4]$ \\
    $\pTmis$\hl & $[20,\infty]\,\text{GeV}$ & $[20,\infty]\,\text{GeV}$ & -- \\
    $\mT$\hl & $[40,\infty]\,\text{GeV}$ & $[40,\infty]\,\text{GeV}$ & -- \\
    $m_{\ell\ell}$\hl & -- & -- & $[80,100]\,\text{GeV}$
  \end{tabular}
  \caption{
    Fiducial phase space.
    \label{tab:results:setup:fdips}
  }
\end{table}

We compute our results in a fiducial phase space that is loosely modelled 
after the respective ATLAS and CMS measurement regions.
It differs slightly depending on the process and the physics 
objects. 
Leptons are required to have a transverse momentum \pT\ of larger 
than 20\,GeV and an absolute pseudo-rapidity of less 
than 2.4. 
In the dilepton channel we require an opposite-sign same-flavour 
pair with an invariant mass $m_{\ell\ell}$ of more than 80\,GeV 
and less than 100\,GeV to effectively restrict it to the $Z$ pole. 
In both lepton-plus-missing-transverse-momentum channels we require 
one such lepton of the respective charge and a missing transverse 
momentum \pTmis\ of more than 20\,GeV. 
For simplicity, the missing transverse momentum is equated with 
transverse momentum of the neutrino. 
In addition, the missing transverse momentum and the charged lepton have 
to have a transverse mass 
$m_\text{T}=\sqrt{2\,\pT\pTmis(1-\cos\Delta\phi)}$ of more 
than 40\,GeV, where \dphi\ is the opening angle between the 
two in the transverse plane.
These definitions are summarised in Tab.\ \ref{tab:results:setup:fdips}.

As we work at strict leading order in the electroweak sector, 
questions about the precise lepton definition do not arise. 
While the resummation part of the matched result is calculated 
for the specific observable under consideration, the fixed-order 
part is generated as conventional collider events and analysed 
using \Rivet \cite{Buckley:2010ar,Bierlich:2019rhm} 

\begin{figure}[t!]
  \centering
  \includegraphics[width=.32\textwidth]{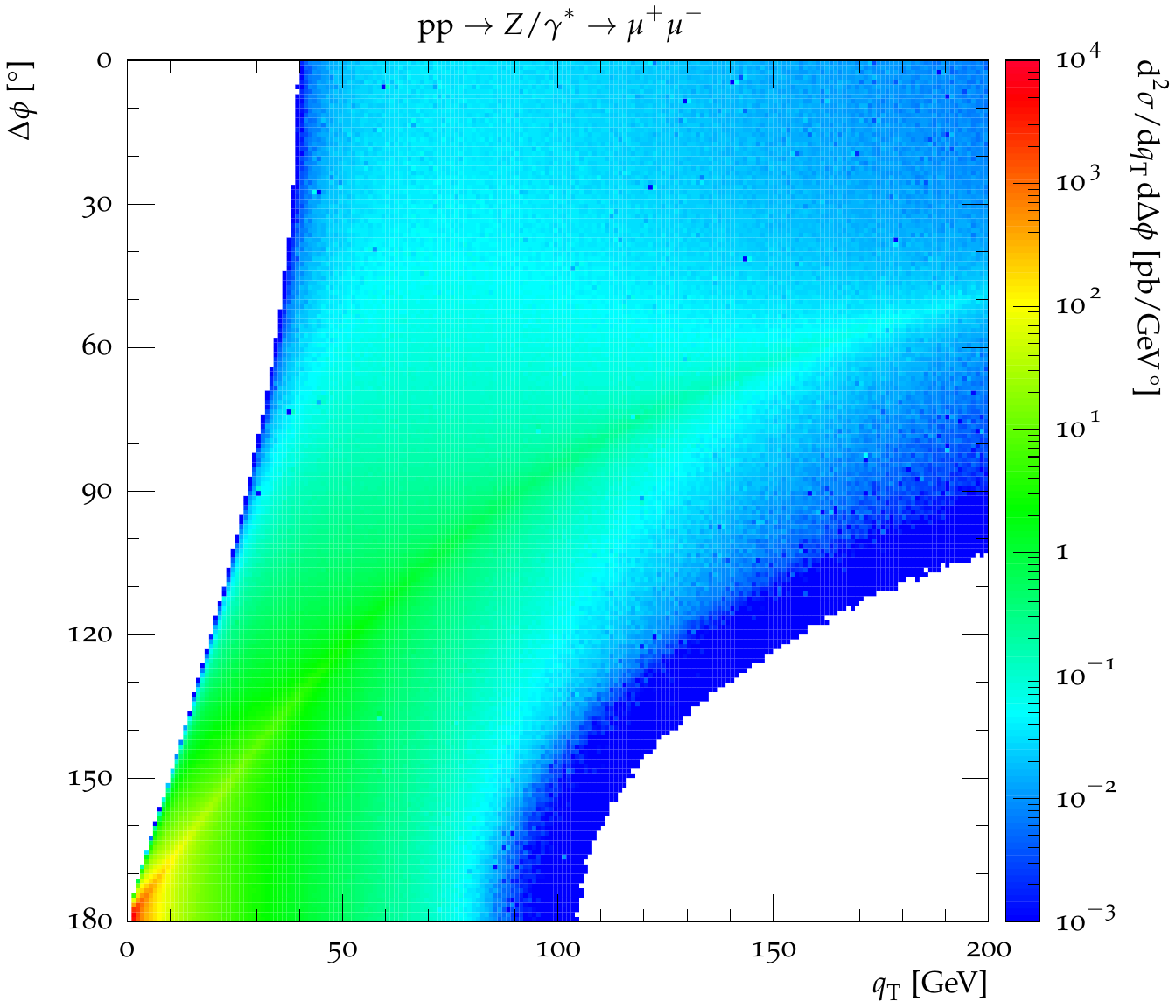}\hfs
  \includegraphics[width=.32\textwidth]{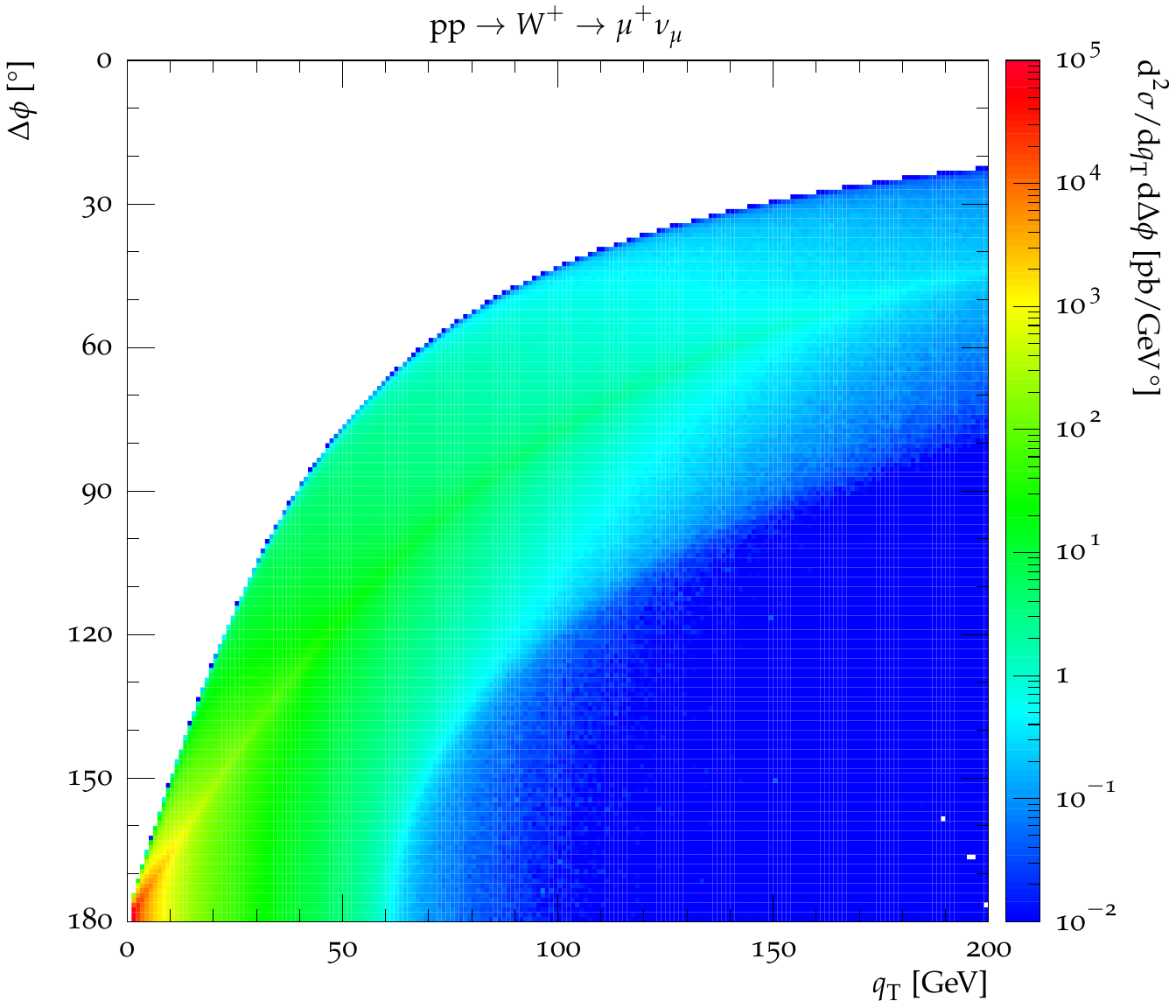}\hfs
  \includegraphics[width=.32\textwidth]{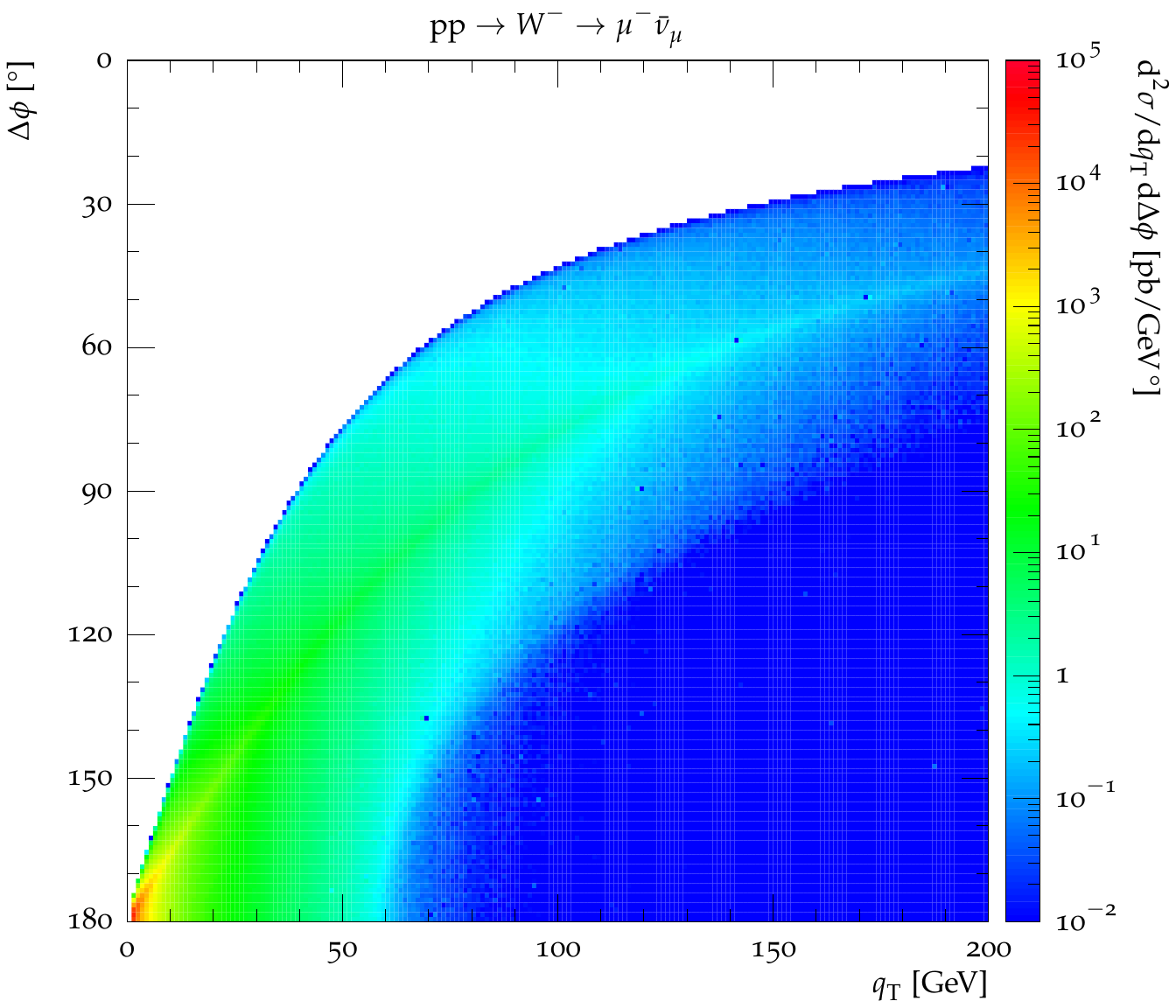}
  \caption{
    N$^2$LO distributions in the transverse momentum of the leptonic 
    system \qT\ and the opening angle between the two leptons in the 
    transverse plane \dphi.
  }
  \label{fig:results:setup:qt_dphi}
\end{figure}

Fig.\ \ref{fig:results:setup:qt_dphi} now details the distribution at 
N$^2$LO, that is including terms up to $\order(\alpha_s^2)$, in the 
transverse momentum of the leptonic system \qT\ and the opening angle 
between the two leptons in the transverse plane \dphi. 
A regularising cut of $\qT>1\,\text{GeV}$ is applied.
As can be seen, the different sets of fiducial cuts necessitated by 
the differing measurable physics objects in each channel, induce 
differing coverages of the \qT-\dphi-plane. 
Although the fiducial regions overlap well at very small \qT\ and \dphi,
they illustrate that ratios between $Z$ and $W$ channels for the 
purpose of reweighting one channel to the other should not be 
taken single-differentially in \qT\ but need to take the constrained 
internal dynamics of the lepton system into account.

\begin{figure}[t!]
  \centering
  \includegraphics[width=.47\textwidth]{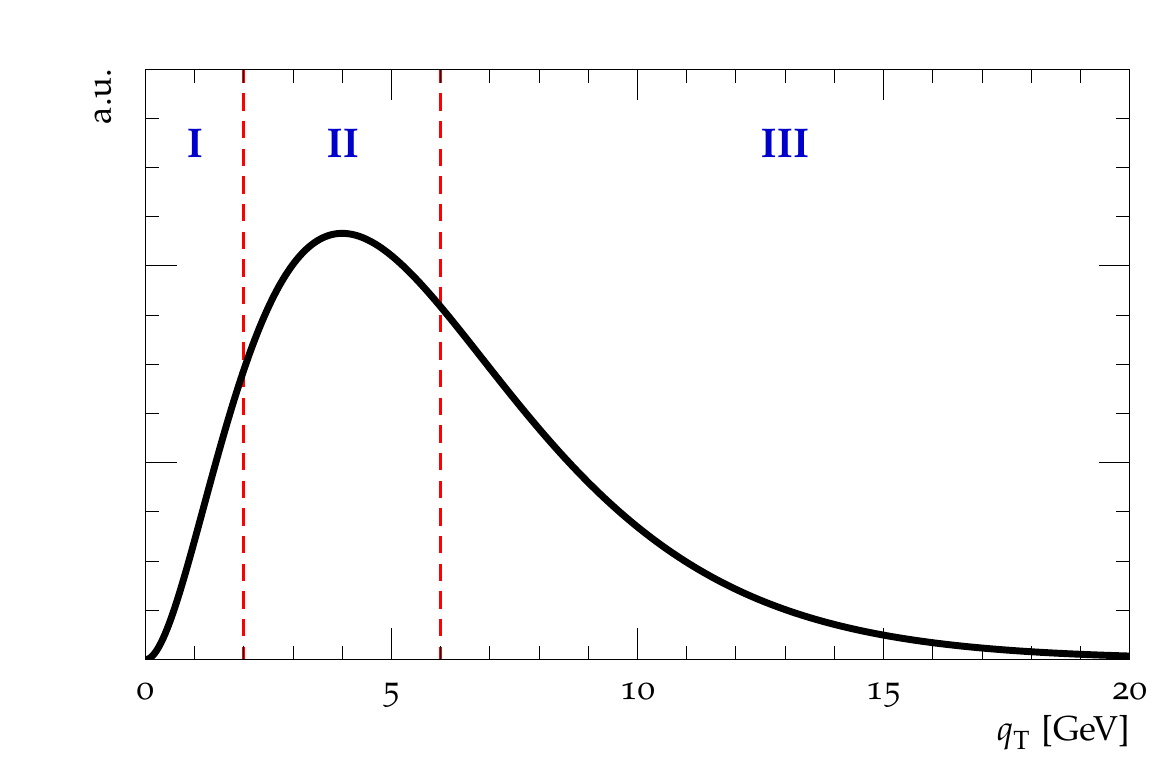}\hfill
  \includegraphics[width=.47\textwidth]{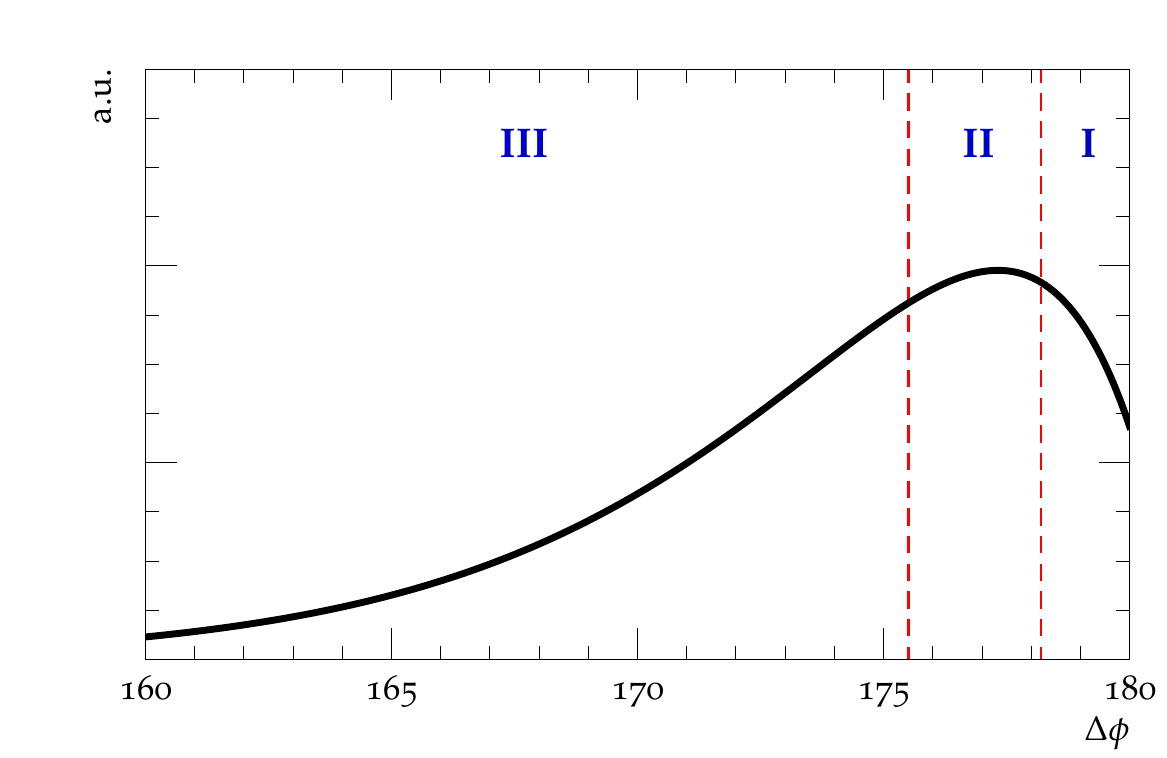}
  \caption{
    Illustration of the position of the three chosen slices 
    in \qT\ and \dphi\, encompassing a region below (I), containing (II), 
    and above (III) the Sudakov peak in each distribution.
  }
  \label{fig:results:setup:slices}
\end{figure}

Finally, Fig.\ \ref{fig:results:setup:slices} shows the qualitative 
behaviour of the single-differential behaviour of the matched cross 
section in \qT\ and \dphi. 
In both distributions the typical resummed shape can be observed, 
exhibiting a Sudakov peak close to the singular point $(\qT,\dphi)=(0,180^\circ)$.
It must be noted though, that the expression of the Sudakov peak 
in the \dphi\ spectrum strongly depends on the precise definition 
of the fiducial region.
With these observations at hand, we define three different regions 
of interest for both spectra: (I) a region between the singular point 
and the Sudakov peak, (II) a region containing the Sudakov peak, and 
(III) a region beyond the Sudakov peak that includes the part of the 
spectrum where the resummation becomes unimportant.
For the present study we therefore examine the \qT\ spectrum in three 
slices of \dphi, $180^\circ>\dphi>178.2^\circ$, 
$178.2^\circ>\dphi>175.5^\circ$, and $175.5^\circ>\dphi$, 
and the \dphi\ spectrum in three slices of \qT, $1\,\text{GeV}<\qT<2\,\text{GeV}$, 
$2\,\text{GeV}<\qT<6\,\text{GeV}$, and $6\,\text{GeV}<\qT$.

\subsection{Validation}
\label{sec:results:val}
%
To begin with the examination of our result we first 
present a comparison of the resummed result, expanded to 
next-to-leading order (\NLOs) and 
next-to-next-to-leading order (\NNLOs), 
i.e.\ truncating at $\order(\alpha_s)$ and 
$\order(\alpha_s^2)$, respectively, to the exact full 
QCD result at the same order, \NLO\ and \NNLO.
It needs to be noted that, contrary to the complete resummed 
result, the (expanded) fixed-order results diverge 
as $\qT\to 0$. 
We thus, impose a cut of $\qT>1\,\text{GeV}$ for the 
validation in this section, both for the \qT\ and 
\dphi\ spectra discussed in the following. 
The N$^3$LO$_s$ expansion of the resummed result will 
be compared with the corresponding full QCD computation in 
a future study.

\begin{figure}[t!]
  \centering
  \includegraphics[width=.32\textwidth]{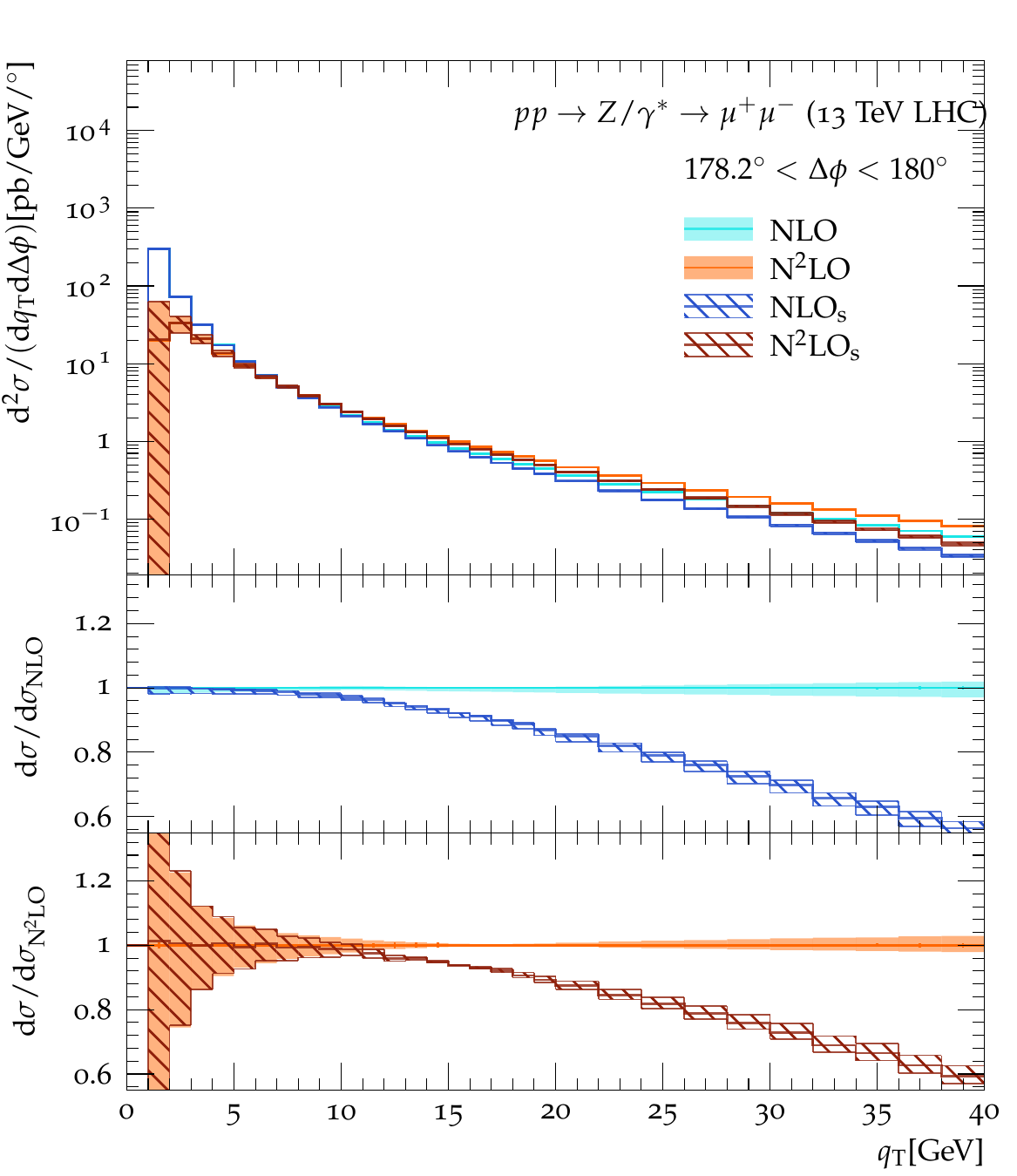}
  \includegraphics[width=.32\textwidth]{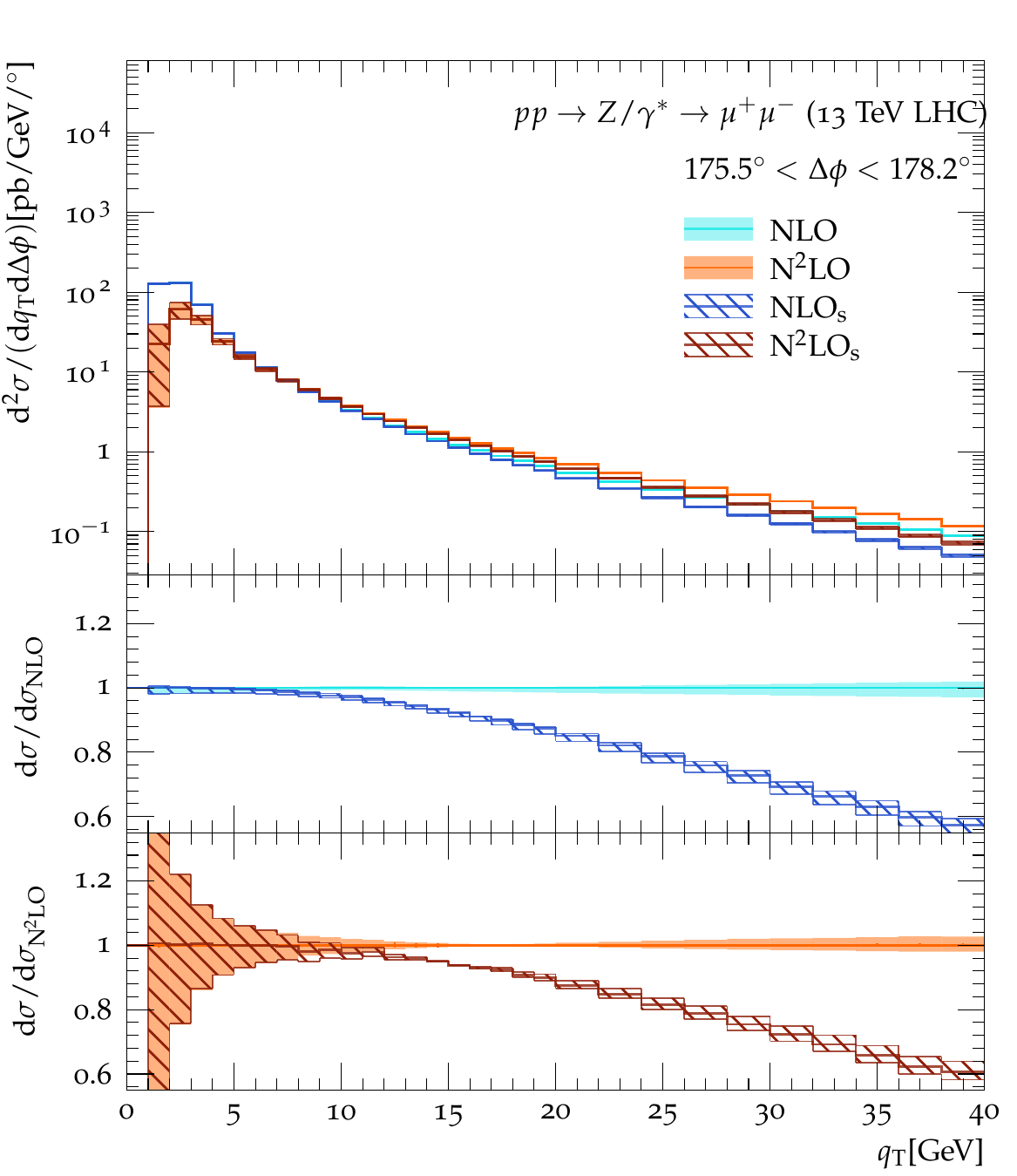}
  \includegraphics[width=.32\textwidth]{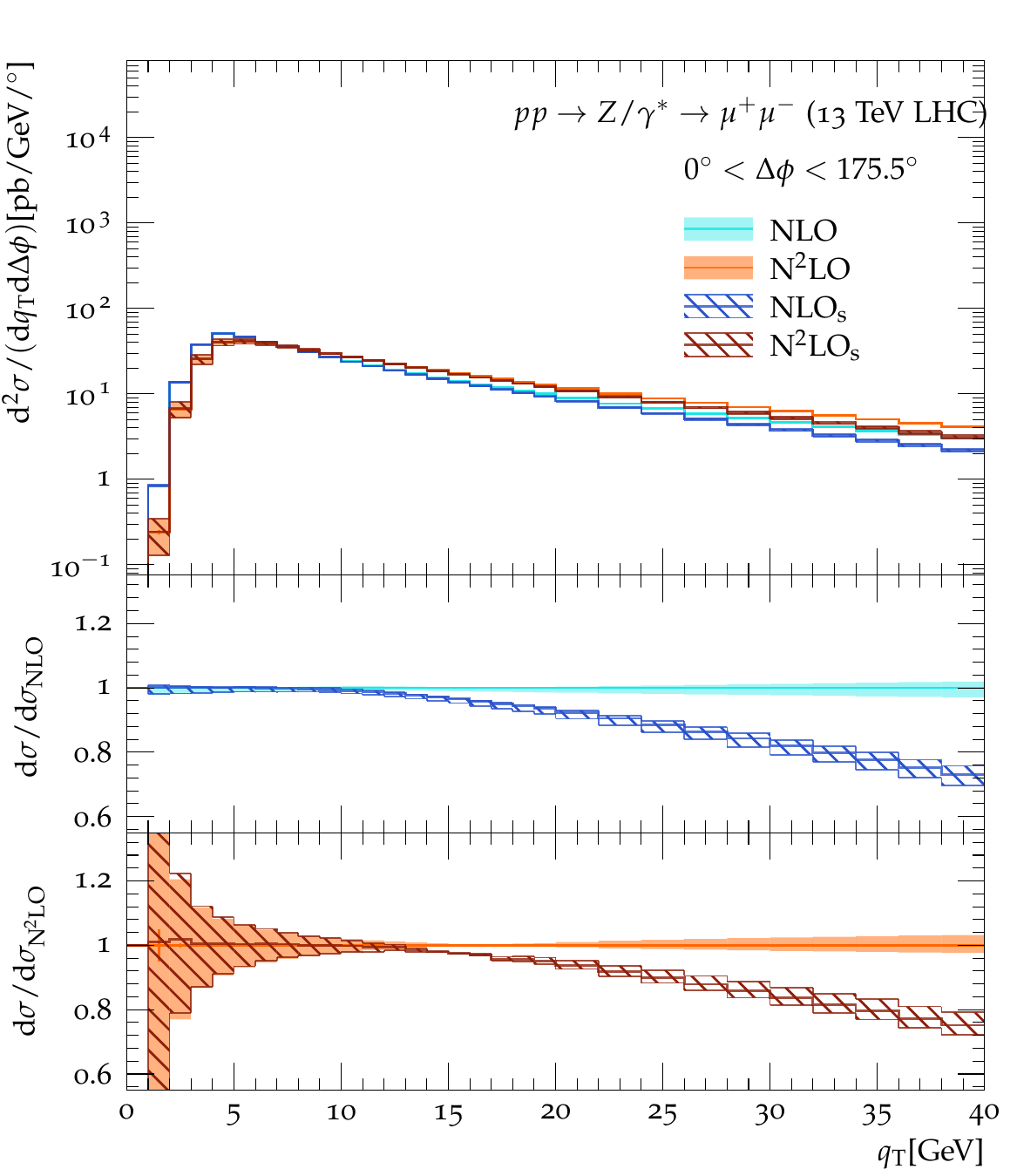}\\[1mm]
  \includegraphics[width=.32\textwidth]{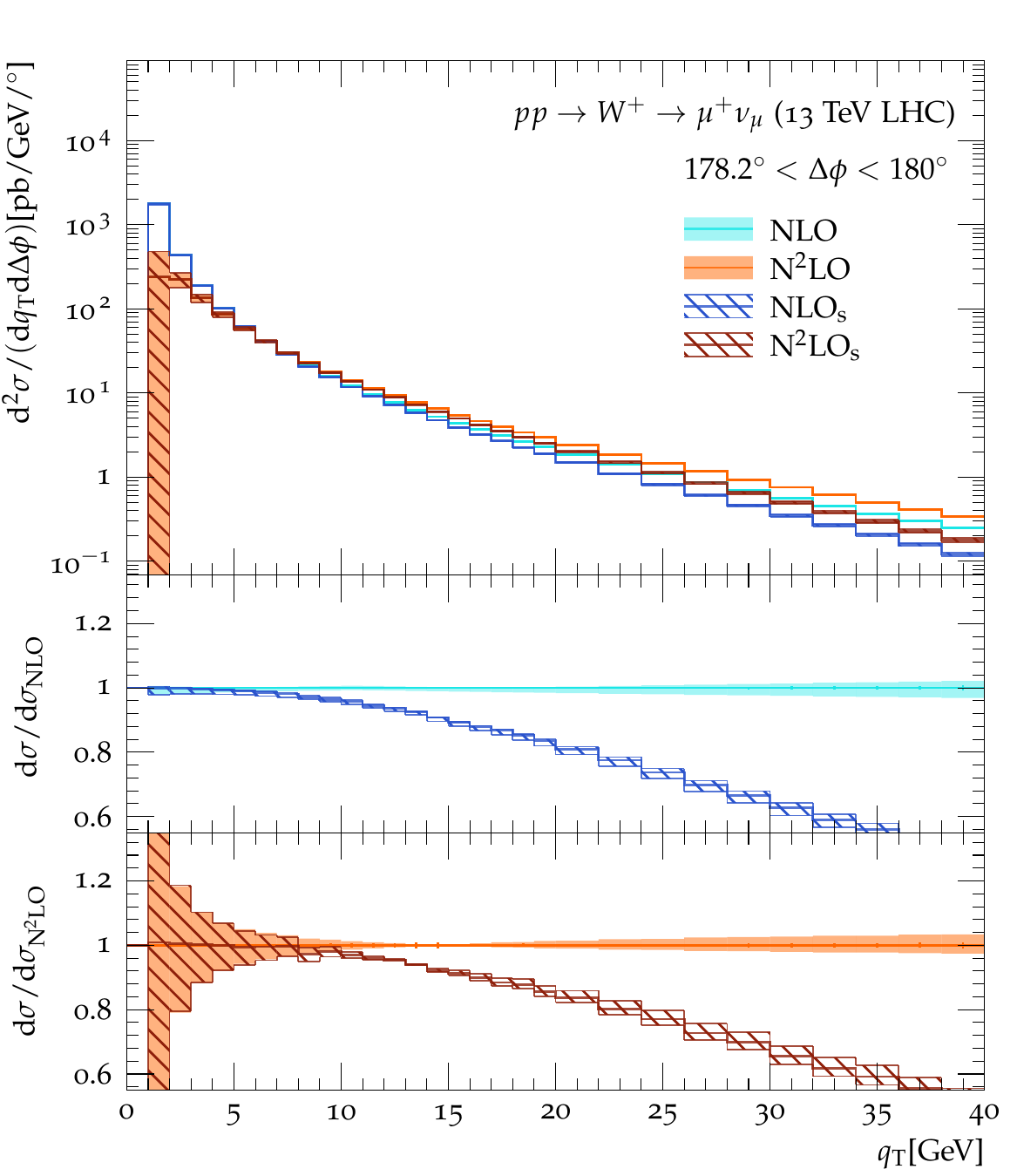}
  \includegraphics[width=.32\textwidth]{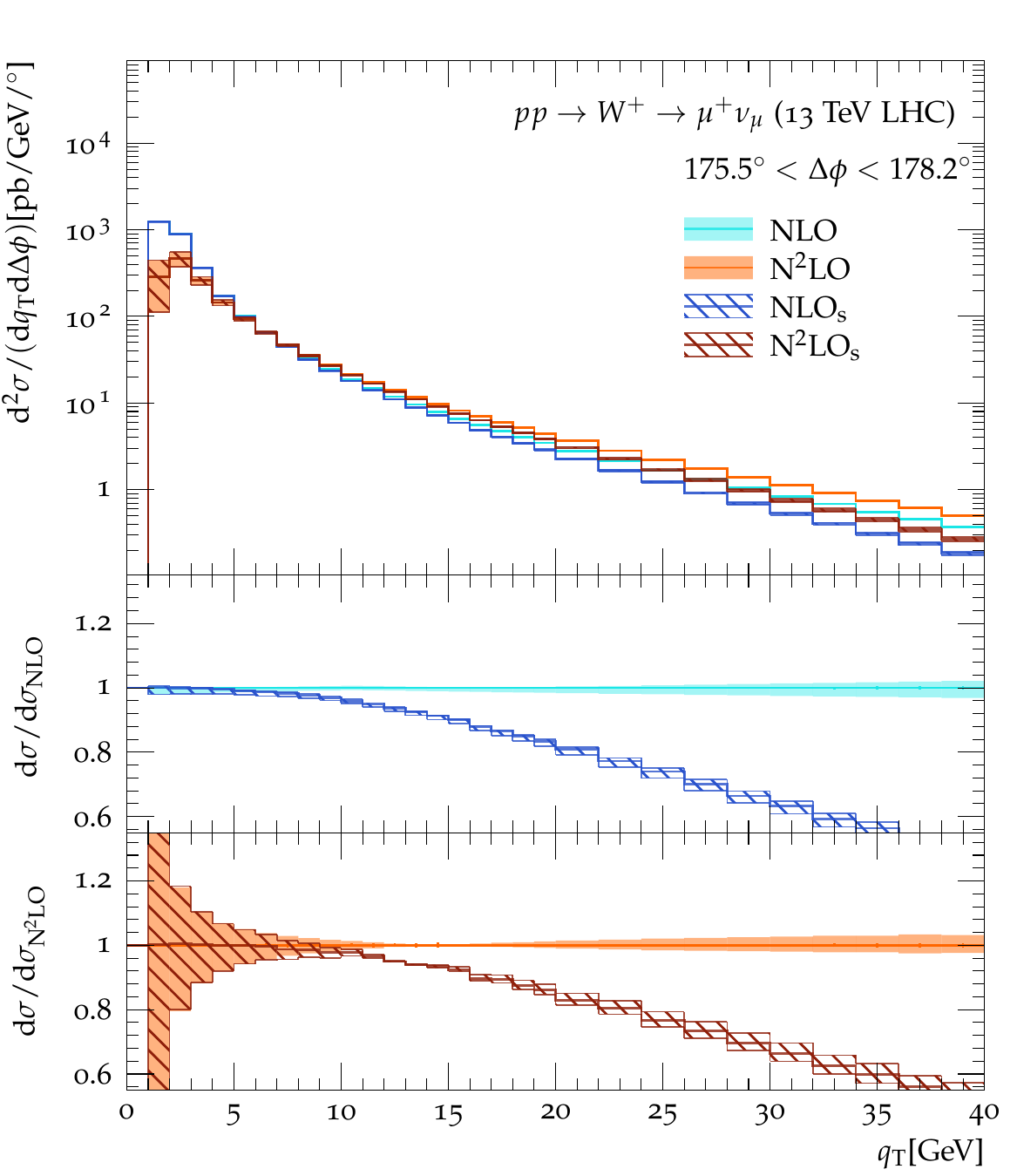}
  \includegraphics[width=.32\textwidth]{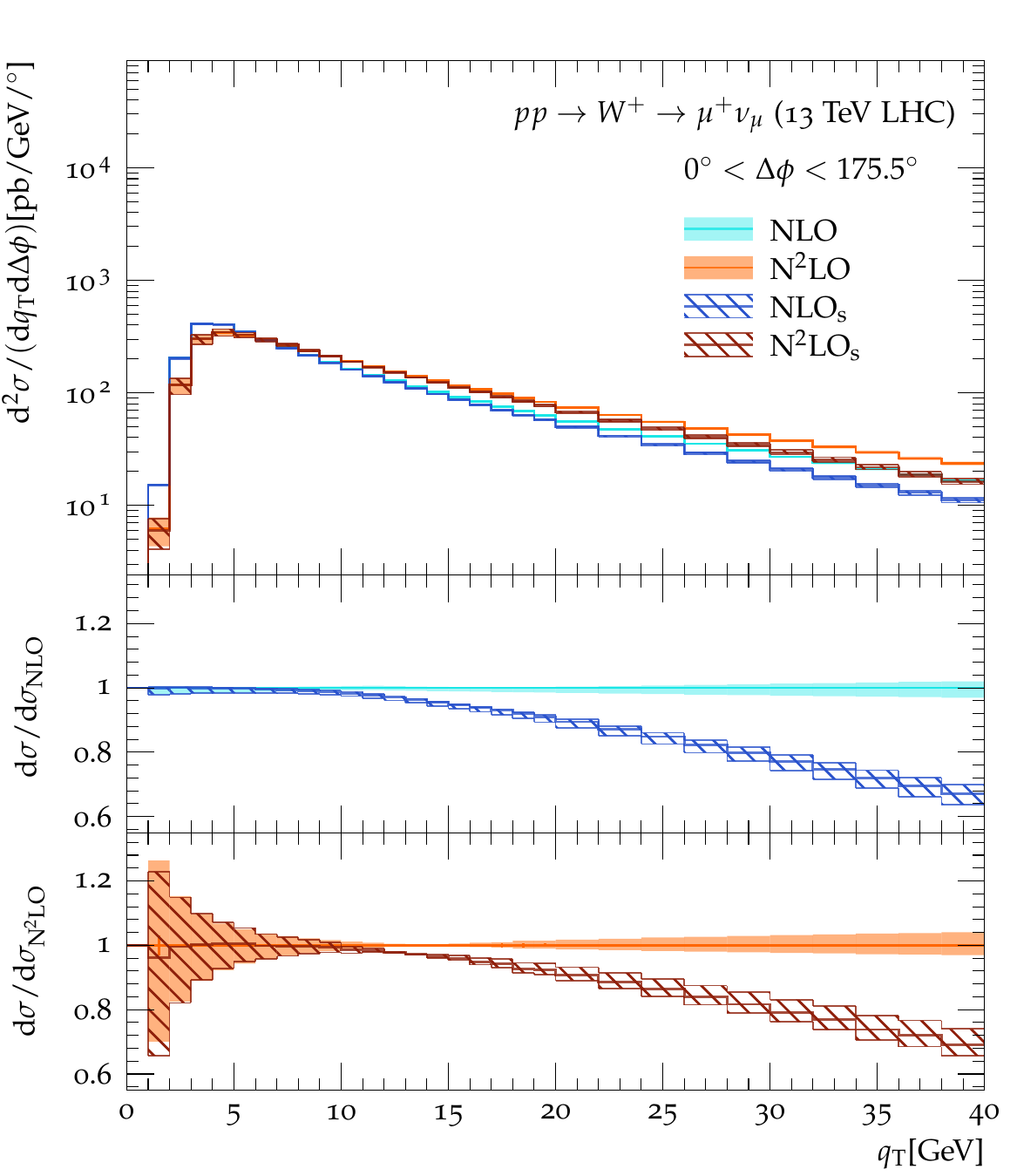}\\[1mm]
  \includegraphics[width=.32\textwidth]{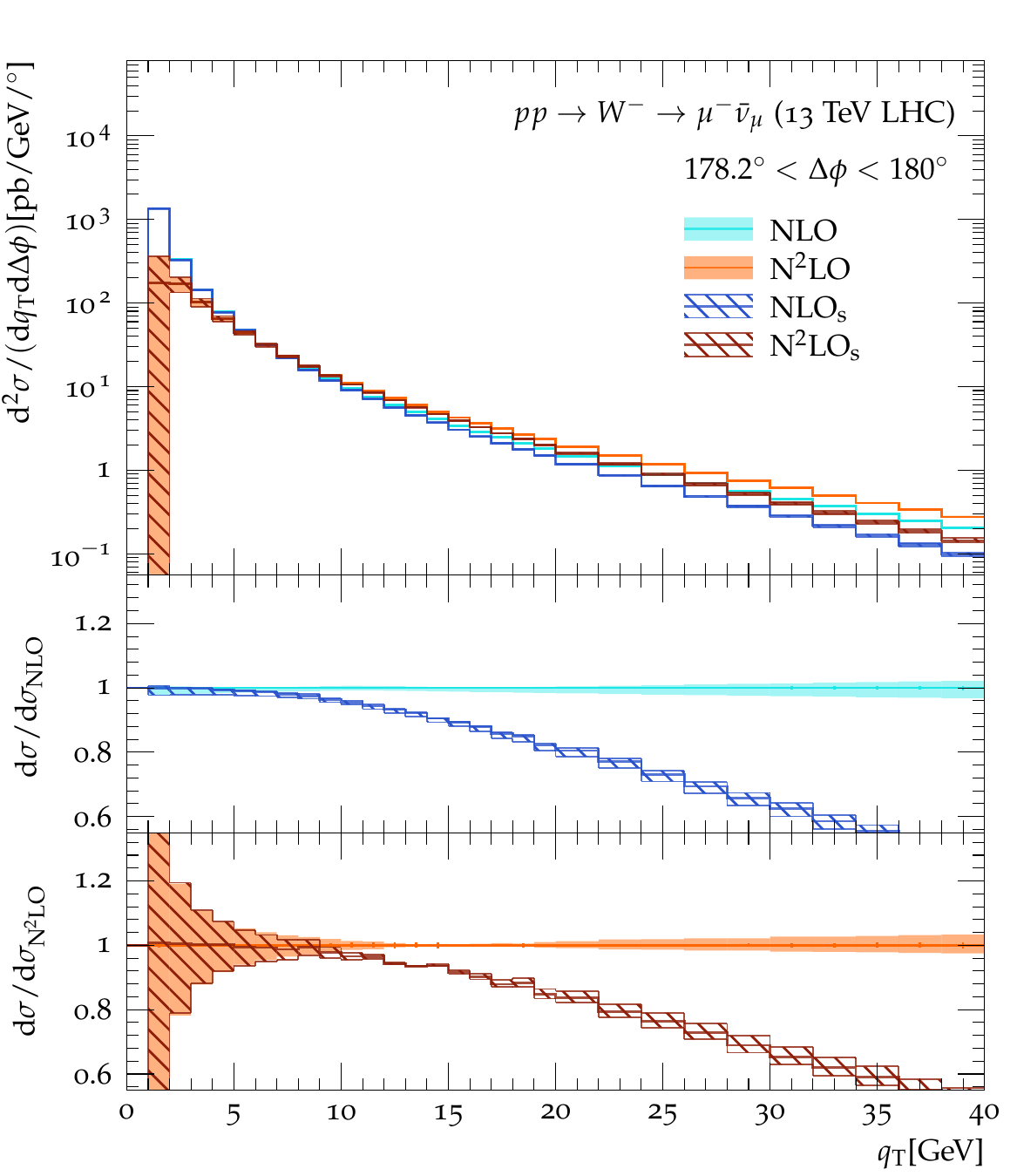}
  \includegraphics[width=.32\textwidth]{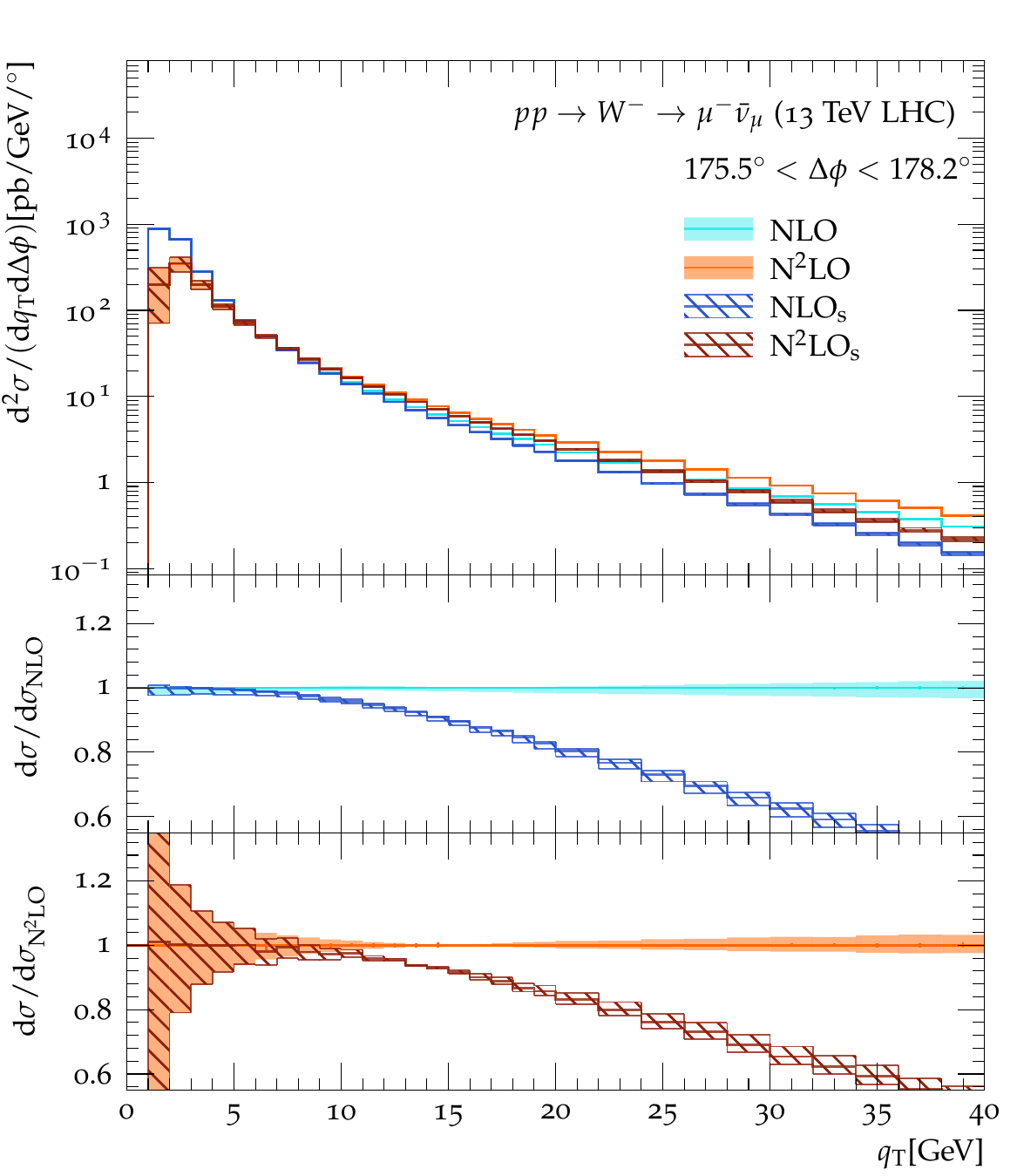}
  \includegraphics[width=.32\textwidth]{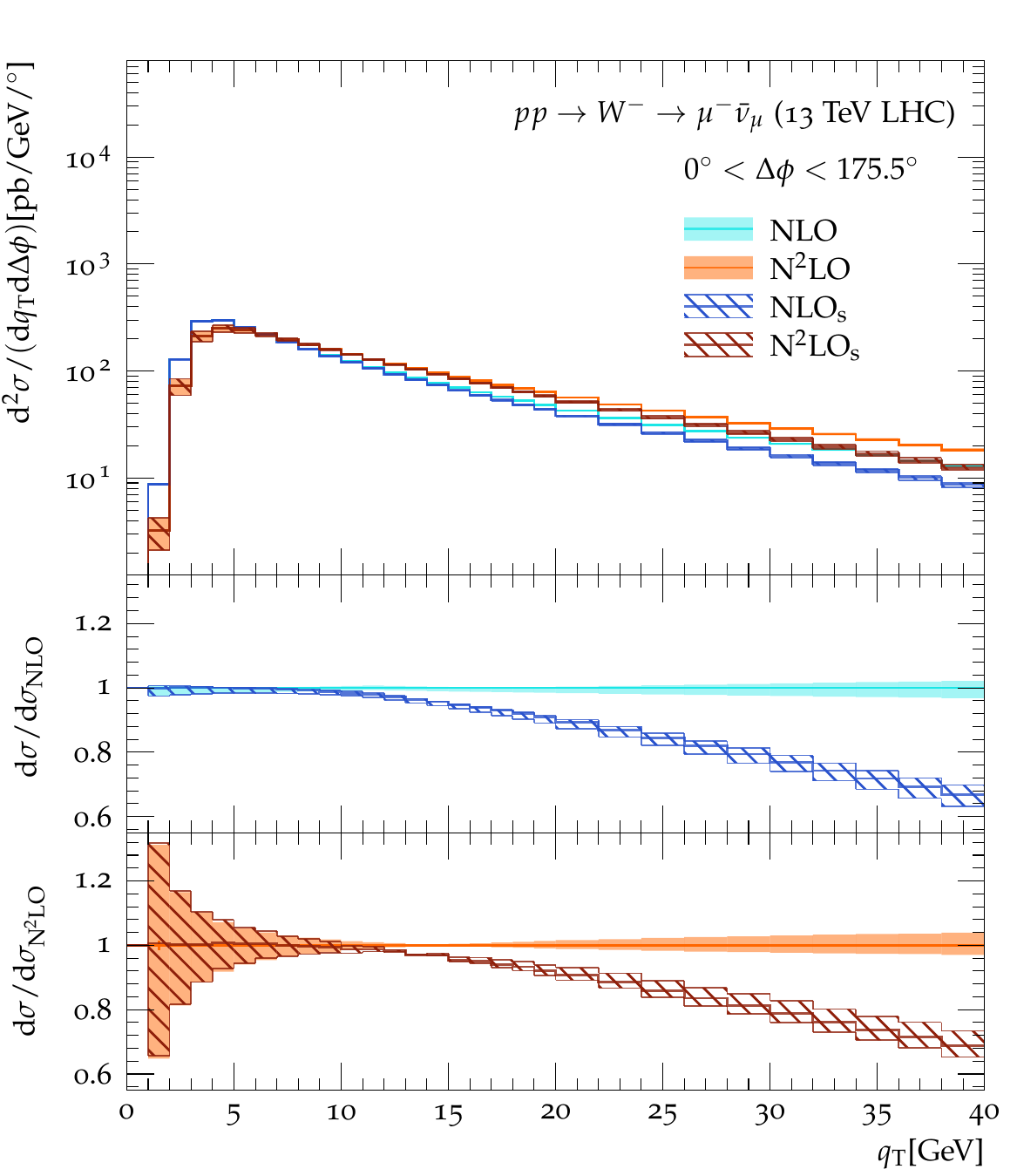}
  \caption{
    Comparison of the fixed-order \qT\ spectra in all three processes. 
    N$^{(2)}$LO denotes the fixed-order full QCD perturbative result, 
    while N$^{(2)}$LO$_\text{s}$ is the fixed-order expansion of the 
    SCET-based resummation.
  }
  \label{fig:results:val:qt}
\end{figure}

Fig.\ \ref{fig:results:val:qt} shows this comparison for
the \qT\ spectra in the three slices in \dphi\ introduced 
earlier. 
In all three slices the agreement between the exact QCD 
result and the expanded resummed result derived from the 
soft-collinear effective theory is excellent for all 
three processes, $Z$, $W^+$, and $W^-$ production, under 
investigation. 
It is interesting to note that within a few percent 
the approximation in the soft-collinear effective theory 
holds to sizeable distances from the singular point at 
$(\qT,\dphi)=(0\,\text{GeV},180^\circ)$. 
In particular, deviations do not exceed 2\% for \qT\ up to 
10\,GeV for all \dphi\ ranges, neither at \NLO\ nor \NNLO.
These observations hold independent of the factorisation 
and renormalisation scale, as is indicated by the 
coincidence of the shown scale variation bands. 

\begin{figure}[t!]
  \centering
  \includegraphics[width=.32\textwidth]{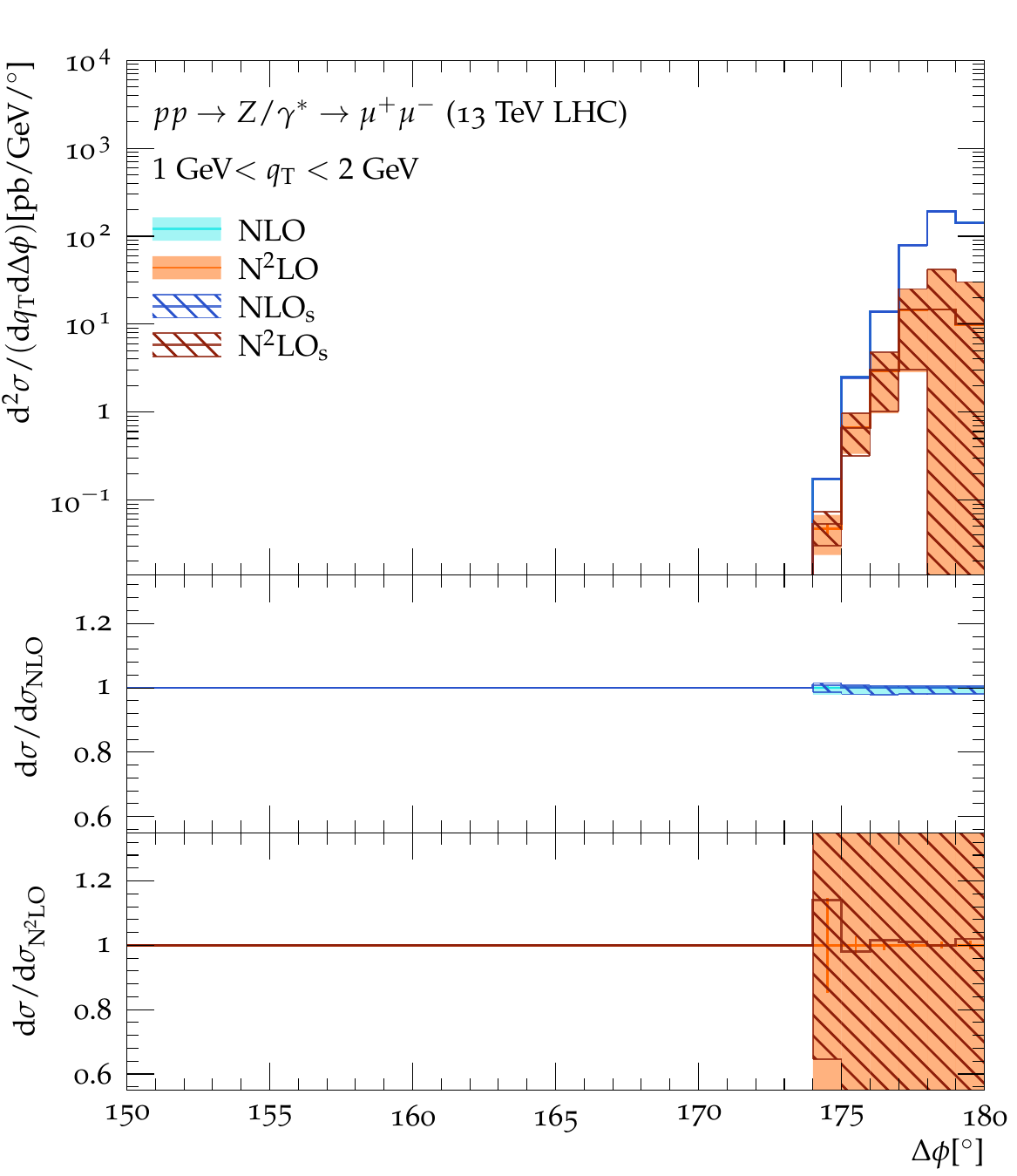}
  \includegraphics[width=.32\textwidth]{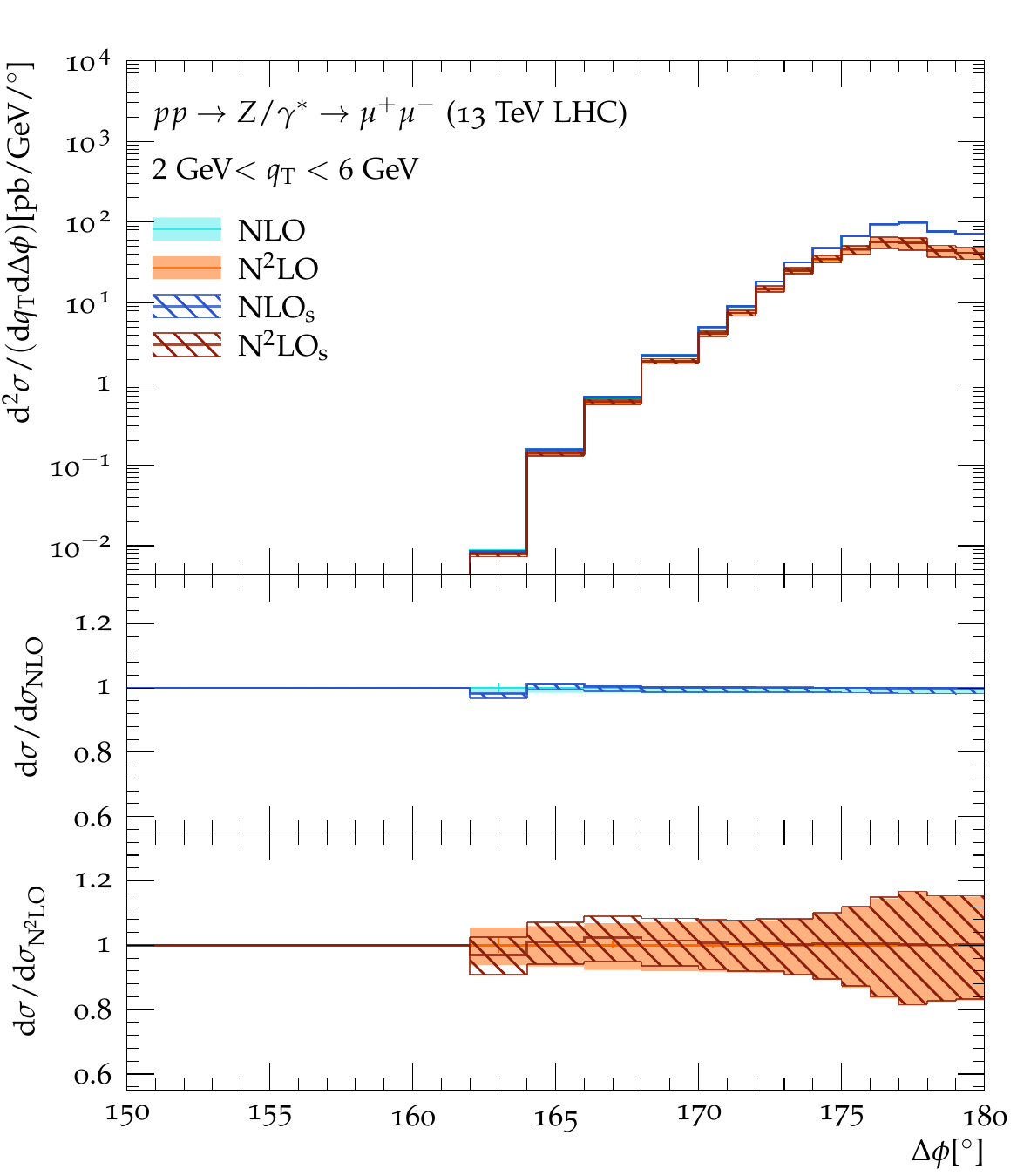}
  \includegraphics[width=.32\textwidth]{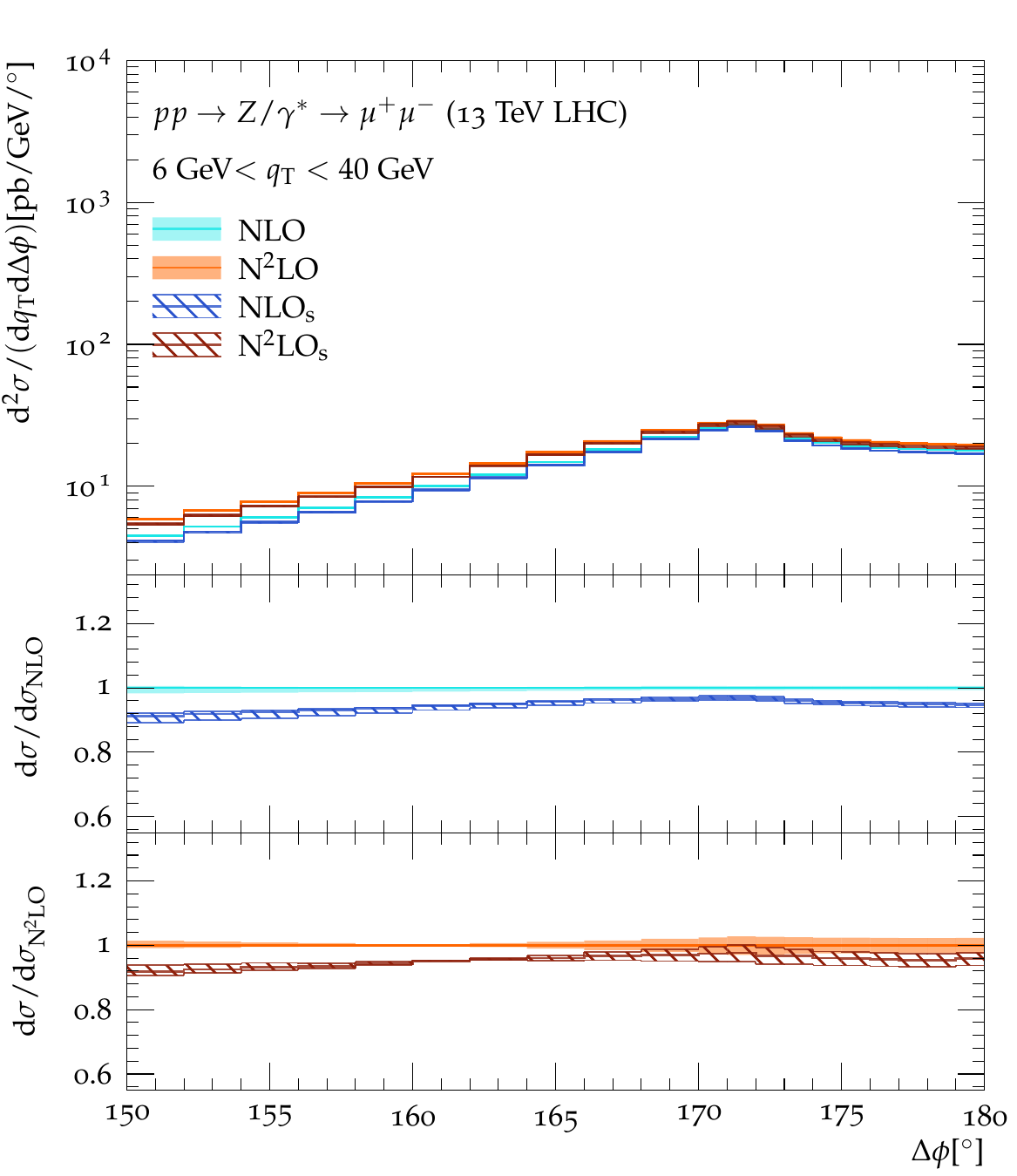}\\[1mm]
  \includegraphics[width=.32\textwidth]{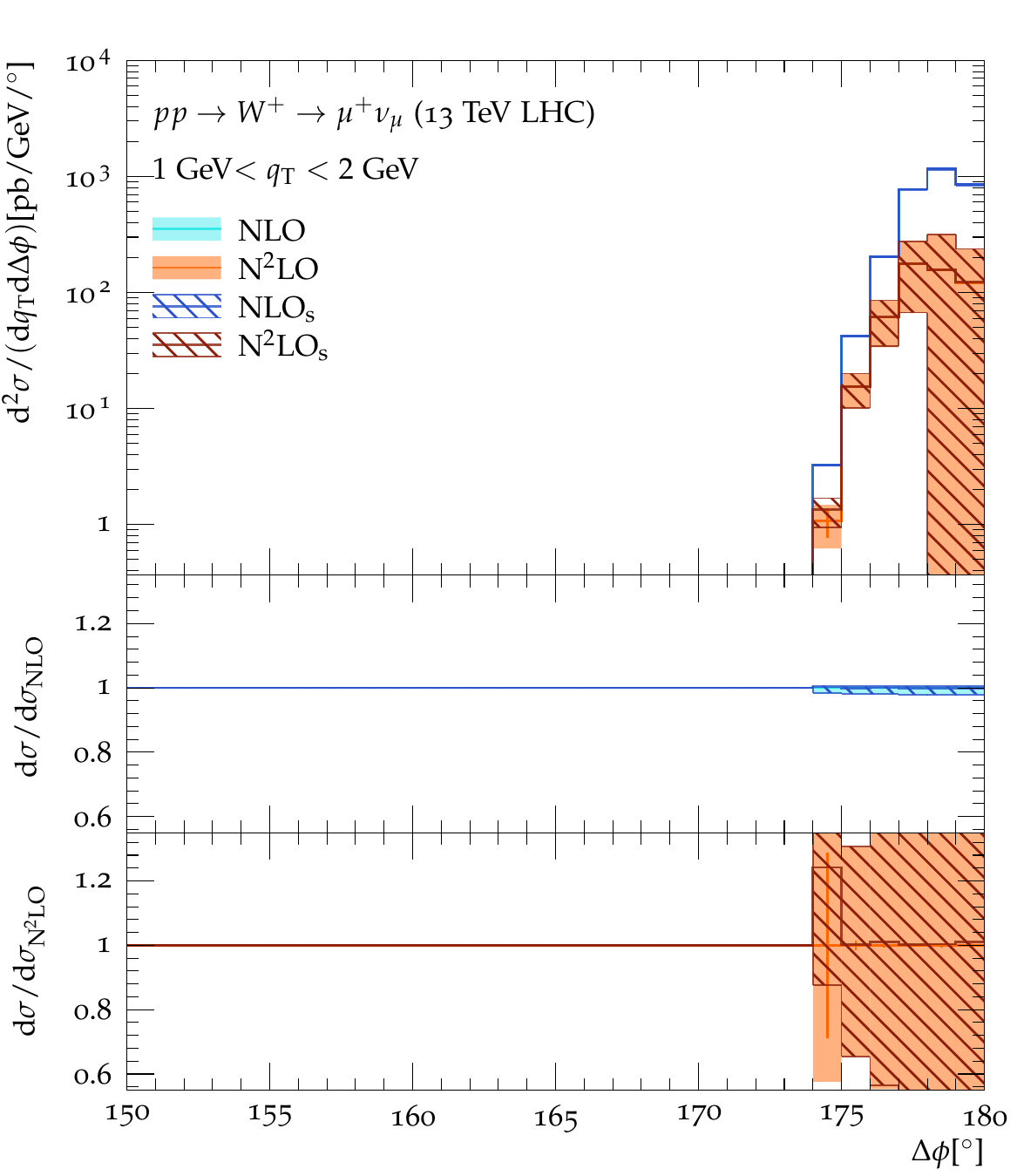}
  \includegraphics[width=.32\textwidth]{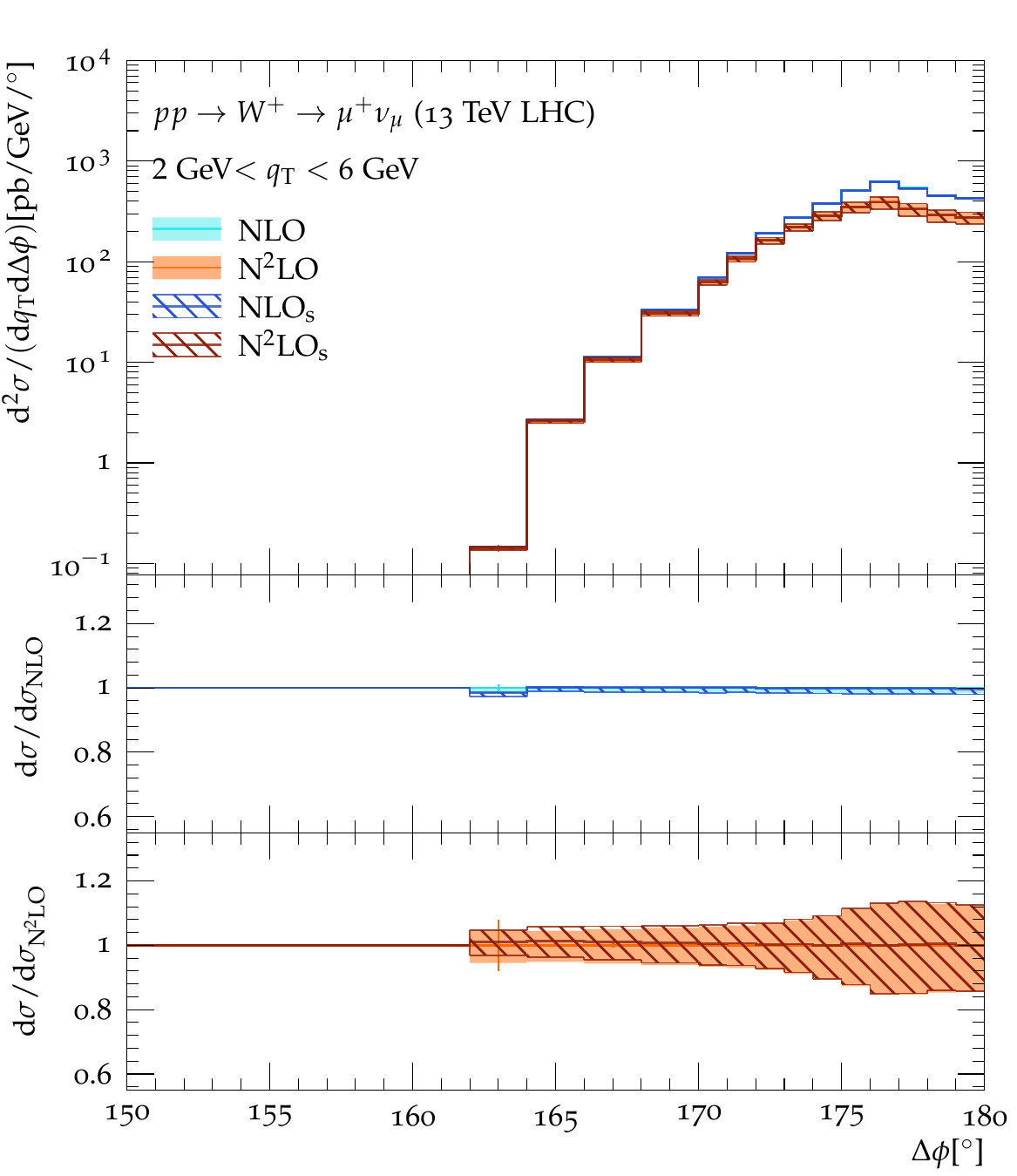}
  \includegraphics[width=.32\textwidth]{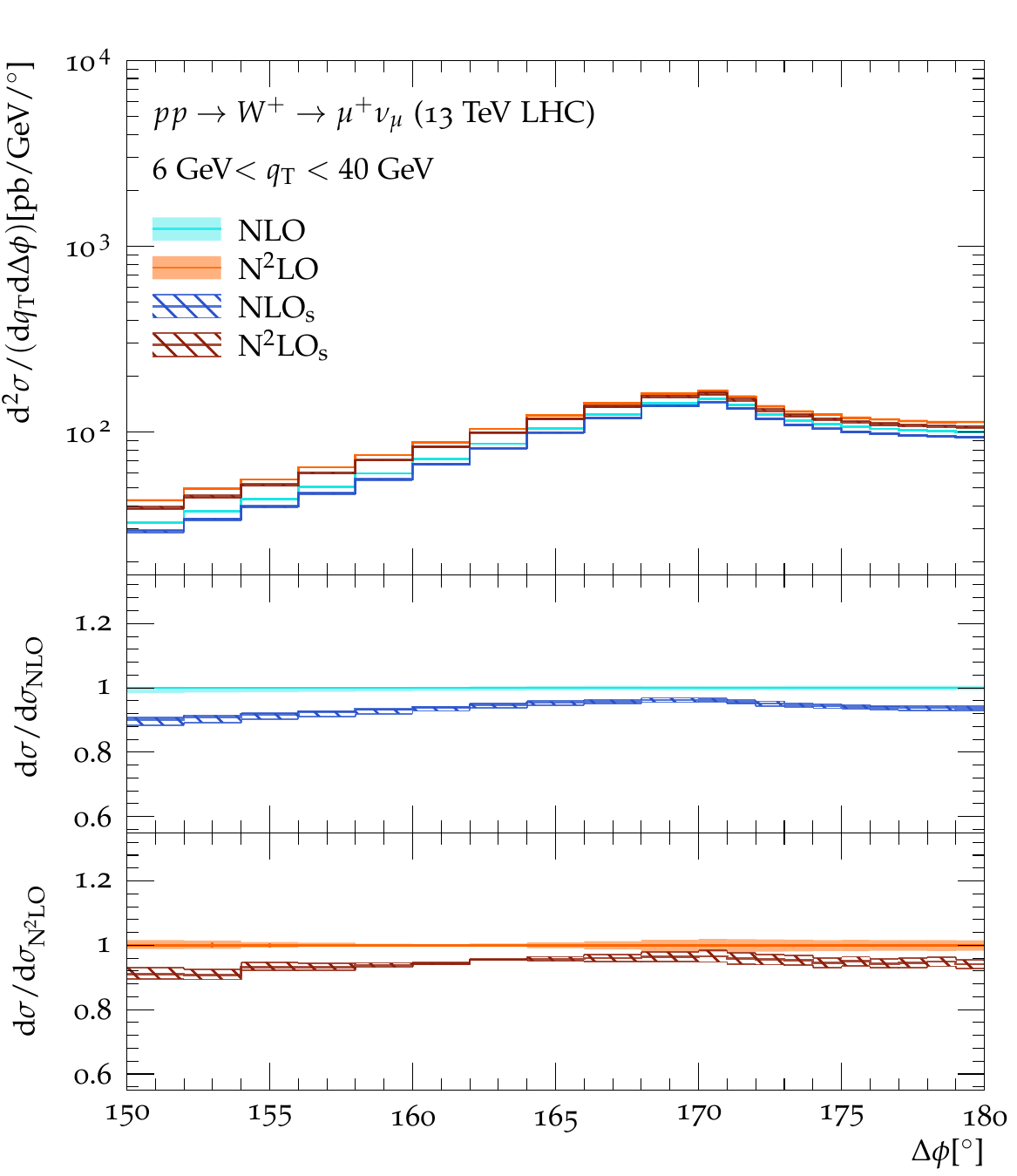}\\[1mm]
  \includegraphics[width=.32\textwidth]{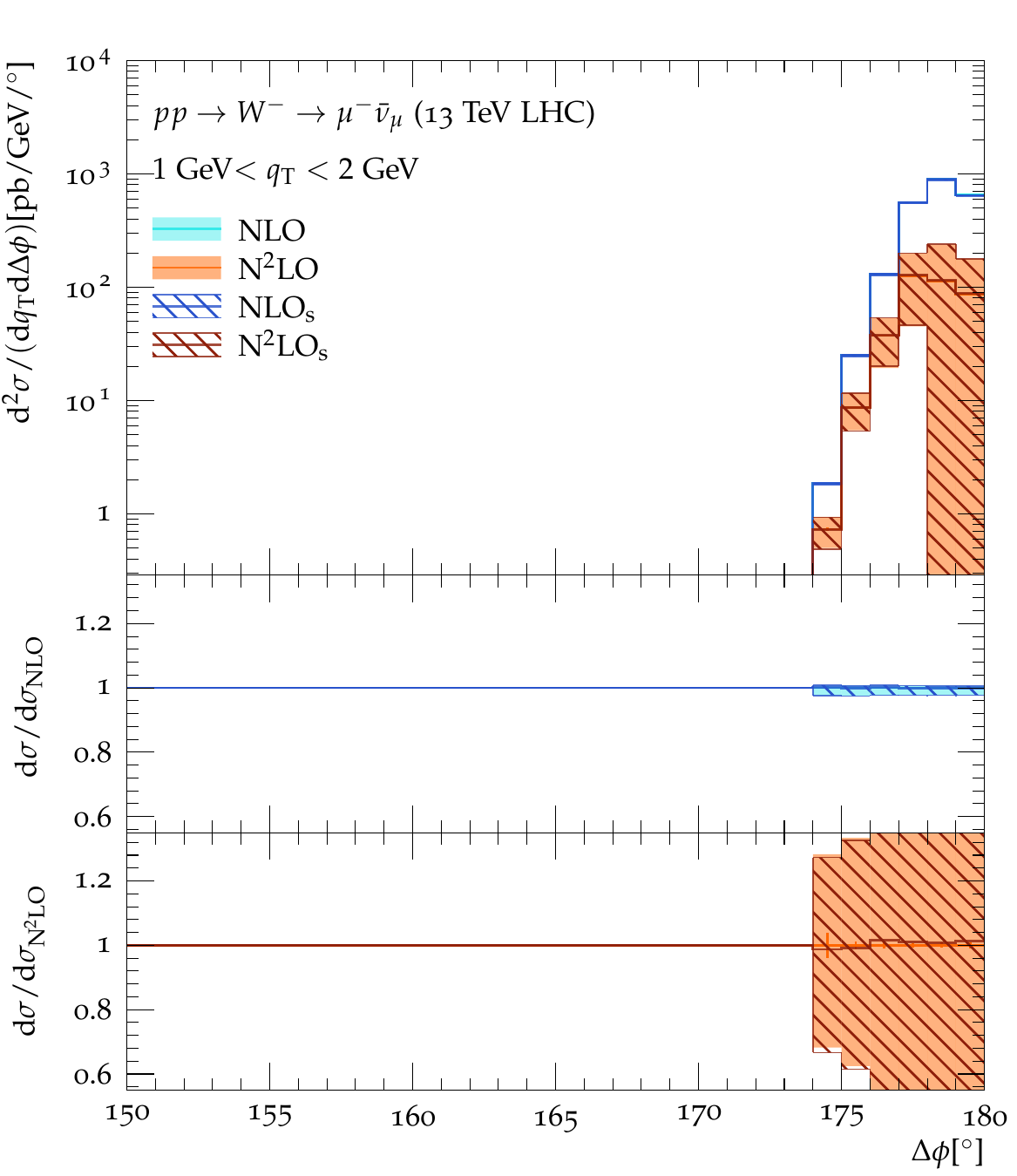}
  \includegraphics[width=.32\textwidth]{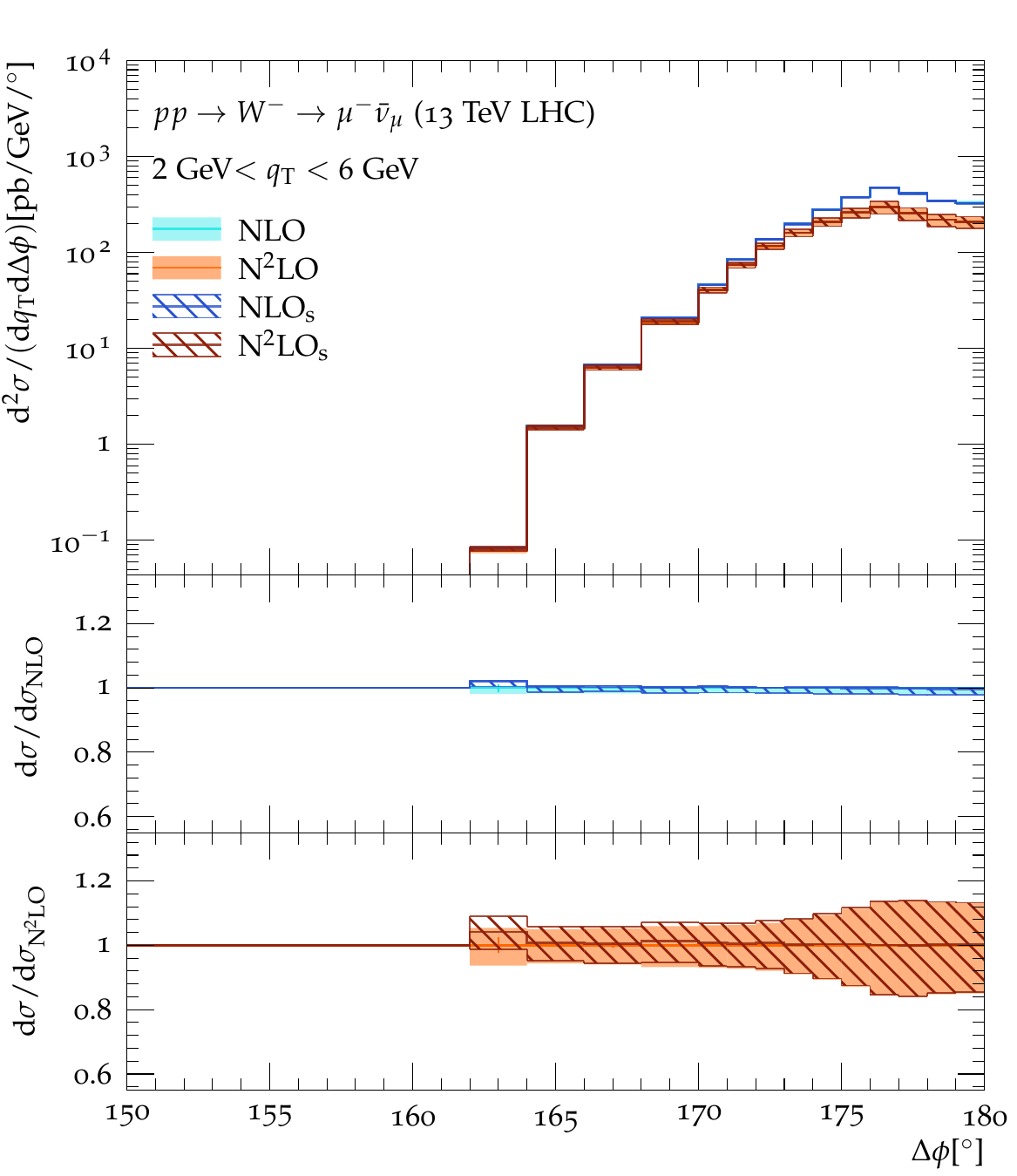}
  \includegraphics[width=.32\textwidth]{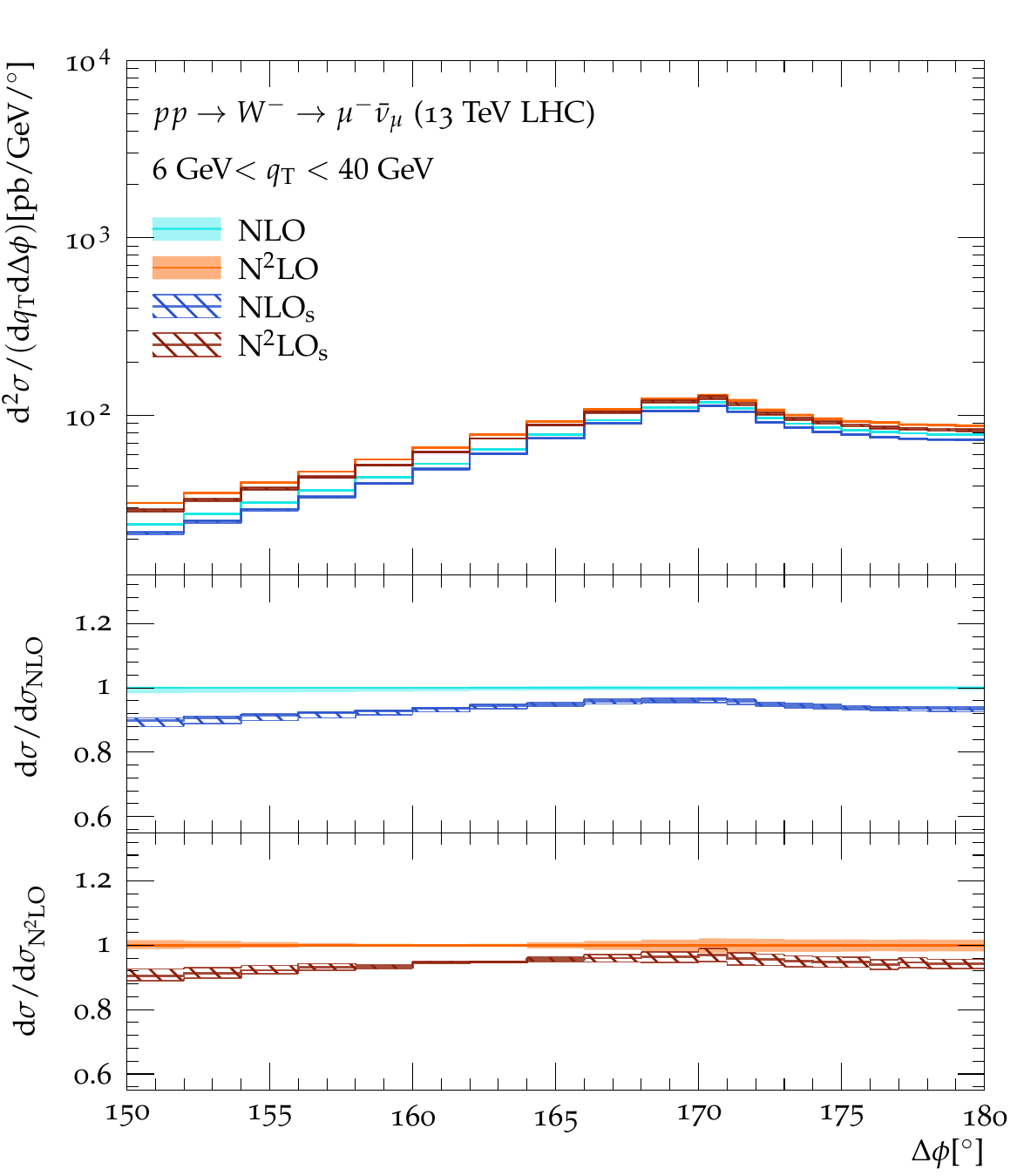}
  \caption{
    Comparison of the fixed-order \dphi\ spectra in all three processes. 
    N$^{(2)}$LO denotes the fixed-order full QCD perturbative result, 
    while N$^{(2)}$LO$_\text{s}$ is the fixed-order expansion of the 
    SCET-based resummation.
  }
  \label{fig:results:val:dphi}
\end{figure}

The situation is slightly different for the \dphi\ 
distributions in the three \qT\ slices depicted in 
Fig.\ \ref{fig:results:val:dphi}.
Here, the fixed-order expansion of the resummation 
very well coincides with the exact QCD calculation 
only in the first two slices for \qT\ smaller than 6\,GeV. 
In the last \qT\ slice, well away from the singular point, 
the agreement is worse, ranging from around $5\%$ to 10\%, 
with the best agreement unsurprisingly found on the Sudakov 
peak.
Small effects due to limited statistics can be observed 
near the phase space boundaries, but do not impact 
our findings.

\subsection{Resummation improved results}
\label{sec:results:res}

We now turn to examine our full resummation improved results, 
calculated according to eq.\ \eqref{eq:methods:extra:obs_calc}. 
Contrary to the fixed-order evaluation of the previous section, 
the behaviour of the cross section as we approach the singular 
point at $(\qT,\dphi)=(0,180^\circ)$ is regular and we can 
evaluate the complete $(\qT,\dphi)$ plane. 
To arrive at our resummation improved results of this section, however, 
the governing eq.\ \eqref{eq:methods:extra:TMD_matched} still 
contains two terms that are separately diverging as the singular 
point is approached, originating in the behaviour of the full QCD 
fixed-order calculation and the fixed-order expansion of the 
resummation in the soft-collinear effective theory. 
To regulate their behaviour, following the results of 
Sec.\ \ref{sec:results:val}, we set them to be exactly equal 
for $\qT<1\,\text{GeV}$, leaving only the resummed result 
similar to the treatment in Refs.~\cite{Becher:2019bnm,Becher:2020ugp}. 
Having said that, the most important question to answer, 
is the choice of matching scale \qTcut. 
For this, we follow two arguments to guide us to our choice of matching scale.

Firstly, we want to restrict the resummation to the asymptotic regime 
where its intrinsic approximations are valid. 
Recalling that the resummation in eq.\ \eqref{eq:methods:res:TMD_Resummation} 
is derived from the factorisation of eq.\ \eqref{eq:methods:TMD_factorisation} 
where only the leading contributions in an expansion in $\qT/M_L$ 
are kept, eq.\ \eqref{eq:methods:res:TMD_Resummation} is valid only 
in the small \qT\ regime and thus should be disabled beyond it. 
To this end, the transition function $f(x)$ of eq.~\eqref{eq:methods:extra:f} 
is employed to provide a smooth transition from the resummed 
spectra to the fixed-order contribution in the matched result 
of eq.\ \eqref{eq:methods:extra:TMD_matched} within the interval 
$\qTcut\mp\dqT$ with $\dqT\approx4.6\,\text{GeV}$ with our choice of scales.  
We note that the value of \qTcut\ is related 
to the effective range of the $\qT/M_L$ expansion.  
Following the spirit of \cite{Becher:2019bnm} we determine 
that range by comparing in 
Fig.\ \ref{fig:results:val:qt} the \NNLO\ result with the 
corresponding expansion of the resummation \NNLO$_s$. 
To be precise, we require the fixed-order expansion of the SCET 
approximation to deviate from the exact result by no more than 20\%. 
For all three processes and \dphi\ slices we extract similar 
values, leading to a common choice of matching scale of 
$\qTcut=16\,\text{GeV}$. 

\begin{figure}[t!]
  \centering
  \includegraphics[width=.32\textwidth]{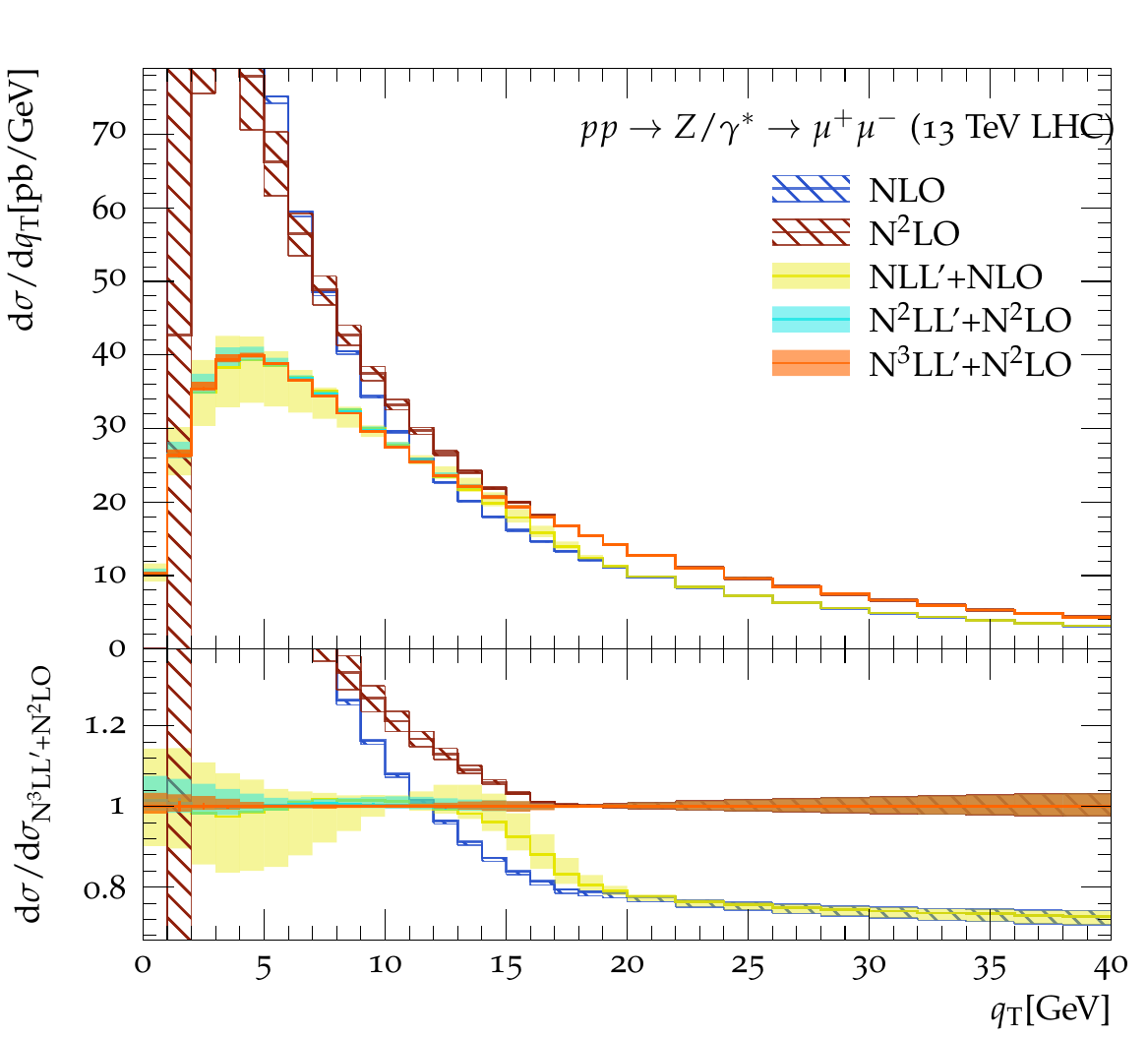}\hfs
  \includegraphics[width=.32\textwidth]{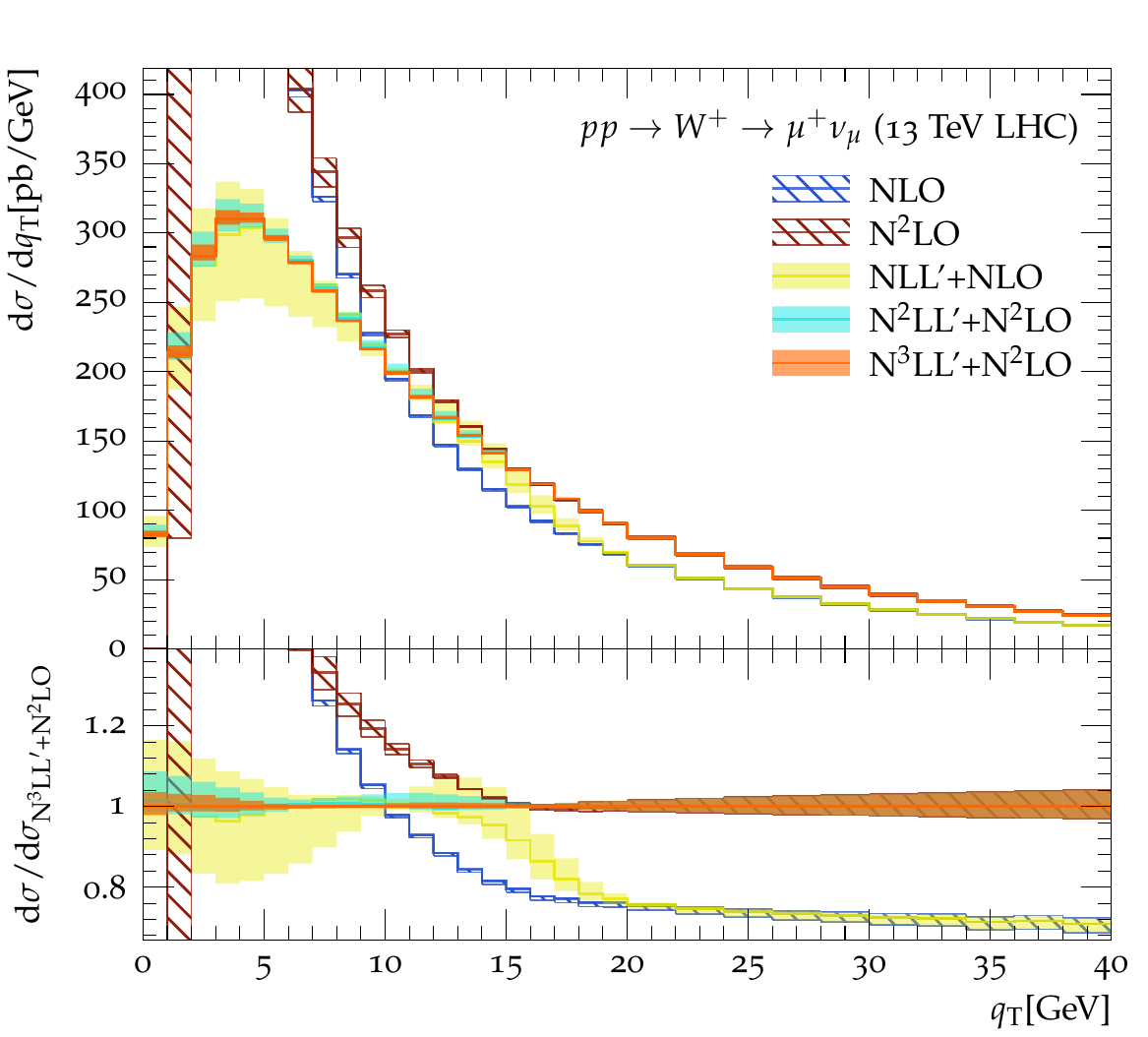}\hfs
  \includegraphics[width=.32\textwidth]{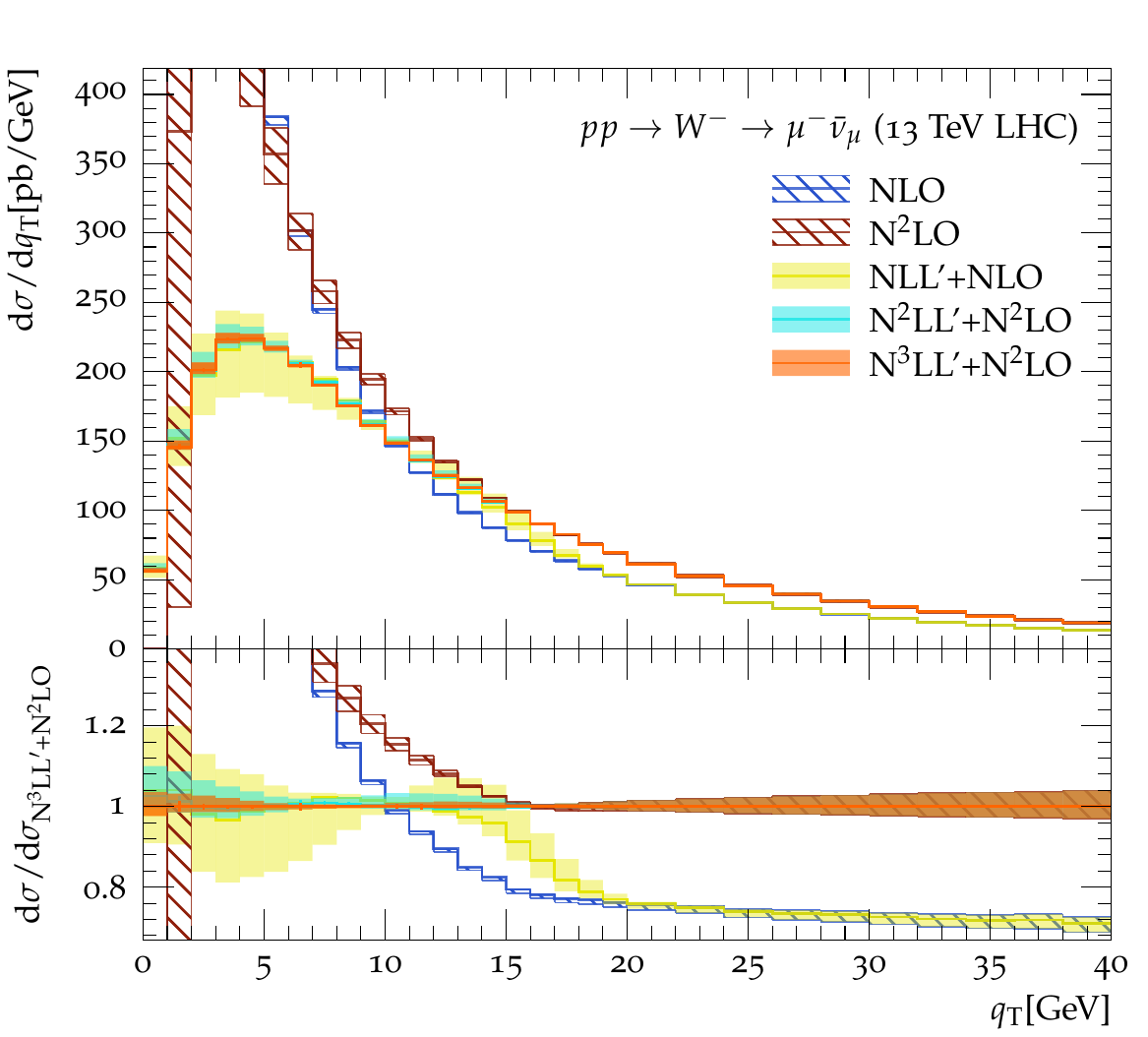}
  \caption{
    Single-differential cross section in \qT\ for all three processes. 
    We compare the full QCD \NLO\ and \NNLO\ distributions to the resummation 
    improved results, \NLLNLO, \NNLLNNLO, and \NNNLLNNLO.
  }
  \label{fig:results:res:qt:inclusive}
\end{figure}

Secondly, we want the additional corrections introduced by the 
resummation with respect to the fixed-order calculation to be 
small or negligible at the matching scale.
To evaluate this requirement, Fig.\ \ref{fig:results:res:qt:inclusive} 
is of particular interest. 
Here we observe that at values around the chosen matching scale 
$\qTcut=16\,\text{GeV}$ the resummation improved results coincide 
with the pure fixed-order one to better than 3\%. 
At this point, the reader is reminded that, although we are not 
comparing the pure resummation with the full QCD result but 
instead a result where the resummation at the scale \qTcut\ 
is already subjected to the suppression function $f(\qT)$ of 
eq.\ \eqref{eq:methods:extra:f}, the suppression function 
has the value $f(\qTcut)=0.5$. 
Therefore, we still find that the resummation and the fixed-order 
result still agree to better than 5\% and resummation effects are 
no longer important. 
It is interesting to note that this observation holds for both 
\NNLLNNLO\ and \NNNLLNNLO.
The situation is slightly different for \NLLNLO. 
However, since this result is mainly included to illustrate 
the progression of the increased accuracy of our calculation 
we choose the same value for \qTcut. 

Analytically this can be understood in the following. 
In the above argument we are essentially evaluating the relative 
size of the contribution the resummation is supplying beyond the 
accuracy of the fixed-order calculation. 
These terms are of $\order(\alpha_s^2L^4+\alpha_s^2L^3+\ldots)$ 
for the \NLLNLO\ matched result, while they are of 
$\order(\alpha_s^3L^6+\alpha_sL^5+\ldots)$ for 
the \NNLLNNLO\ and \NNNLLNNLO\ calculations, $L=\log(\qT/M_L)$. 
Now while \qT\ is small, these contributions are of the same order. 
Choosing a \qTcut\ sufficiently removed from the singular point, 
such that the ratio $\qT/M_L$ is of $\order(1)$, $L$ follows a different 
power counting. 
Thus, the additional terms induced by the resummation with 
respect to the fixed-order calculation are indeed of higher-order 
in \NNLLNNLO\ and \NNNLLNNLO\ than in \NLLNLO.

\begin{figure}[t!]
  \centering
  \includegraphics[width=.32\textwidth]{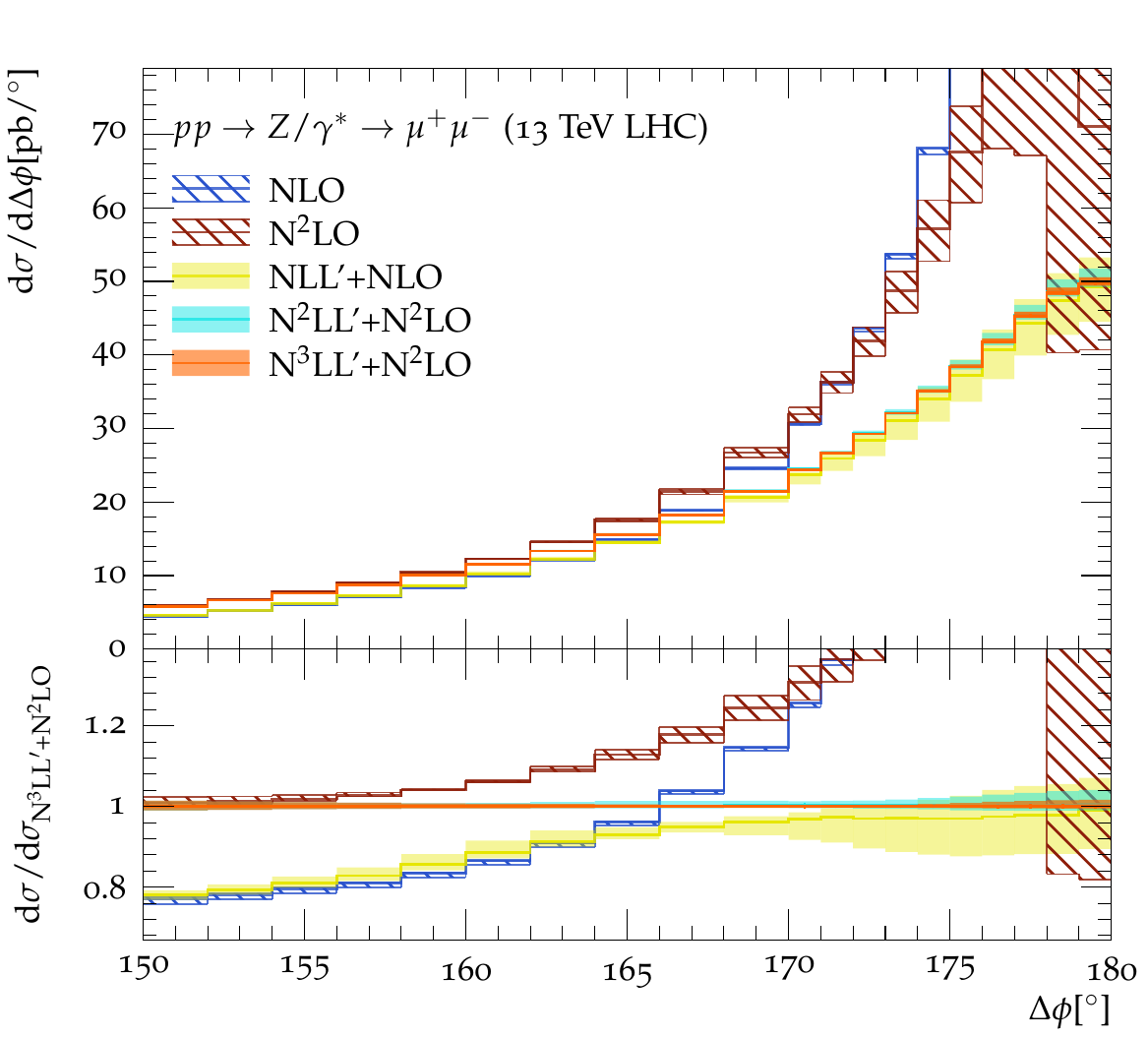}\hfs
  \includegraphics[width=.32\textwidth]{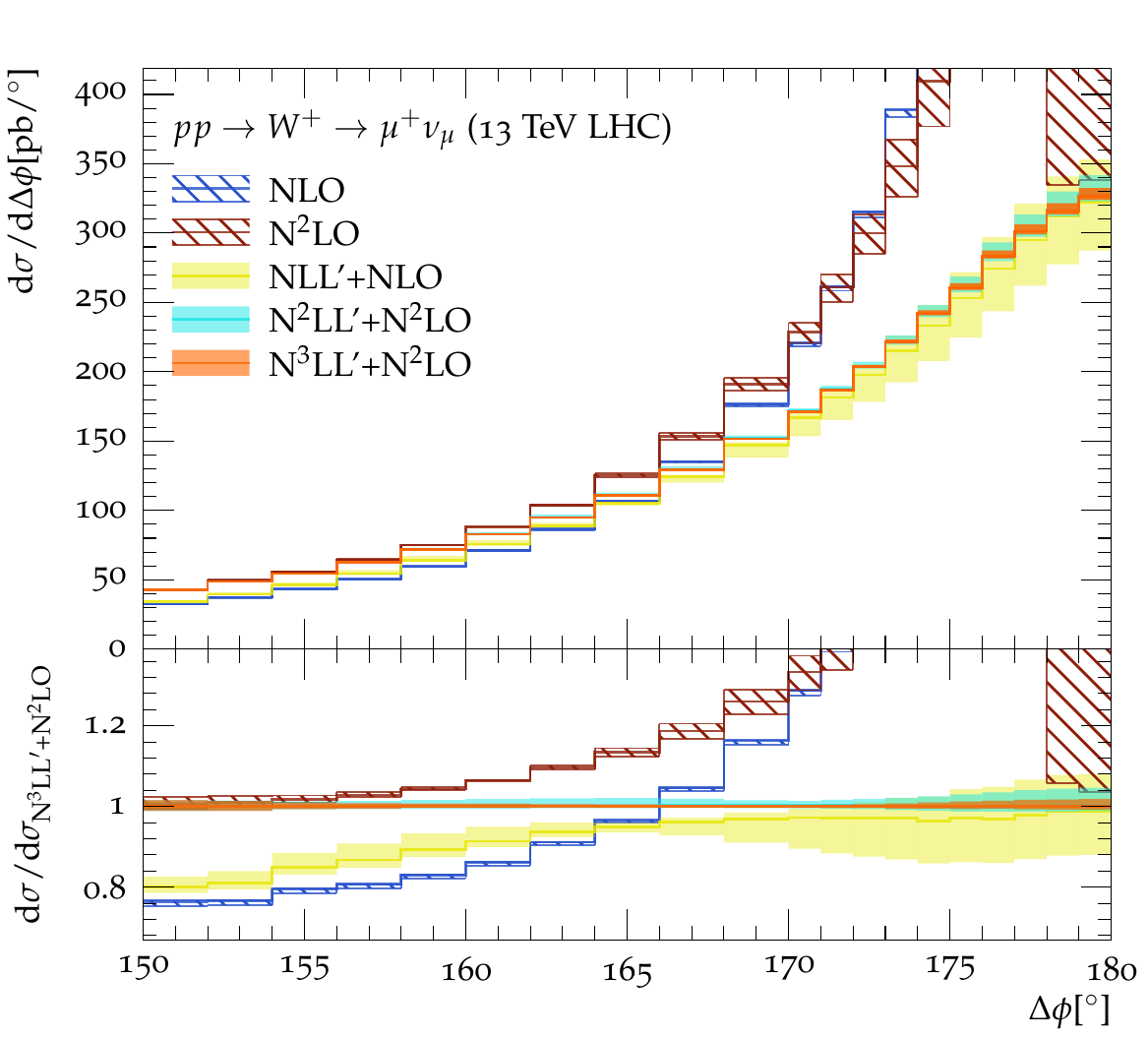}\hfs
  \includegraphics[width=.32\textwidth]{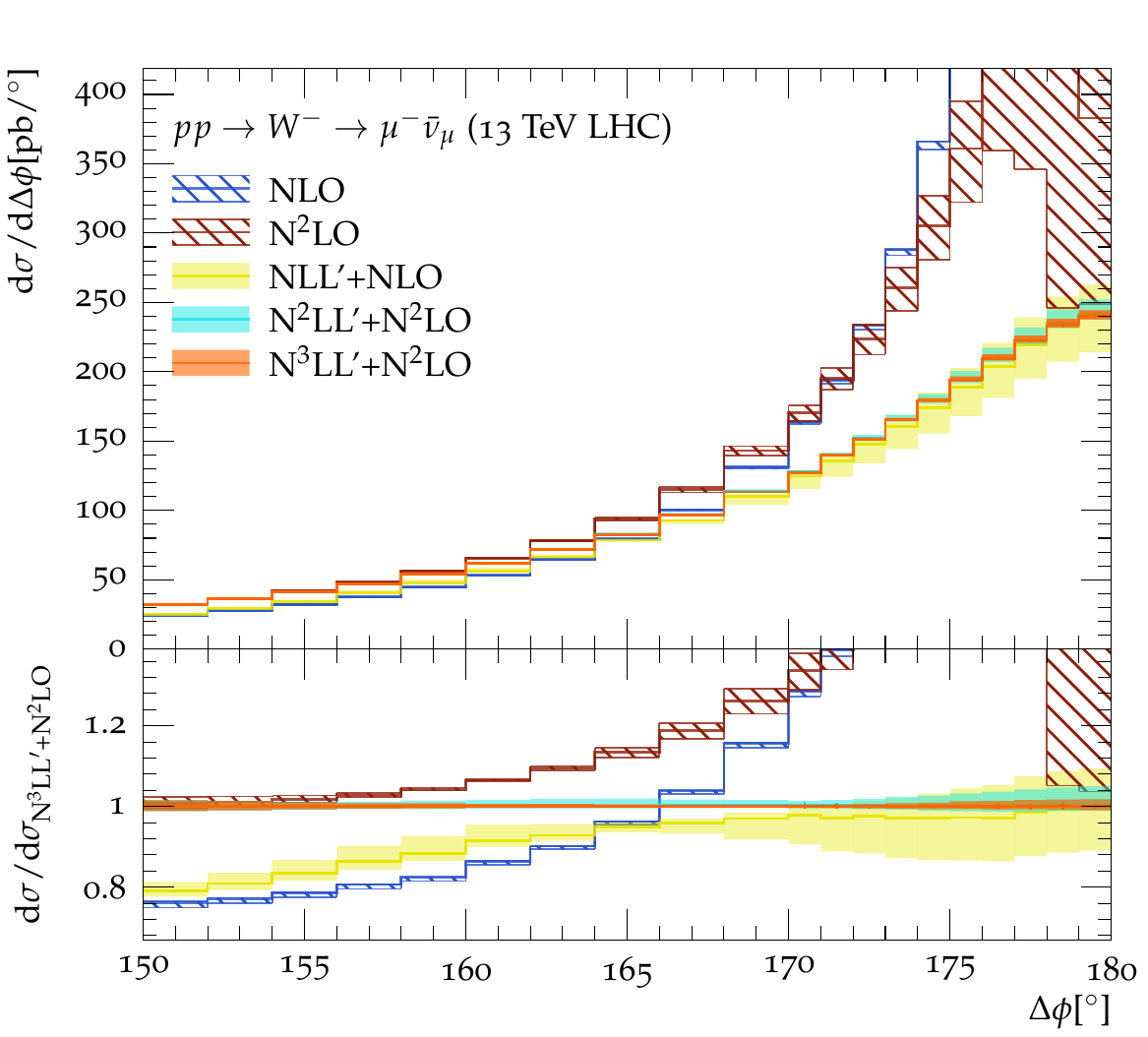}
  \caption{
    Single-differential cross section in \dphi\ for all three processes. 
    We compare the full QCD \NLO\ and \NNLO\ distributions to the resummation 
    improved results, \NLLNLO, \NNLLNNLO, and \NNNLLNNLO.
  }
  \label{fig:results:res:dphi:inclusive}
\end{figure}

With this choice of resummation scale the single-differential 
distributions in the leptonic transverse opening angle \dphi\ 
similarly receive substantial resummation effects, as shown 
in Fig.\ \eqref{fig:results:res:dphi:inclusive}. 
Since the suppression function $f$ acts in another variable, 
no clear transition from one regime to the other can be observed 
at any order. 
We observe, however, that while at our highest order, \NNNLLNNLO, 
all three processes behave extremely similarly and receive very 
similar resummation induced corrections to the \NNLO\ result, 
this is markedly different at \NLLNLO\ accuracy. 
Here, the $W$ production channels receive somewhat larger 
corrections in the region between $\dphi=150^\circ$ and 
$\dphi=165^\circ$ than in $Z$ production. 

\begin{figure}[t!]
  \centering
  \includegraphics[width=.32\textwidth]{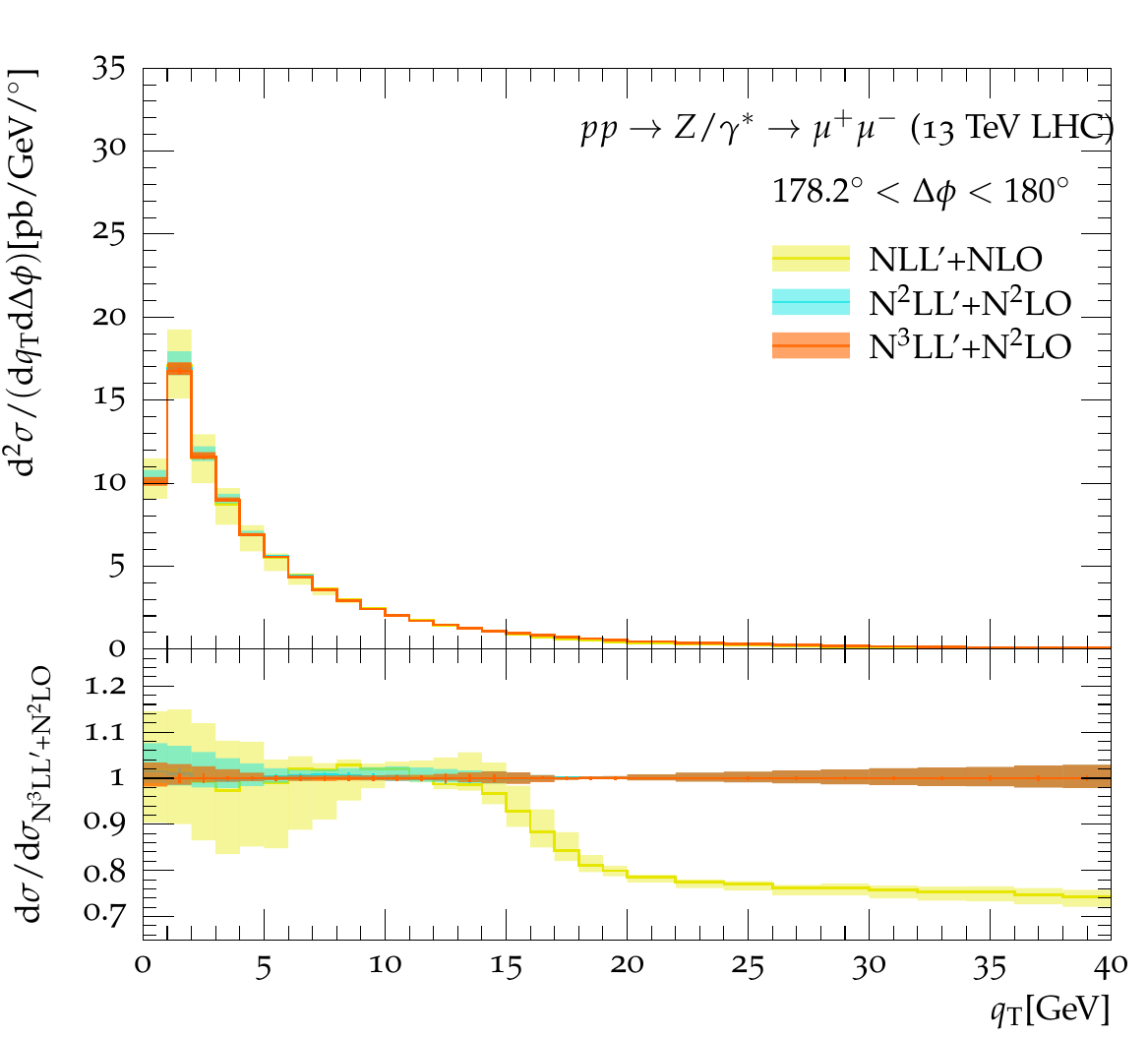}\hfs
  \includegraphics[width=.32\textwidth]{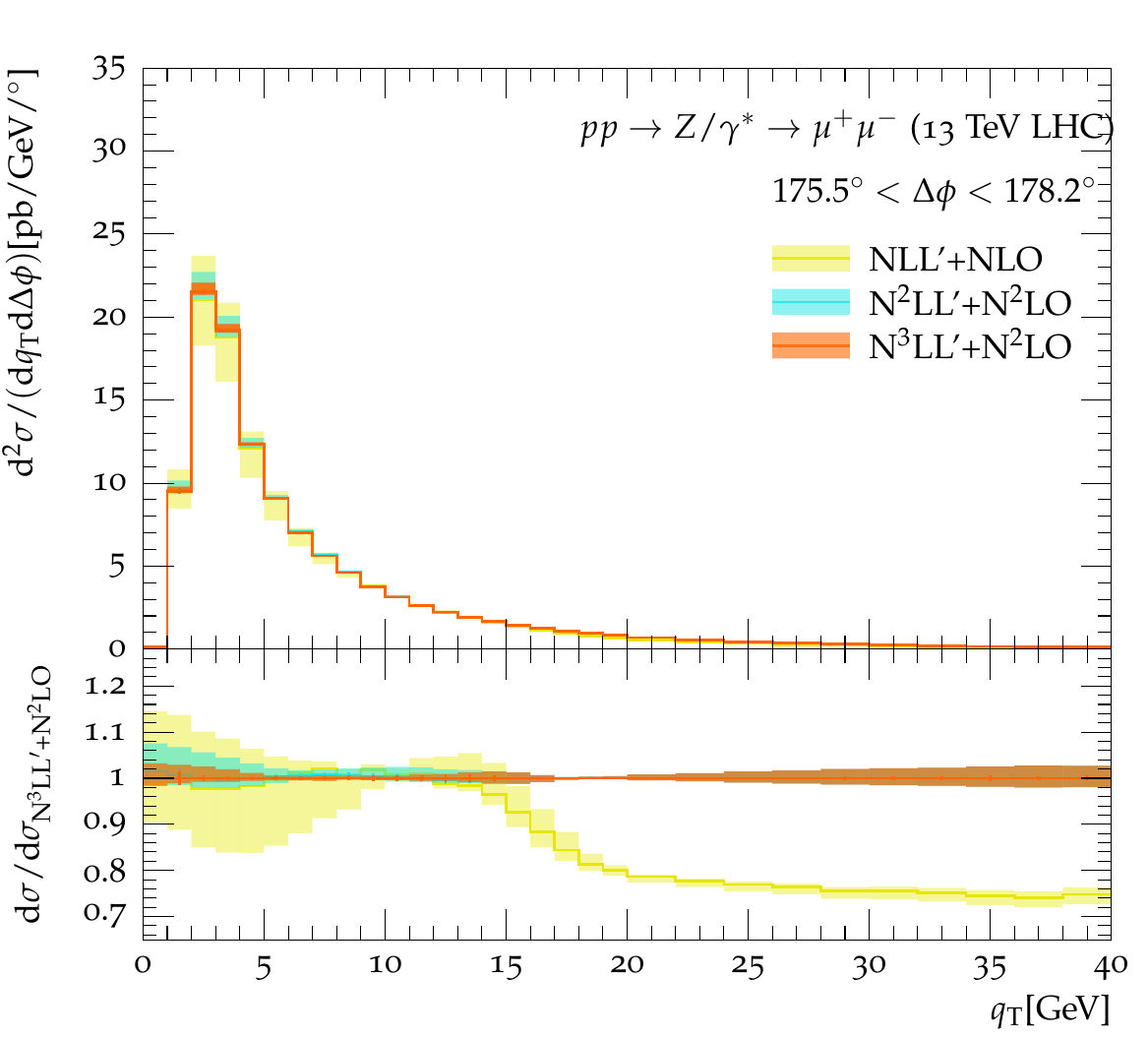}\hfs
  \includegraphics[width=.32\textwidth]{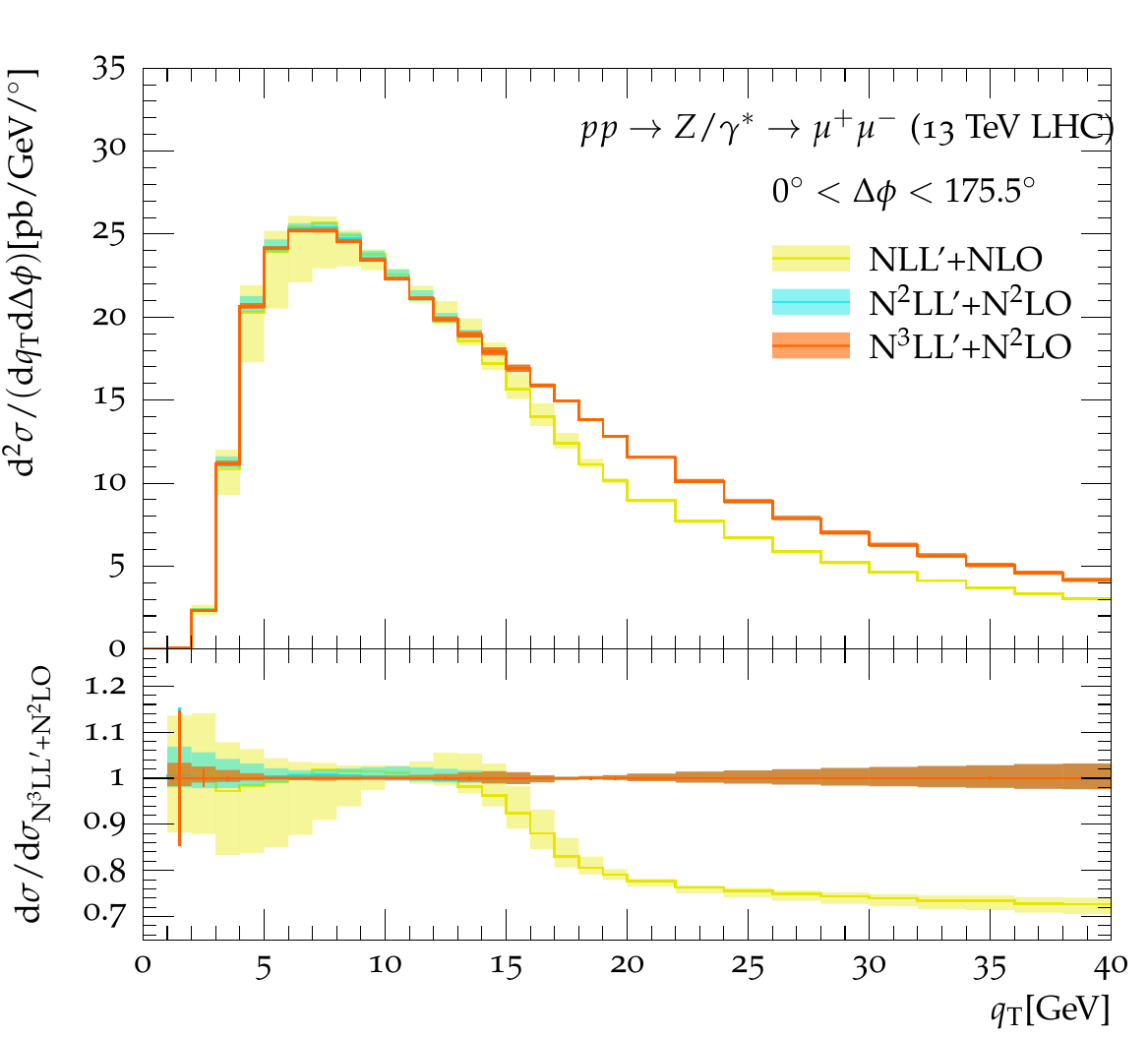}\\[1mm]
  \includegraphics[width=.32\textwidth]{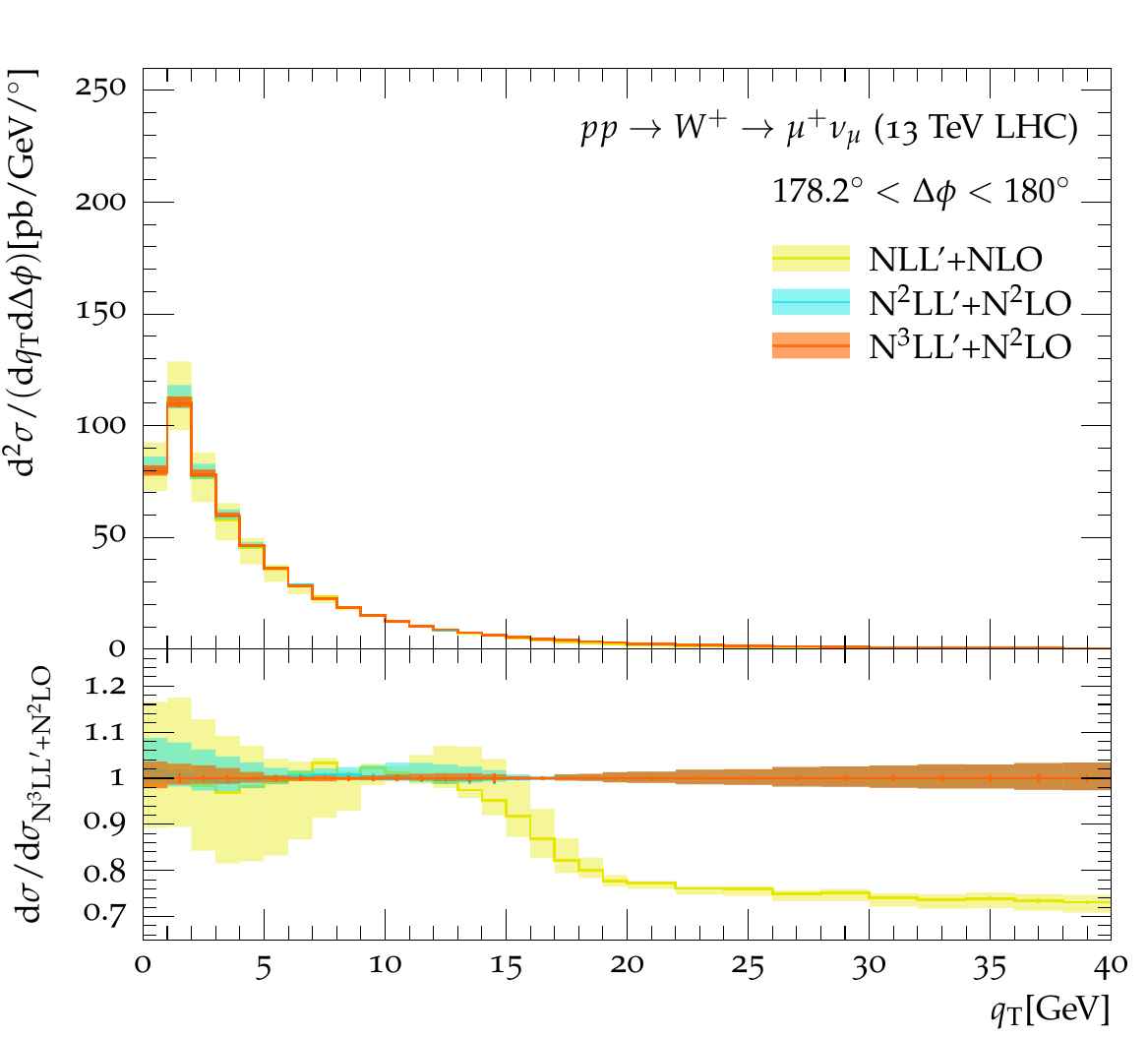}\hfs
  \includegraphics[width=.32\textwidth]{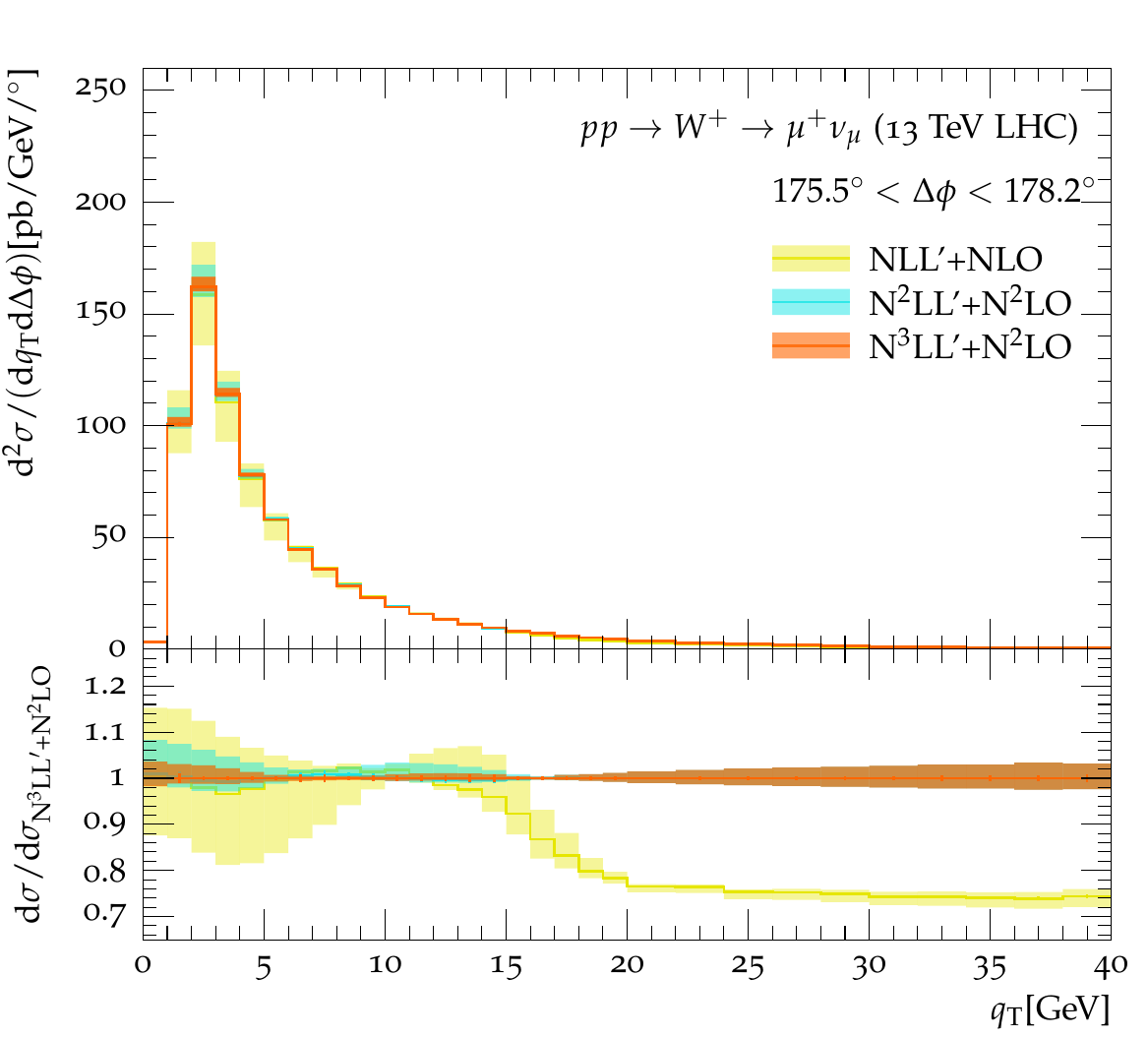}\hfs
  \includegraphics[width=.32\textwidth]{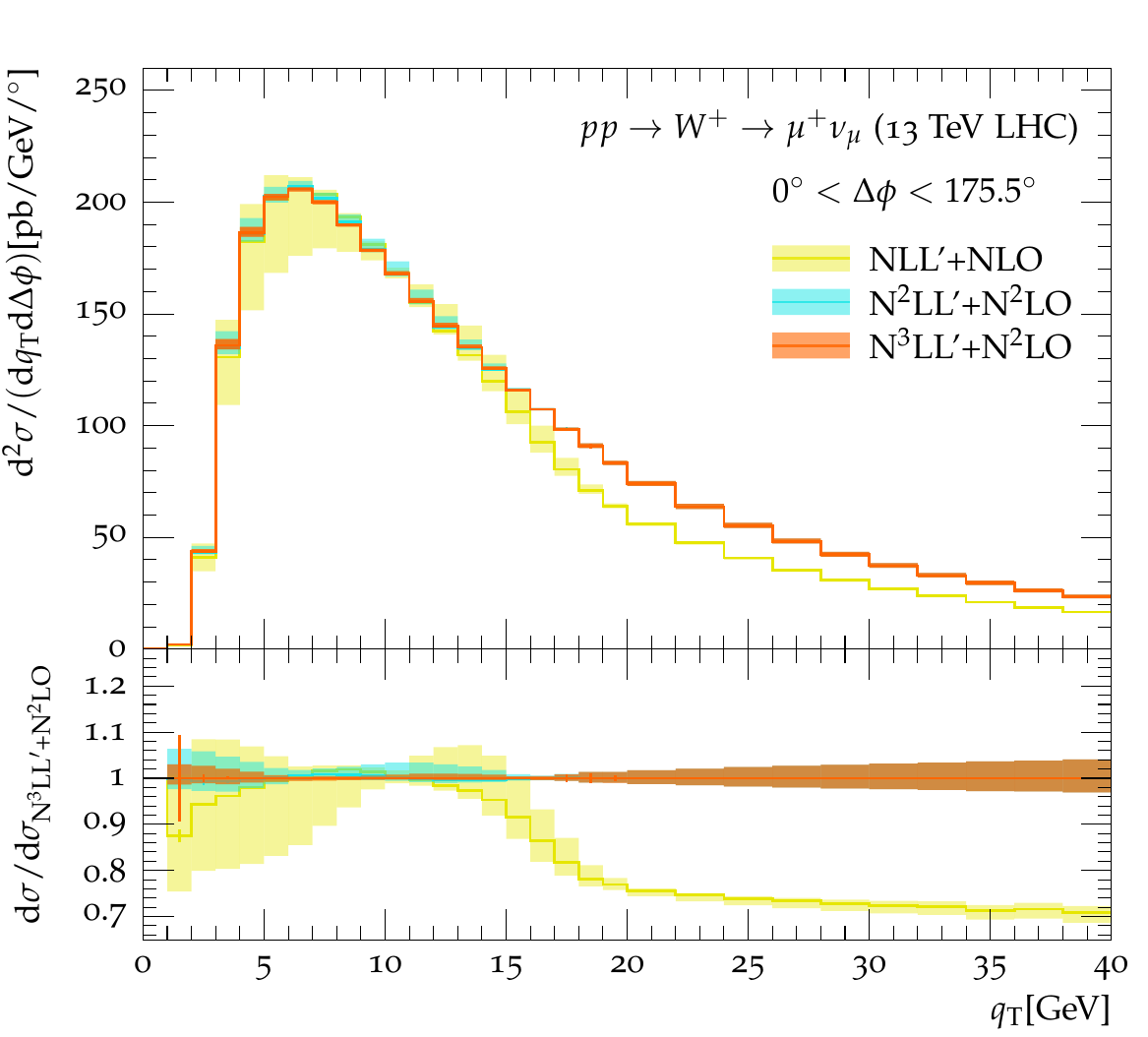}\\[1mm]
  \includegraphics[width=.32\textwidth]{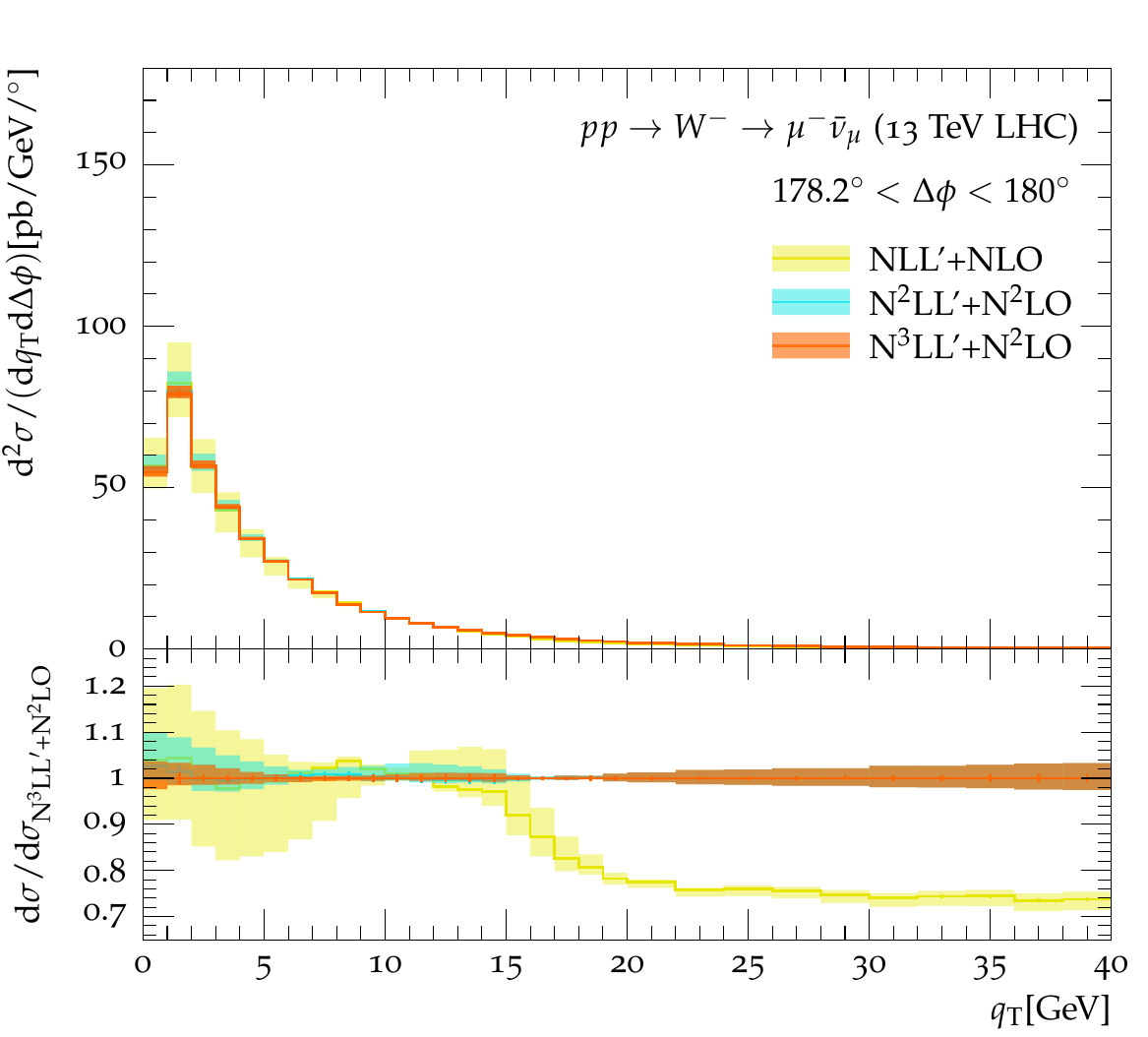}\hfs
  \includegraphics[width=.32\textwidth]{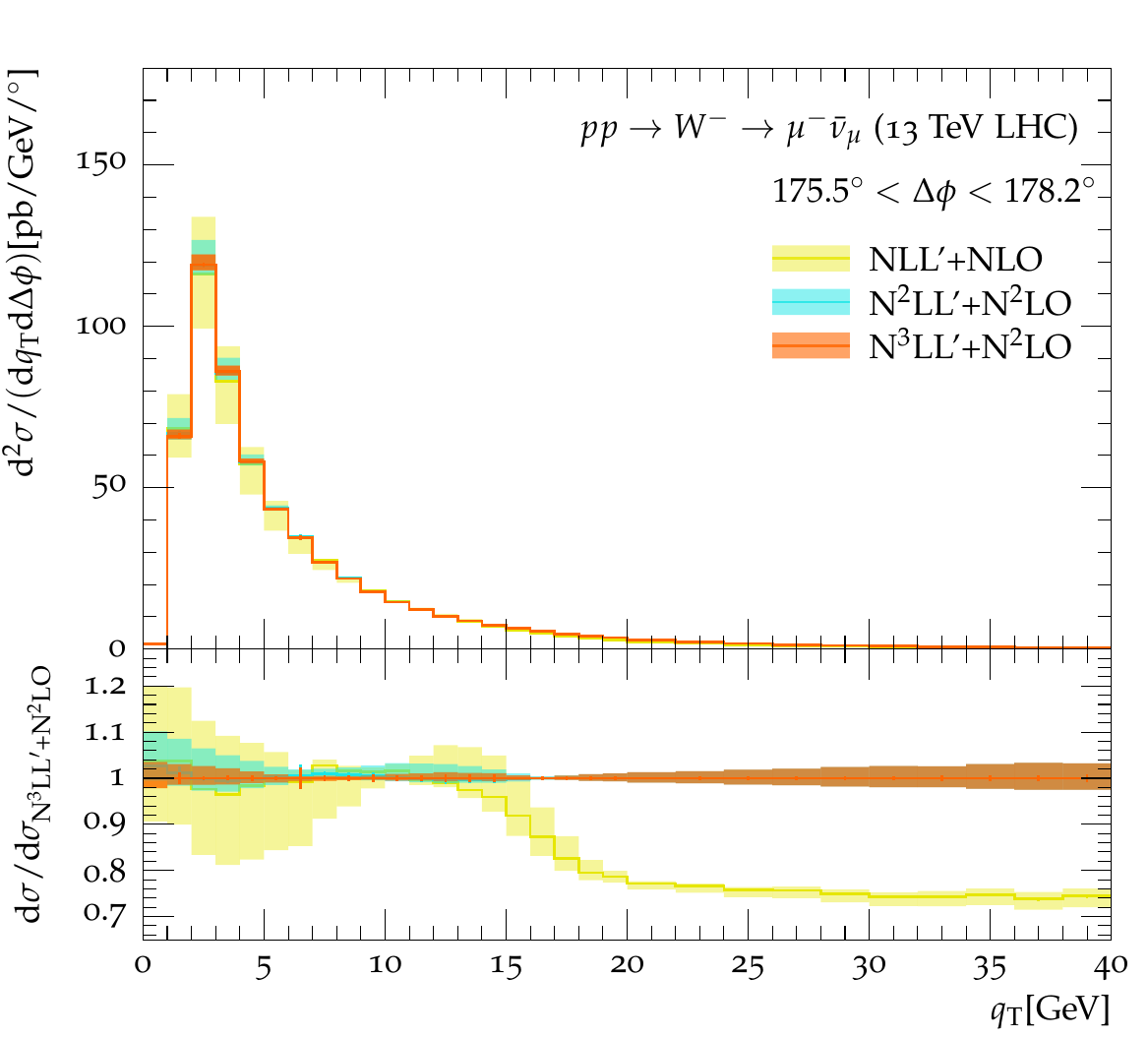}\hfs
  \includegraphics[width=.32\textwidth]{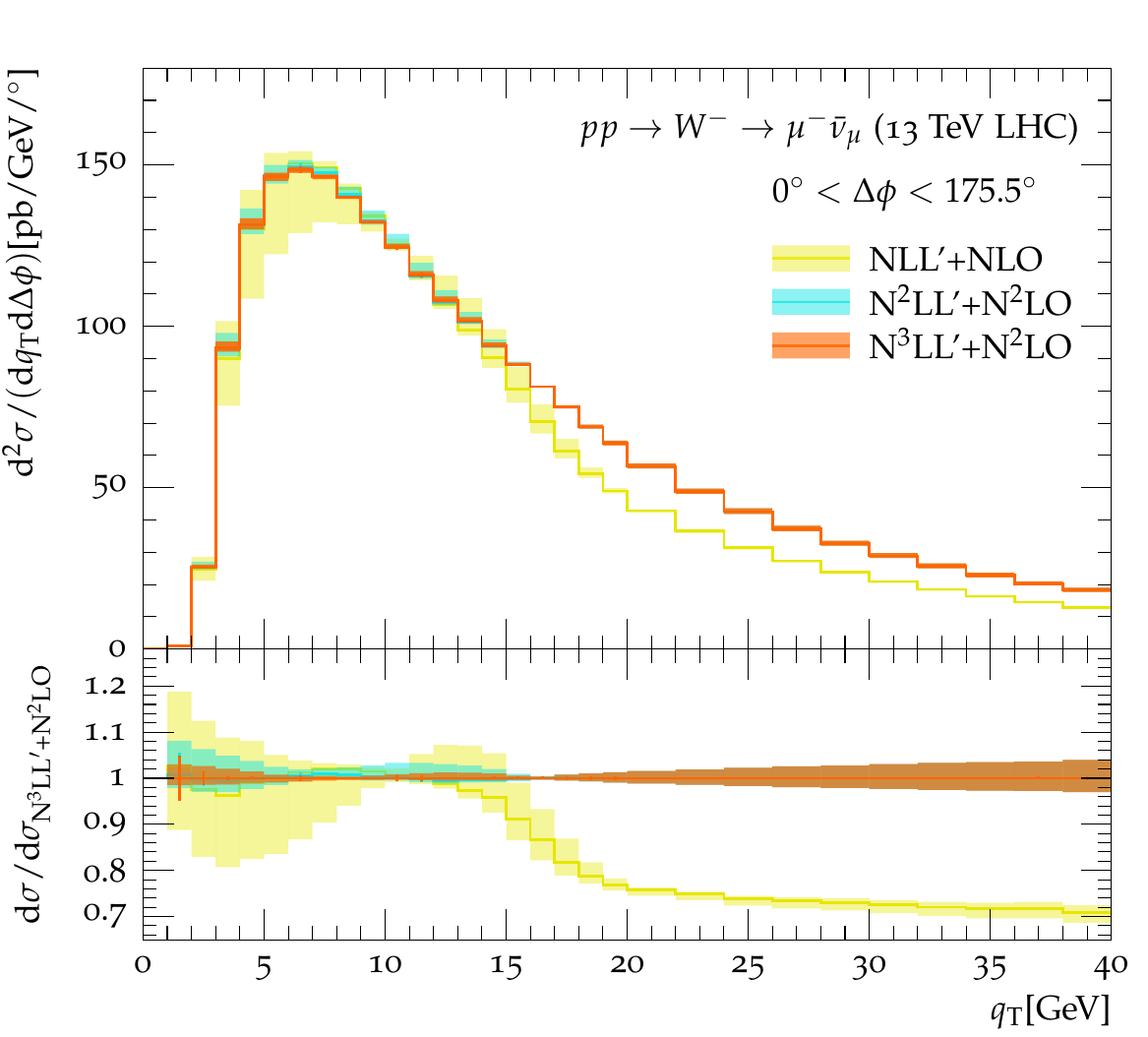}
  \caption{
    Double-differential cross section in \qT\ and three slices of \dphi\ for 
    all three processes. 
    We present the resummation improved results at \NLLNLO, \NNLLNNLO, and 
    \NNNLLNNLO\ accuracy.
  }
  \label{fig:results:res:qt}
\end{figure}

With this global picture examined, we now turn to the aim of 
our study, the double-differential distributions. 
We again begin with the \qT\ spectra in our three chosen 
slices of \dphi\ in Fig.\ \ref{fig:results:res:qt}. 
While only the first slice for $\dphi>175.5^\circ$ contains 
the singular point, all slices are close enough to the singularity 
that a Sudakov peak is formed. 
The description of this resummation region depends 
strongly on the order of logarithmic corrections 
included. 
While the central values do not change significantly 
order-by-order, validating our choice for the central 
scales involved in the resummation, the estimated 
uncertainty steadily decreases when going to higher 
orders. 
To be definite, the uncertainty in our lowest accuracy 
calculation (\NLL) ranges from $-20\%$ to $+15\%$ around 
the central value around the Sudakov peak, independent 
of the \dphi\ slice, and involves a sizeable shape 
uncertainty as well. 
This shape uncertainty in particular differs in the 
region below the Sudakov peaks depending on \dphi. 
Both uncertainties are reduced greatly when higher-order 
logarithmic corrections are included. 
At \NNLL\ they amount to $-4\%$ to $+7\%$ while at \NNNLL\ 
they are reduced to $-2\%$ to $+4\%$.

Moving towards larger \qT, the matching uncertainty 
tends to become comparable to the perturbative uncertainties 
in the resummation, dominating in particular \NLLNLO\ 
calculation around \qTcut. 
As discussed above, the matching scale was chosen based on 
arguments for the highest precision calculation in this 
study and significant contributions of the resummation 
beyond the \NLO\ fixed-order accuracy were found. 
Hence, this finding is not surprising. 
On the contrary, the matching uncertainty is substantially 
reduced in both the \NNLLNNLO\ and \NNNLLNNLO\ calculations, 
not exceeding $\pm 1\%$ in the latter. 
At even higher transverse momenta the fixed-order calculation 
dominates the spectrum and its usual behaviour and 
uncertainties are recovered.

\begin{figure}[t!]
  \centering
  \includegraphics[width=.32\textwidth]{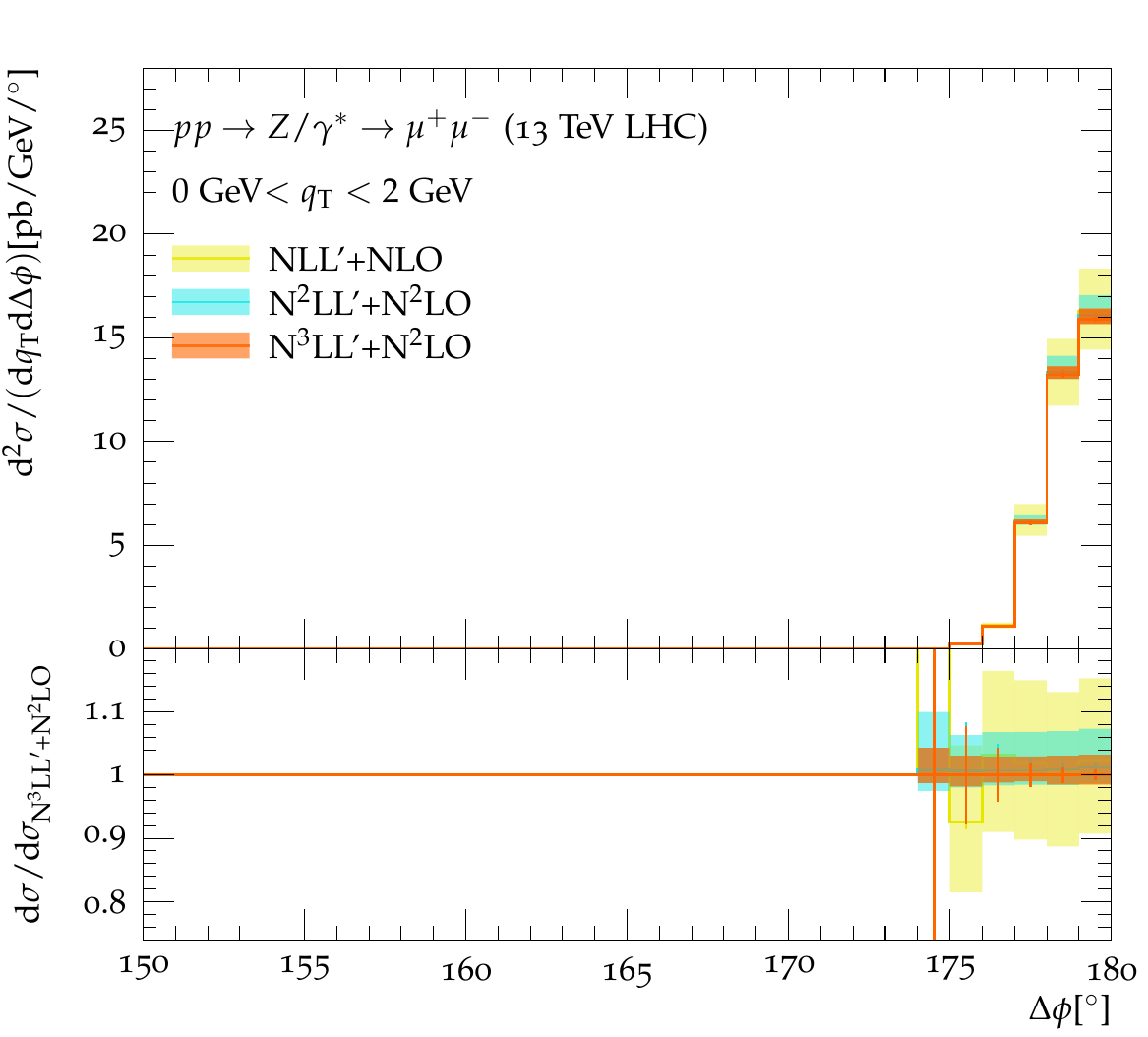}\hfs
  \includegraphics[width=.32\textwidth]{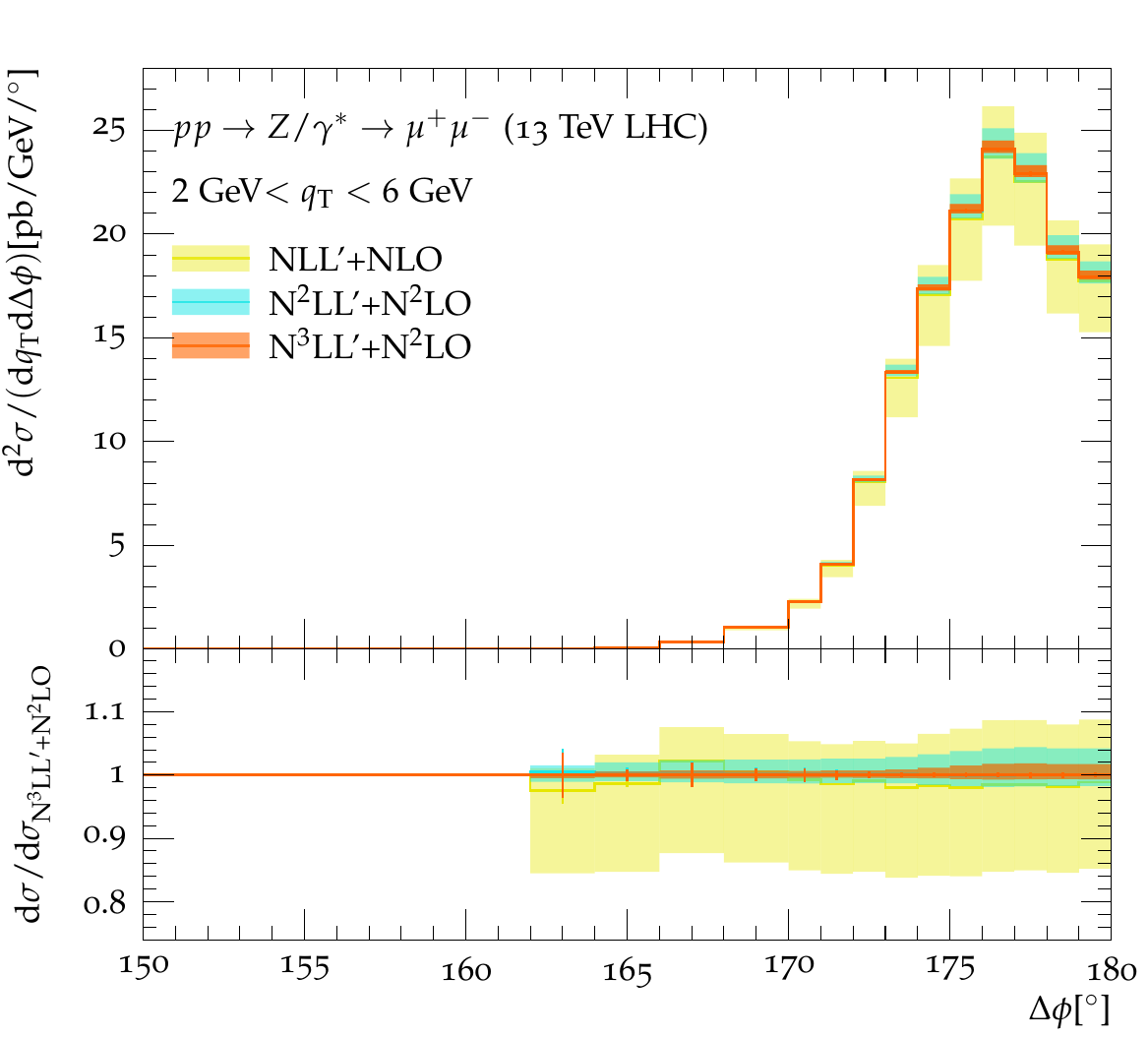}\hfs
  \includegraphics[width=.32\textwidth]{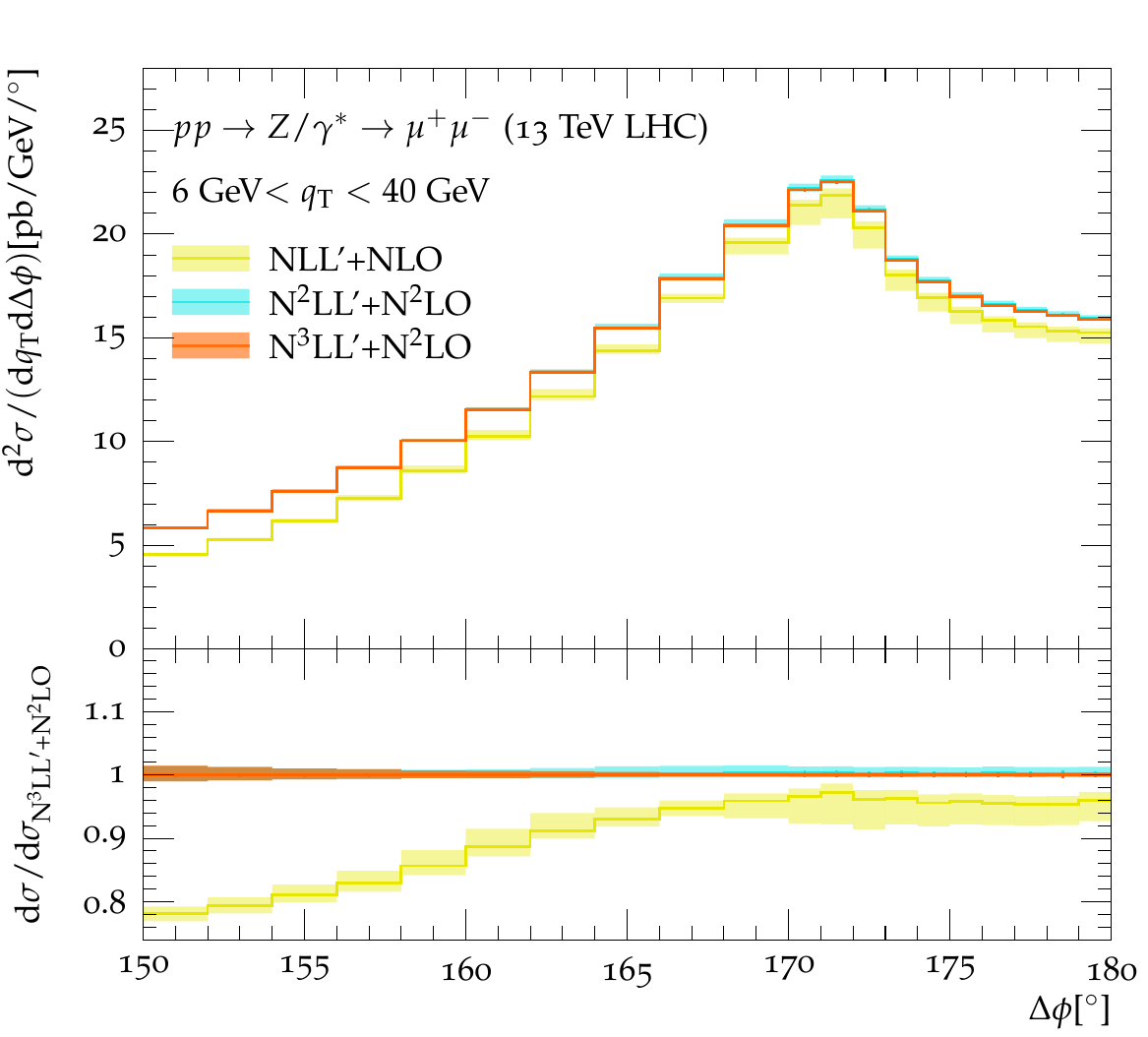}\\[1mm]
  \includegraphics[width=.32\textwidth]{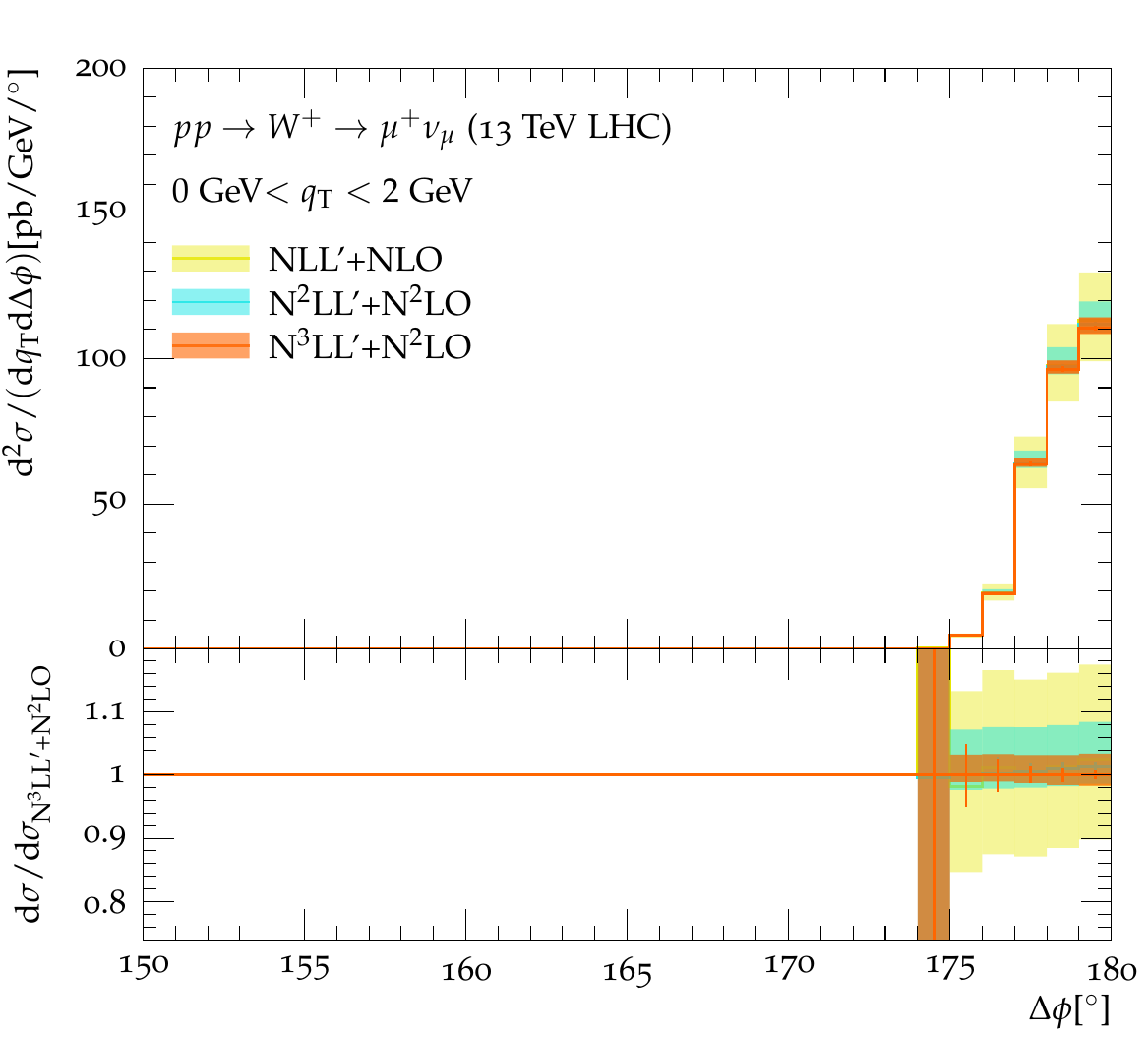}\hfs
  \includegraphics[width=.32\textwidth]{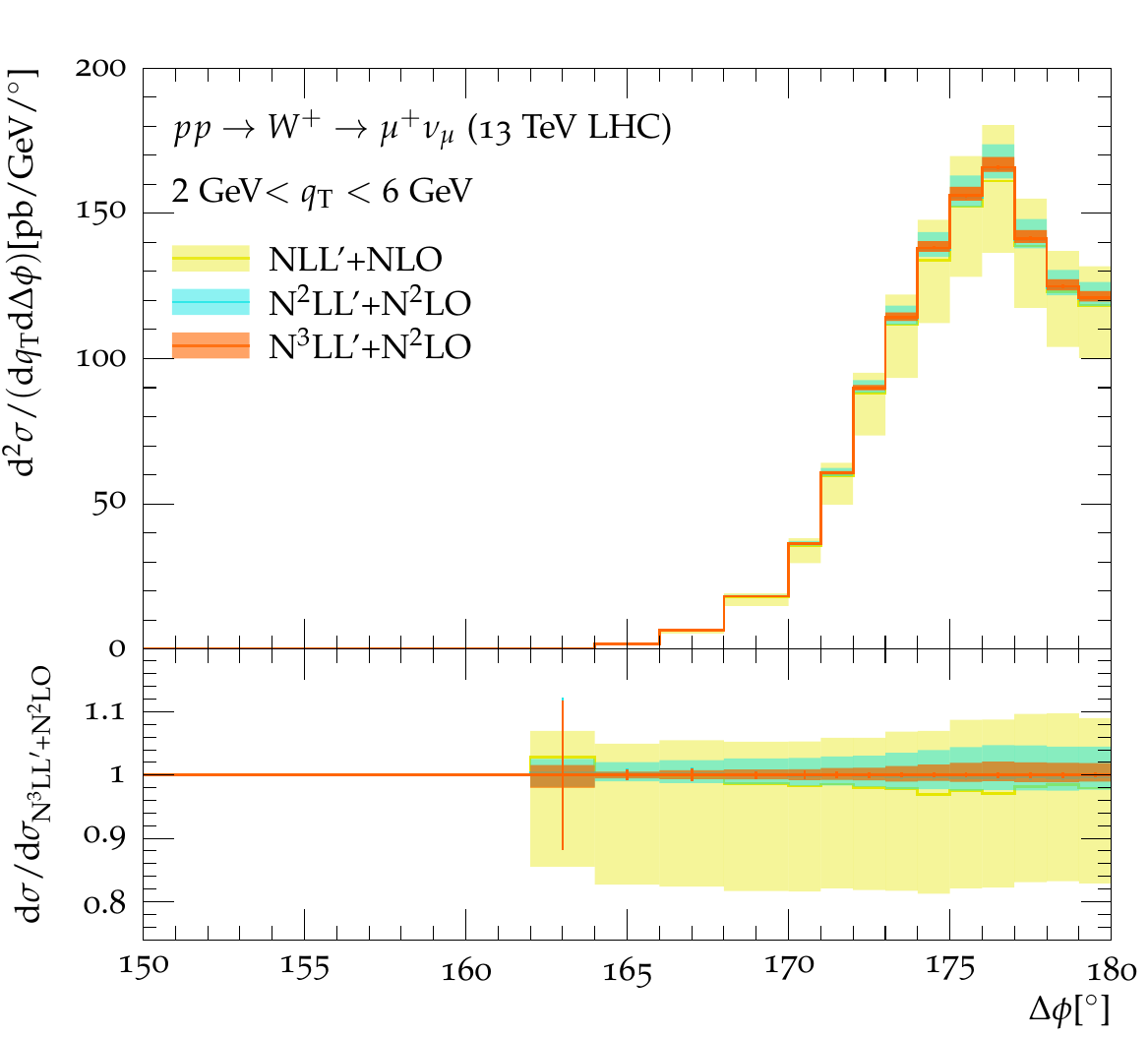}\hfs
  \includegraphics[width=.32\textwidth]{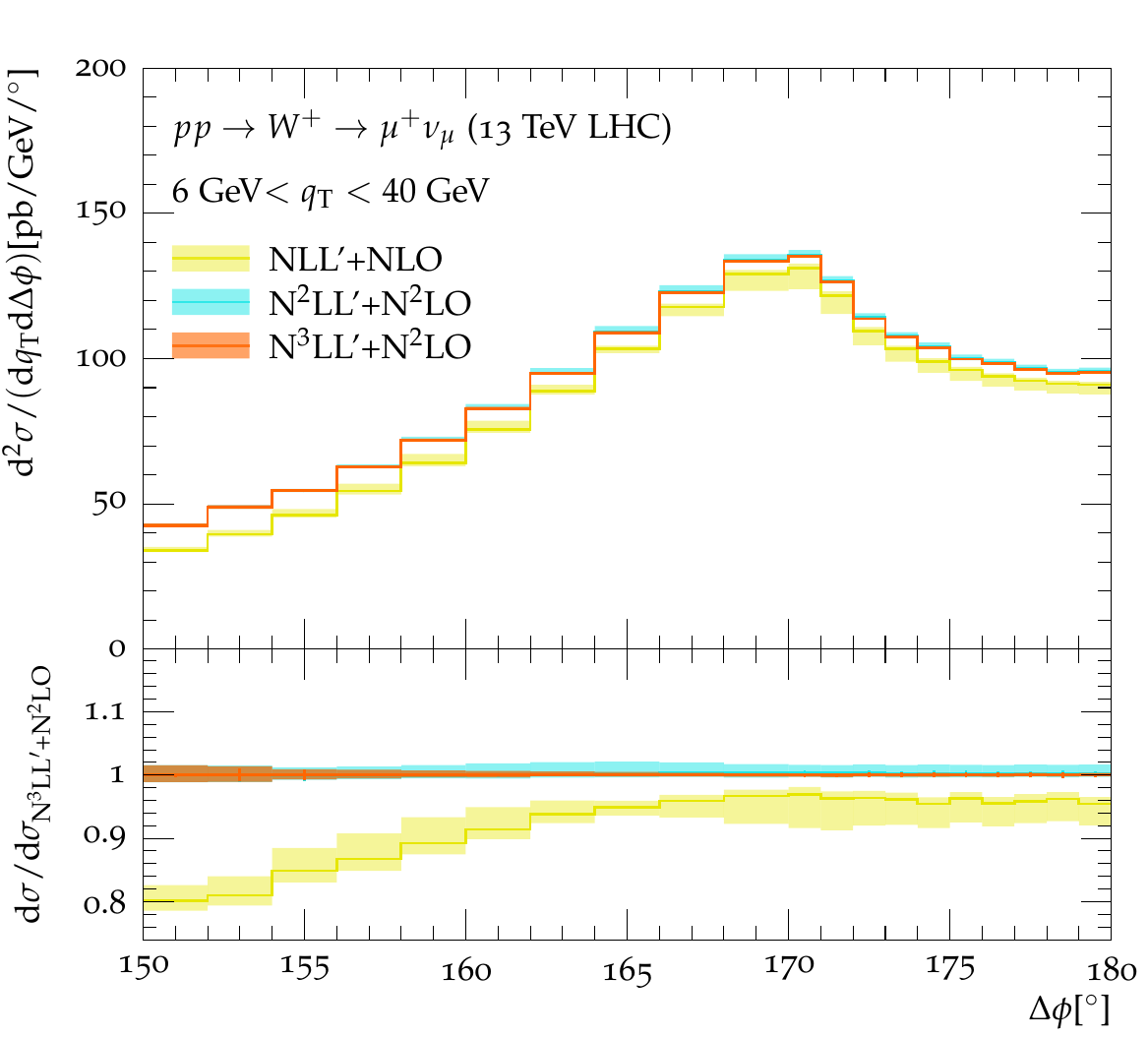}\\[1mm]
  \includegraphics[width=.32\textwidth]{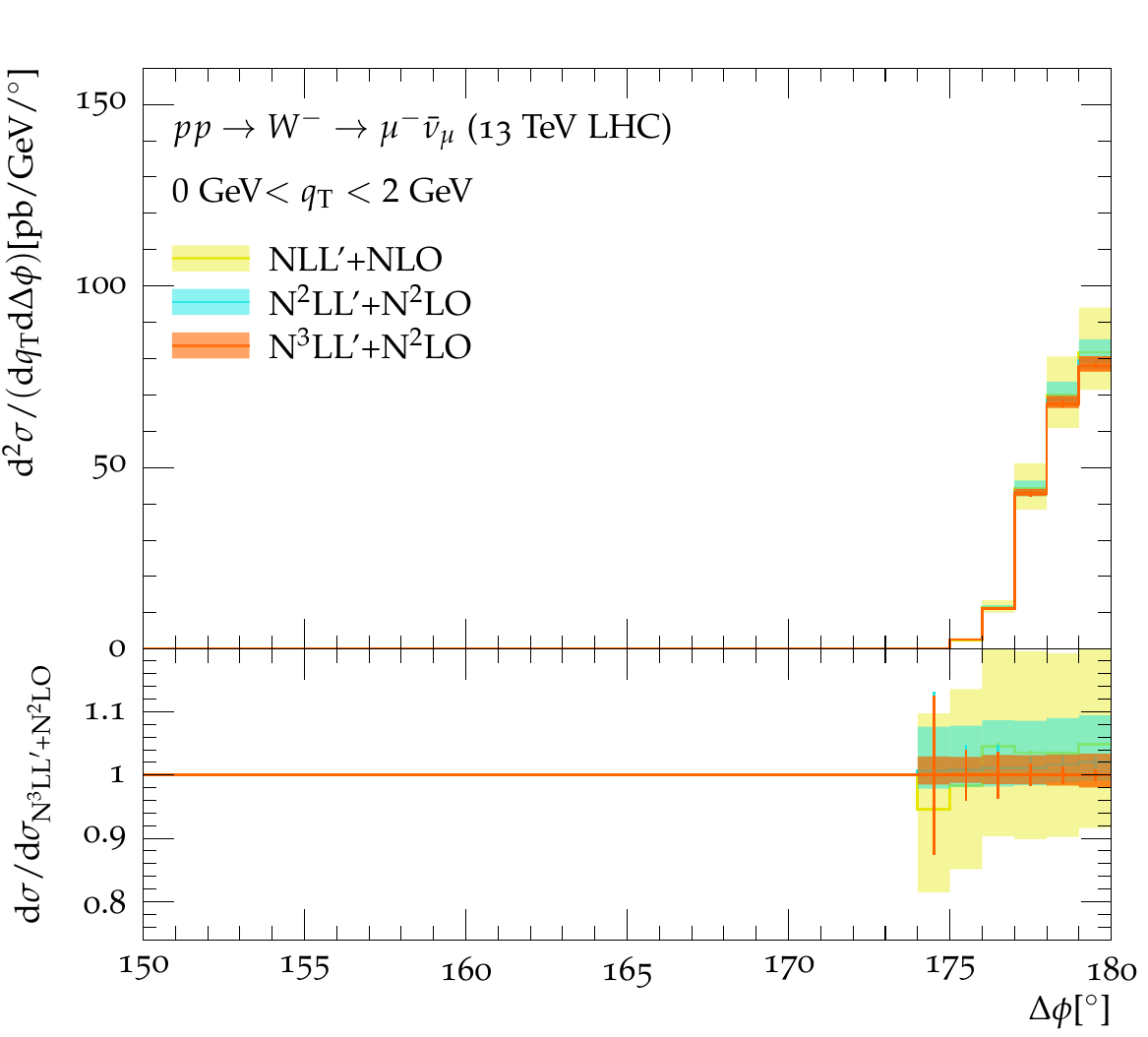}\hfs
  \includegraphics[width=.32\textwidth]{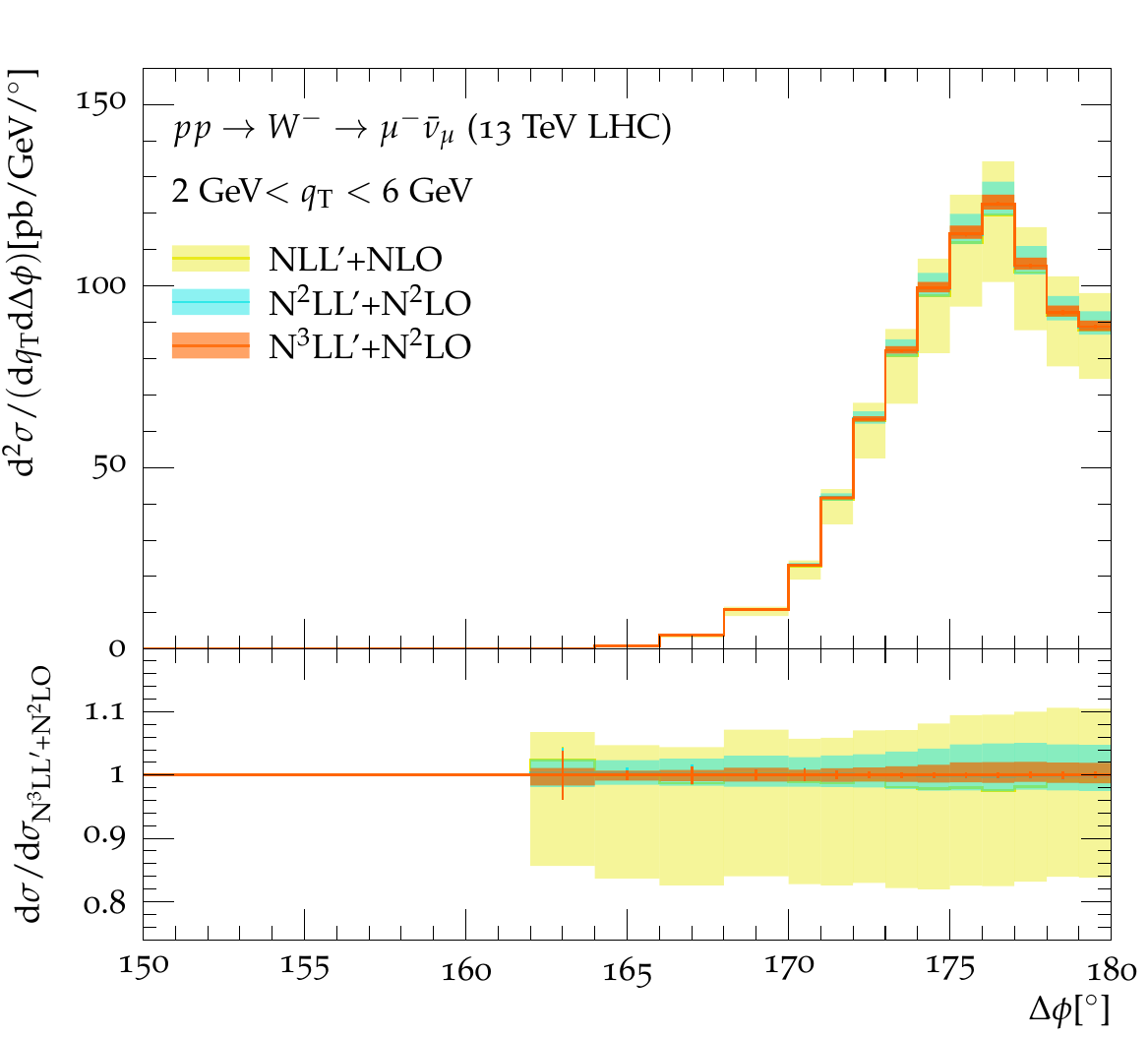}\hfs
  \includegraphics[width=.32\textwidth]{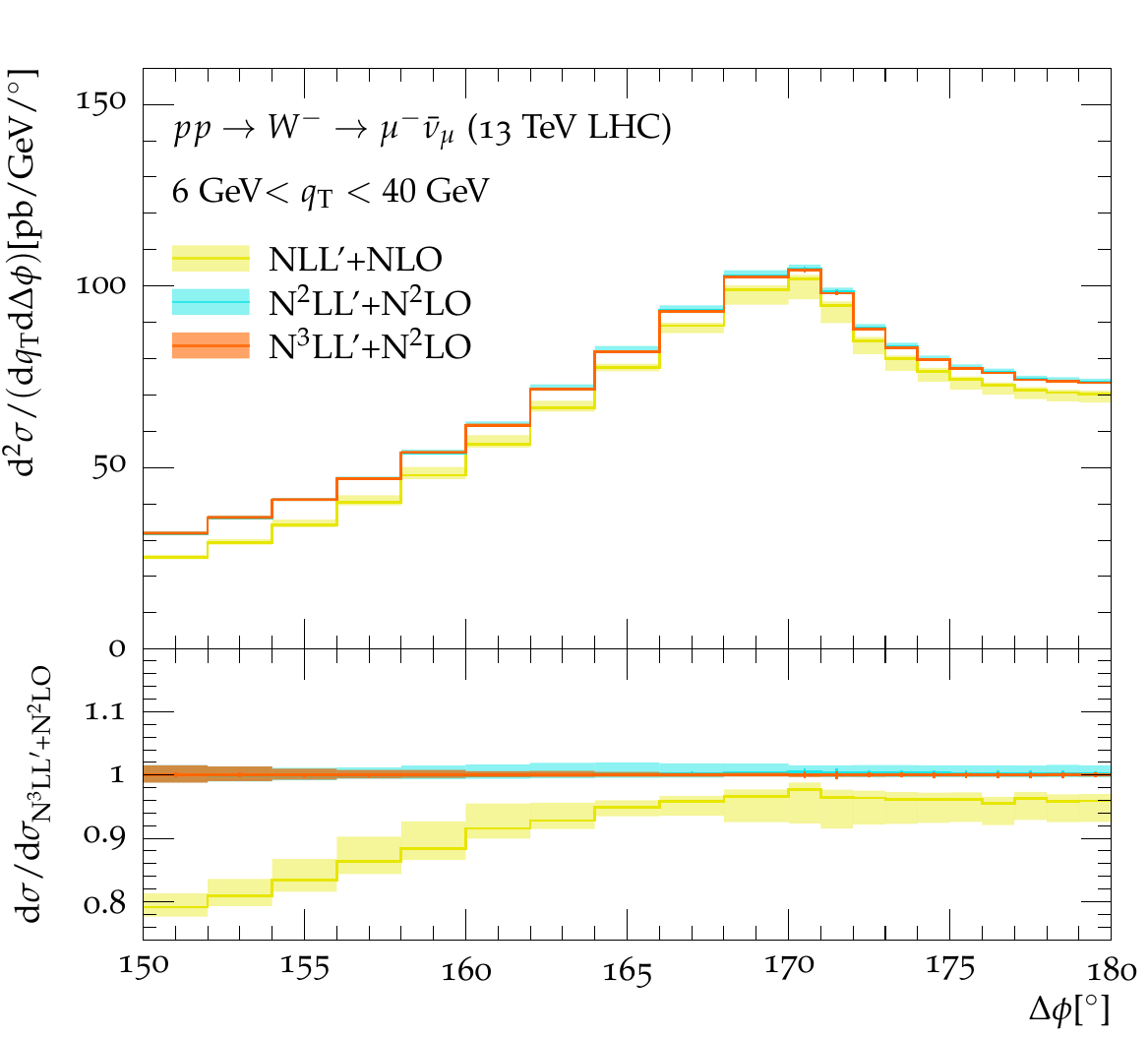}
  \caption{
    Double-differential cross section in \dphi\ and three slices of \qT\ for 
    all three processes. 
    We present the resummation improved results at \NLLNLO, \NNLLNNLO, and 
    \NNNLLNNLO\ accuracy.
  }
  \label{fig:results:res:dphi}
\end{figure}

Finally, we examine the resummation improved results 
for the \dphi\ spectra in the three chosen \qT\ regions. 
Recall that the first region with transverse momenta 
smaller than 2\,GeV resides entirely below the Sudakov 
peak in the transverse momentum spectrum, while the 
second one contains the peak, and the third region with 
$\qT>6\,\text{GeV}$ resides entirely beyond the Sudakov peak. 
Consequently, very different behaviour can be observed 
despite the per-bin-cross sections being of similar orders 
of magnitude. 
The results of our computation for the \dphi\ spectra 
are shown in Fig.\ \ref{fig:results:res:dphi}.
As all three processes exhibit a very similar 
behaviour we continue to discuss them simultaneously. 

In the lowest \qT\ region, which 
probes the region below the Sudakov peak in the \qT\ 
spectrum, near back-to-back topologies are favoured 
unsurprisingly and resummation effects dominate the 
calculation throughout. 
Consequently, as observed before, the central values 
of our three predictions of increasing logarithmic 
accuracies agree very well. 
Their uncertainties have nearly no \dphi\ dependence 
and are steadily decreasing with the increasing accuracy of 
the resummation, reaching $-1\%$ to $+2\%$ in the \NNNLL\ 
case. 

In the intermediate \qT\ slice, focussing on the region 
around the Sudakov peak in the \qT\ spectrum, a peaked 
structure is developing. 
Still, the distribution is dominated by resummation effects, 
leading again very well agreeing central values, their 
uncertainties being dictated by 
the order of the logarithms included in the exponent. 
We, thus, find a variation of $\pm 1\%$ in our most 
accurate calculation. 
This time, however, there is a slight shape to these 
scale uncertainties, predominantly in the back-to-back 
region.

The third \qT\ slice, located entirely above the \qT\ 
Sudakov peak, now sees a stronger impact of the full 
QCD fixed-order calculations. 
Hence we find the central values no longer agree, 
partially reflecting the difference in cross section 
between the \NLO\ and \NNLO\ predictions for the spectrum. 
The increasing importance of the full QCD fixed-order 
part for smaller \dphi\ likewise explains the shape 
corrections at higher order.
Still, the uncertainties are much reduced in our highest 
precision calculation at \NNNLLNNLO\ accuracy, being 
smaller than $\pm 1\%$ throughout.

Finally, please note that the smallest \dphi\ bin in the lower two 
\qT\ slices carries almost no cross section and, thus, 
suffers from larger statistical uncertainties.

\subsection{The \texorpdfstring{$W^{\pm}/Z$}{W/Z} and \texorpdfstring{$W^+/W^-$}{W/W} correlations}
\label{sec:results:ratios}

\begin{figure}[p]
  \centering
  \includegraphics[width=.31\textwidth]{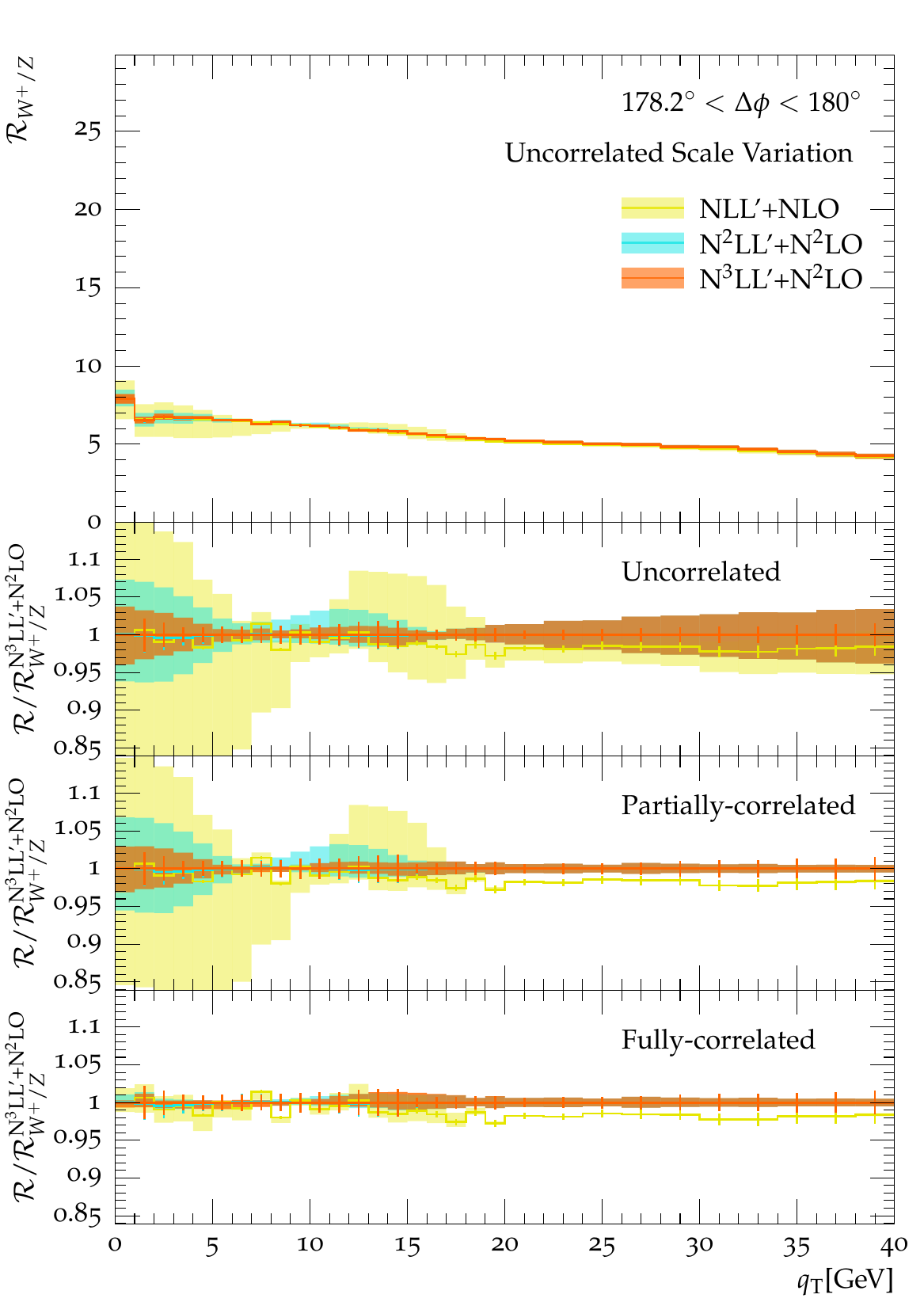}\hfs\hfs
  \includegraphics[width=.31\textwidth]{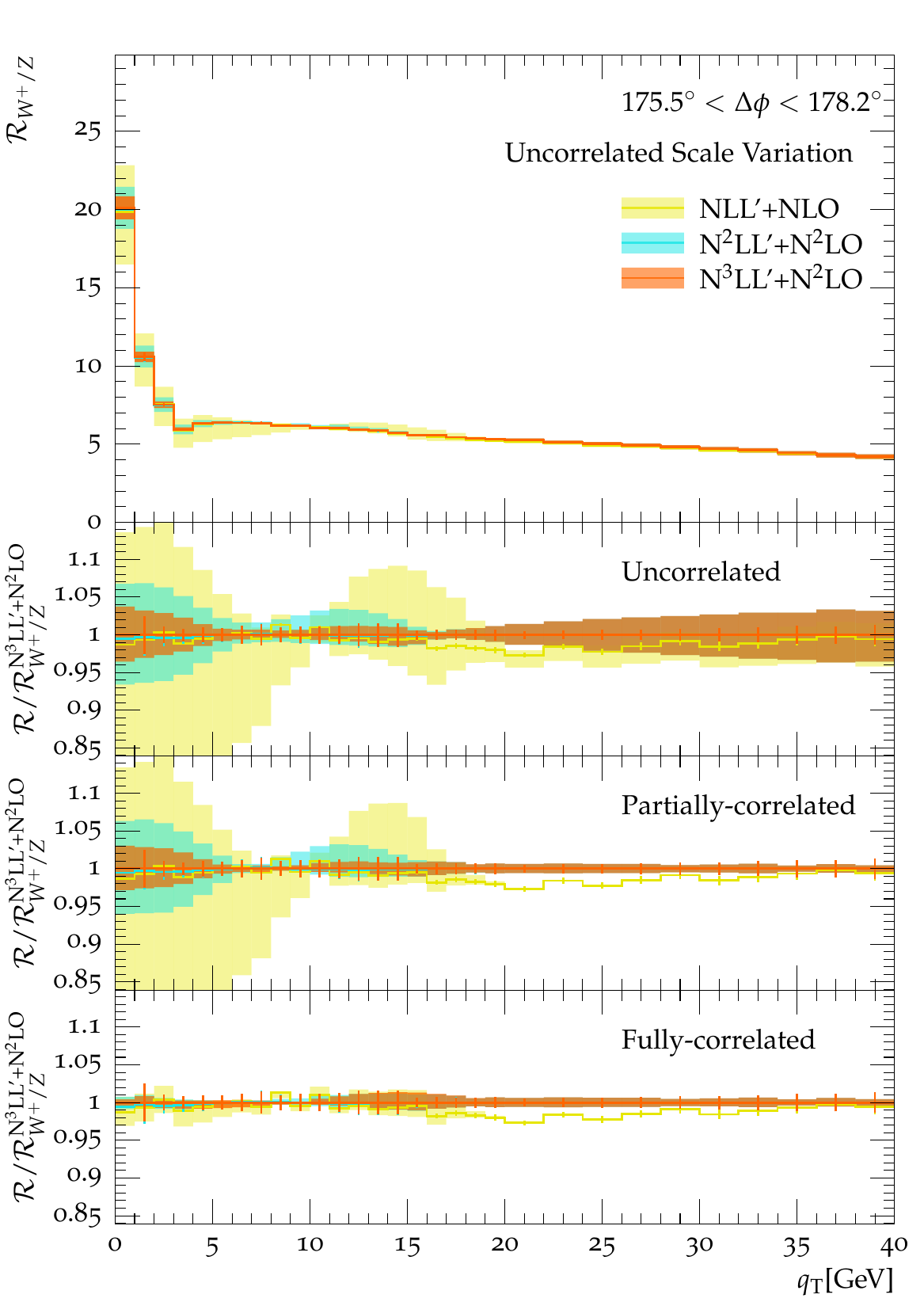}\hfs\hfs
  \includegraphics[width=.31\textwidth]{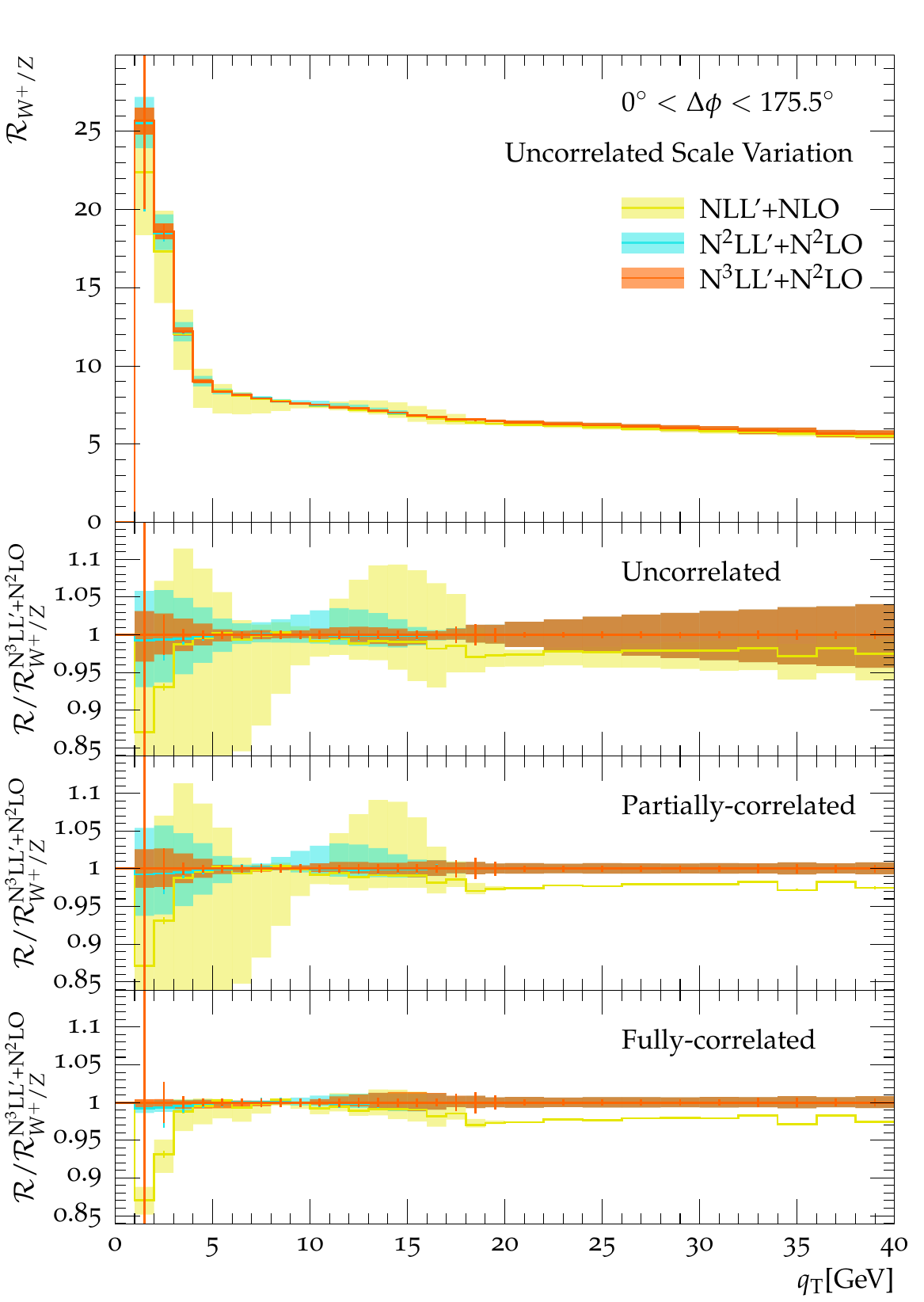}\\[1mm]
  \includegraphics[width=.31\textwidth]{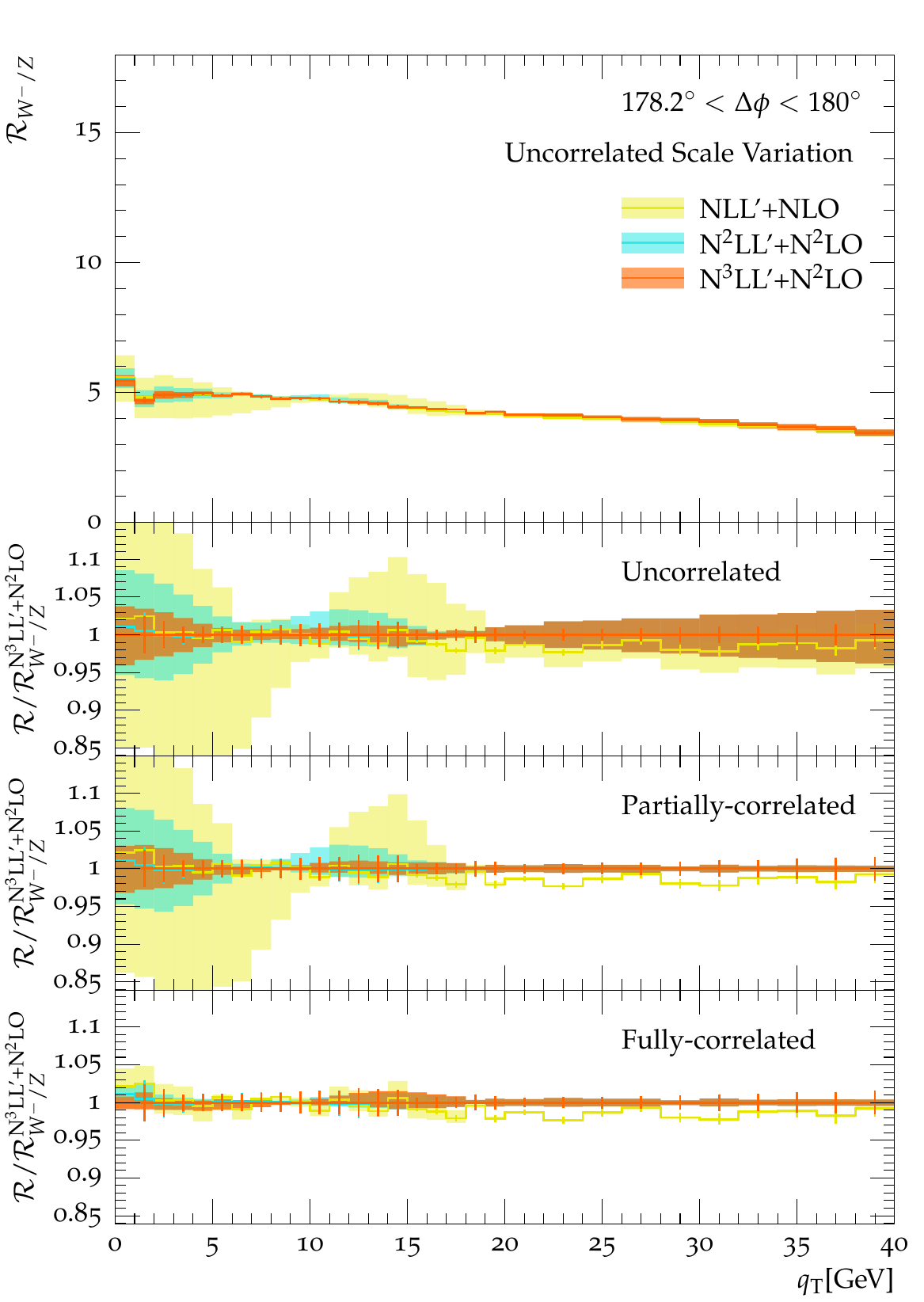}\hfs\hfs
  \includegraphics[width=.31\textwidth]{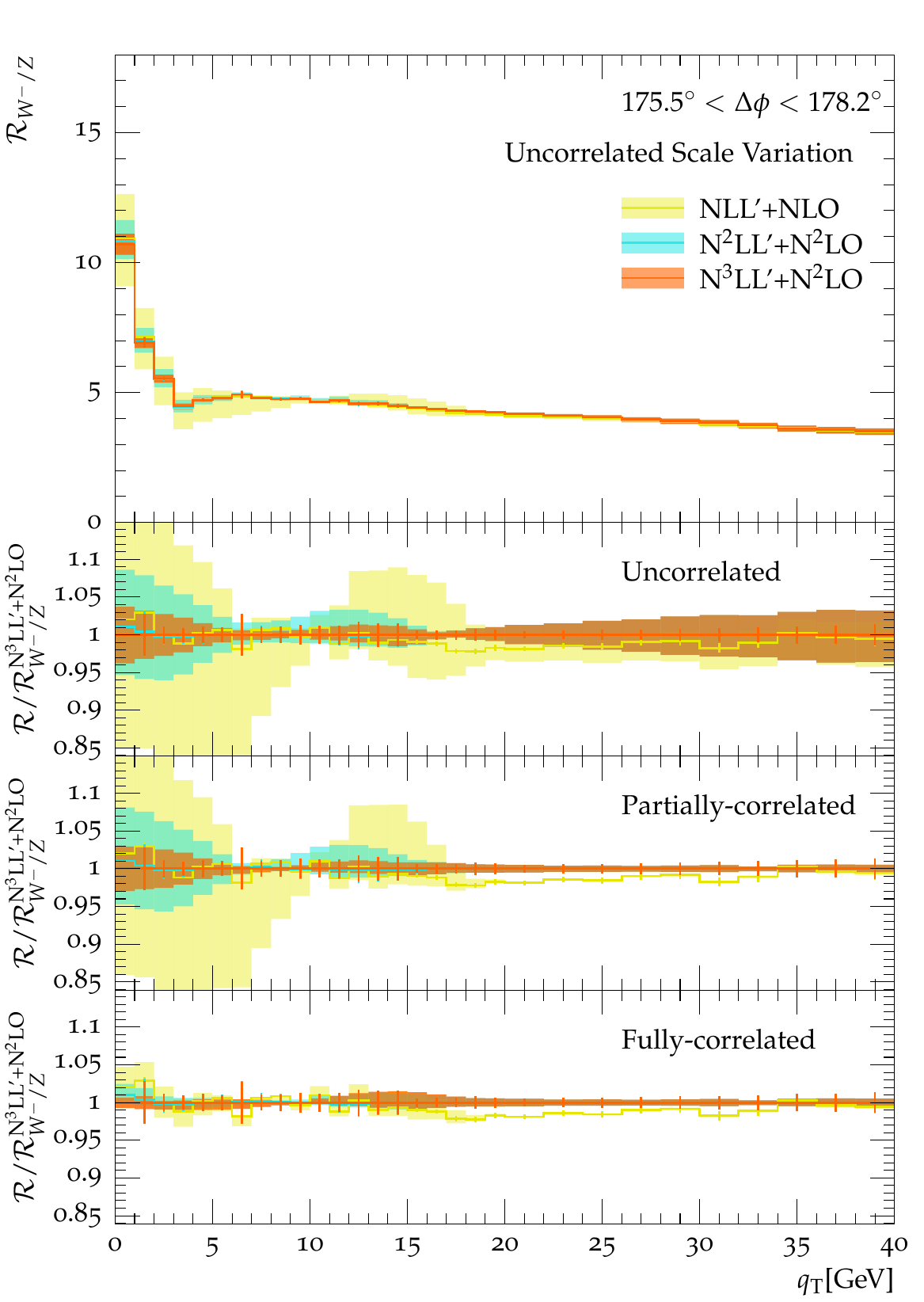}\hfs\hfs
  \includegraphics[width=.31\textwidth]{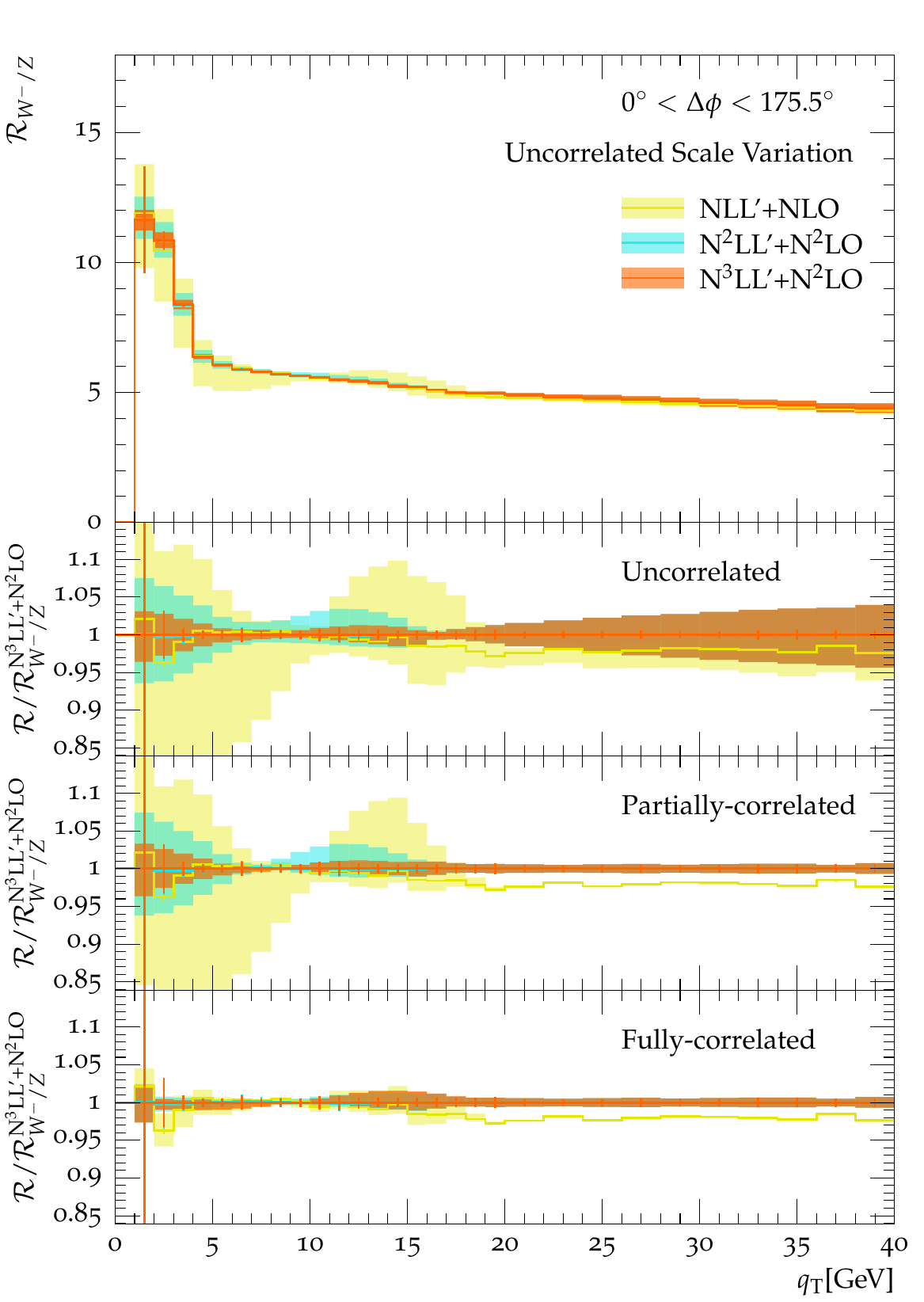}\\[1mm]
  \includegraphics[width=.31\textwidth]{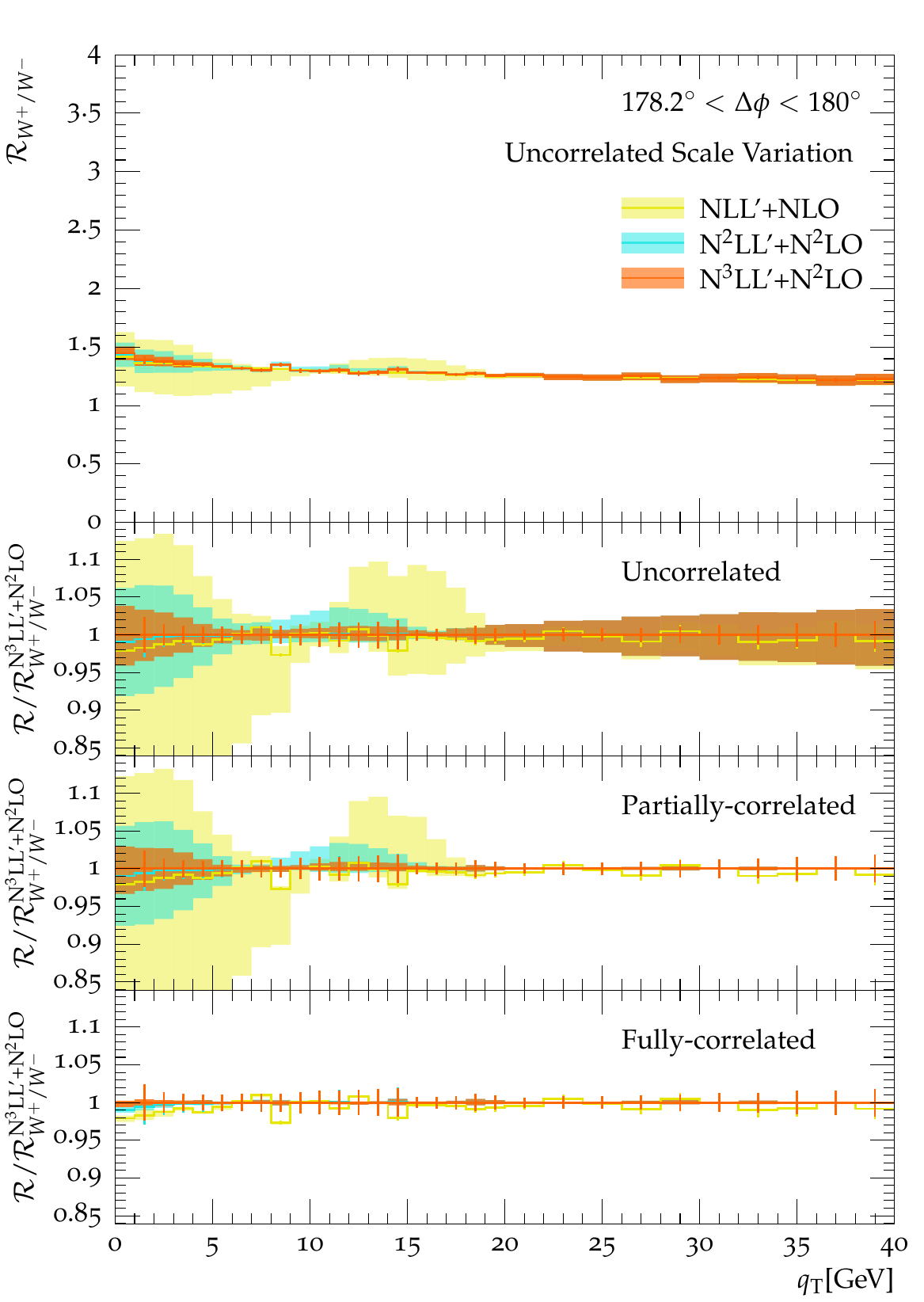}\hfs\hfs
  \includegraphics[width=.31\textwidth]{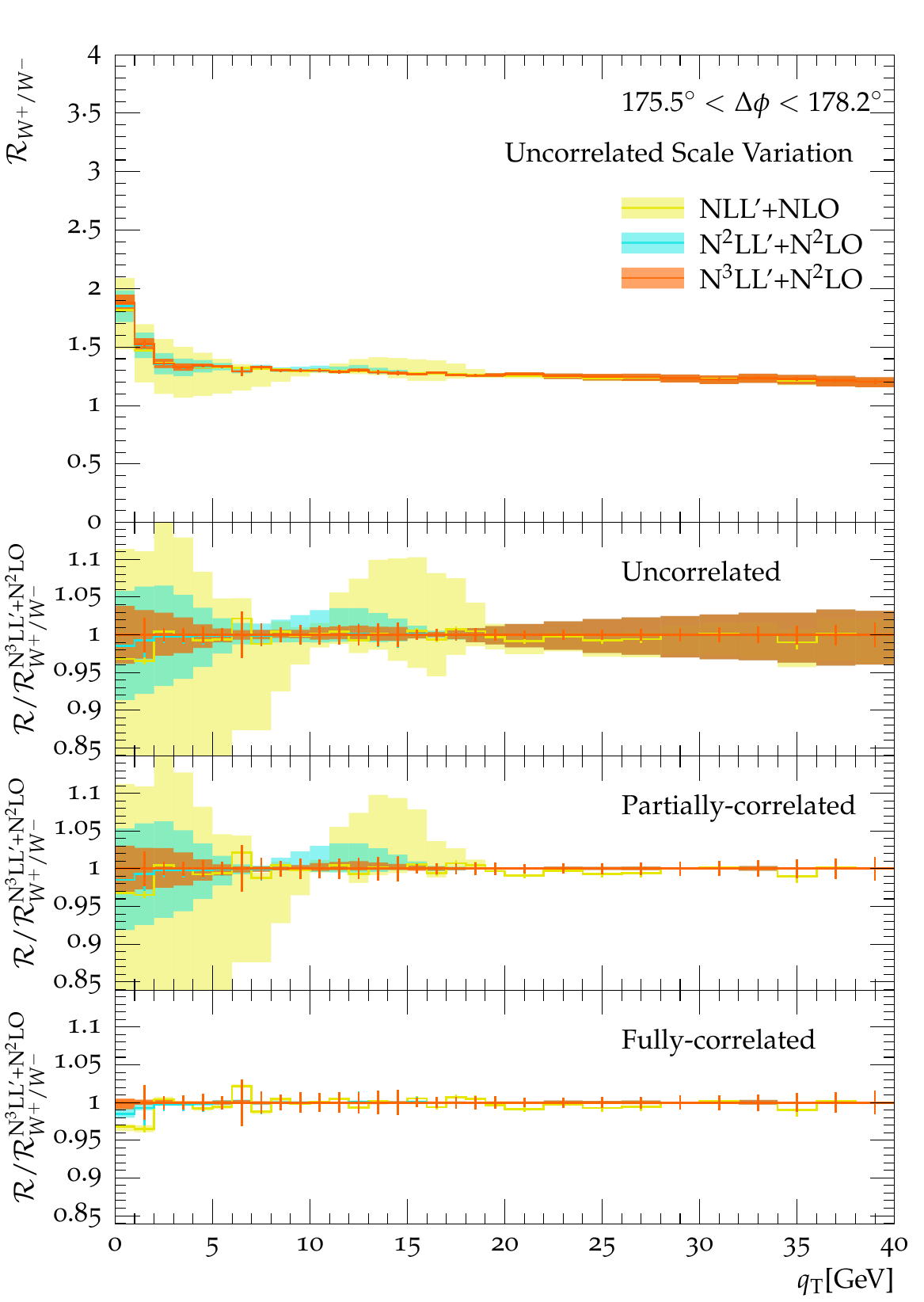}\hfs\hfs
  \includegraphics[width=.31\textwidth]{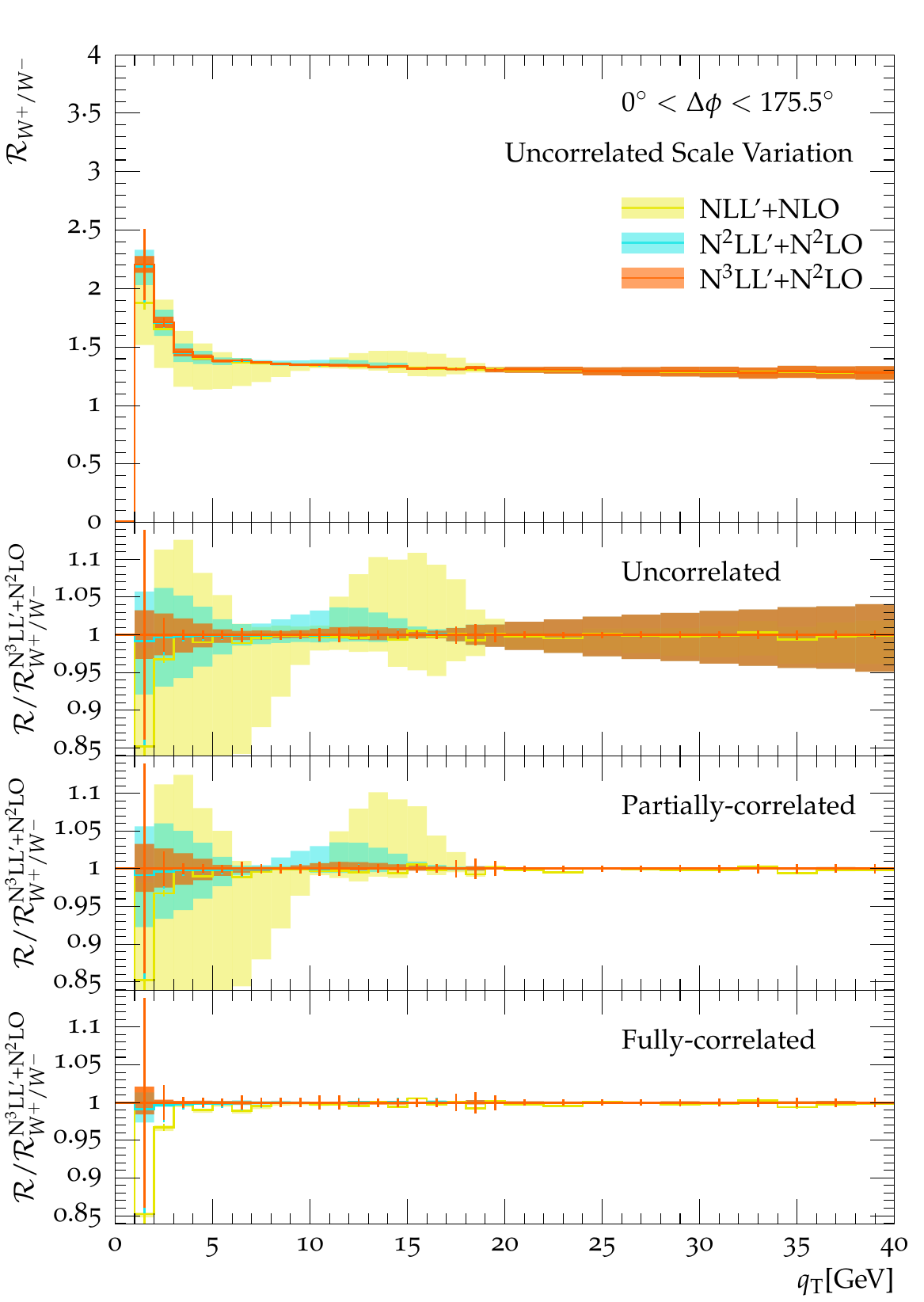}
  \caption{
    Double-differential cross section ratios in \qT\ and three slices of \dphi\ for 
    all three processes. 
    We present the resummation improved results at \NLLNLO, \NNLLNNLO, and 
    \NNNLLNNLO\ accuracy.
  }
  \label{fig:results:ratios:qt}
\end{figure}

\begin{figure}[p]
  \centering
  \includegraphics[width=.31\textwidth]{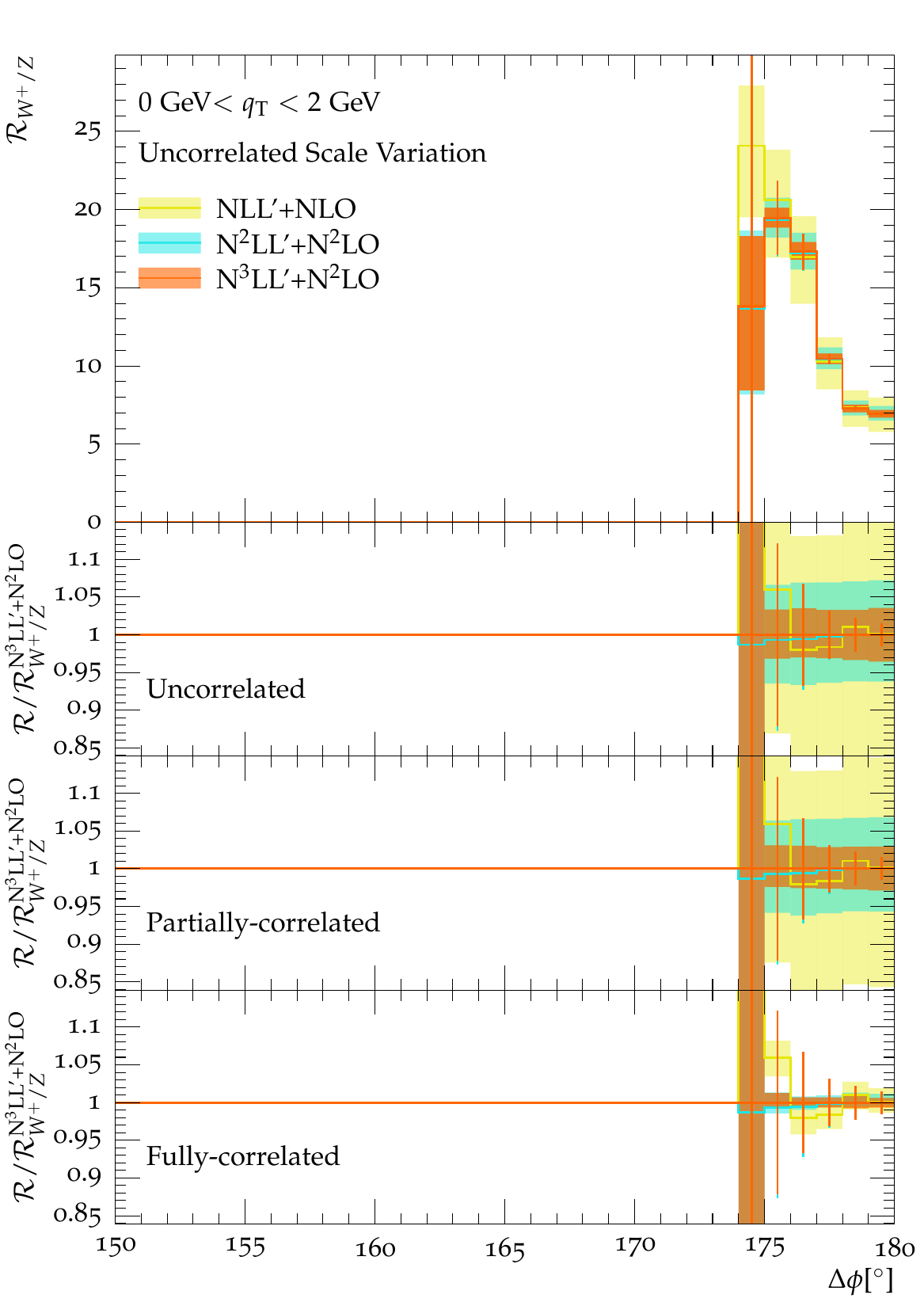}\hfs\hfs
  \includegraphics[width=.31\textwidth]{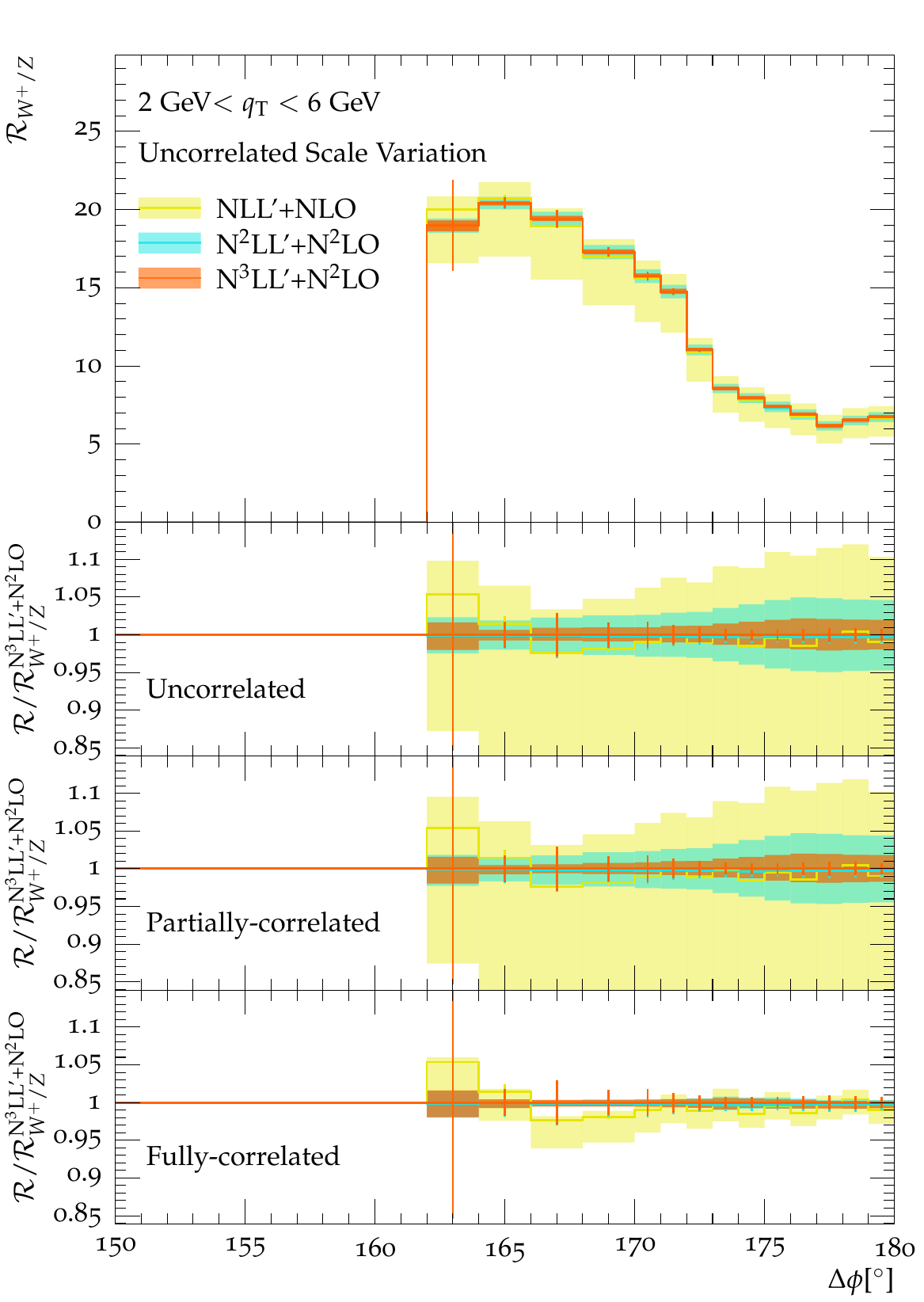}\hfs\hfs
  \includegraphics[width=.31\textwidth]{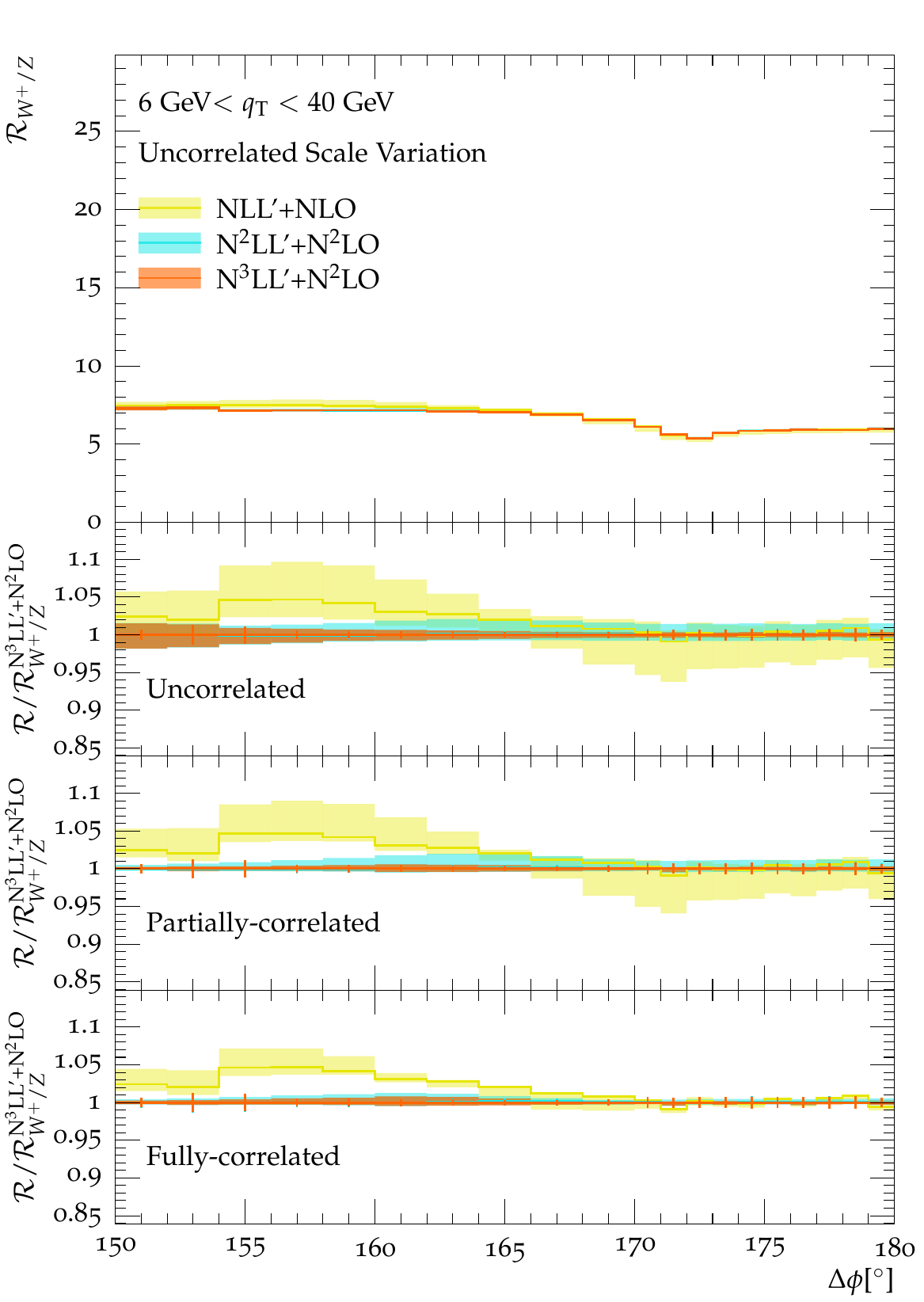}\\[1mm]
  \includegraphics[width=.31\textwidth]{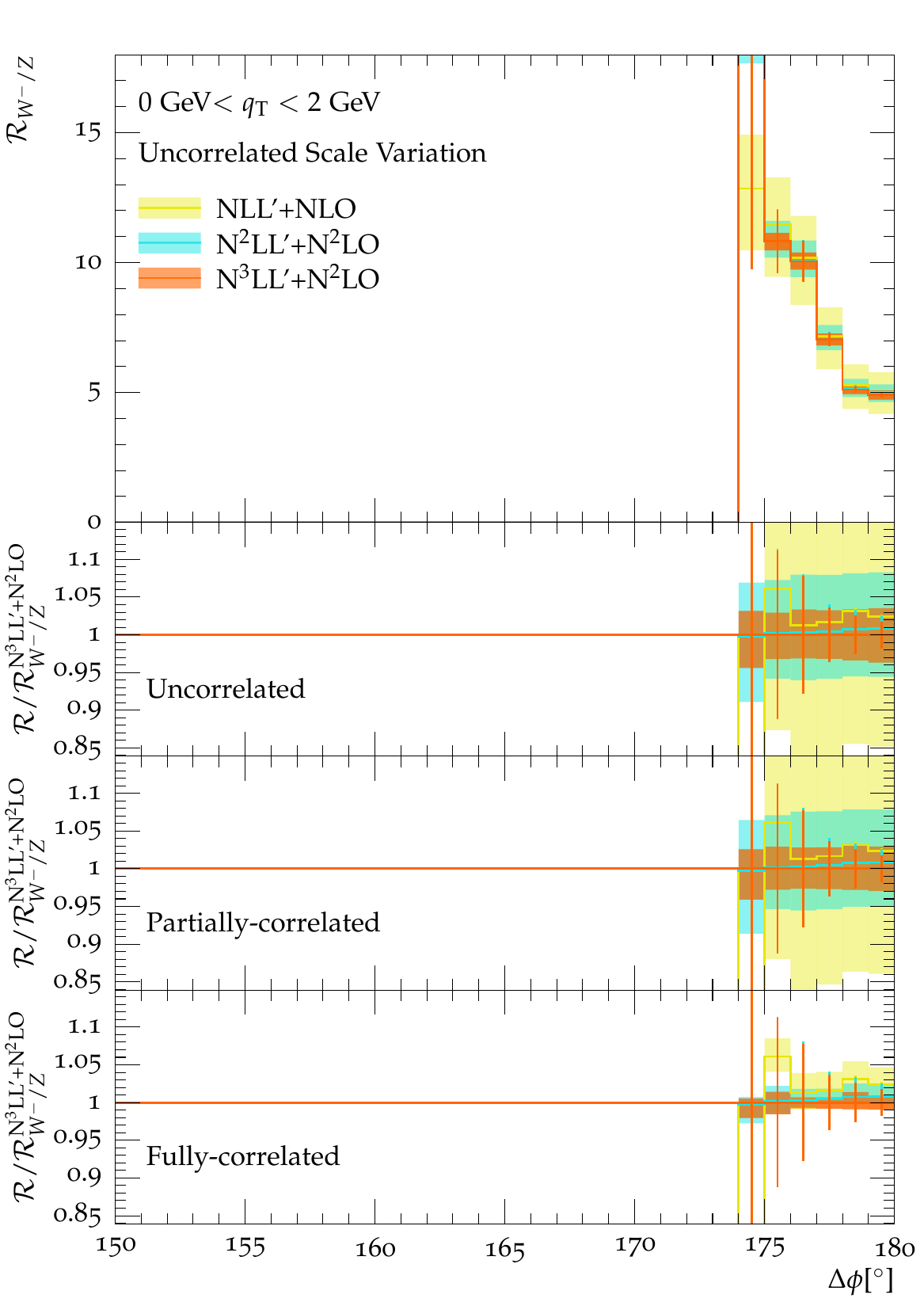}\hfs\hfs
  \includegraphics[width=.31\textwidth]{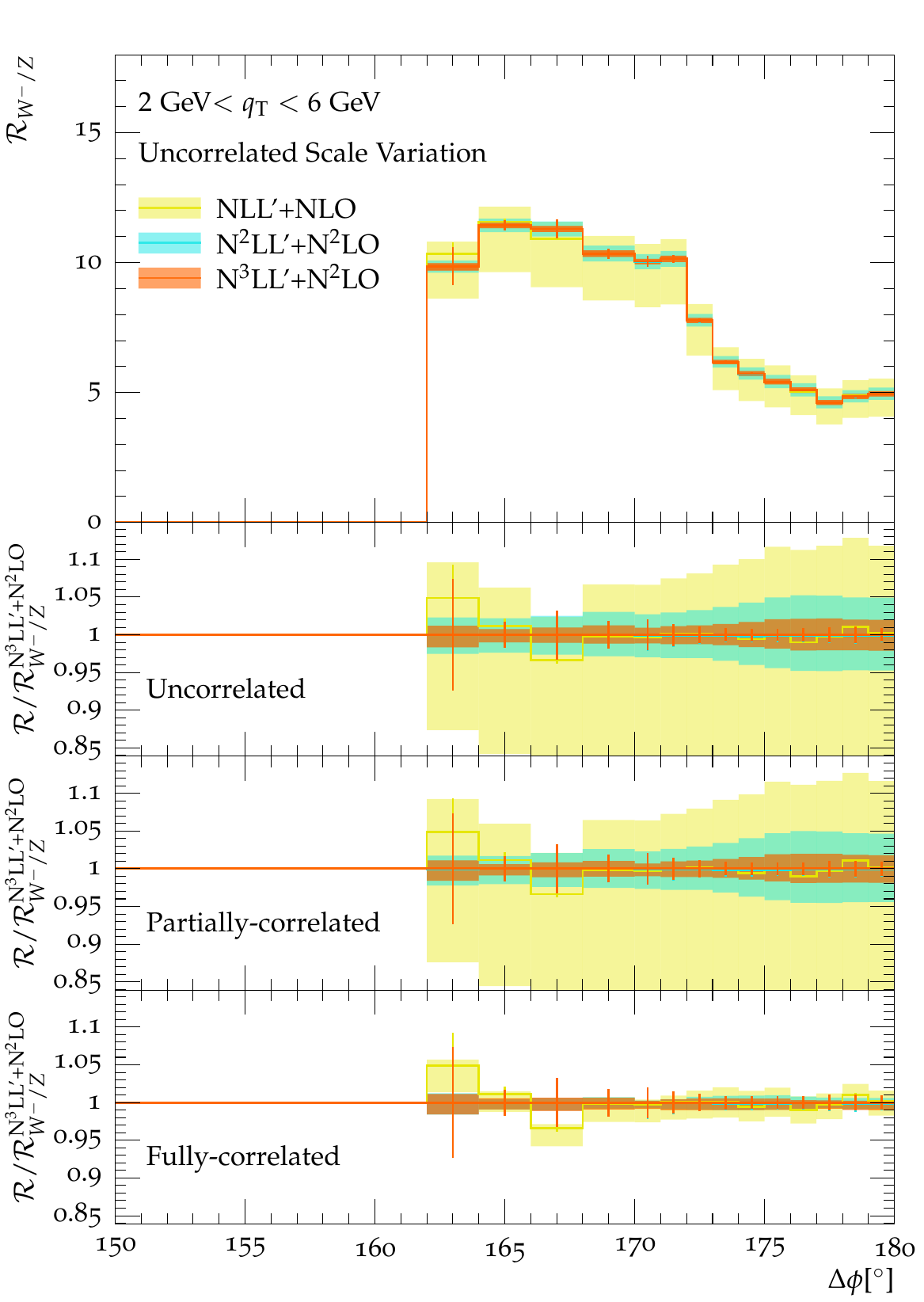}\hfs\hfs
  \includegraphics[width=.31\textwidth]{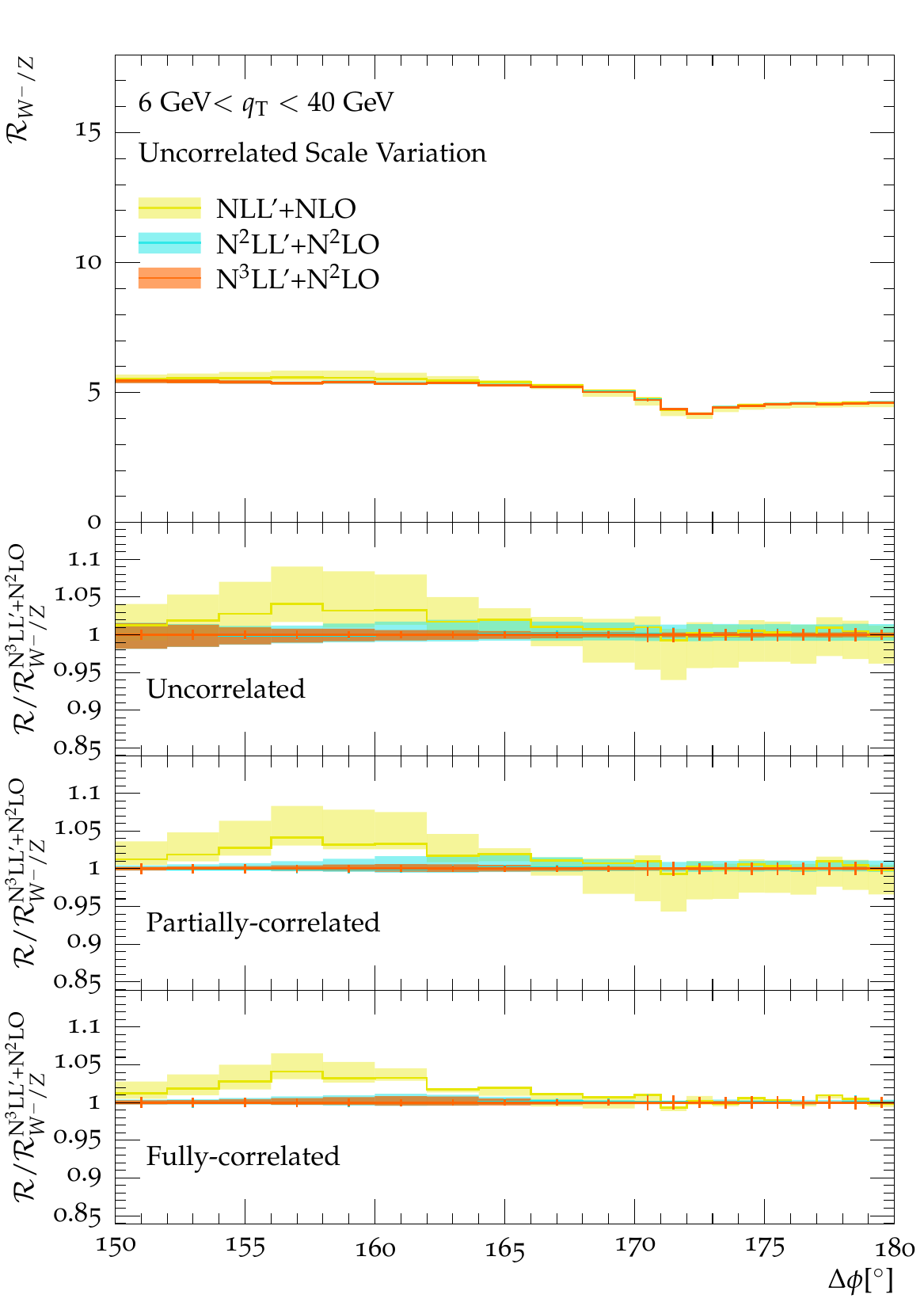}\\[1mm]
  \includegraphics[width=.31\textwidth]{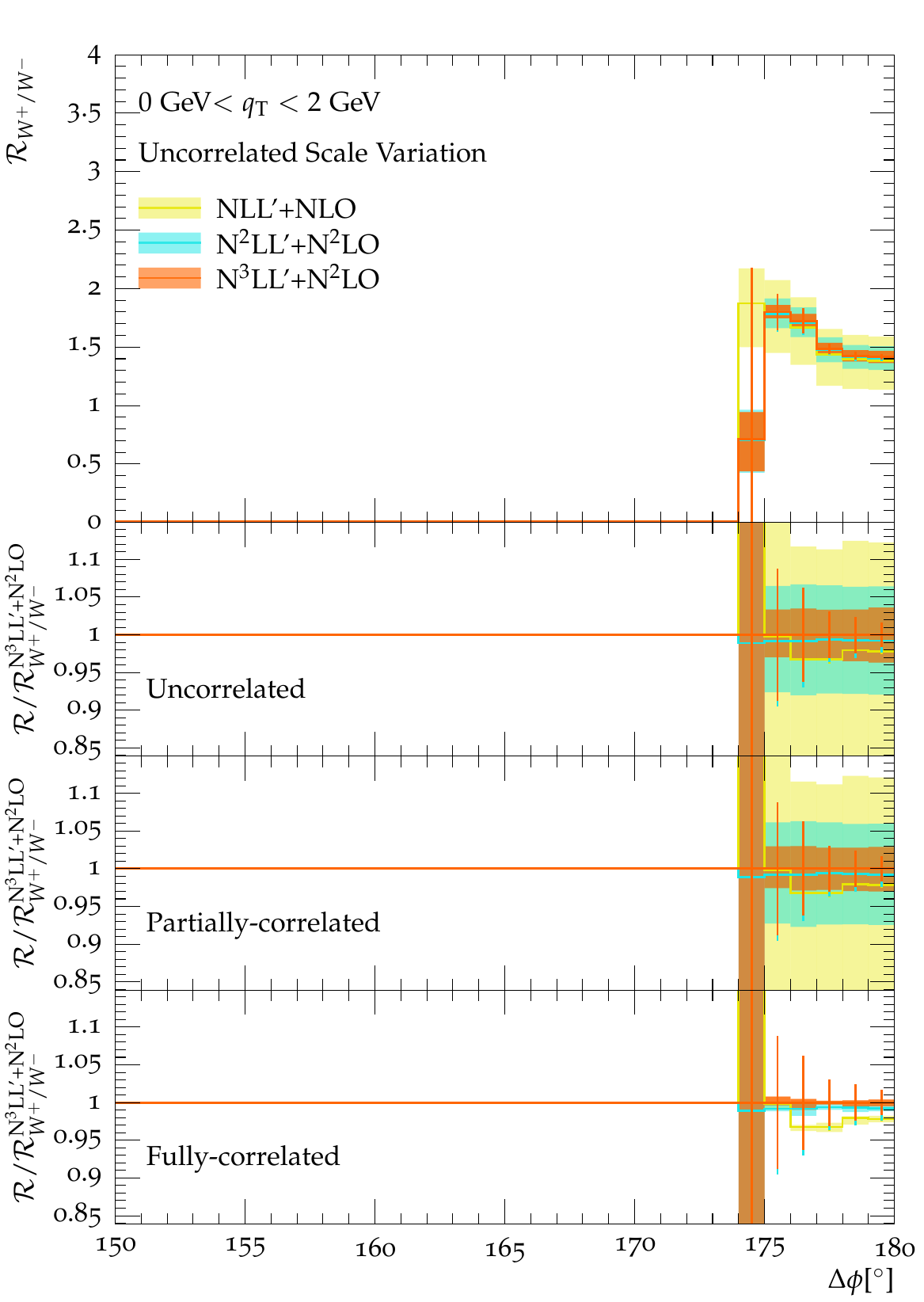}\hfs\hfs
  \includegraphics[width=.31\textwidth]{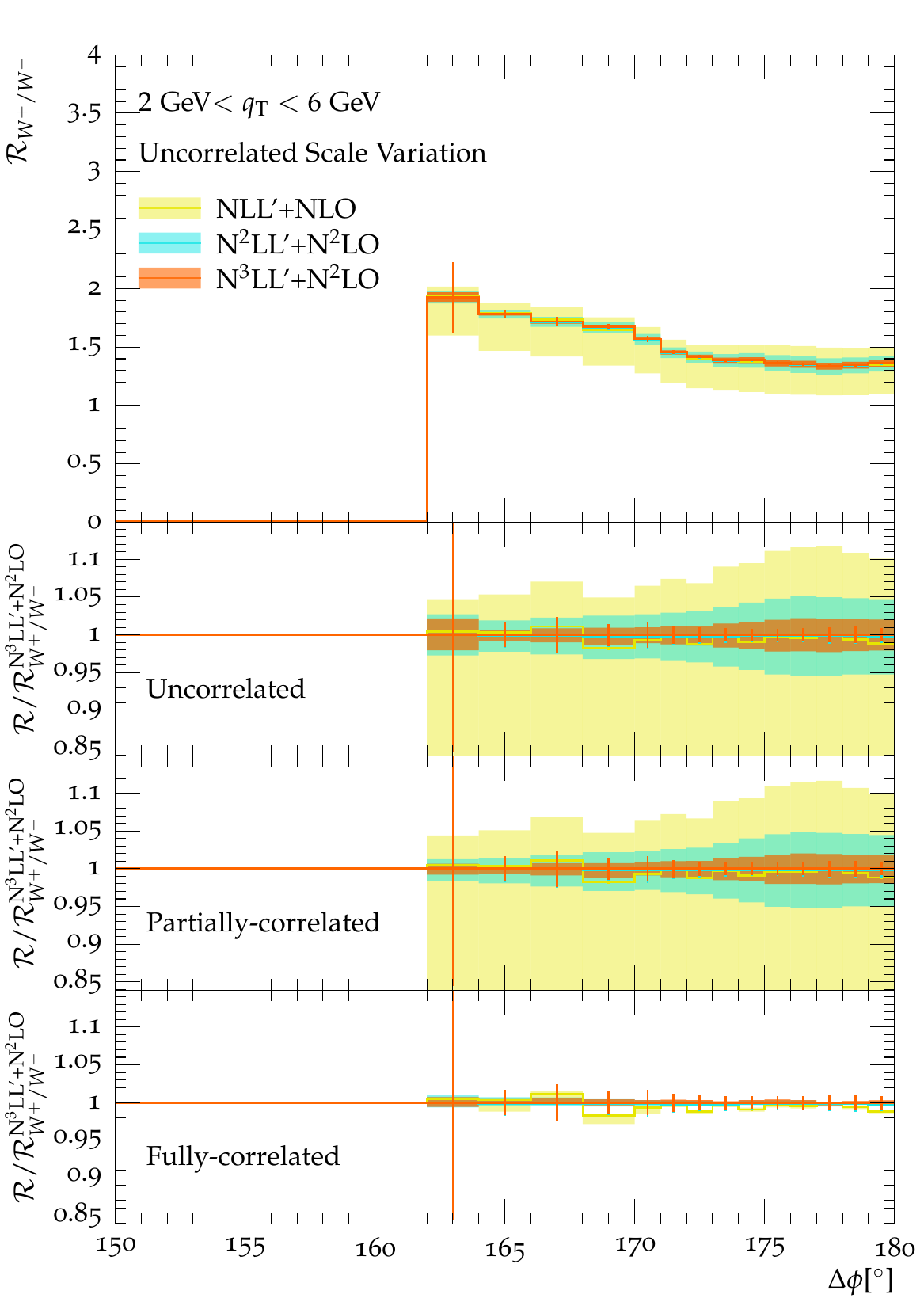}\hfs\hfs
  \includegraphics[width=.31\textwidth]{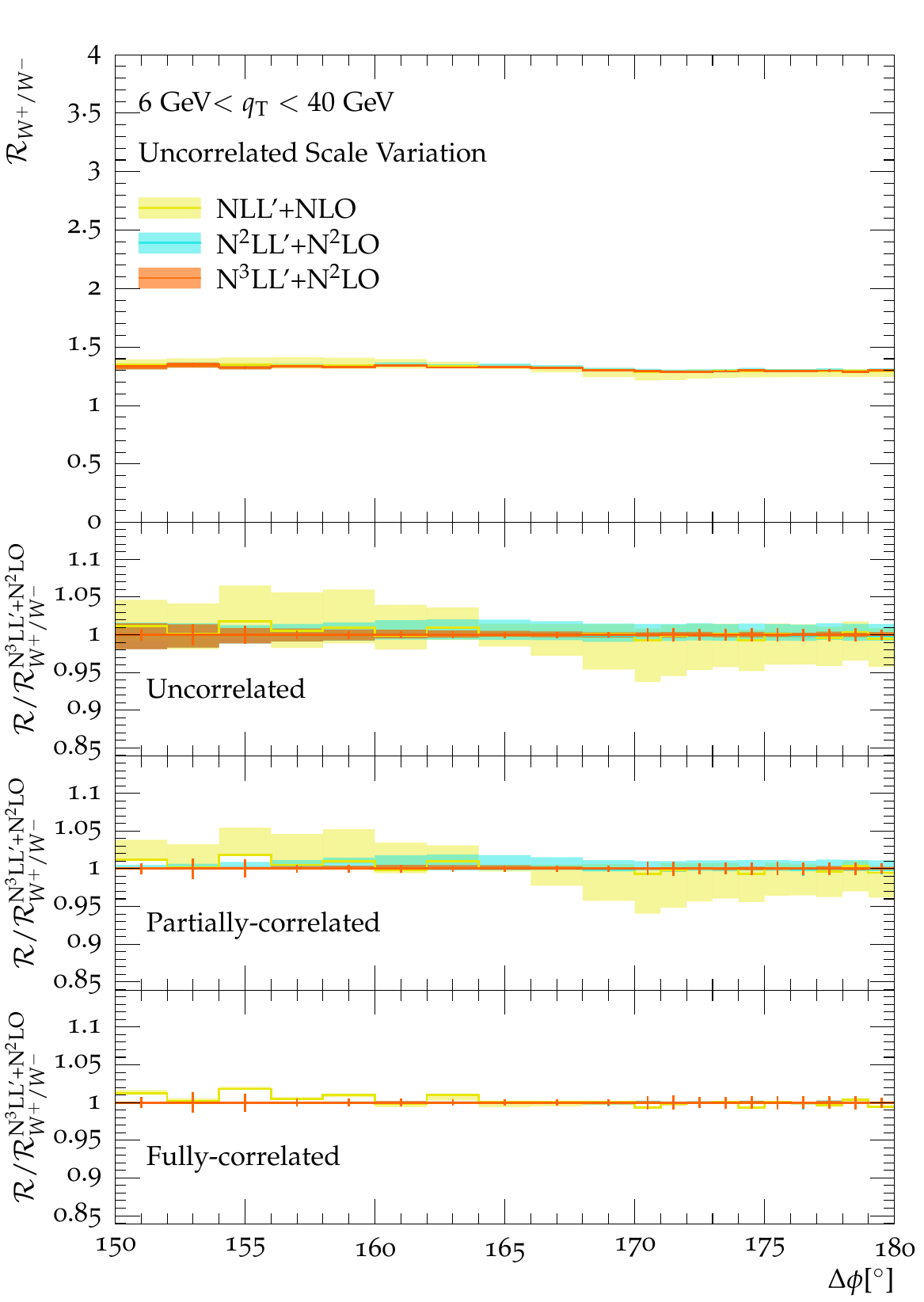}
  \caption{
    Double-differential cross section ratios in \dphi\ and three slices of \qT\ for 
    all three processes. 
    We present the resummation improved results at \NLLNLO, \NNLLNNLO, and 
    \NNNLLNNLO\ accuracy.
  }
  \label{fig:results:ratios:dphi}
\end{figure}

With the double-differential resummation-improved cross
sections of the previous section at hand we finally 
turn towards cross section ratios. 
They are useful to obtain high-precision $W^+$ and 
$W^-$ production data, by measuring the fully 
and precisely reconstructible cross sections in 
$Z$ production and applying the following theory 
predictions, see e.g.\ Refs.\ \cite{Aaboud:2017svj}.  
Similarly, the $W^+$ to $W^-$ ratio is of interest 
for PDF extractions, see e.g.\ \cite{Ball:2017nwa}.
A key question, however, is how to determine the 
uncertainties of such a ratio. 
The main bottleneck is the fundamental lack of 
statistical interpretation of the theoretical 
uncertainties on (multi-)differential cross sections 
as presented so far. 
Thus, in particular various assumptions about their
(non-)correlation in the numerator and denominator 
of the $R_{W^\pm/Z}$ have to be made. 
In the following, we present results for the following 
three correlation assumptions:
\begin{itemize}
  \item \textbf{Uncorrelated.} 
        The scales of the $W^\pm$ and $Z$ processes are 
        assumed to be completely uncorrelated. 
        All scales in the numerator and denominator are 
        varied independently. 
        This corresponds to the assumption that both 
        processes have no common structure in the form 
        of their higher-order corrections or input functions, 
        such as the PDFs. 
        As this is known not to be the case, the uncertainties 
        on the ratio obtained this way are likely to be severely 
        overestimated. 
  \item \textbf{Fully correlated.}
        All scales of the $W^\pm$ and $Z$ processes are 
        assumed to be completely correlated. 
        They are thus varied by a common factor in the 
        numerator and denominator simultaneously. 
        This corresponds to the assumption that both 
        processes have exactly the same structure in the form 
        of their higher-order corrections and input functions, 
        such as the PDFs. 
        As this is known not to be the case, the uncertainties 
        on the ratio obtained this way are likely to be severely 
        underestimated.
  \item \textbf{Partially correlated.}
        A careful analysis of the internal structure of the 
        higher-order corrections to $W^\pm$ and $Z$ production 
        allows to carefully assess which corrections have 
        identical (or at least very similar) structures and 
        which differ.
        This allows, to first approximation, to select a subset 
        of scales to fully correlate, uncorrelating the rest. 
        Following a detailed analysis of the derivations of 
        Sec.\ \ref{sec:methods}, we find that the singlet 
        contributions to $Z$ production are numerically small, 
        see App.\ \ref{app:numeric:singlet}.
        In consequence, both the hard and soft functions 
        for $W^\pm$ and $Z$ decays show the same dependence on 
        the respective scales. 
        Differences, however, occur for the beam functions, 
        arising in the different composition of initial 
        states in all three processes. 
        We thus choose to fully correlate the variation of all 
        scales except for $\mu_b$ and $\nu_b$, which we fully 
        uncorrelate.
\end{itemize}

Figs.\ \ref{fig:results:ratios:qt} and \ref{fig:results:ratios:dphi} 
now show the ratios \RWpZ, \RWmZ, and \RWpWm\ for all three 
definitions of their uncertainties detailed above. 
In the following, we will discuss the different features of 
both their central values and uncertainties separately.

\begin{figure}[t!]
  \centering
  \includegraphics[width=.31\textwidth]{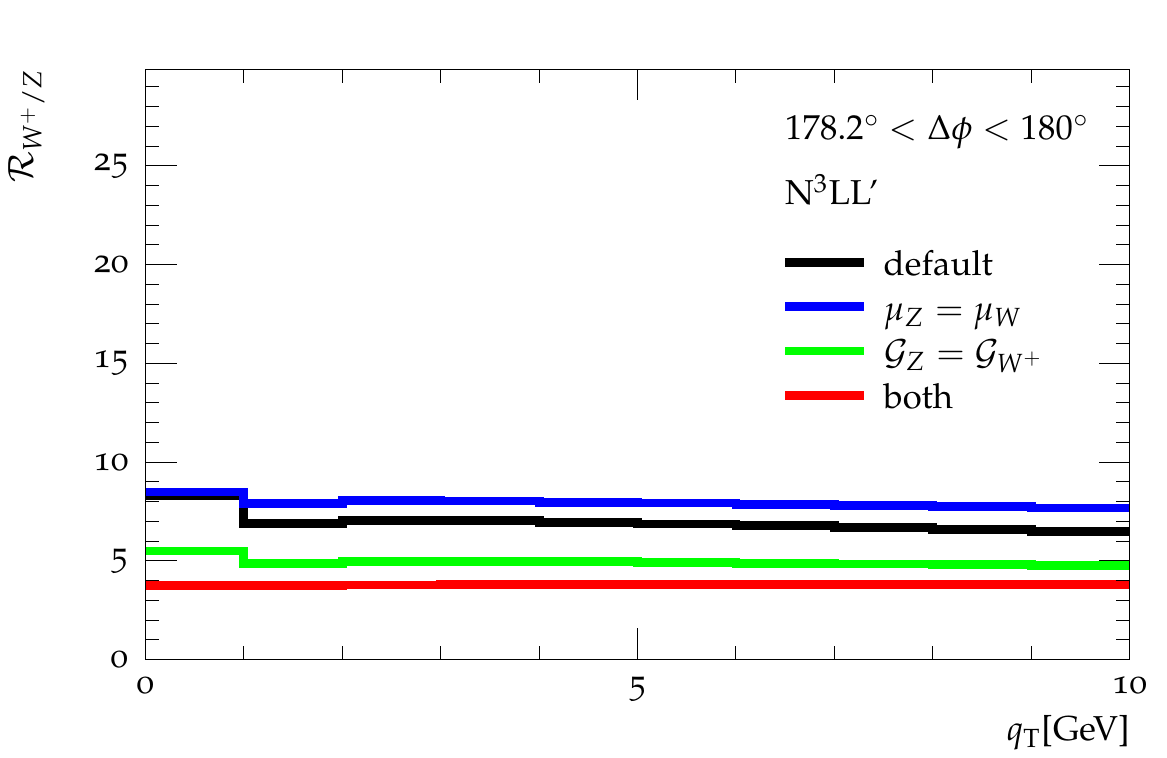}\hfs\hfs
  \includegraphics[width=.31\textwidth]{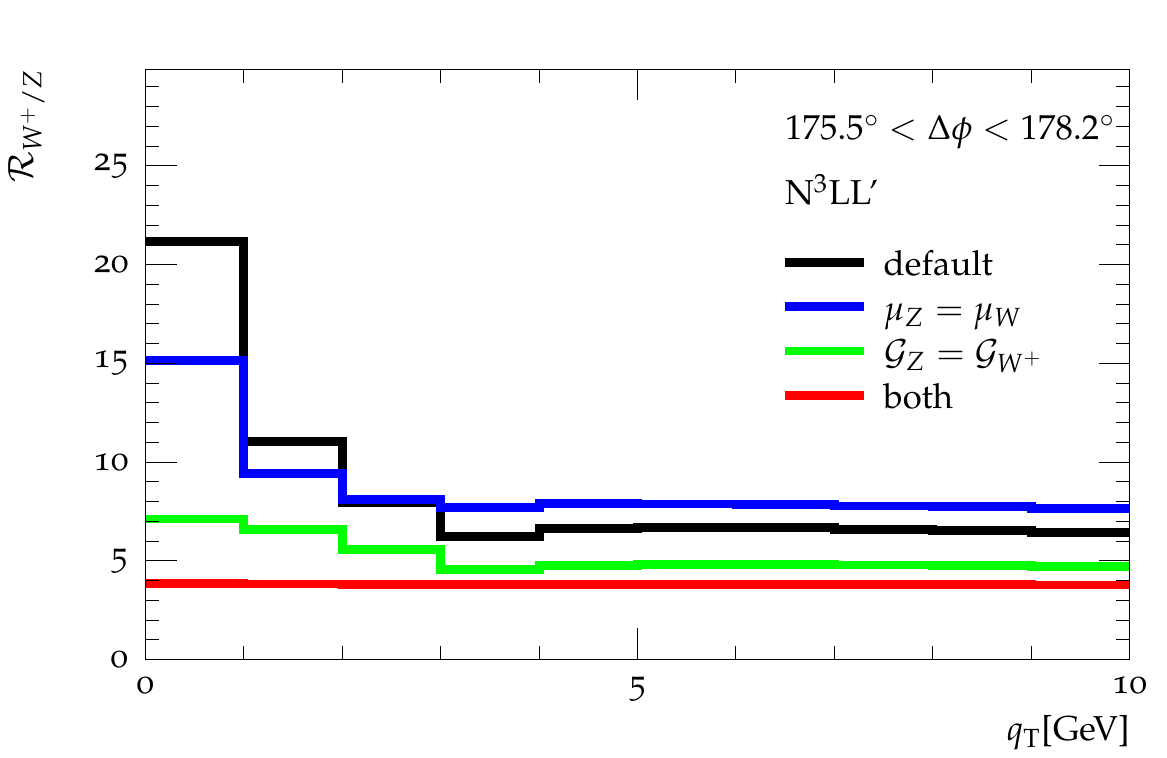}\\[1mm]
  \includegraphics[width=.31\textwidth]{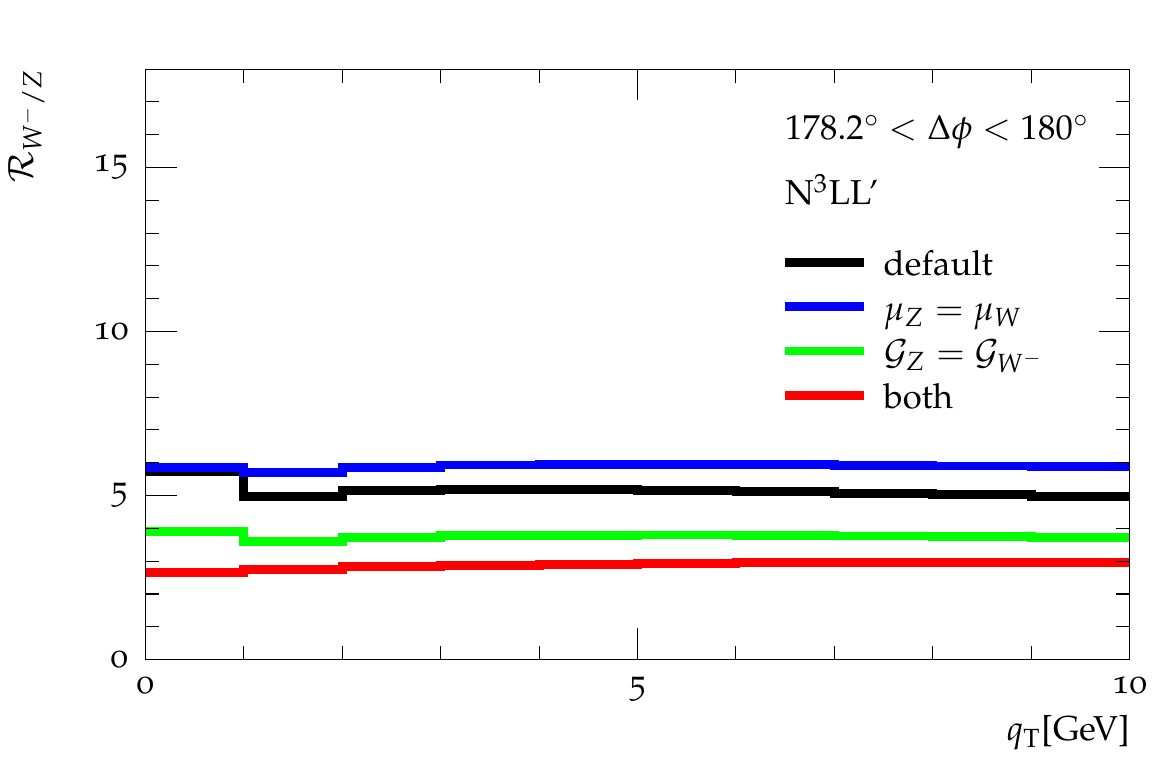}\hfs\hfs
  \includegraphics[width=.31\textwidth]{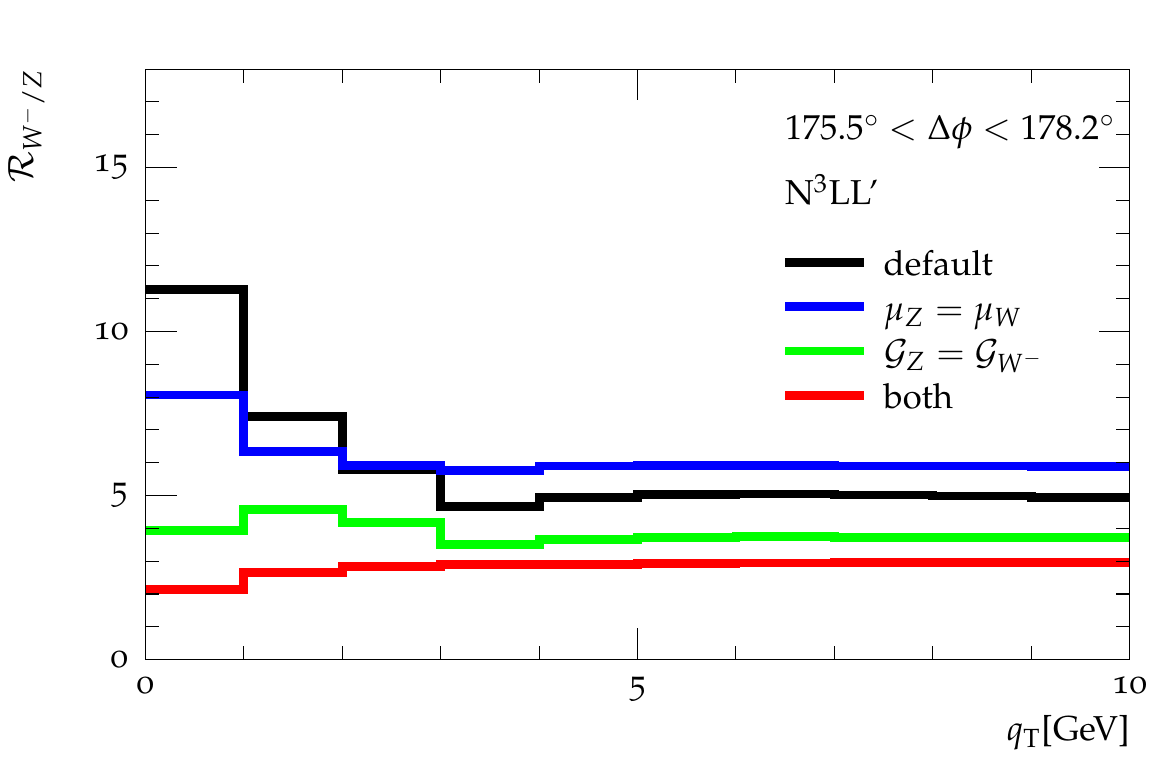}
  \caption{
    Double-differential cross section ratios \RWpZ\ and \RWmZ\ in \qT\ 
    and the two \dphi\ slices closest to the singular region. 
    We present the resummed results at \NNNLL\ accuracy for the 
    default parameters and fiducial regions used in all physical 
    predictions (black), for the default parameters and the $Z$ 
    production phase space $\mathcal{G}_Z$ adapted to the 
    $W^\pm$ production phase space $\mathcal{G}_{W^\pm}$ (green), for
    the $Z$ boson mass set equal to the $W$ boson mass and default 
    fiducial regions (blue), and for both adaptations of the $Z$ 
    production parameters (red), see text for details.
  }
  \label{fig:results:ratios:Comparing:qt}
\end{figure}

\paragraph*{Central values.}
The central values of the cross section ratios are of 
course unaffected by the precise definition of the uncertainty 
band. 
Instead, they are largely determined by the slightly different 
location of the Sudakov peak induced by the mass difference, 
the different $x$-dependence of the contributing parton 
distributions, and the slightly different fiducial phase spaces. 
In addition, they only exhibit a small dependence on the 
perturbative order at which they are calculated. 
In fact, with respect to the perturbative stability of these ratios, 
we observe only minor corrections on the level of up 
to 2\% in \RWpZ\ and \RWmZ, and much smaller in \RWpWm,
when increasing the intrinsic accuracy of the 
resummation-improved calculation from \NLLNLO\ to \NNNLLNNLO. 

Further, the cross section ratios depend only weakly on 
the transverse momentum of the reconstructed vector boson 
for $\qT>5\,\text{GeV}$. 
Below that value, in the vicinity and left flank of the Sudakov 
peak, all three ratios exhibit a marked increase. 
As alluded to earlier, this increase is induced by the differing 
precise locations of the Sudakov peak in each process. 
Fig.\ \ref{fig:results:ratios:Comparing:qt} investigates 
this phenomenon more closely, by  
\begin{itemize}
  \item[a)] setting complex mass of the $Z$ boson to that of 
            the $W$ boson in the propagator only, keeping all 
            other parameters at the default values, 
  \item[b)] replacing the fiducial phase space definition for 
            the $Z$ production channel, $\mathcal{G}_Z$, by that 
            of the $W^\pm$ channel, $\mathcal{G}_{W^\pm}$ with 
            the $\ell^\mp$ taking the role of the neutrino, and
  \item[c)] applying both modifications a) and b).
\end{itemize}
This leaves the differences due to the participating parton 
fluxes and the different spin-structures in the underlying 
EW couplings. 
We find that in the \qT\ spectra the  majority of the effect 
is induced by the differing fiducial regions, with smaller 
additional corrections stemming from the different $W$ and $Z$ 
boson masses. 
With both effects accounted for, the ratios are nearly \qT\ 
and \dphi\ independent, with the remaining small deviations 
attributed to the PDFs and the different spin-structures of 
the underlying EW coupling. 

It needs to be noted, though, that the difference in fiducial 
phase spaces between $W^\pm$ and $Z$ measurements in the 
$W$-boson mass measurement \cite{Aaboud:2017svj} 
was larger than the one used here. 
Hence, the effect can be estimated to have been larger 
in that phase space as well.

\begin{figure}[t!]
  \centering
  \includegraphics[width=.31\textwidth]{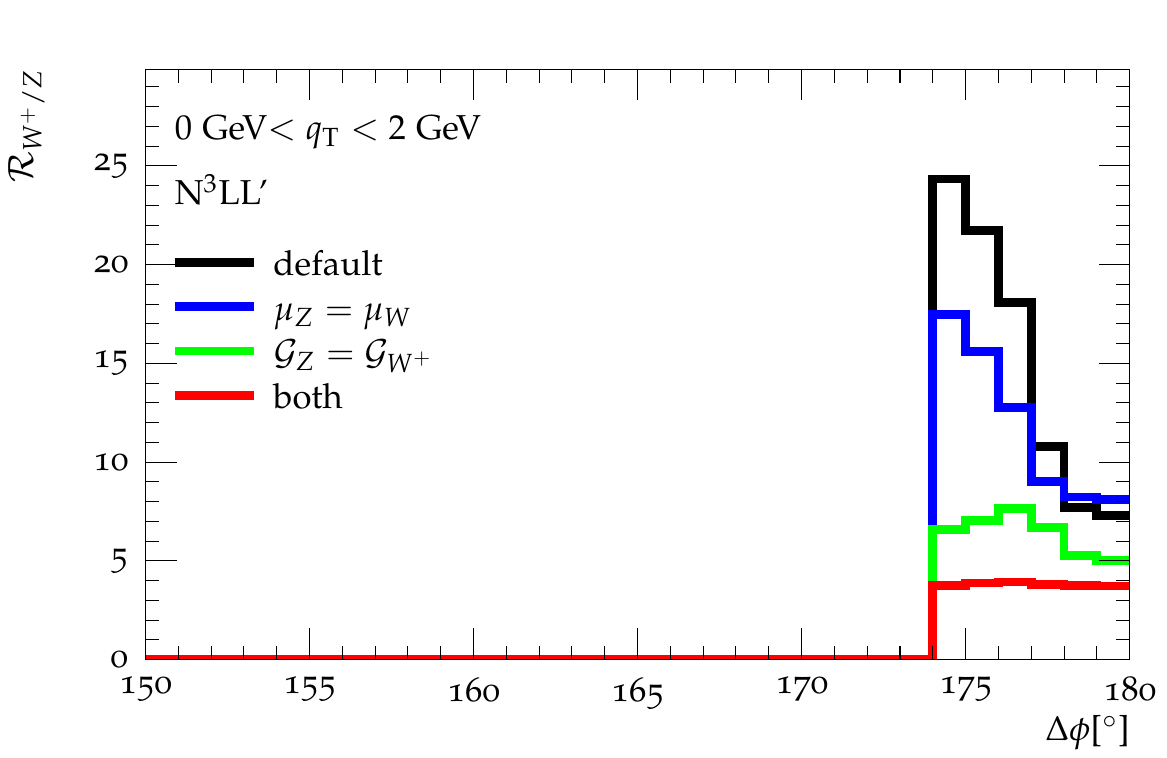}\hfs\hfs
  \includegraphics[width=.31\textwidth]{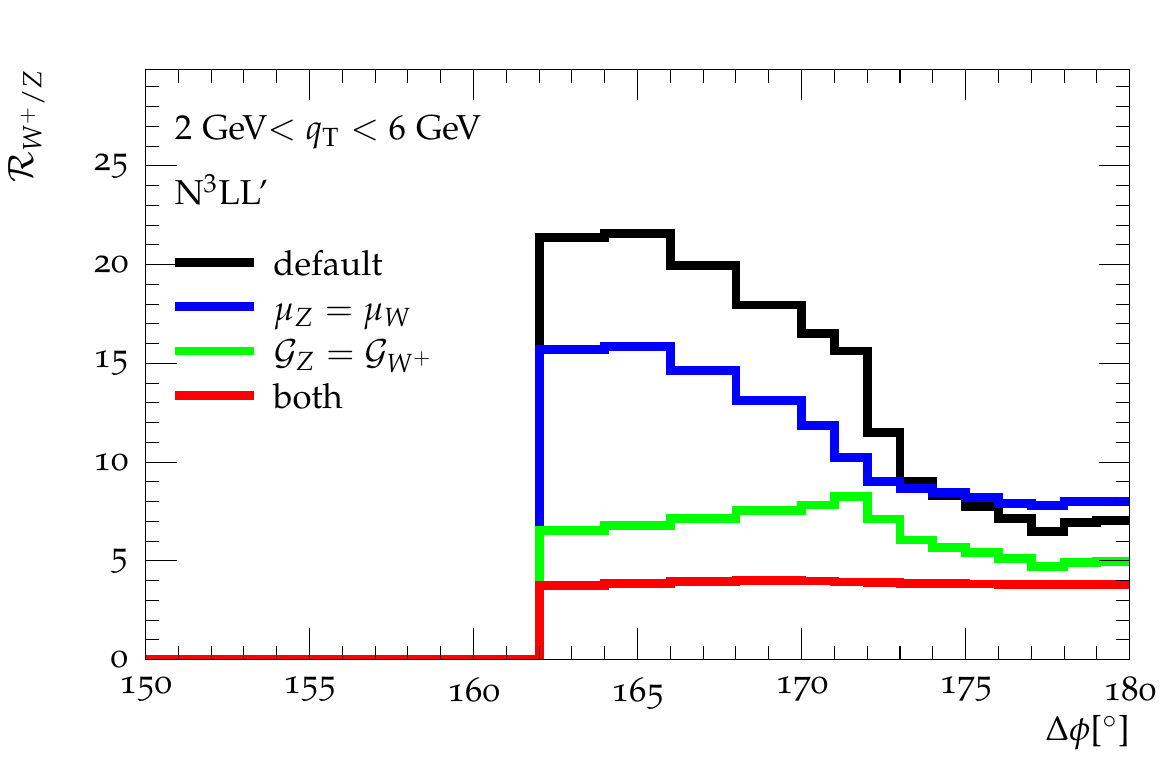}\\[1mm]
  \includegraphics[width=.31\textwidth]{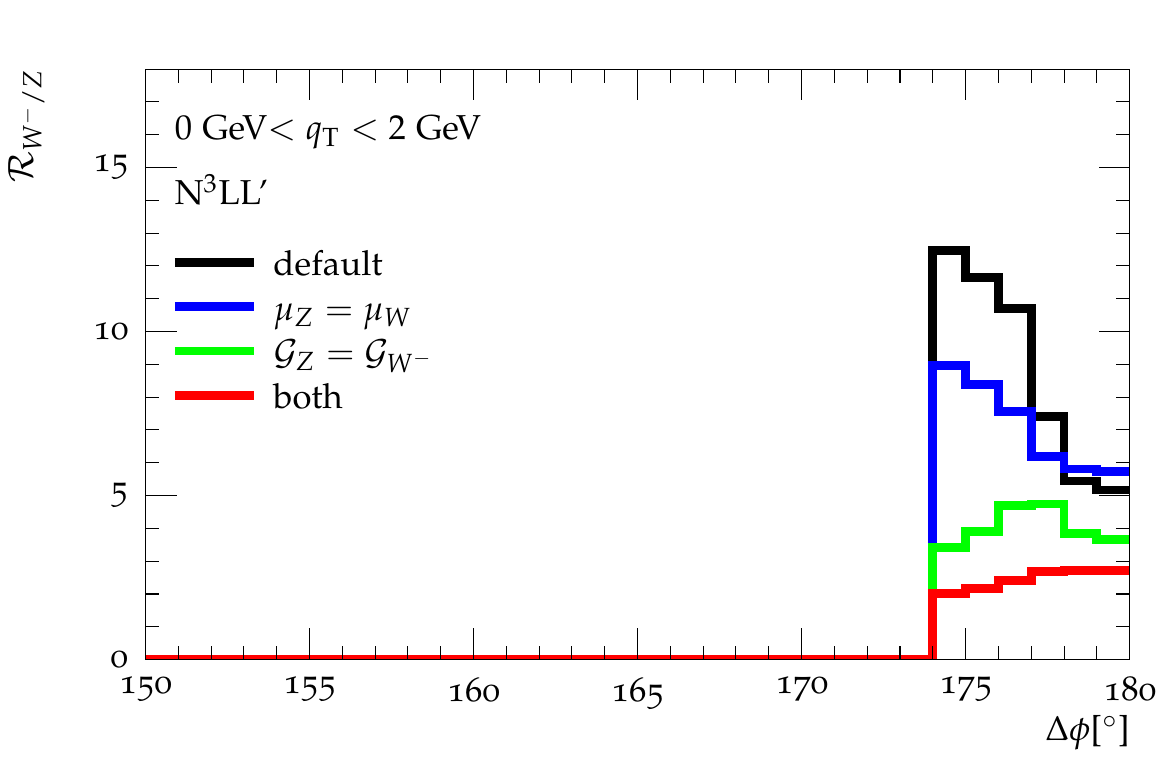}\hfs\hfs
  \includegraphics[width=.31\textwidth]{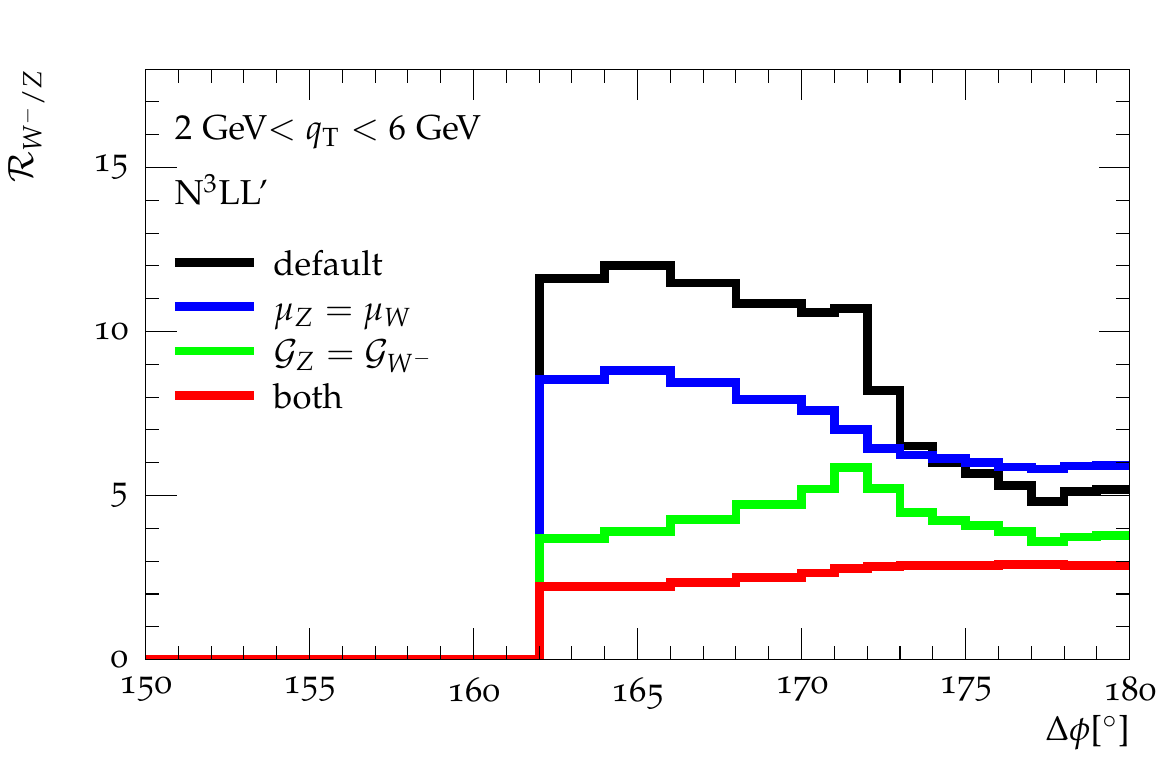}
  \caption{
    Double-differential cross section ratios \RWpZ\ and \RWmZ\ in \dphi\ 
    and the two \qT\ slices closest to the singular region. 
    We present the resummed results at \NNNLL\ accuracy for the 
    default parameters and fiducial regions used in all physical 
    predictions (black), for the default parameters and the $Z$ 
    production phase space $\mathcal{G}_Z$ adapted to the 
    $W^\pm$ production phase space $\mathcal{G}_{W^\pm}$ (green), for
    the $Z$ boson mass set equal to the $W$ boson mass and default 
    fiducial regions (blue), and for both adaptations of the $Z$ 
    production parameters (red), see text for details.
  }
  \label{fig:results:ratios:Comparing:dphi}
\end{figure}

The \dphi\ spectra show a stronger variation of the central 
value of the cross section ratio, increasing to up to a factor 
of three above their value far away from the Sudakov peak in 
both \qT\ and \dphi, in particular in the first two regions.
This increase, however, appears on the far side of the peak 
away from the back-to-back region, 
in contrast to the increase observed in the \qT\ spectra. 
Nonetheless, its origin can be traced to the same factors 
as for the \qT\ spectra 
in Fig.\ \ref{fig:results:ratios:Comparing:dphi}, 
the different definitions of the fiducial phase space in $W$ 
and $Z$ production and the different $W$ and $Z$ boson 
masses.
To be specific,  
we observe that when both effects are accounted for the ratio is nearly independent of 
both \qT\ and \dphi.
This also means that the remaining PDF dependence is small. 

At this point it is important to note that although both the 
fiducial phase spaces in $W^+$ and $W^-$ appear to be the same, 
they are not. 
The reason is that, in terms of spin-correlation the anti-neutrino 
produced in the decay of the $W^-$ takes the role of the charged 
lepton in the decay of the $W^+$, but not in the observable 
definition. 
This is compounded by the fact that, out of the three processes 
under consideration here, $W^+$ and $W^-$ show the largest 
divergence of the contributing partonic fluxes, imparting 
differing rapidity distributions on the produced boson and, 
thus, slightly different effects of the fiducial cuts. 
These factors add up to explain the remaining small, but 
non-negligible \qT\ and \dphi\ dependence of \RWpWm.
 
Finally, the ratios \RWpZ, \RWmZ, and \RWpWm\ are nearly 
flat in the third region containing events with $\qT>6\,\text{GeV}$, 
its only structure being induced mostly by the difference 
in the fiducial phase space in $W$ and $Z$ boson production.

\paragraph*{Uncertainties.}
The uncertainty of the cross section ratios follows the pattern laid out 
in their definition: while the fully-correlated case leads to vanishingly 
small uncertainties, smaller than 1\% in most regions 
(in fact, the largest surviving uncertainty is related to the 
matching procedure), the fully-uncorrelated case lies on the 
opposite end of the spectrum with uncertainties of $\pm 4\%$ 
for our best calculation at \NNNLLNNLO\ accuracy both in 
the fixed-order region and the resummation region. 
The partially-correlated ansatz so far yields the, in our 
judgement, most reliable result, 
ranging from $\pm 1\%$ in the fixed-order region where the 
respective scales are correlated and $\pm 3\%$ in the 
resummation region. 
In particular, it is interesting to note that the beam 
function uncertainties are the driving force of the 
resummation uncertainties overall, reinforcing the difference 
in contributing parton fluxes as a driving factor for the 
details of the ratio overall. 
Similarly, we observe that, apart from the fully-correlated 
uncertainty estimate, the uncertainty for \RWpWm\ largely 
follows the pattern of \RWpZ\ and \RWmZ\, both qualitatively 
and quantitatively.

%% file: text/conclusions.tex
\section{Conclusions}
\label{sec:conclusions}

In this paper we have computed the single-differential 
\qT\ and \dphi\ as well as the double-differential 
$(\qT,\dphi)$ spectra for inclusive $Z$, $W^+$, and 
$W^-$ production in the experimentally accessible 
fiducial phase space 
up to \NNNLLNNLO\ accuracy resumming small transverse 
momentum logarithms. 
Besides the essential inclusion of the third-order soft 
and beam functions, the resummation features the incorporation 
of leptonic power corrections and the singlet contribution 
into the hard sector. 
The leptonic power corrections have been found to extend 
the region of validity of the approximate SCET result. 
The singlet contributions, on the other hand, characterised by 
topologies where the external quarks do not directly couple to 
the electroweak gauge boson, enter the $Z$ boson production 
process at second and third order in $\alpha_s$, and have been found to yield 
corrections of similar size as the non-singlet third order ones, 
and are thus non-negligible at \NNNLL. 

In our numerical evaluation we first confronted the approximate 
results derived from the SCET with the exact ones and 
excellent agreement has been observed in the asymptotic regime.
We then computed the resummation-improved single- and multidifferential 
distributions at \NLLNLO, \NNLLNNLO\ and \NNNLLNNLO\ and found 
excellent perturbative convergence in the asymptotic regime, 
i.e.\ the higher order predictions and their estimated 
uncertainties are fully contained in the lower order 
uncertainty band. 
Further, the respective uncertainties themselves are 
systematically reduced to the level around 4\% or 
less at \NNNLLNNLO. 

In addition, we computed the ratios \RWpZ, \RWmZ, and 
\RWpWm\ of these calculations and estimated their 
uncertainty assuming no correlation, full correlation, 
and, as our best prediction, a partial correlation 
of the scale variations in the numerator and denominator making up the uncertainty. 
For the partial correlation case in particular, a careful 
assessment of the internal structure of our calculation 
allowed to identify identical (or very similar) 
components and structures that differed between the 
three different processes. 
Consequently, the scales used to estimate the uncertainties 
originating in similar components were correlated 
while scales used to estimate the uncertainties 
originating in differing structures were varied independently. 
The main driver for the differences were identified to 
be related to the incident beams and the partonic 
fluxes initiating the respective processes. 

In summary, we determine the ratios with relative 
uncertainties of less than 
1\%, rising to 3-4\% in the resummation region 
at \NNNLLNNLO\ accuracy. 
The shape of the ratios, although largely perturbatively 
stable, is not constant but shows a strong impact 
of the fact that the Sudakov peak 
is located at slightly different positions in all three
processes. 
This observation is the consequence of three main factors: 
\begin{itemize}
  \item[1)] the difference in the $W$ and $Z$ boson masses,
  \item[2)] the different partonic fluxes contributing to the three 
            processes and the different $(x,Q)$-dependence of the $u$ 
            and $d$ valence quarks in particular, and 
  \item[3)] the difference in the fiducial regions for $\ell^+\ell^-$ 
            and $\ell^\pm+\pTmis$ final states.
\end{itemize}
The latter is, in fact, the dominating factor in the \dphi\- and 
low-\qT-dependence of the \RWpZ\ and \RWmZ\ ratios.

With the presented calculation at hand, precise predictions 
for both absolute fiducial multidifferential cross sections 
and their ratios that are vital for the LHC's precision 
measurements programme can be made. 
Nonetheless, important contributions from higher-order 
corrections originating in the electroweak sector of the 
Standard Model are not included in our calculation yet, 
and we leave their investigation to a future publication. 
Similarly, we have for now not included non-perturbative 
corrections that are expected to give non-negligible 
contributions in particular at small transverse momenta 
or large azimuthal separations. 
Their reliable modelling is intricate, including both 
flavour and $x$-dependent contributions, and goes beyond 
the scope of this paper. 
It will also be addressed in a future publication.

\subsection*{Acknowledgements}

WJ is grateful to Xuan Chen and Markus Ebert for helpful discussions on the 
matching procedure. WJ also thanks Konstantin G. Chetyrkin and Thomas Gehrmann for constructive discussions on the singlet hard functions.
MS is funded by the Royal Society through a
University Research Fellowship (URF\textbackslash{}R1\textbackslash{}180549). 
WJ and MS are supported by a Royal Society Enhancement Award 
(RGF\textbackslash{}EA\textbackslash{}181033 and CEC19\textbackslash{}100349).
This work has received funding from the European Union's 
Horizon 2020 research and innovation programme as part of 
the Marie Sklodowska-Curie Innovative Training Network 
MCnetITN3 (grant agreement no.\ 722104).

%% file: text/Appendixes.tex
\section{Impact of the singlet contributions}
\label{app:numeric:singlet}

\begin{figure}[t!]
  \centering
  \includegraphics[width=.45\textwidth]{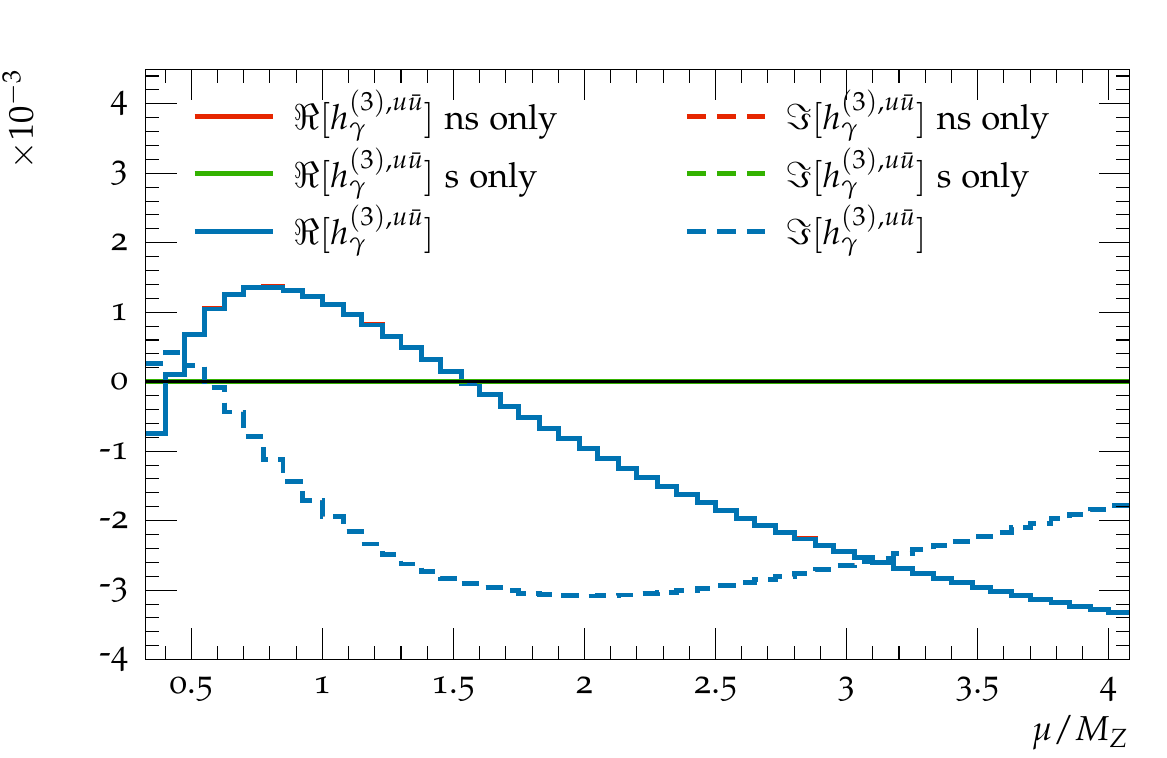}\hfs\hfs
  \includegraphics[width=.45\textwidth]{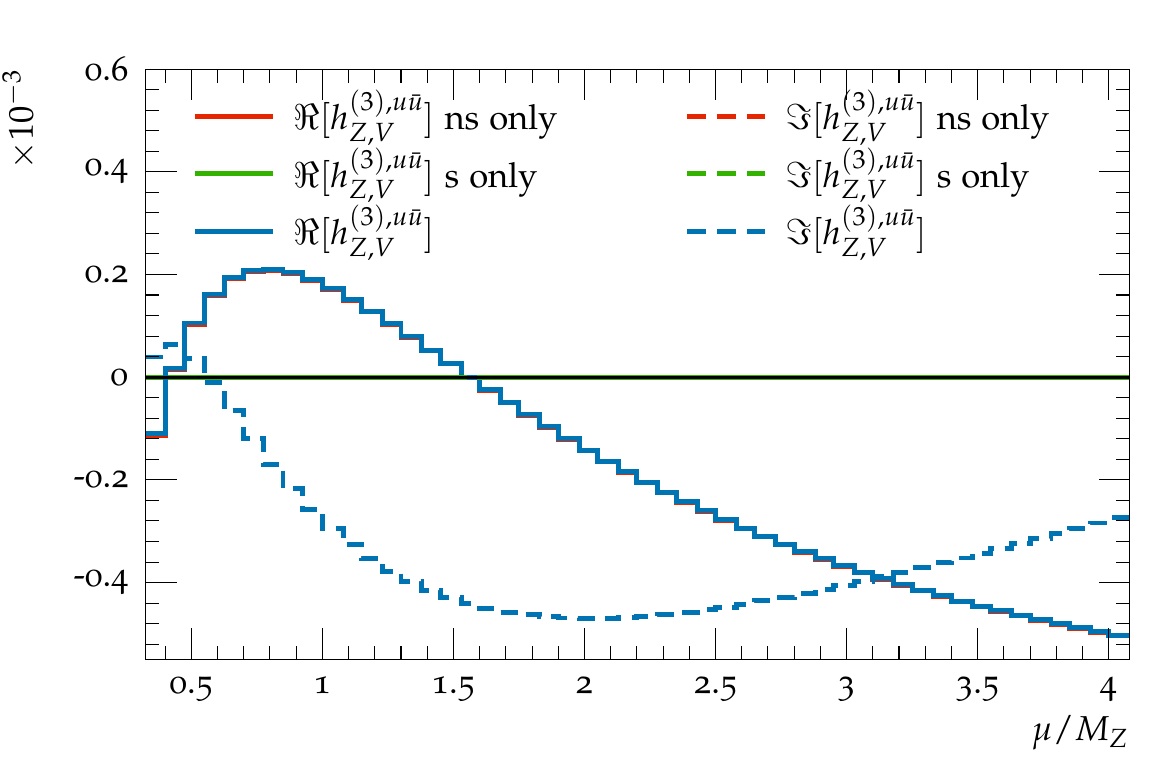}\hfs\\[1mm]
  \includegraphics[width=.45\textwidth]{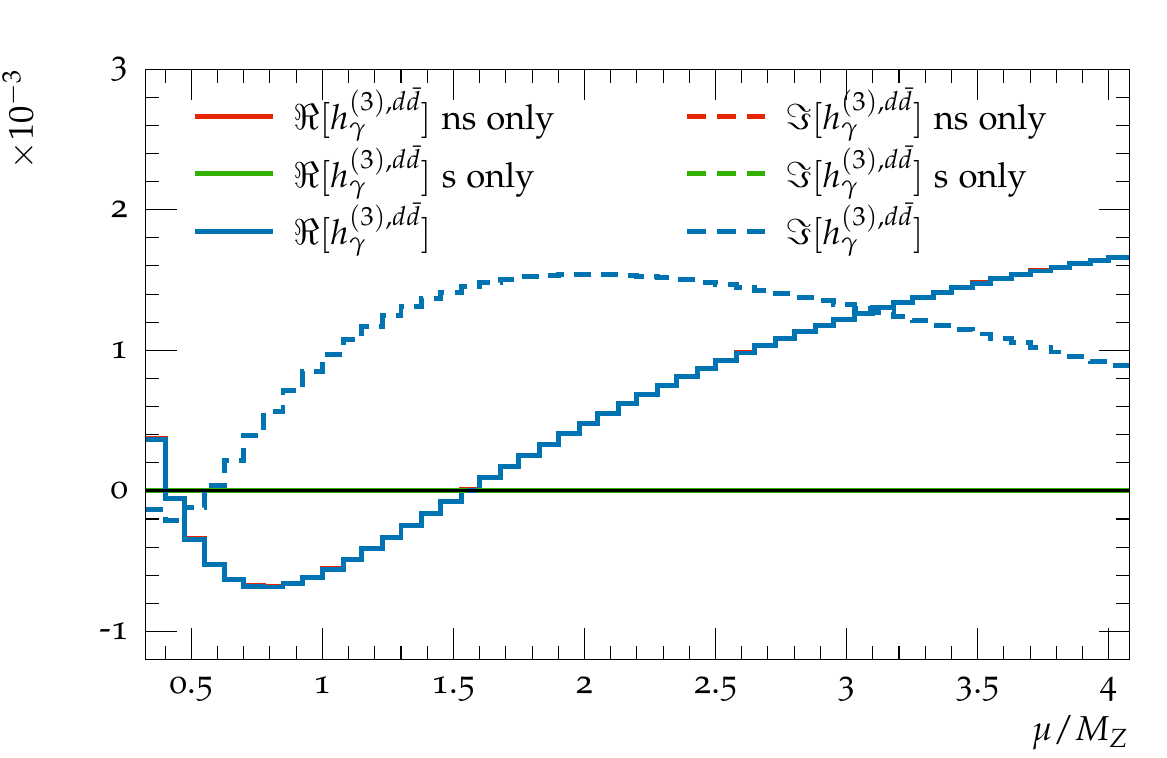}\hfs\hfs
  \includegraphics[width=.45\textwidth]{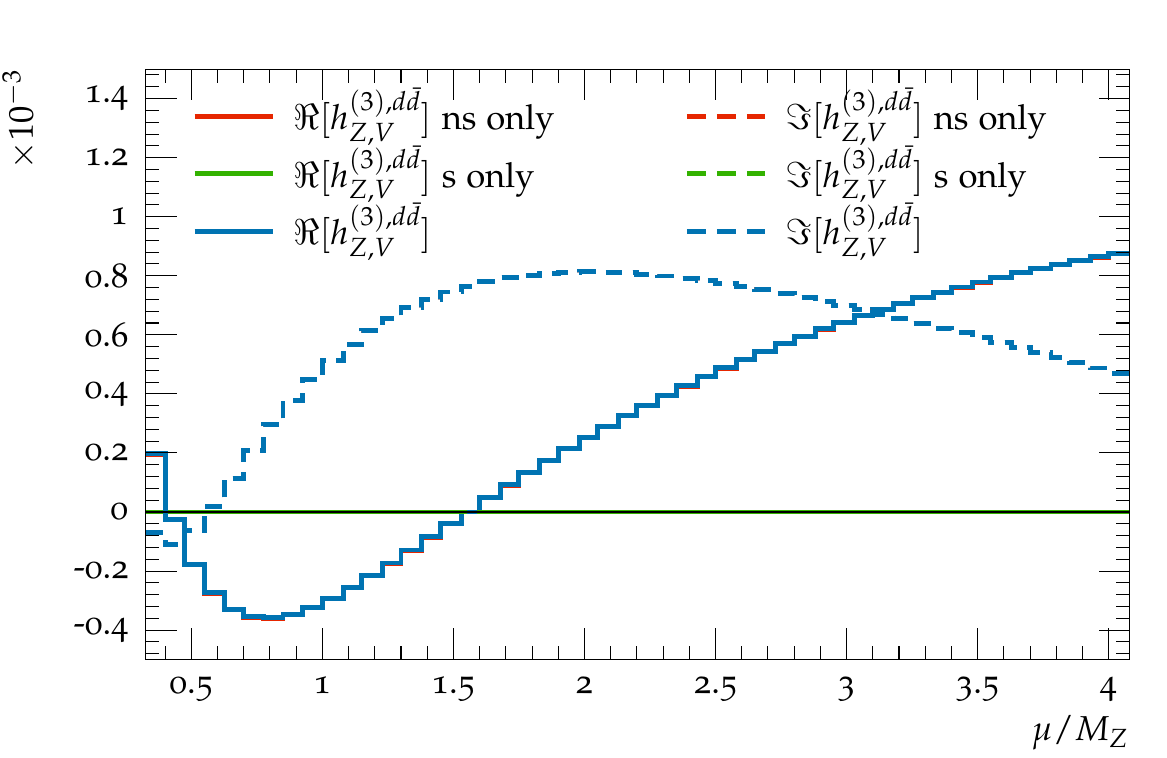}\hfs\\[1mm]
  \caption{
    Numeric results for the hard coefficients $h_{\gamma}^{(3)}$ (left) and 
    $h_{Z,V}^{(3)}$ (right) in the vector current.
  }
  \label{fig:apps:singNum:Vec}
\end{figure}

\begin{figure}[t!]
  \centering
  \includegraphics[width=.45\textwidth]{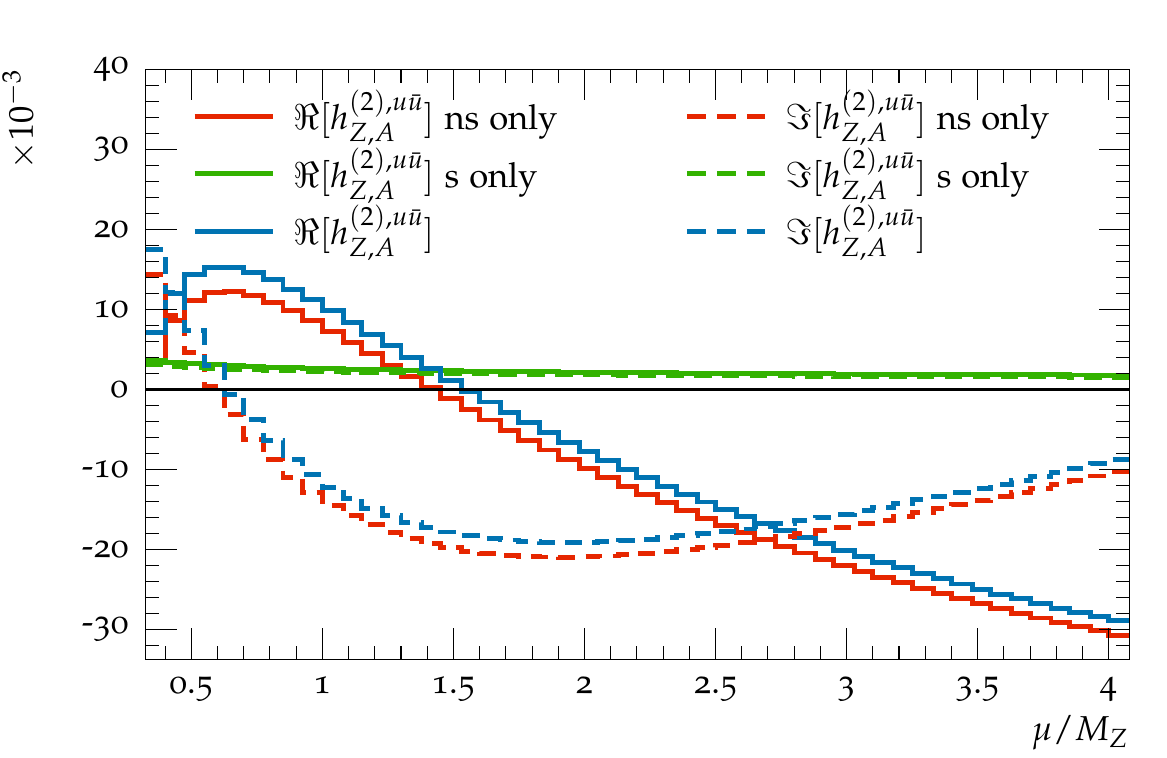}\hfs\hfs
  \includegraphics[width=.45\textwidth]{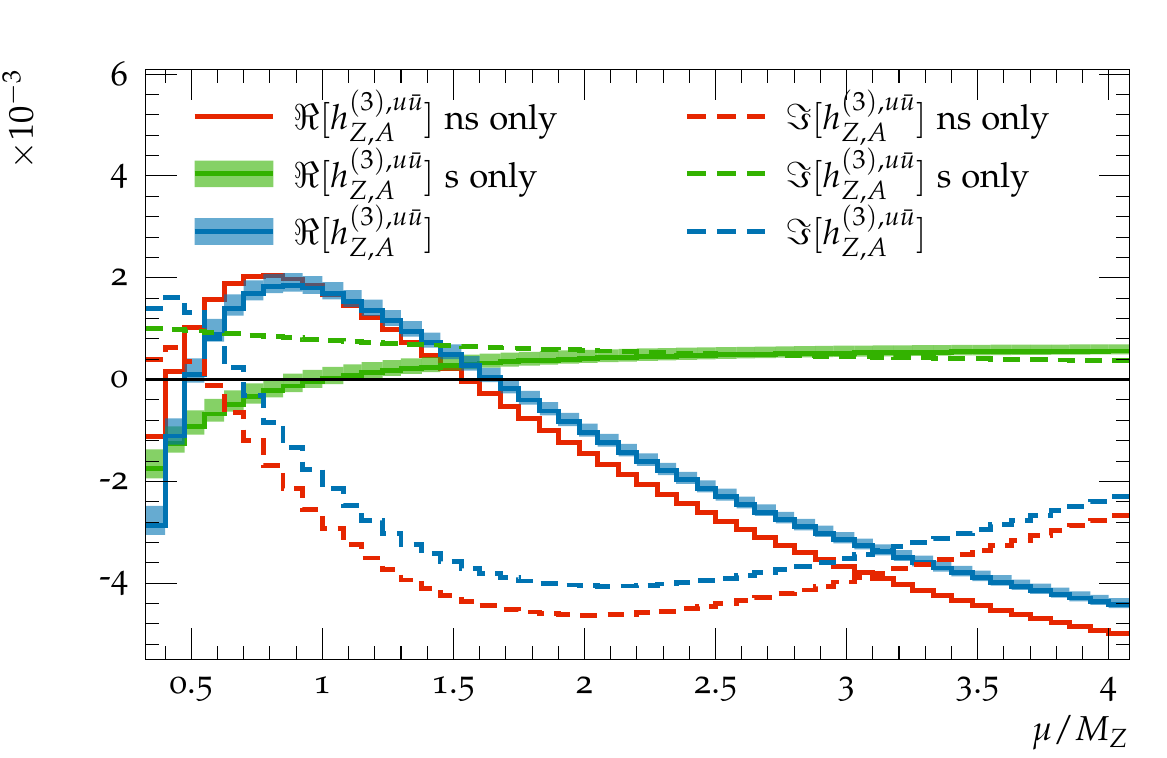}\hfs\\[1mm]
  \includegraphics[width=.45\textwidth]{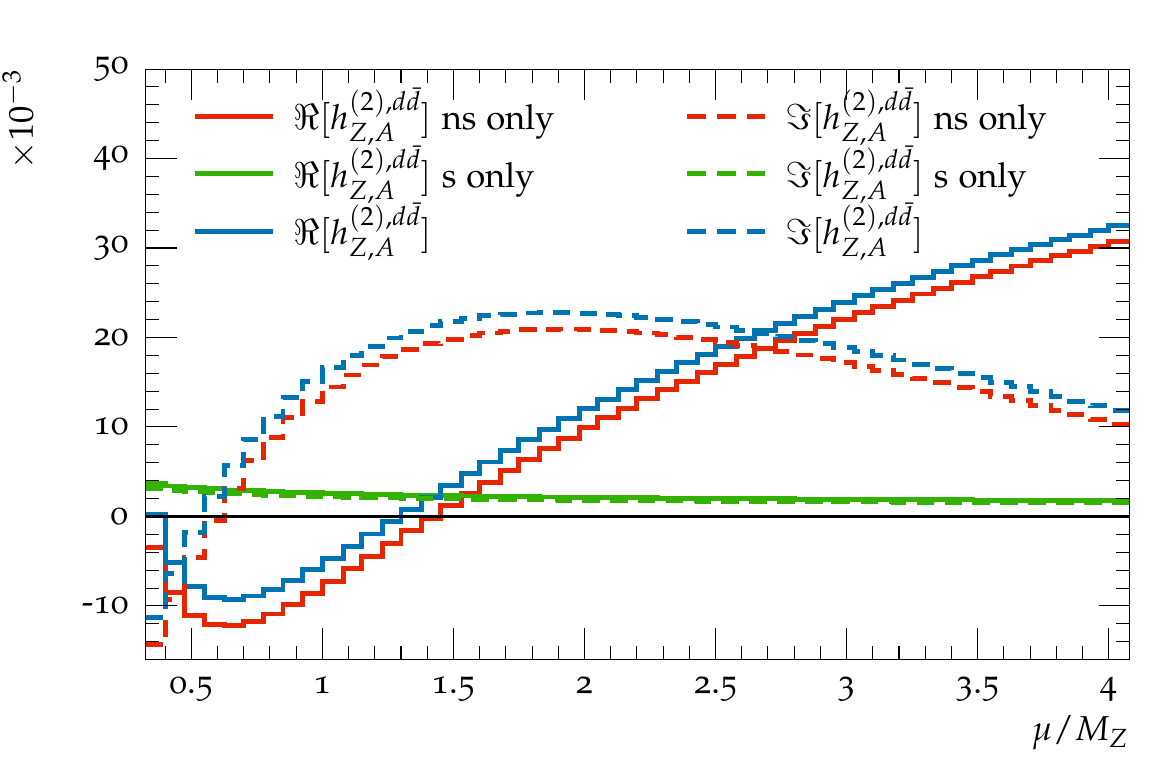}\hfs\hfs
  \includegraphics[width=.45\textwidth]{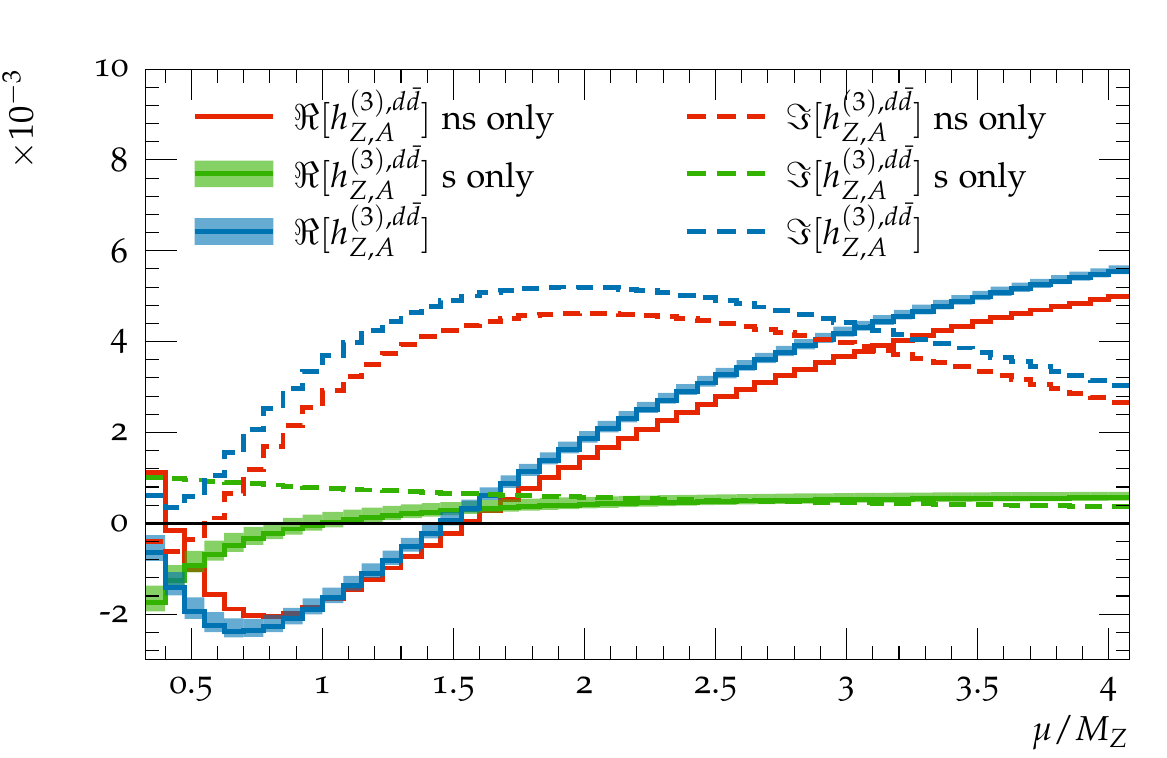}\hfs\\[1mm]
  \caption{
    Numeric results for the hard coefficients $h_{Z,A}^{(2)}$ (left) and 
    $h_{Z,A}^{(3)}$ (right) in the axial-vector current. 
  }
  \label{fig:apps:singNum:AVec1}
\end{figure}

\begin{figure}[t!]
  \centering
  \includegraphics[width=.45\textwidth]{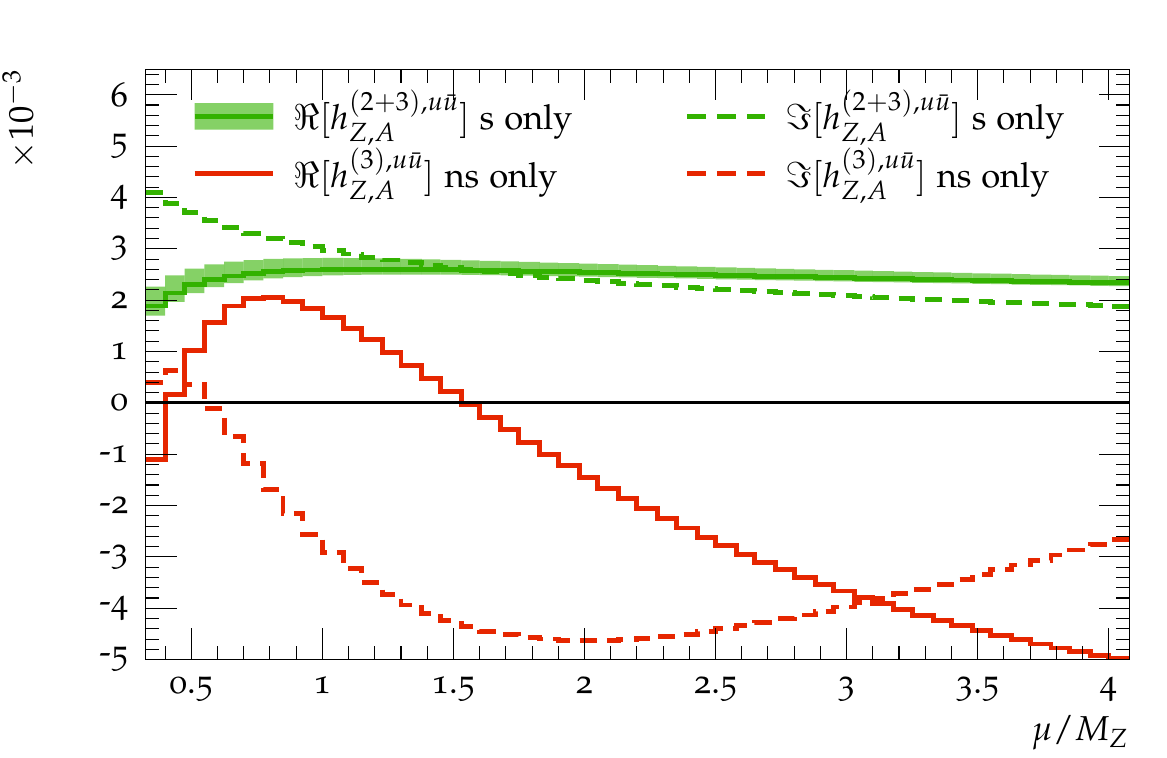}\hfs\hfs
  \includegraphics[width=.45\textwidth]{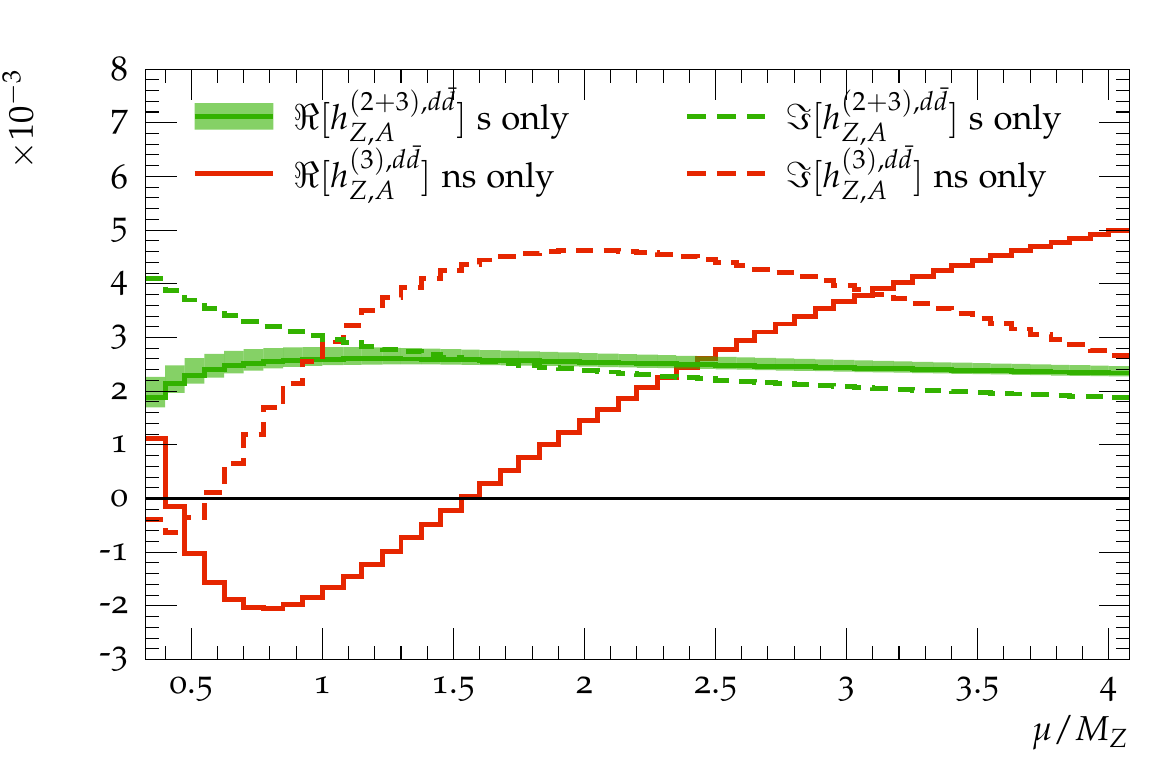}\hfs\\[1mm]
  \caption{
    Comparison of the total singlet contribution to the third-order 
    non-singlet contribution. 
  }
  \label{fig:apps:singNum:AVec2}
\end{figure}

In this appendix we present a quantitative discussion of the 
singlet contributions to the hard function. 
To this end, we re-express the hadronic current $H^{\mu,ij}$ 
of \eqref{eq:method:def:hadcur} in the following form,
\begin{equation}\label{eq:method:def:hadcur:2}
  \begin{split}
    H^{\mu,ij}_{\gamma}
    =&\;
      \left(g_{\gamma}^{q_i}C_{\mathrm{ns}}+g_{\gamma}^\sssS C_{\mathrm{s}}^V \right)
      \mathcal{V}^{\mu}_{ij}\equiv    \,e\, \mathcal{V}^{\mu}_{ij}\,\sum_{m}\, h_{\gamma}^{(m),ij}    \,,\\
    H^{\mu,ij}_{Z,V}
    =&\;
      \left(g_{V}^{q_i} C_{\mathrm{ns}}+ g_{V}^\sssS C_{\mathrm{s}}^V \right)
      \mathcal{V}^{\mu}_{ij}\equiv  \,\frac{e}{s_w\,c_w}\,\mathcal{V}^{\mu}_{ij} \sum_{m} \, h_{Z,V}^{(m),ij}\,,\\
    H^{\mu,ij}_{Z,A}
    =&\;
      g_{A}\left[\left(2T_{q_i}+\frac{1}{N_F}\right)C_{\mathrm{ns}}+C_tC_{\mathrm{s}}^A\right]
      \mathcal{A}^{\mu}_{ij}\equiv \,g_{A}\,\mathcal{A}^{\mu}_{ij} \sum_{m}\, h_{Z,A}^{(m),ij} \,,
  \end{split}
\end{equation}
where the coefficients $ h_{\gamma}^{(m),ij}$, $ h_{Z,V}^{(m),ij}$ 
and $h_{Z,A}^{(m),ij}$ represent the $\order(\alpha_s^m)$ corrections 
to the hadronic currents.
During our calculations here, the invariant mass $M_L$ is specifically 
fixed to $m_Z$, while the scale $\mu$ is left as a variable.  

Fig.~\ref{fig:apps:singNum:Vec} exhibits the size of the non-singlet 
\NNNLO\ coefficients $h_{\gamma}^{(3)}$ and $h_{Z,V}^{(3)}$ for the 
$u\bar{u}$ and $d\bar{d}$ partonic channels. 
As the singlet contribution to the vector current is forbidden at two 
loop level by $C$-parity, we plot only the third order therein. 
The numerical results computed in three different ways are displayed: 
1) the $\order(\alpha_s^3)$ correction with both the non-singlet 
   coefficient $C_{\mathrm{ns}}$ and the singlet coefficient 
   $C_{\mathrm{s}}^V$ included; 
2) the same $\order(\alpha_s^3)$ correction including only the 
   non-singlet contribution $C_{\mathrm{ns}}$; and
3) the $\order(\alpha_s^3)$ singlet contribution $C_{\mathrm{s}}^V$.
It is seen that for the entire $\mu$ range, the both the real 
and the imaginary part of the full $\order(\alpha_s^3)$ correction 
(in blue lines) almost coincide with the pure non-singlet ones 
(in red lines), and the singlet contributions (in green lines) 
are negligible. 
This indicates that at least up to $\mathcal{O}(\alpha_s^3)$, 
the role played by the singlet terms in the vector hadronic current 
is of little phenomenological impact. 
Moreover, it can also be observed that the full and non-singlet 
results of $u\bar{u}$ initial state are curved in the opposite 
direction to the $d\bar{d}$ ones. 
This is caused by the different charges and weak isospins 
of the $u$ and $d$ quarks.  
As shown in eq.~\eqref{eq:method:def:hadcur}, given the negligible 
singlet vector terms, the vector current hard coefficients between 
different initial states can be related as follows  
\begin{equation}
  \begin{split}
    \frac{h_{\gamma}^{(3),u\bar{u}}}{h_{\gamma}^{(3),d\bar{d}}}
    &\sim\frac{Q_u}{Q_d}=-2\,,\\
    \frac{h_{Z,V}^{(3),u\bar{u}}}{h_{Z,V}^{(3),d\bar{d}}}
    &\sim\frac{T_u^3-2Q_us_w^2}{T_d^3-2Q_ds_w^2}\sim-\frac{1}{2}\,.\\
\end{split}
\end{equation}
Here the flipping signs account for the different curvatures as observed, 
and the proportions derived in the last step give the relationship of 
the magnitudes of the $u\bar{u}$ and $d\bar{d}$ initiated results.

In Fig.~\ref{fig:apps:singNum:AVec1}, the magnitudes of the real and 
imaginary parts of axial-vector current hard coefficients are depicted. 
As the singlet contribution here starts at $\order(\alpha_s^2)$, we 
show both $h_{Z,A}^{(2)}$ and $ h_{Z,A}^{(3)}$.
In complete analogy to the vector current coefficients, we here also 
graph the three types of outputs therein: 
1) the full result calculated as eq.~\eqref{eq:method:def:hadcur}; 
2) the non-singlet one including $C_{\mathrm{ns}}$ only; and 
3) the singlet one which is the difference between the previous two cases.
Contrary to Fig.~\ref{fig:apps:singNum:Vec}, the axial-vector singlet 
terms give an unignorable contribution here for all values of $\mu$, 
but in particular for values of $\mu$ close to 0.5 or 1.5 where either 
the real or imaginary part of the non-singlet contribution vanishes.
In the region $\mu\sim m_Z$, which is the default hard scale taken in 
the resummation in this paper (see eq.\ \eqref{eq:results::setup:scales_II}), 
the singlet terms can account for around $20\%$ of the real contributions 
of the full results in $\mathcal{O}({\alpha_s^2})$, and nearly $40\%$ in 
$\mathcal{O}({\alpha_s^3})$. 
In the imaginary parts, it is also noted that more than $10\%$ of 
$\Im[h_{Z,A}^{(2),u\bar{u}}]$ and $\Im[h_{Z,A}^{(2),d\bar{d}}]$ is made 
up of the singlet terms for the majority of the $\mu$ range, and although 
at the third order accuracy the singlet terms experience zeros in the 
vicinity of $\mu=m_Z$, its proportions can recover to $\sim10\%$ when 
$\mu/m_Z$ exceeds $2$. 
Furthermore, one interesting phenomenon in Fig.~\ref{fig:apps:singNum:AVec1} 
is that the singlet contributions remain the same in the $u\bar{u}$ and 
$d\bar{d}$ initial states, but the non-singlet lines are curved in the 
different directions. 
The reason can be found in eq.~\eqref{eq:method:def:hadcur}. 
Therein the non-singlet contribution $C_{\mathrm{ns}}$ is directly 
proportional to the weak isospin of the initial parton, 
$T_{q_i}$, whilst the singlet terms 
$C_{t}C_{\mathrm{s}}^A+C_{\mathrm{ns}}/N_F$ are universal for all 
initial states.
Therefore, after excluding the singlet terms, the $h_{Z,A}^{(2,3),u\bar{u}}$ 
results are actually taking the opposite numbers of those for the $d\bar{d}$ 
channel, while the singlet terms hold. 
Please note that we have added an uncertainty band to the results of 
$C_{\mathrm{s}}^{A,(3)}$ to account for the uncertainty arising 
from the finite terms, see eq.\ \eqref{eq:CAS per order:3} and 
discussion thereafter. 
The error estimation is carried out by varying $c_{\text{s}}^{A,(3)}$ 
around the default choice as $[\tfrac{1}{2},2]\cdot C_{\text{ns}}^{(3)}|_{_{L_H=0}}$.

To further explore the properties of the axial-vector singlet terms, 
we confront the whole singlet contribution (the sum of the 
$\mathcal{O}(\alpha_s^2)$ and $\mathcal{O}(\alpha_s^3)$ corrections) 
with the non-singlet $ h_{Z,A}^{(3)}$s in Fig.~\ref{fig:apps:singNum:AVec2}. 
It is seen that for the majority of the $\mu$ range, the singlet 
contribution takes the comparable magnitude to the third-order 
non-singlet result. 
Especially in the $\mu\sim m_Z$ region, the singlet terms approach 
$\Im[h_{Z,A}^{(3),d\bar{d}}]$ but are even greater than 
$\Re[h_{Z,A}^{(3),u\bar{u}}]$, $\Im[h_{Z,A}^{(3),u\bar{u}}]$ as well 
as $\Re[h_{Z,A}^{(3),d\bar{d}}]$ in magnitude.  
In fact, this observation can substantially highlight the importance 
of the singlet terms in the resummation.  
As shown in Table.~\ref{tab:methods:res_accuracy}, the resummation at 
\NNLL\ requires the $\mathcal{O}(\alpha_s^2)$ hard function, whilst 
the \NNNLL\ accuracy needs those up to the third order. 
So if only a precision of the order of $10\%$ of the $\mathcal{O}(\alpha_s^2)$ 
correction is needed, the singlet terms could be neglected at the 
\NNLL\ accuracy in a numerically approximate sense. 
Nevertheless, as the complete singlet corrections are of the same 
magnitude as, if not larger than, the third-order non-singlet corrections,  
their inclusion is mandatory for a meaningful and robust \NNNLL\ resummation.

\section{Impact of leptonic power corrections}
\label{sec:app_lpc}

\begin{figure}[t!]
  \centering
  \includegraphics[width=.22\textwidth]{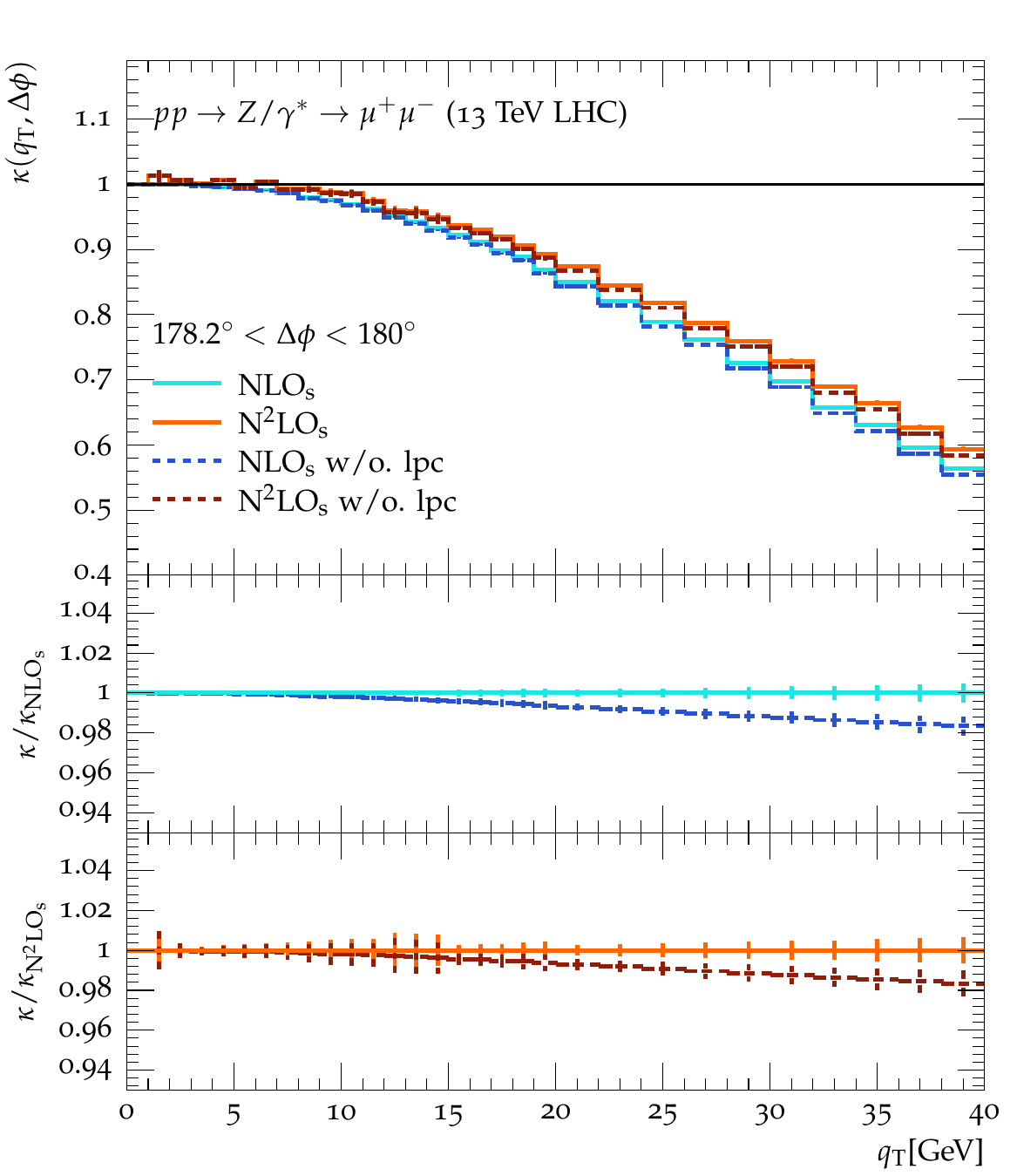}\hfs\hfs
  \includegraphics[width=.22\textwidth]{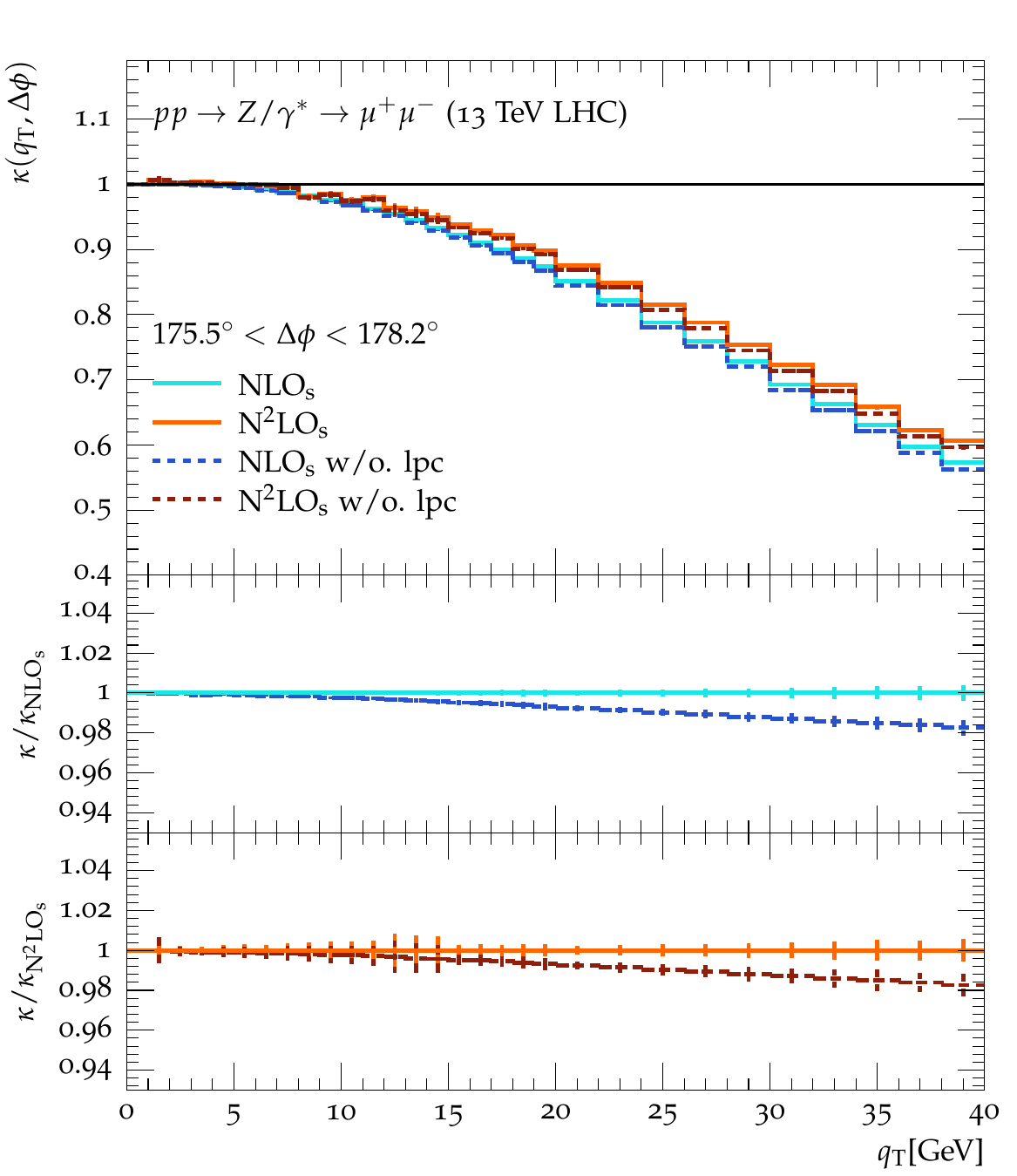}\hfs\hfs
  \includegraphics[width=.22\textwidth]{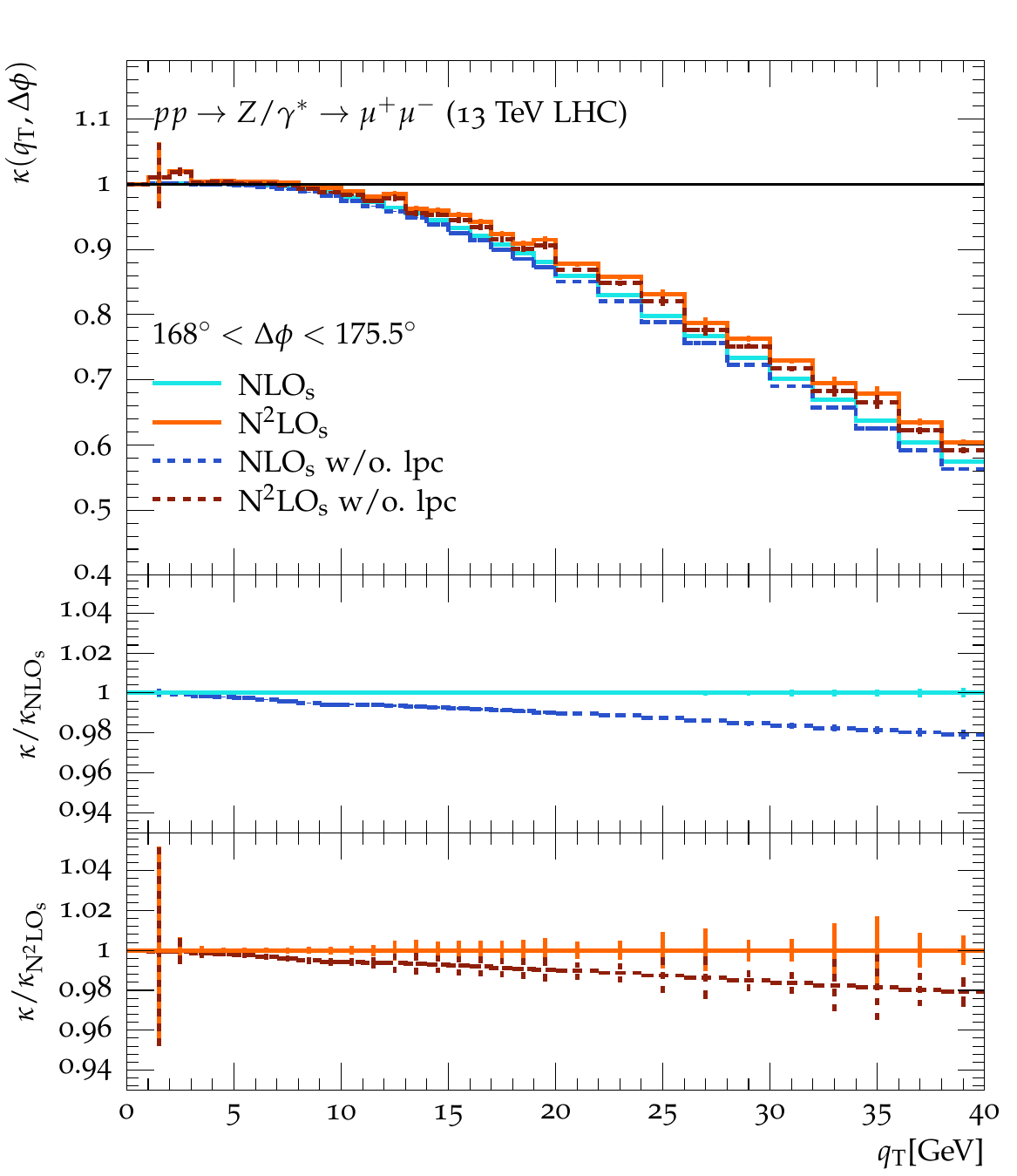}\hfs\hfs
  \includegraphics[width=.22\textwidth]{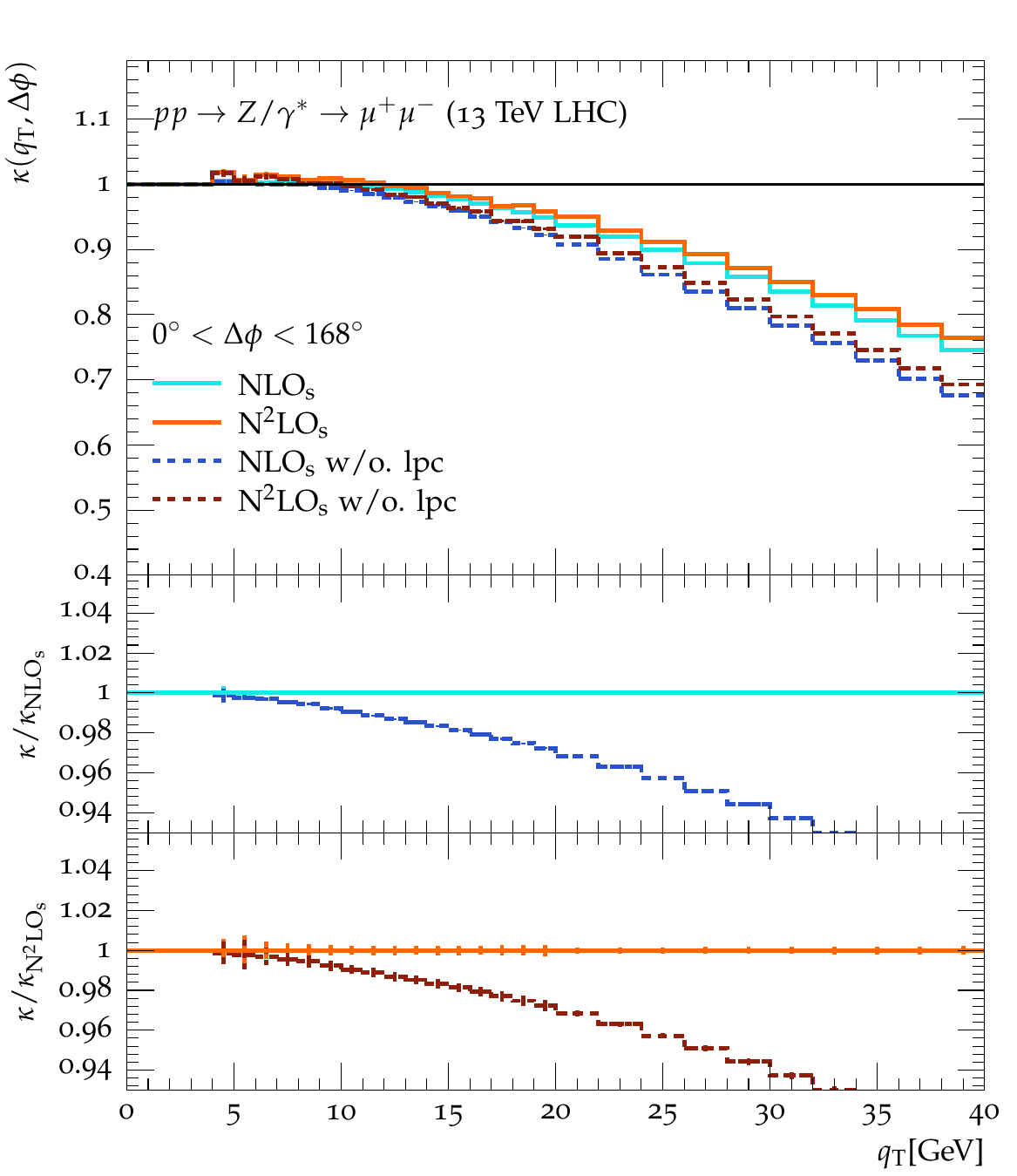}\\[1mm]
  \includegraphics[width=.22\textwidth]{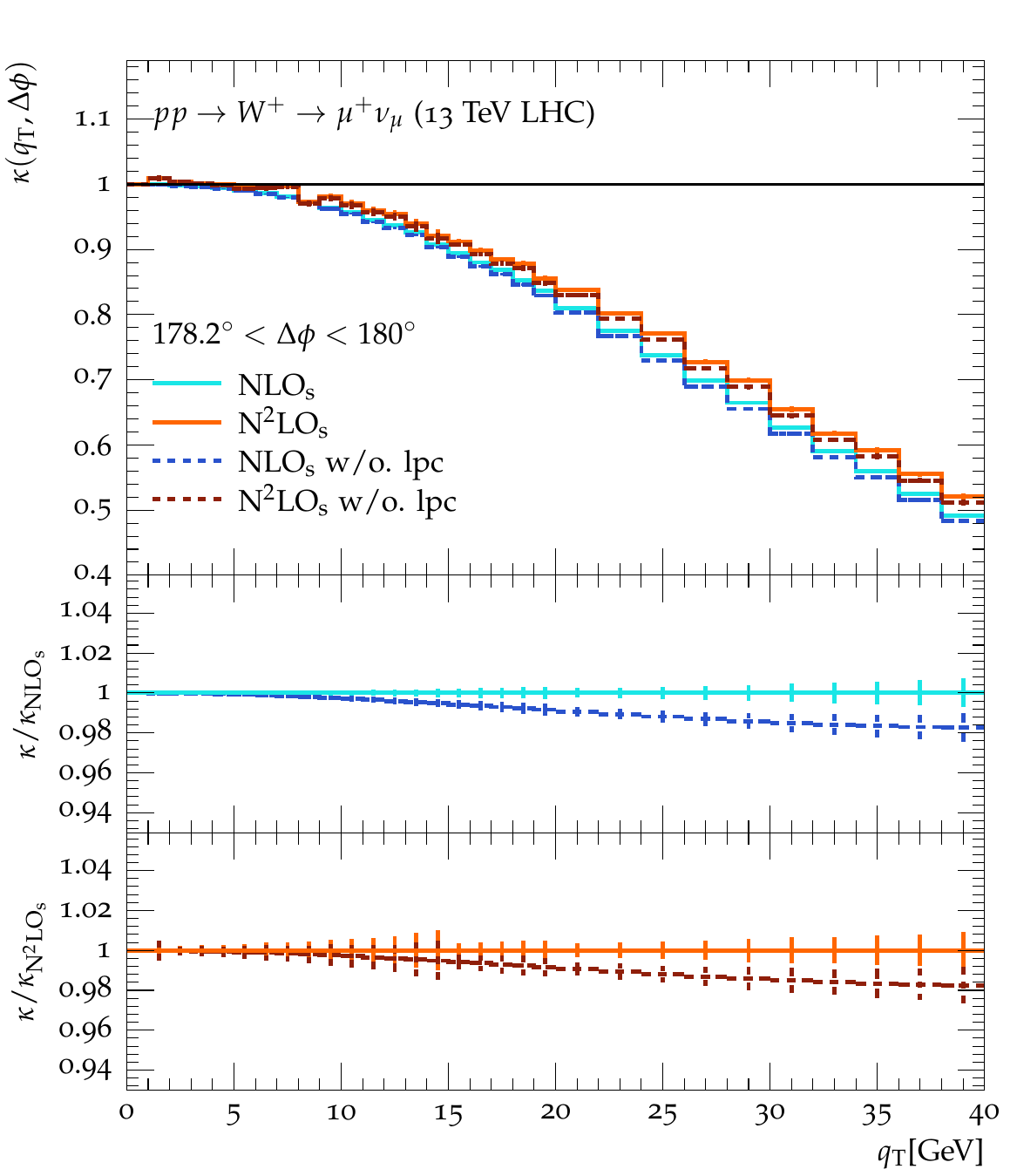}\hfs\hfs
  \includegraphics[width=.22\textwidth]{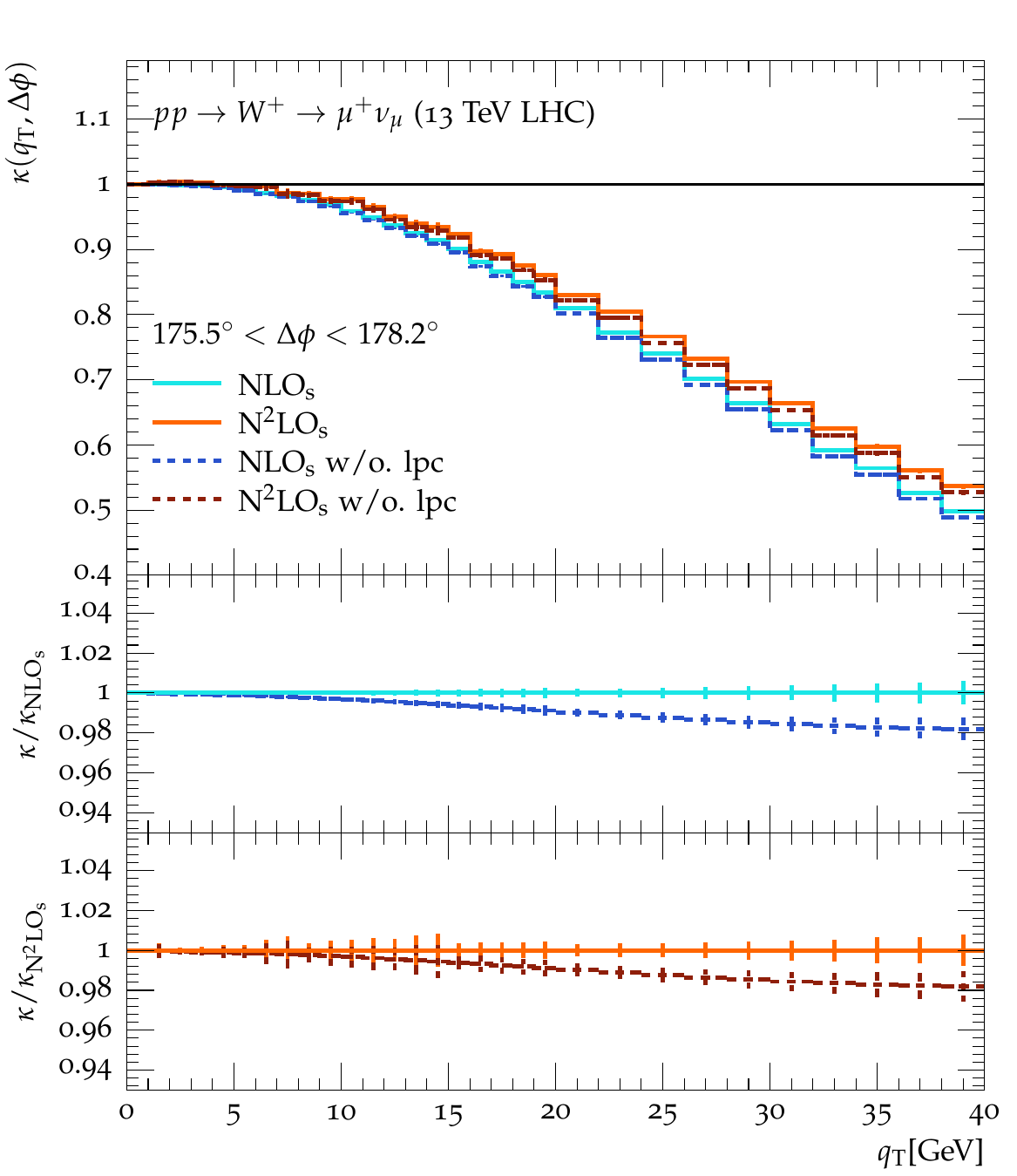}\hfs\hfs
  \includegraphics[width=.22\textwidth]{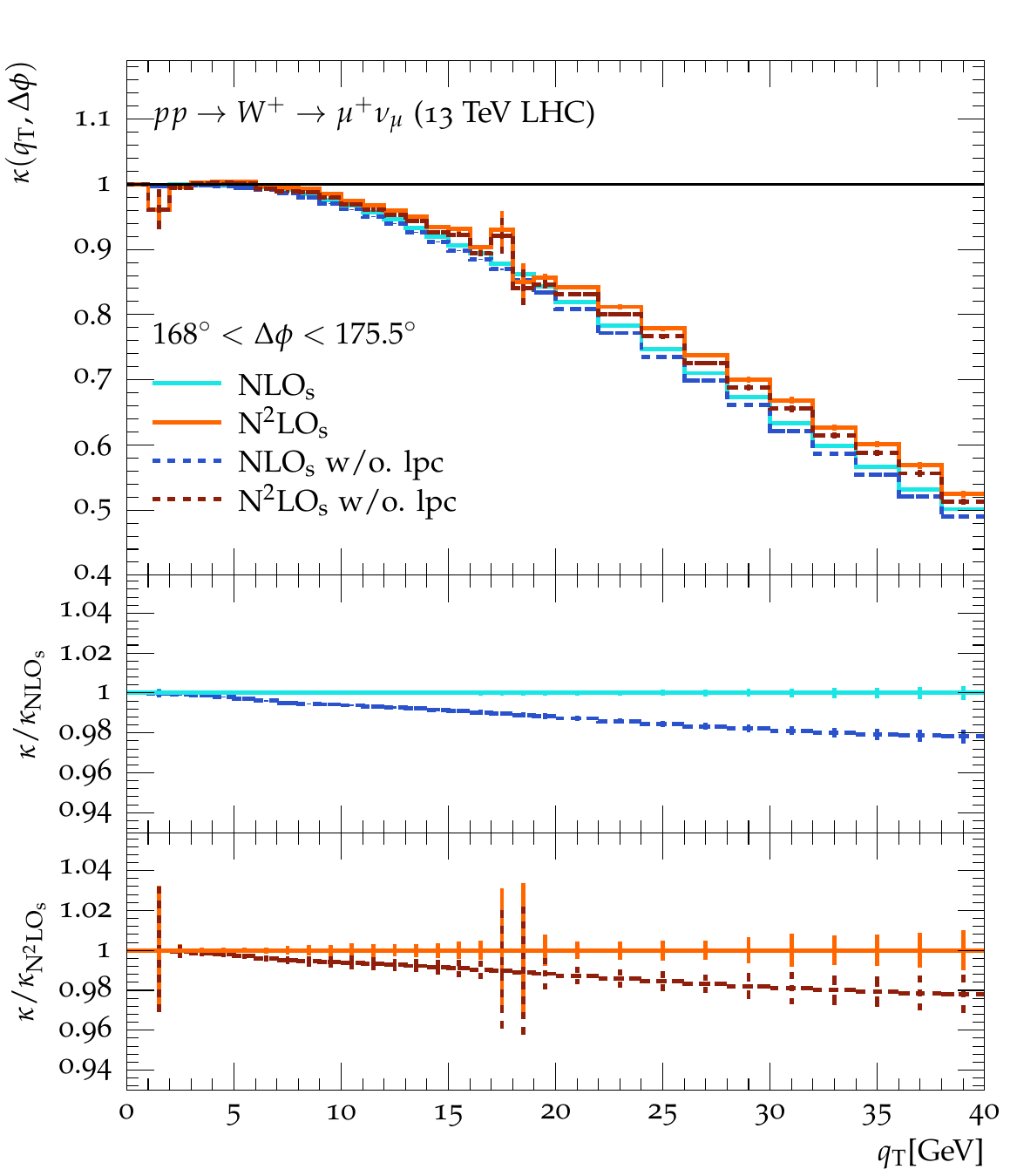}\hfs\hfs
  \includegraphics[width=.22\textwidth]{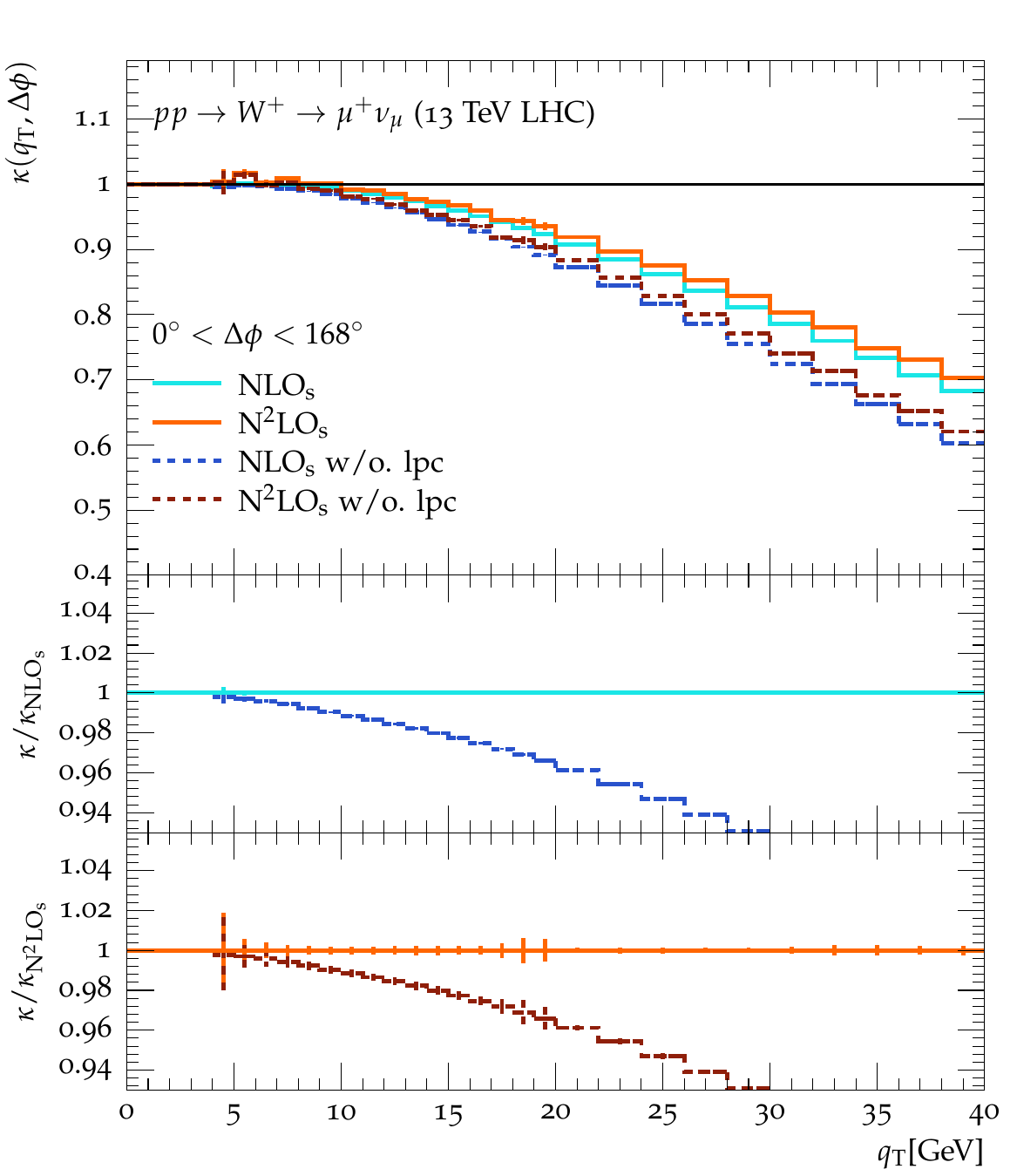}\\[1mm]
  \includegraphics[width=.22\textwidth]{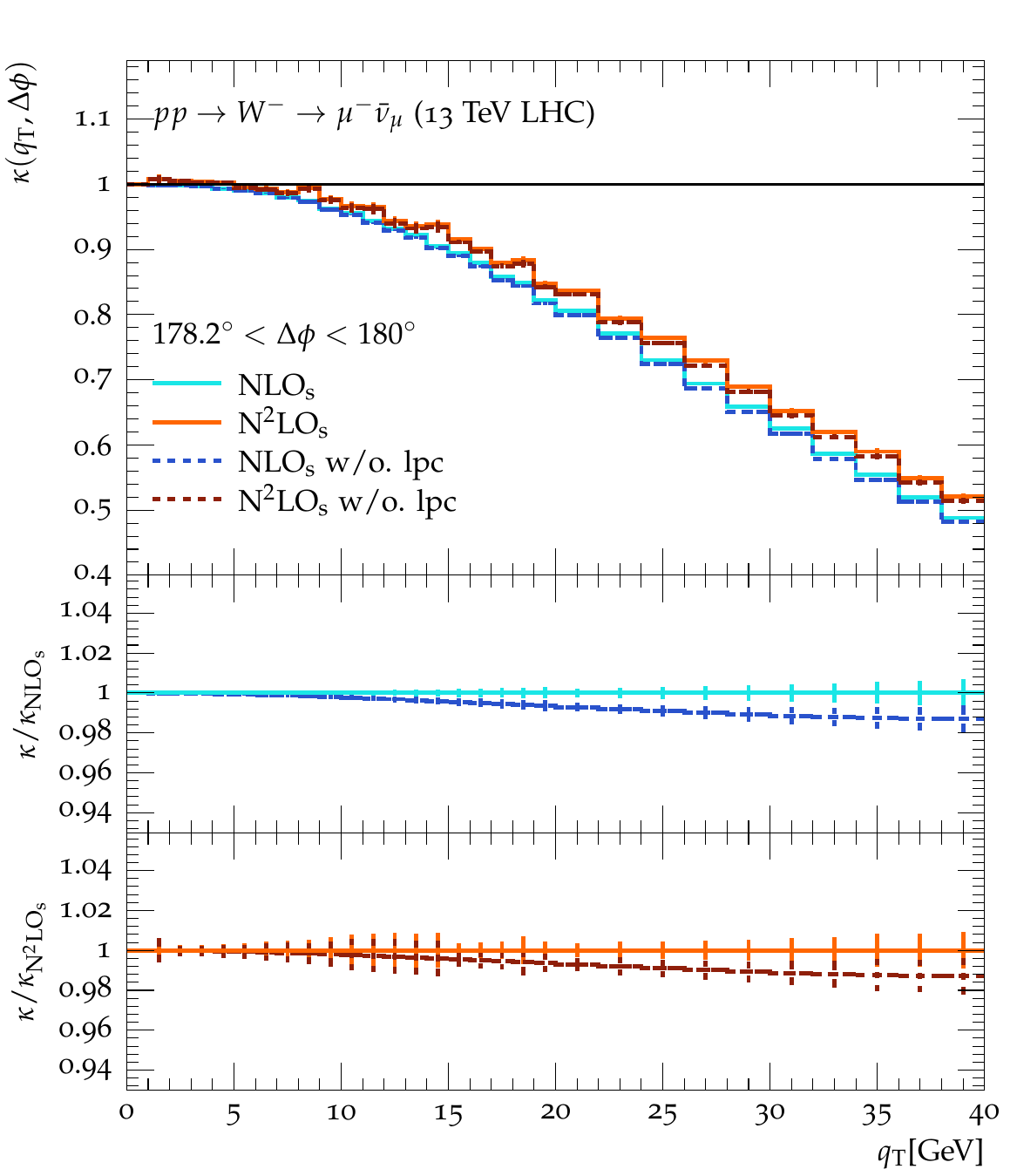}\hfs\hfs
  \includegraphics[width=.22\textwidth]{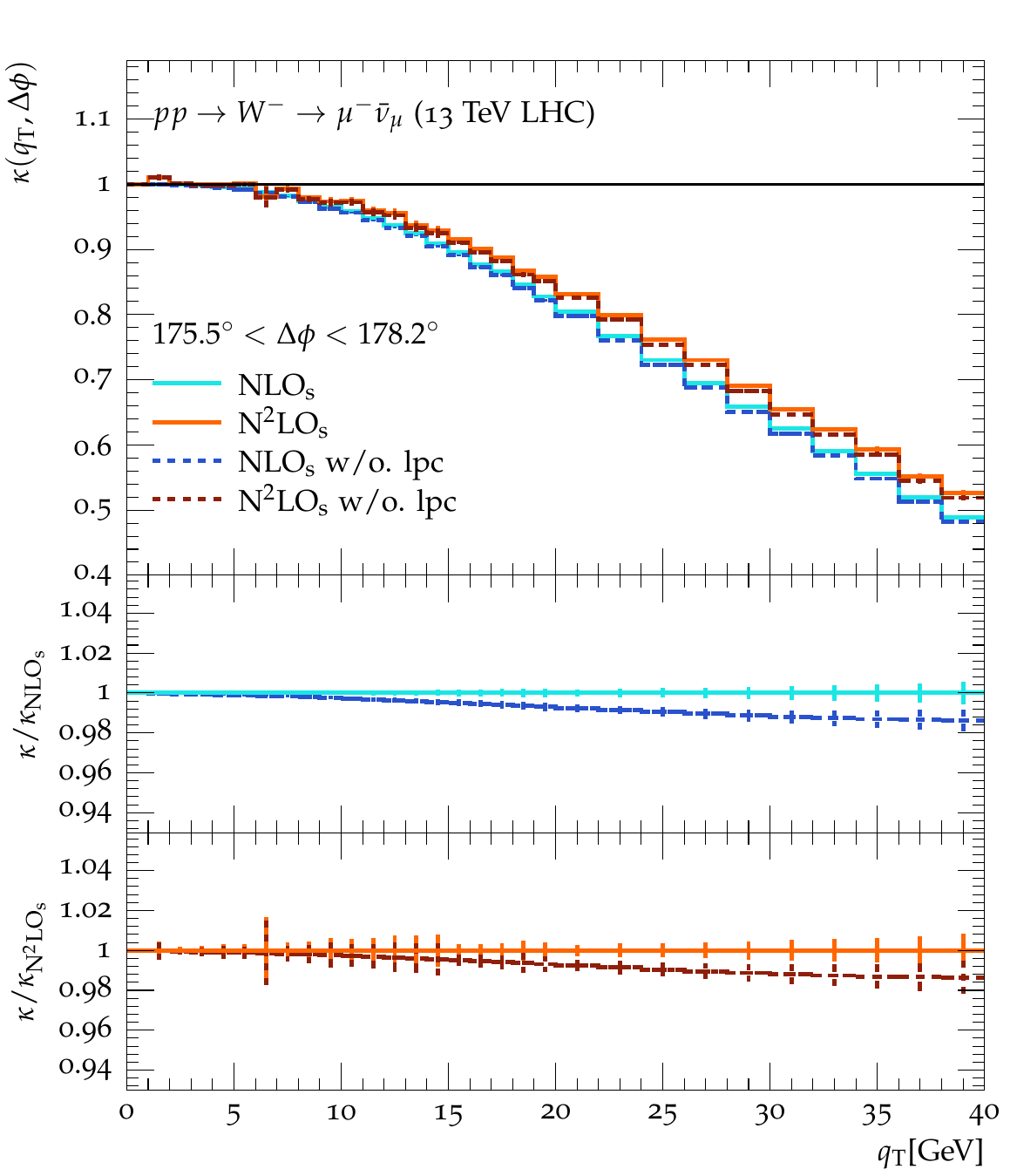}\hfs\hfs
  \includegraphics[width=.22\textwidth]{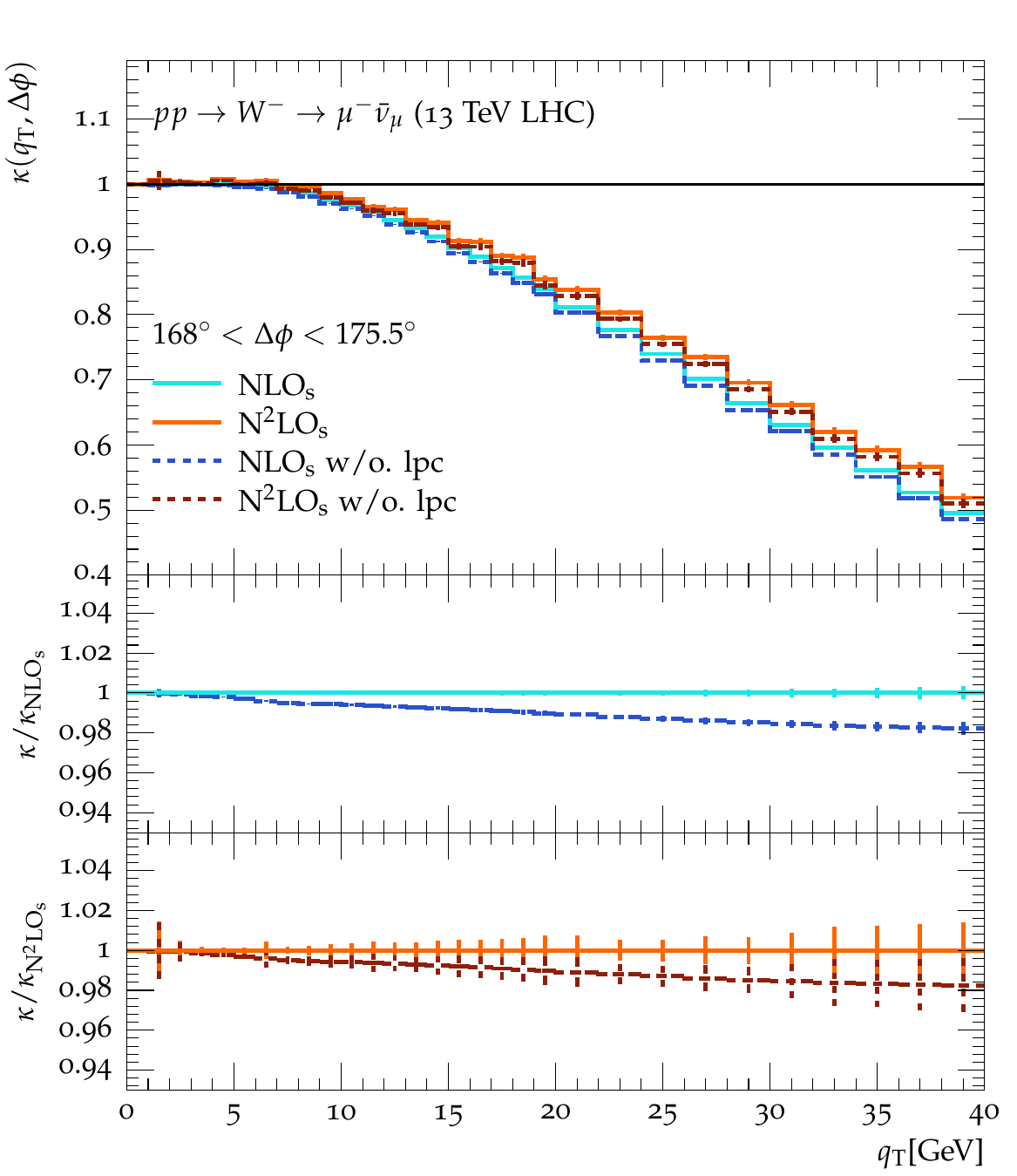}\hfs\hfs
  \includegraphics[width=.22\textwidth]{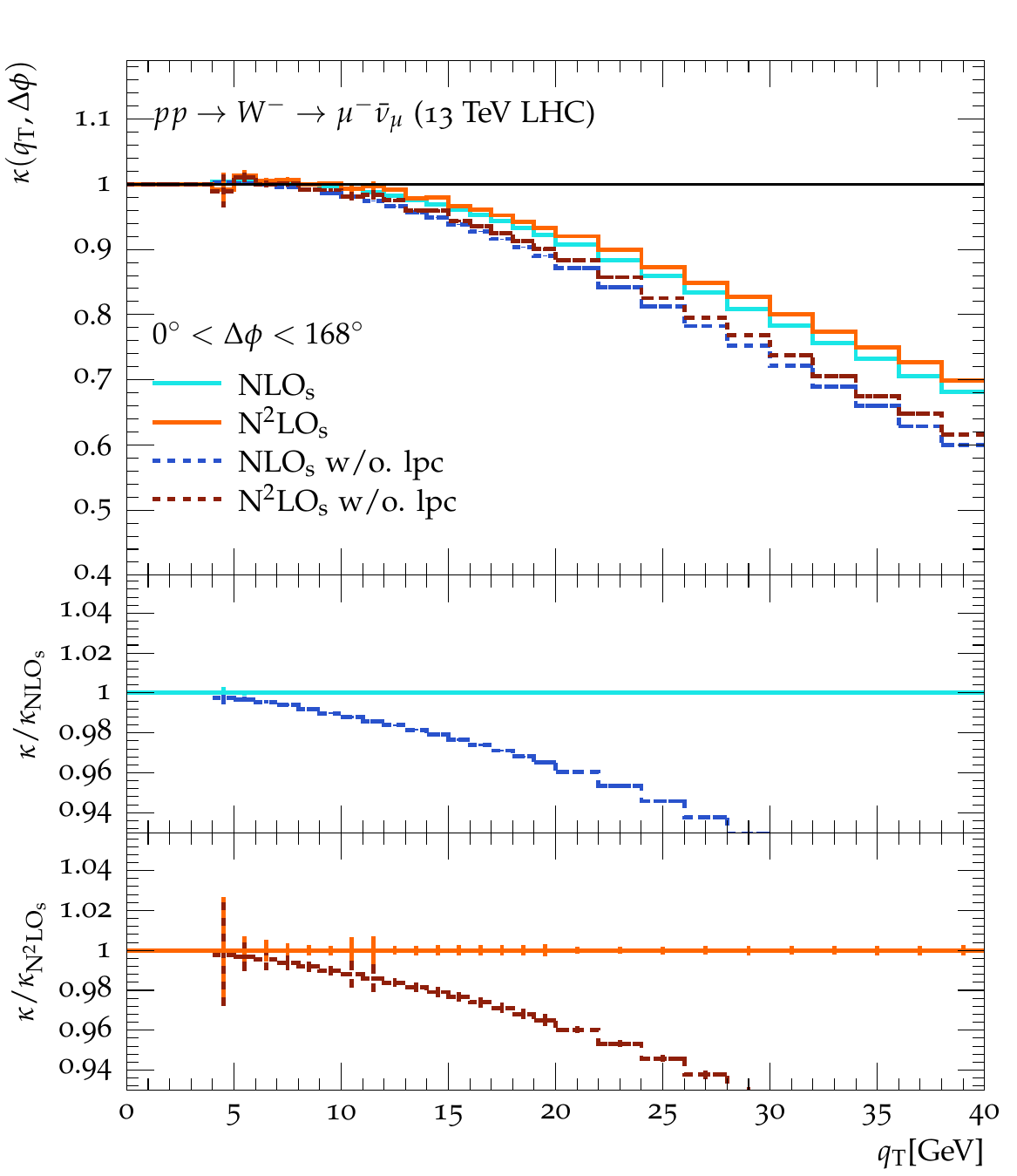}\\[1mm]
  \caption{Numeric impacts of the leptonic power corrections on the $\qT$ spectra.        }
  \label{fig:apps:lpc:tmd}
\end{figure}

\begin{figure}[t!]
  \centering
  \includegraphics[width=.22\textwidth]{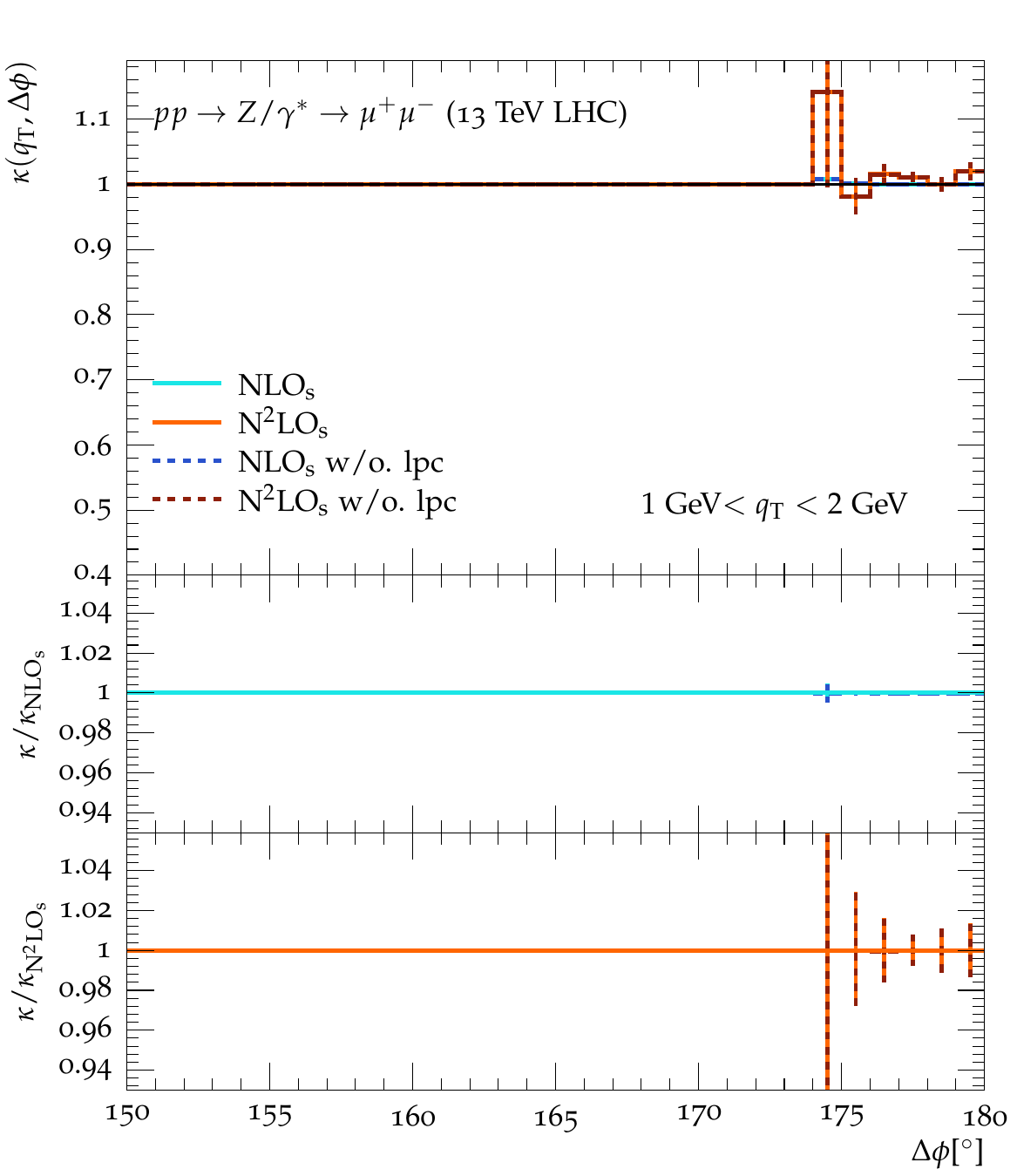}\hfs\hfs
  \includegraphics[width=.22\textwidth]{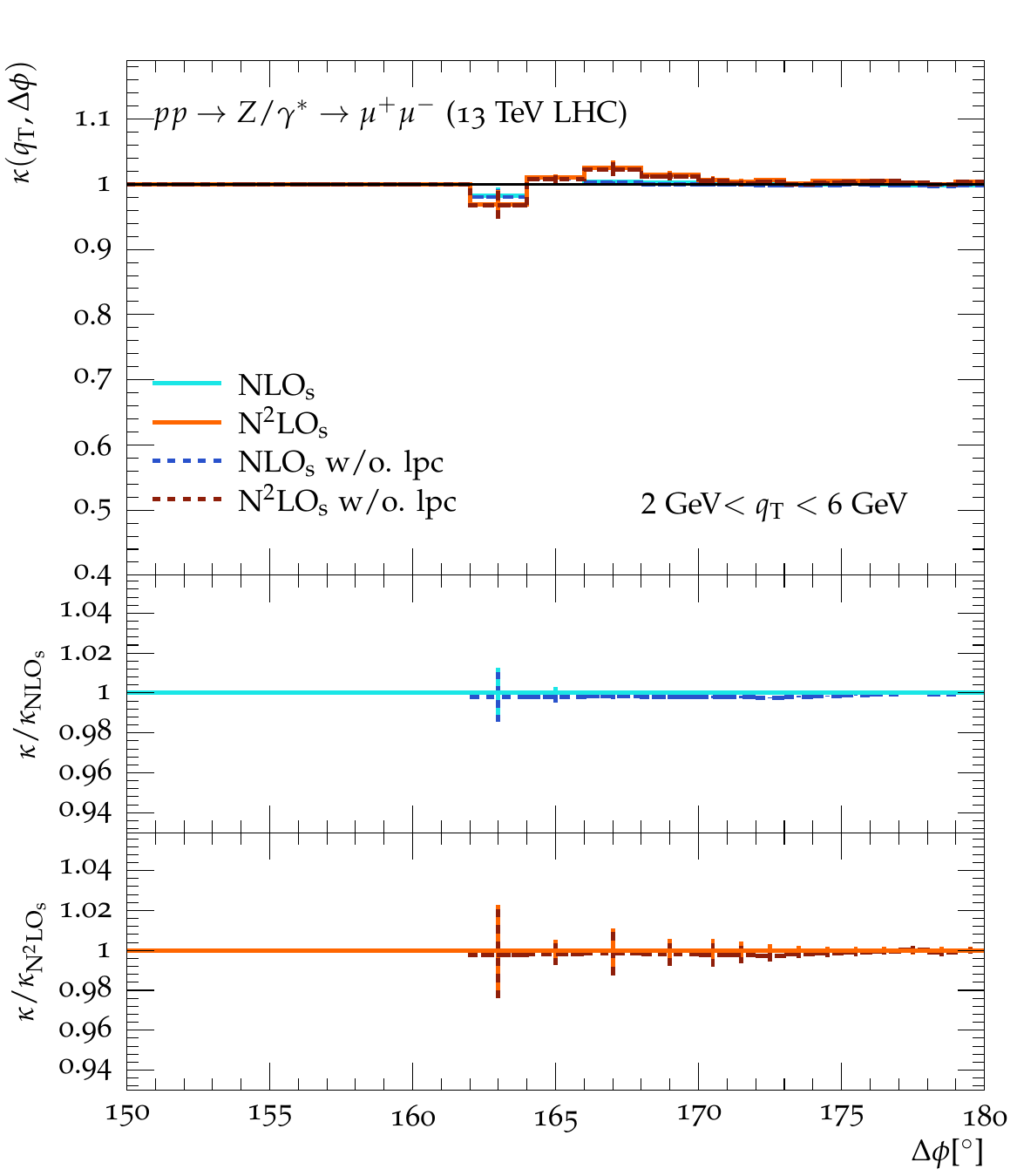}\hfs\hfs
  \includegraphics[width=.22\textwidth]{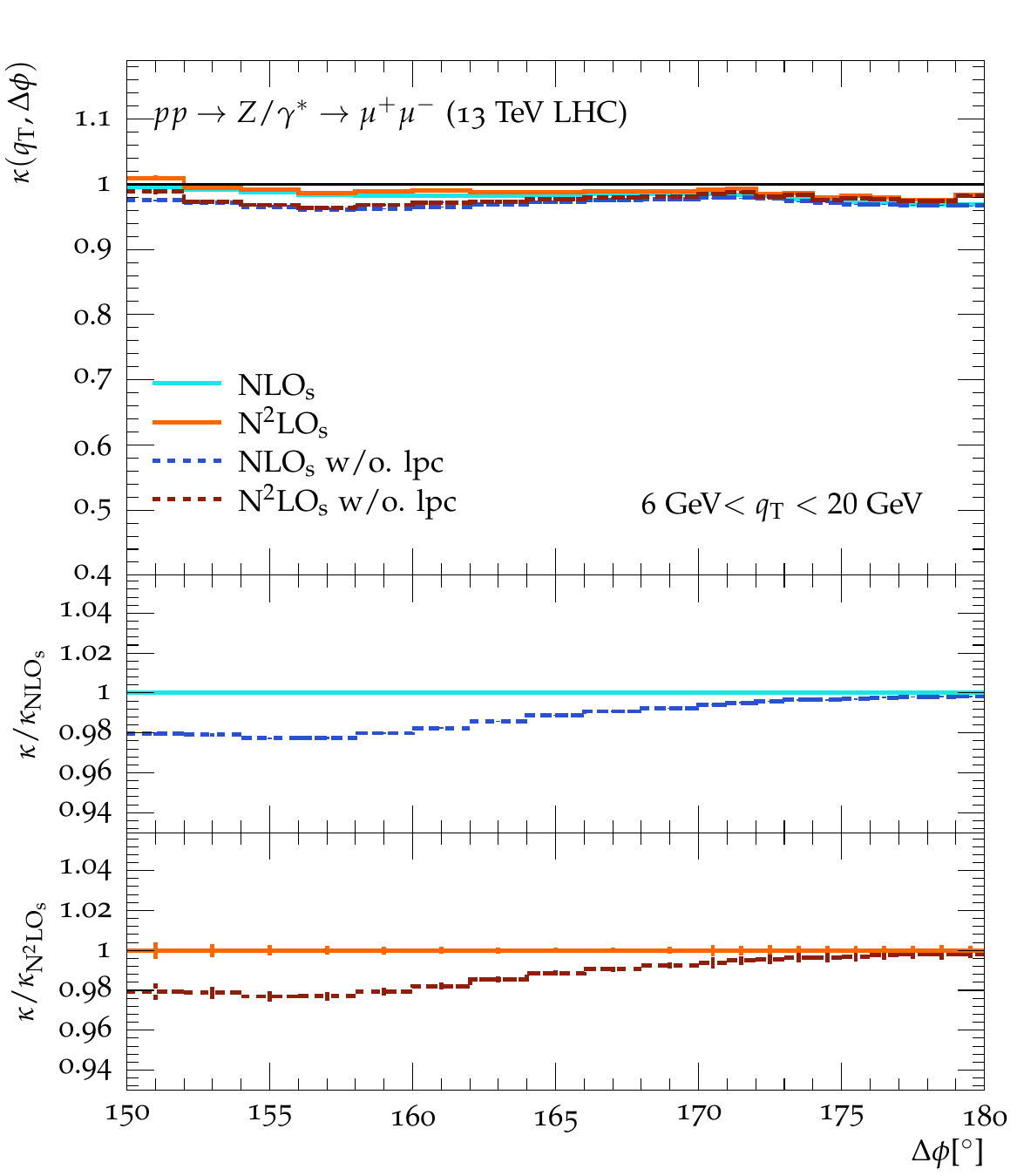}\hfs\hfs
  \includegraphics[width=.22\textwidth]{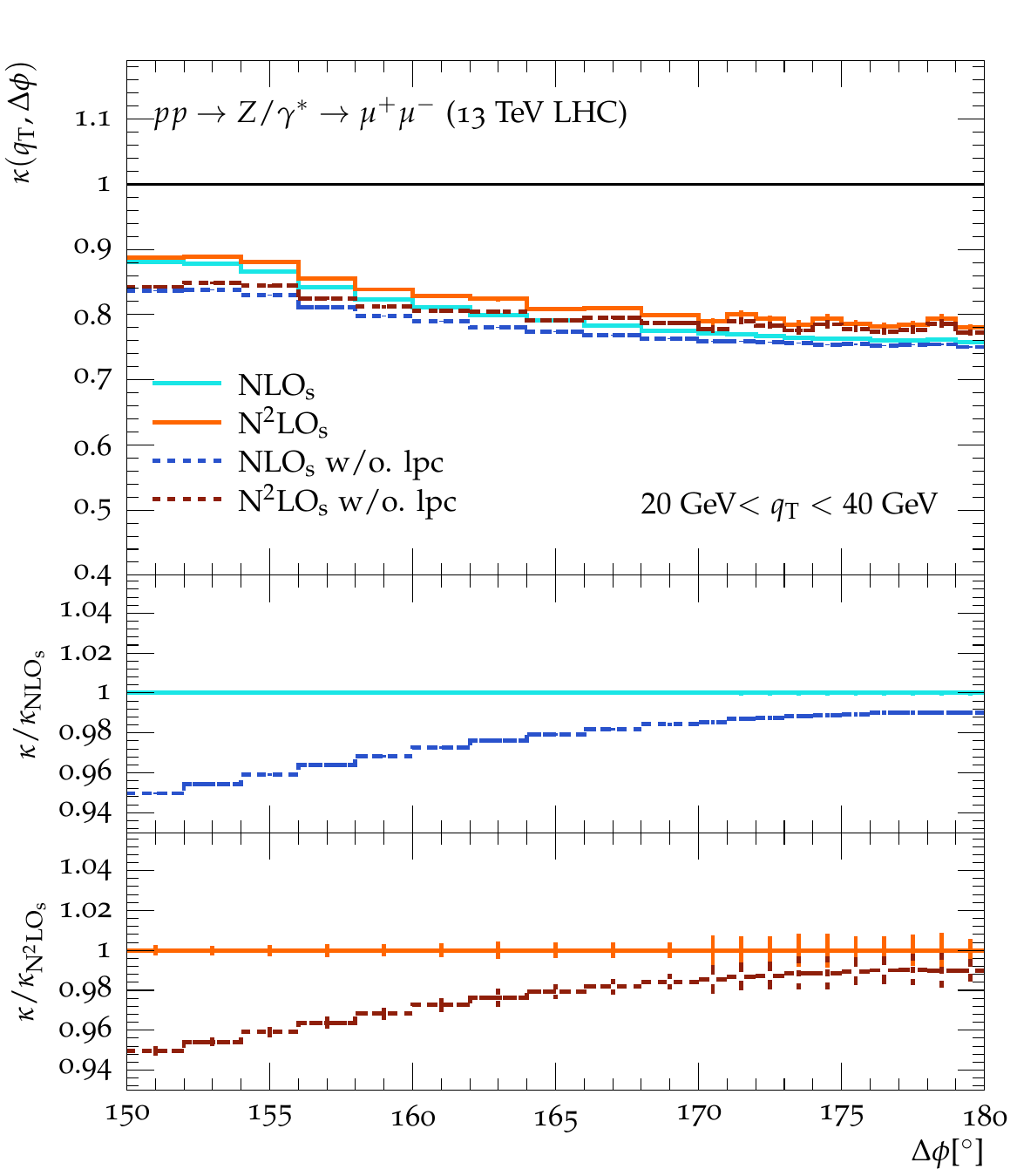}\\[1mm]
   \includegraphics[width=.22\textwidth]{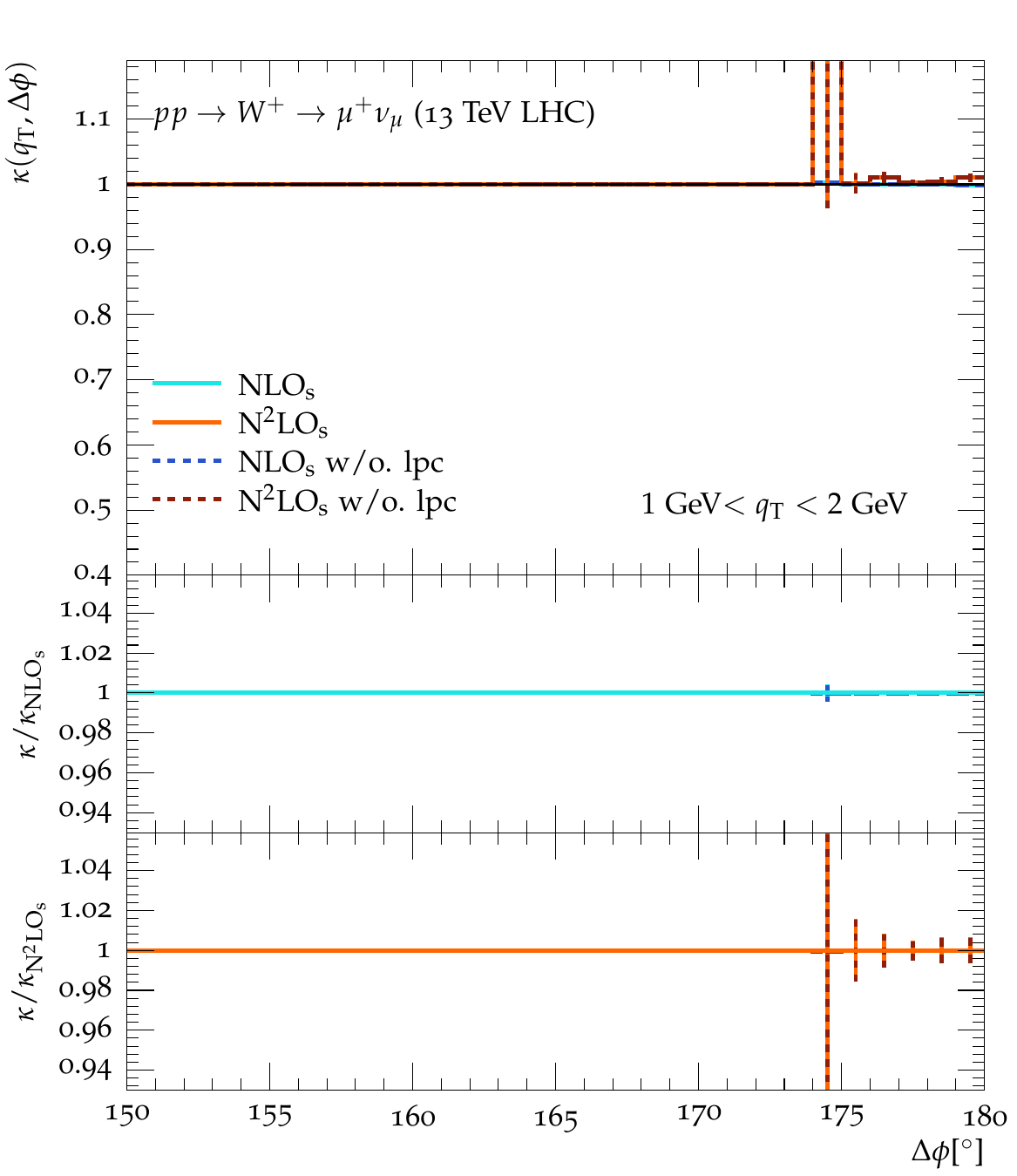}\hfs\hfs
  \includegraphics[width=.22\textwidth]{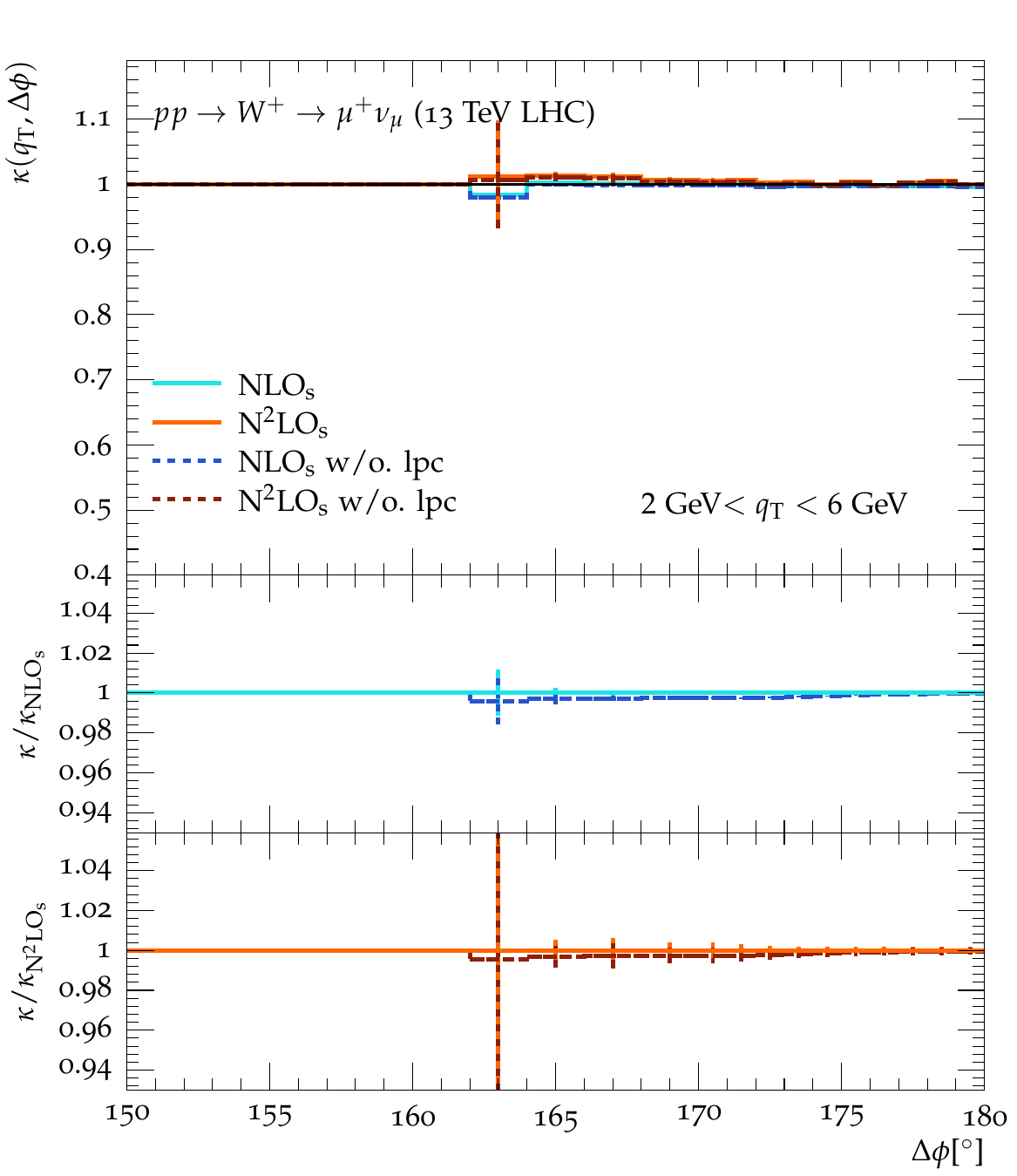}\hfs\hfs
  \includegraphics[width=.22\textwidth]{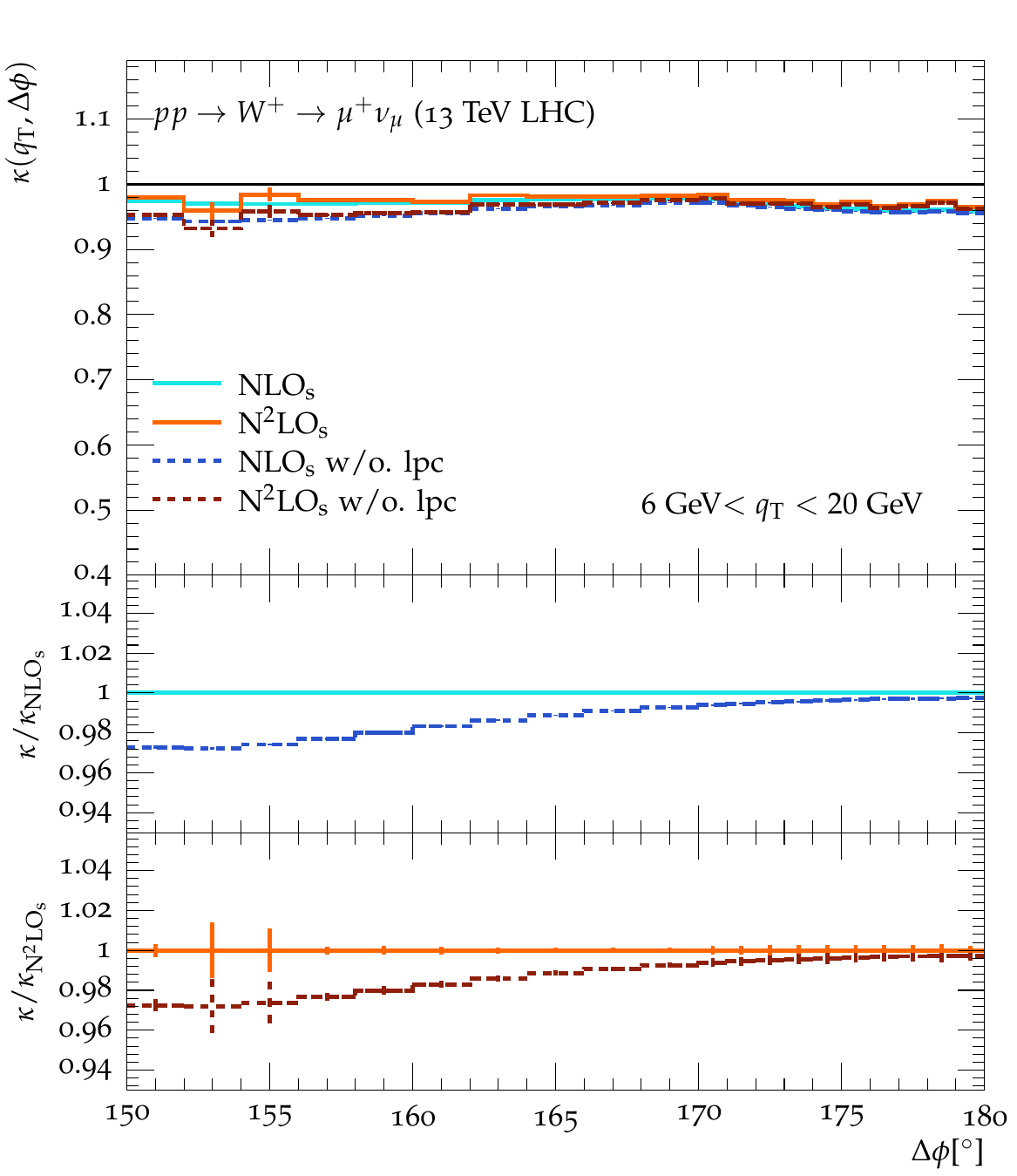}\hfs\hfs
  \includegraphics[width=.22\textwidth]{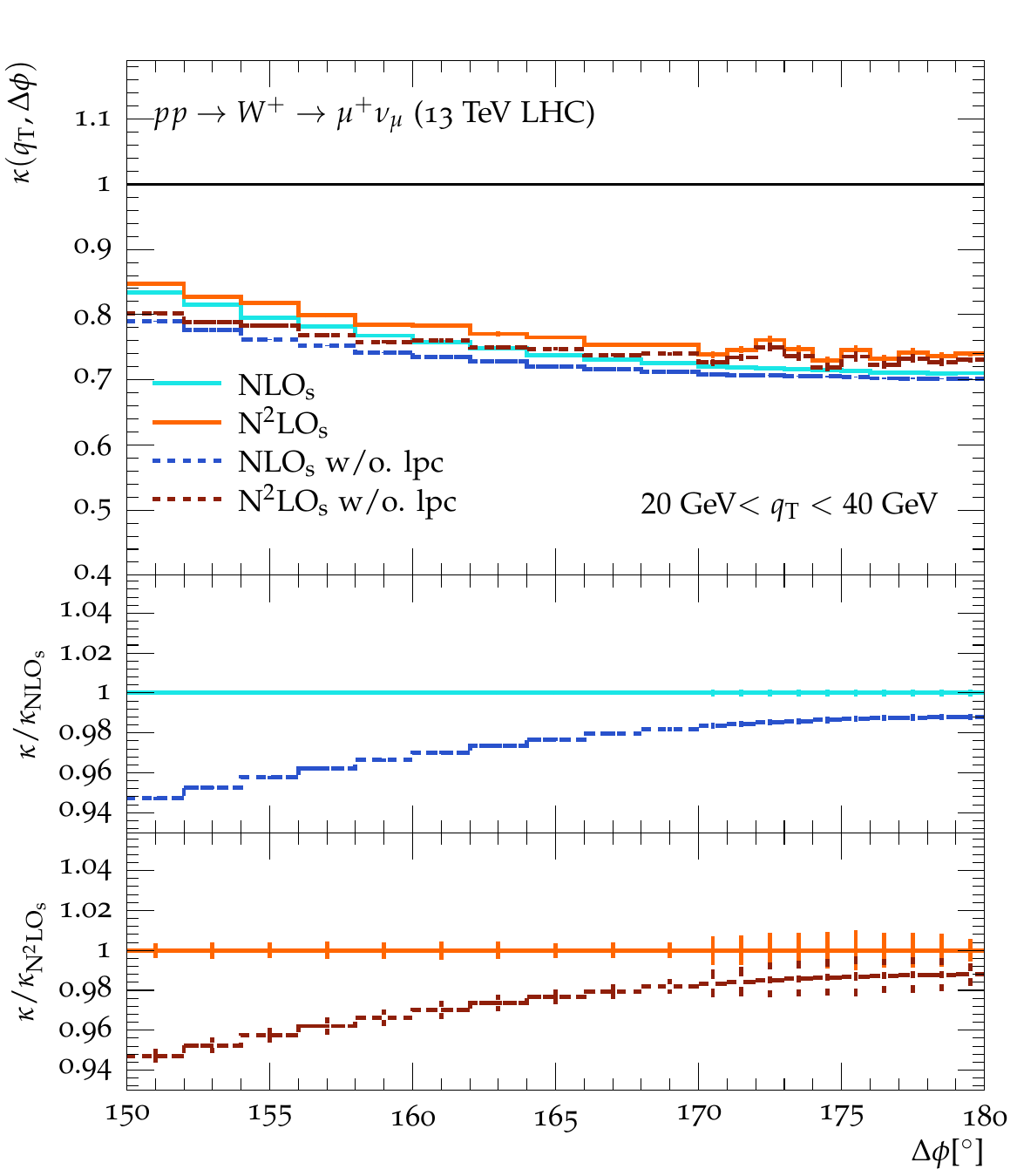}\\[1mm]
  \includegraphics[width=.22\textwidth]{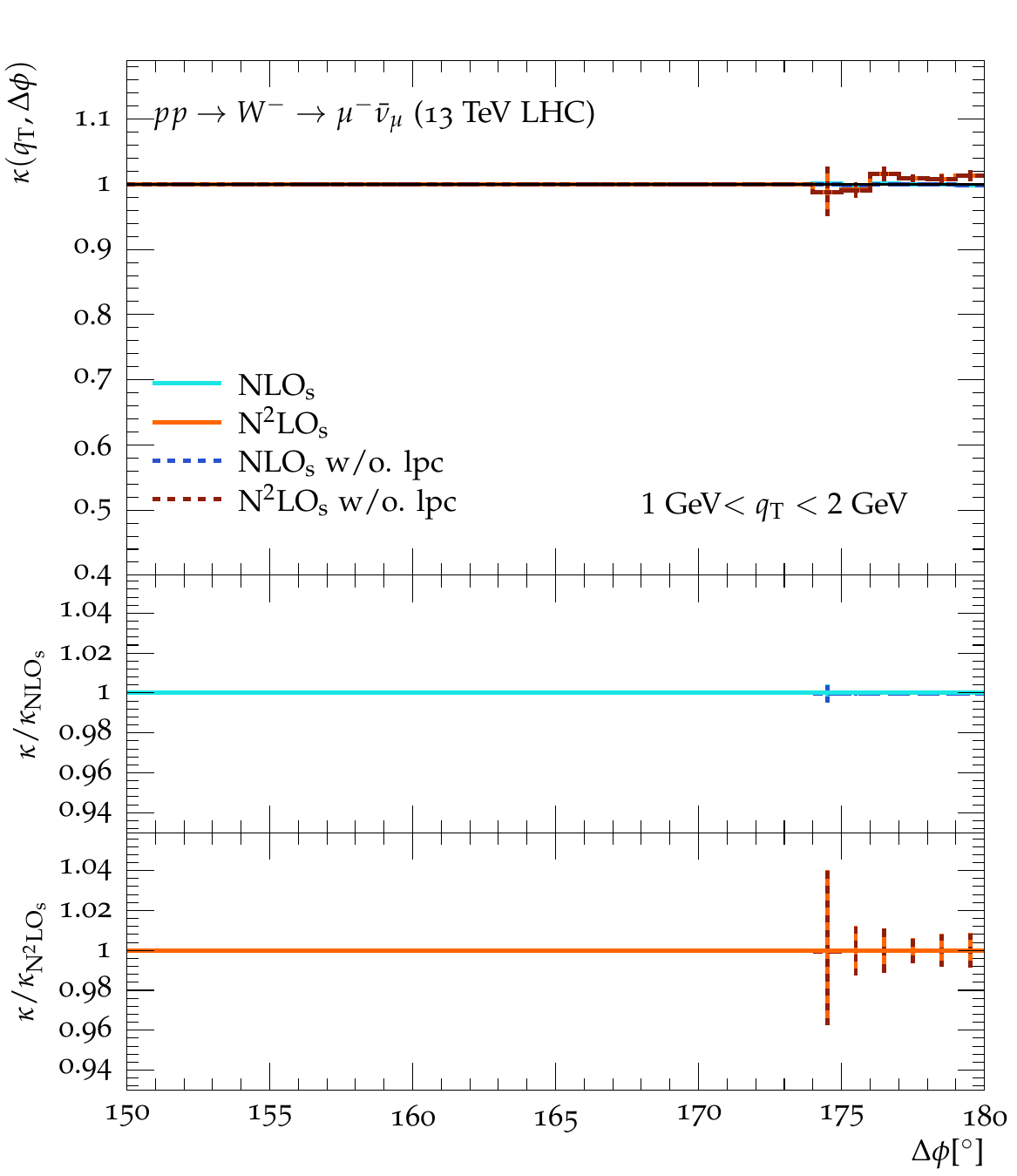}\hfs\hfs
  \includegraphics[width=.22\textwidth]{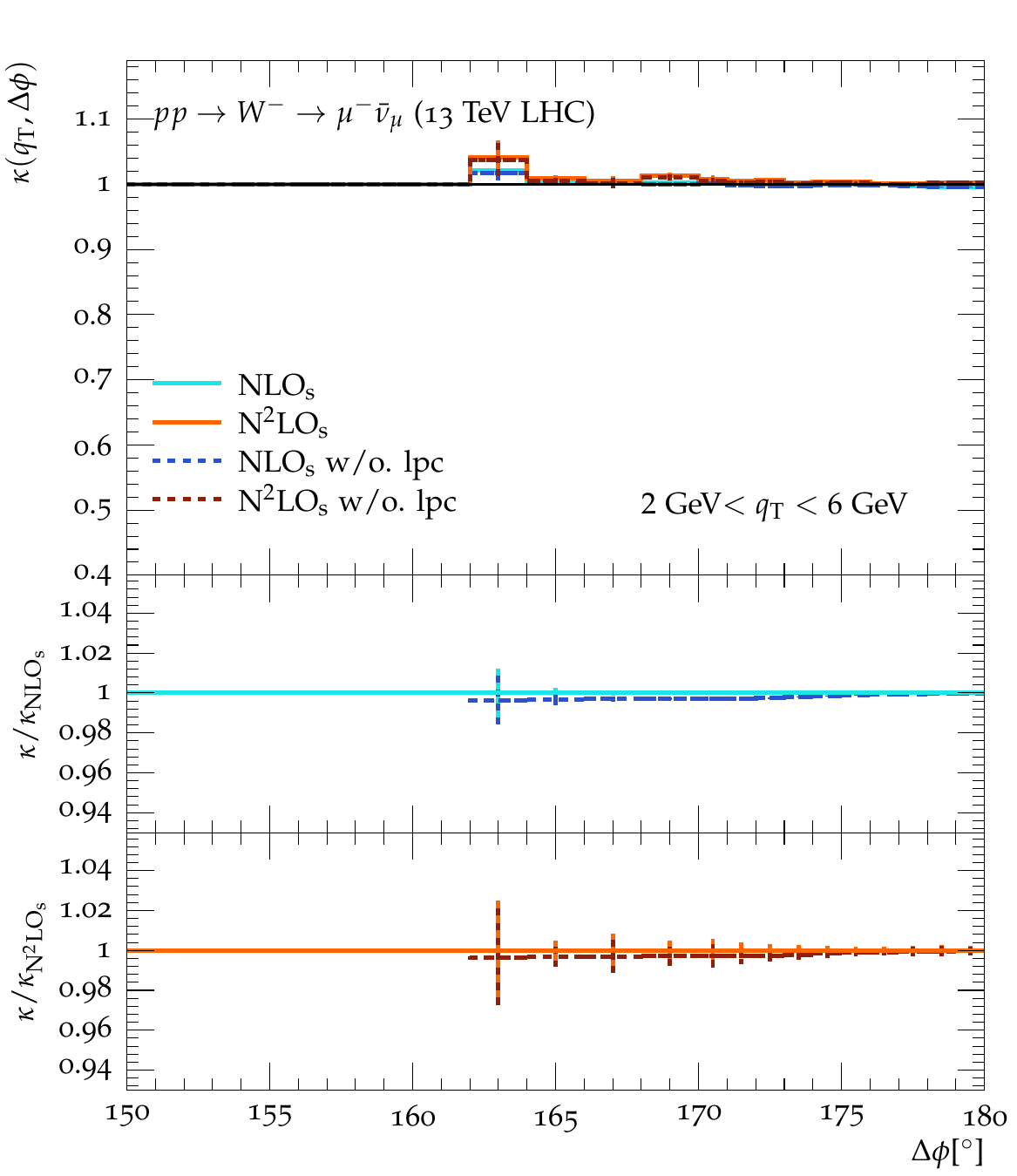}\hfs\hfs
  \includegraphics[width=.22\textwidth]{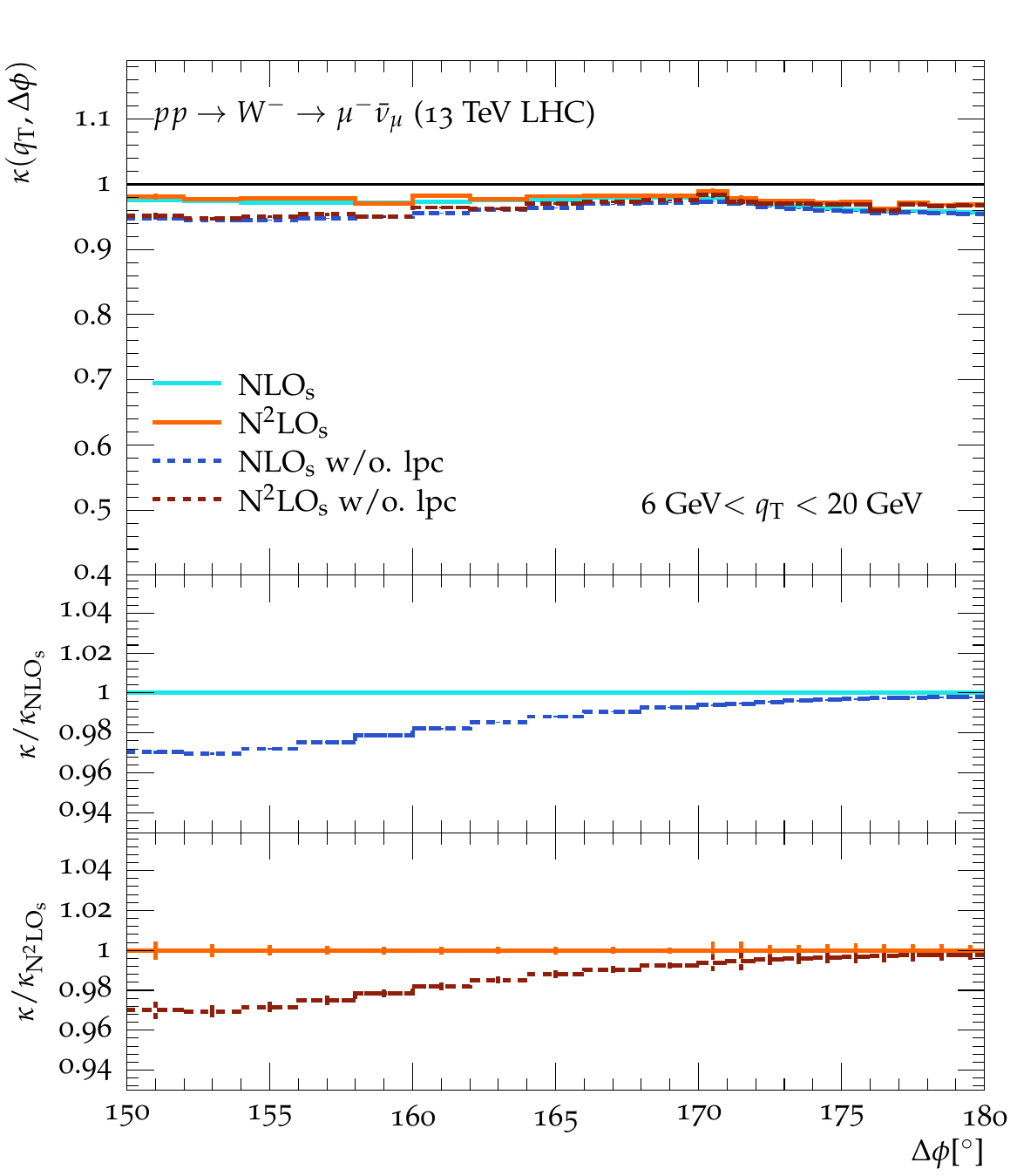}\hfs\hfs
  \includegraphics[width=.22\textwidth]{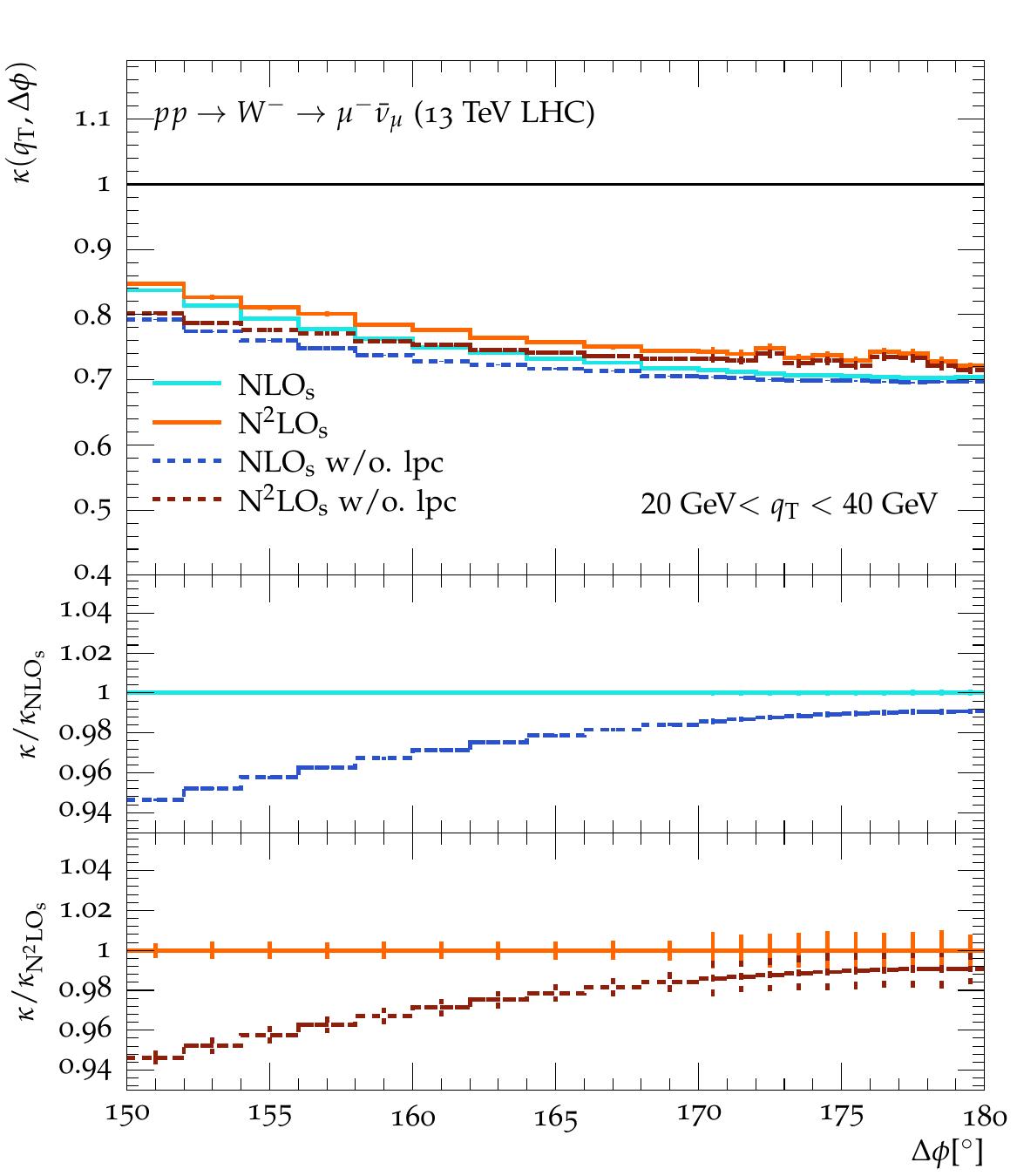}\\[1mm]
  \caption{Numeric impacts of the leptonic power corrections on the $\dphi$ spectra. 
  }
  \label{fig:apps:lpc:dphi}
\end{figure}

In eq.~\eqref{eq:methods:extra:leppowcorr},  the Lorenz-transformation matrix  $\Lambda_{\ell}(\qTvec)$ has been incorporated into the hard functions for including the leptonic power corrections. In this part, we will investigate its numerical influences on the double-differential observable $\done^2 \sigma/(\done \qT\done\dphi)$. To this end, we define the ratio of the approximate results to the exact ones as follows, 
\begin{equation}
\kappa( \qT,\,\dphi)\equiv\frac{\done^2 \sigma_{\mathrm{exp}}/(\done \qT\done\dphi) }{\done^2 \sigma_{\mathrm{f.o.}}/(\done \qT\done\dphi) }, 
\end{equation}
where $\sigma_{\text{f.o.}}$ denotes the exact fixed-order perturbative 
result.  $\sigma_{\mathrm{exp}}$ represents the perturbative expansions of the resummed distribution in eq.~\eqref{eq:methods:res:TMD_Resummation}. For comparison, we will calculate $\sigma_{\mathrm{exp}}$ in two different ways: with or without $\Lambda_{\ell}(\qTvec)$ contributions.  While the result experienced eq.~\eqref{eq:methods:extra:leppowcorr} is still named $\mathrm{N^{(m)}LO}_{\mathrm{s}}$, same as those in Figs.~(\ref{fig:results:val:qt}-\ref{fig:results:val:dphi}), those excluding $\Lambda_{\ell}(\qTvec)$ effects are labelled as  ``$\mathrm{N^{(m)}LO}_{\mathrm{s}}$ w/o. lpc''. Here the superscript ``m'' specifies the  expansion order in $\alpha_s$.  Throughout this section, the results in former case will be depicted in the solid lines, and to distinguish, the dashed ones illustrate the later case.

 In Fig.~\ref{fig:apps:lpc:tmd}, we exhibit the numerical results for $\kappa( \qT,\,\dphi)$ after integrating out $\dphi$ over the following four intervals:
 
 \begin{tabular}{lllll}
 1) 178 .2$^{\circ}$&$<\dphi<$&180$^{\circ}$; \\
 2) 175 .5$^{\circ}$&$<\dphi<$&178.2$^{\circ}$;\\
  3) 168$^{\circ}$&$<\dphi<$&175.5$^{\circ}$; \\
  4) 0$^{\circ}$&$<\dphi< $&168$^{\circ}$.  
\end{tabular}

In the small $\qT$ regime, the $\kappa$ distributions in the four slices all approach the unity and the differences due to the leptonic power corrections are insensible. This phenomenon is in agreement with the observations in Sec.~\ref{sec:results:val} and also validates the leading power factorization in eq.~\eqref{eq:methods:TMD_factorisation}. However, with the increase in the $\qT$, the power corrections start to manifest themselves. For instance, in the intermediate region $\qT\sim20$~GeV, one can find that the $\kappa$ spectra obviously deviate from the unity grid  and furthermore the discrepancies arising from the $\Lambda_{\ell}(\qTvec)$ incorporation emerge.  One interesting phenomenon is that differing from the first three slices, the $\kappa$ spectra in the last slice are fairly sensitive to the leptonic power corrections. 
To be specific, around the point $\qT=25$~GeV, the  $\Lambda_{\ell}(\qTvec)$ incorporations in the first three slices improve the dashed lines by nearly $1\%$ towards the ``f.o.'' ones,  whilst it escalates to $5\%$ for the last slice.  

 To interpret this,  note the topological configuration of final leptons in the fourth slice considerably differs from the others. 
For the first three slices,  as required by the small value of $(\pi- \dphi)$, the final leptons are almost in the back to back configuration. So for the $\qT\ge25$~GeV region, the direction of $\qTvec$
tends to be aligned with $\hat{ \vec{p}}_{_{\ell,\mathrm{T}}}$ in the transverse plane, so as to avoid the energetic recoil enhancing the $(\pi- \dphi)$ value.  Here $\hat{ \vec{p}}_{_{\ell,\mathrm{T}}}$ represents the (anti-)lepton momentum in the rest frame of the lepton pair. Given the limit $\qTvec \parallel \hat{ \vec{p}}_{_{\ell,\mathrm{T}}}$, the  $\Lambda_{\ell}(\qTvec)$  matrix acts only on the time-like and the longitudinal components of $L_{V}$, whereas the later case is only able to polarize perpendicularly to $\hat{ \vec{p}}_{_{\ell,\mathrm{T}}}$ in the massless limit.\footnote{Generically, one can decompose the transformation matrix as $\Lambda^{\mu}_{\ell,\nu}(\qTvec)\equiv \Lambda^{\mu}_{z} \Lambda^{\mu}_{x} \Lambda^{\mu}_{y} $. Here $\Lambda^{\mu}_{z}$ represents the boost-transformation along the colliding beam direction. $\Lambda^{\mu}_{x(y)}$ stands for those in the transverse plane.
 Here we only consider the transformation in the transverse plane, since the boost along the beam direction has vanishing impacts after being contracted against the leading power hadronic current. } After the contraction, the $\Lambda_{\ell}(\qTvec)$ matrix is effectively reduced to the metric tensor and therefore gives rise to mild influences in the first three slices. 
The situation in the last slice is however different. A variety of $(\pi- \dphi)$  therein can encourage the stronger recoils against  lepton momenta, so that $\qTvec$ is capable of developing sizable perpendicular component with respect to $\hat{ \vec{p}}_{_{\ell,\mathrm{T}}}$, which in turn permits $\Lambda_{\ell}(\qTvec)$  non-trivially coupled with $L_{V}$. 

With particular attention is that  except for the boundary bins which suffer from statistical issues, the $\kappa$ values are all enhanced towards the unity grid after incorporating the leptonic power correction matrix $\Lambda_{\ell}(\qTvec)$.  This indicates that $\Lambda_{\ell}(\qTvec)$ indeed compensates for the power expansions and therefore demonstrates the effectiveness of our strategy.

 Additionally, Fig.~\ref{fig:apps:lpc:dphi}  illustrates $\kappa( \qT,\,\dphi)$s after integrating out $\qT$ over the four intervals as follows:
 
  \begin{tabular}{lllll}
 1) $1$ GeV$< q_{\mathrm{T}} <2$ GeV; \\
 2) $2$ GeV$<q_{\mathrm{T}} <6$ GeV;\\
  3) $6$ GeV$<q_{\mathrm{T}} <20$ GeV; \\
  4) $20$ GeV$<q_{\mathrm{T}}<40$ GeV.  
\end{tabular}

 For the first two slices, since the $\qT$ value is constrained to be fairly small and thus the power corrections are  significantly suppressed therein, 
 it is challenging to observe any impacts from $\Lambda_{\ell}(\qTvec)$. However, with the value of $\qT$  growing in the next two slices, it emerges that the appreciable discrepancies between  the results of $\mathrm{N^{(m)}LO}_{\mathrm{s}}$ and those without $\Lambda_{\ell}(\qTvec)$ insertions.  Instructively, it is observed that the  $\Lambda_{\ell}(\qTvec)$ influences are gradually corrupted when $\dphi$ moves towards $180^{\circ} $.  For example,  $\Lambda_{\ell}(\qTvec)$ can account for $3\%\sim5\%$ contribution in the $\dphi\sim 150^{\circ}$ region, whilst the proportion  decreases to less than $1\%$ in the vicinity of $\dphi=180^{\circ}$. This phenomenon can be interpreted by the arguments above:  as to the non-vanishing $\qT$, the greater $\dphi$  value, the less    opportunities that $\Lambda_{\ell}(\qTvec)$ is able to couple with the leptonic current.  
 In this way, the $\Lambda_{\ell}(\qTvec)$ contributions in the right end of the graphs are relatively weaker than the left one, which therefore produces the inverse relationships between   the $\Lambda_{\ell}(\qTvec)$ influences  and  $\dphi$  as illustrated in the ratio plots of the last two slices.

\section{Fixed-order functions}

\subsection*{Axial Wilson coefficient \texorpdfstring{$C_t$}{Ct}}
\label{app:Ct}

As illustrated in \eqref{eq:methods:HijV:Ja_EFT}, the axial-vector effective current comprises a novel structure  
$C_t \mathcal{O}_s$ to restore the RGI in LEEFT and encode the contributions induced by the top loops. 
In general,  $C_t$ can be determined by matching the SM amplitudes 
induced by $\bar{t}\gamma^{\mu}\gamma_5t-\bar{b}\gamma^{\mu}\gamma_5b$ 
onto those from $\mathcal{O}_s$ in the limit $M_{L}\ll m_t$. 
However, owing to differences in the $\mathcal{O}_s$ renormalisation, 
the expressions for $C_{t}$ differ. 
Two prescriptions exist in the literatures: 
1) Larin's~\cite{Larin:1993tq} and 
2) Chetyrkin's~\cite{Chetyrkin:1993jm,Chetyrkin:1993hk,Chetyrkin:1993ug}.
In the latter case, the results for $C_t$ up to four-loop accuracy 
have been computed~\cite{Chetyrkin:1993jm,Chetyrkin:1993hk,Kniehl:1989bb,Kniehl:1989qu,Chetyrkin:1993ug,Baikov:2012er,Rittinger:2012bha}.
In the former scheme, the axial-anomaly form factor induced by a top 
quark loop has been calculated in~\cite{Bernreuther:2005rw} at the 
two-loop  level. 
The $C_t$ expressions in different schemes can be related as
\begin{equation}\label{eq:methods:convertor:ct}
  \begin{split}
    \frac{{C_t^{\mathrm{Larin}}}}{{C_t^{\mathrm{Chetyrkin}}}}
    =
      \frac{Z_5^{f,\text{Chetyrkin}}}{Z_5^{f,\text{Larin}}}
    =&\;
        1
        -\left(\frac{\alpha_s}{4\pi}\right)^2
         \left(\frac{3}{2}\,C_F N_F\right)
        +\left(\frac{\alpha_s}{4\pi}\right)^3
         \left(-26\,C_AC_FN_F\,\zeta_3+\frac{163}{27}C_A C_FN_F\right.\\
     & 
       \left.\hspace*{53.5mm}
         +24C_F^2N_F\zeta_3-\frac{35}{2}C_F^2N_F-\frac{88}{27}C_FN_F^2 \right)\,,
  \end{split} 
\end{equation}
where the superscripts represent the schemes of 
$\mathcal{O}_s$ renormalisation. 
This paper employs $\alpha_s$ which is renormalized with 
$N_F=5$ active quarks throughout.
In the Chetyrkin's prescription, the finite renormalisation 
constant $Z_5^{f,\text{Chetyrkin}}$ is actually equal to 
the non-singlet case $Z_{\text{ns}}^{f,\text{Larin}}$ in 
Larin's scheme. 
The $Z_{\text{ns}}^{f,\text{Larin}}$ expression up to 
$\mathcal{O}(\alpha^3_s)$ can be found in 
\cite{Larin:1991tj,Chetyrkin:1993ug}.  
As to the $Z_5^{f,\text{Larin}}$ expression, the first two 
order investigation was carried out some time ago in 
\cite{Larin:1993tq}, while the third order result is given 
in a very recent publication~\cite{Ahmed:2021spj}.  
In this work, we use Larin's scheme. 
The corresponding expression for $C_t$ reads,
\begin{equation}\label{eq:apps:def:ct}
  \begin{split}
    C_t
    =&\;
      -\frac{1}{N_F}
      +\left(\frac{\alpha_s}{4\pi}\right)^2
       \left(-8L_t+4\mhhl\right)
      +\left(\frac{\alpha_s}{4\pi}\right)^3
       \left(-\frac{184}{3}L_t^2-\frac{784}{9}L_t+208\zeta_3-\frac{6722}{27}\right)\;.
  \end{split} 
\end{equation}
where $L_t=\ln(\mu^2/m_t^2)$.  
Here the tree-level result $(-1/N_F)$ balances 
the $\mathcal{O}_s/N_F$ term in $\Delta^{\mathrm{ns}}_3$. 
The singlet contribution starts from the two-loop level and 
the $\mathcal{O}(\alpha^2_s)$ results can be either 
straightforwardly read from the axial-anomaly form factor 
in Ref.~\cite{Bernreuther:2005rw}, or extracted from 
\cite{Collins:1978wz,Chetyrkin:1993jm} with the aid of 
eq.\ \eqref{eq:methods:convertor:ct}. 
We can confirm that both methods result in the same 
$\mathcal{O}(\alpha^2_s)$ expression. 
The third order expression is obtained from 
\cite{Chetyrkin:1993ug} after multiplying the converter 
in eq.\ \eqref{eq:methods:convertor:ct}.
We have checked that the $C_t$ expression here indeed satisfies the RGE in \eqref{eq:methods:HijV:ct_RGE} up to $\mathcal{O}(\alpha_s^3)$, as expected by the
Larin's renormalisation scheme~\cite{Larin:1993tq}.

\subsection*{Non-singlet and singlet functions \texorpdfstring{$C_\text{ns}$, $C_\text{s}^V$, and $C_\text{s}^A$}{Cns, CsV, and CsA}}
\label{app:funcs}

As shown in eq.\ \eqref{eq:method:def:hadcur}, the hadronic currents $H_{\gamma}$,  $H_{Z,V}$ and $H_{Z,A}$ involve a set of coefficients $C_{\text{ns}}$, $C_\text{s}^V$ 
and $C_{\text{s}}^A$ encoding the hard contributions in the loop integrals. 
In practice, they can be extracted from the UV-renormalized and IRC-subtracted quark form factors.  
To cope with the possible ambiguities arising from the $\gamma_5$ manipulation, the form factors with  the Larin's prescription will be adopted throughout, in accordance with the choice in $C_t$.
In the following paragraphs, their expressions will be presented.

First, we specify the non-singlet function $C_{\text{ns}}$. As illustrated in eq.\ \eqref{eq:method:def:hadcur}, $C_{\text{ns}}$ participates  in both the vector and axial-vector hadronic sectors.  Due to the appearance of  $\gamma_5$,  one may expect that the non-singlet vector quark form factor would differ from those in the axial vector case. 
However, since (at least) in Larin's prescription the anticommutativity of $\gamma_5$ is effectively restored for the massless QCD,  the axial-vector form factor coincides with the vector one after the renormalisation~\cite{Larin:1993tq}. Therefore, we  utilize $C_{\text{ns}}$  to represent both cases here. 
Without any loss of generosity, the perturbative expansion for $C_{\text{ns}}$  can be defined as 
\begin{equation}
\begin{split}
C_{\text{ns}}=\sum_{i=0}^{\infty}   \left(  \frac{\alpha_s}{4\pi} \right)^i   C_{\text{ns}}^{(i)}.
\end{split}
\end{equation}
According to the calculations on $\gamma^*q\bar{q}$ amplitudes, the first three coefficients can be given as \cite{Moch:2005id,Baikov:2009bg,Gehrmann:2010ue}
\begin{equation}
\begin{split}
C_{\text{ns}}^{(0)}=&1,\\
C_{\text{ns}}^{(1)}=&-\frac{4 L_H^2}{3}+4 L_H+\frac{2 \pi ^2}{9}-\frac{32}{3} ,\\
C_{\text{ns}}^{(2)}=& \frac{8 L_H^4}{9}-\frac{52 L_H^3}{27}+\left(\frac{28 \pi ^2}{27}-\frac{418}{27}\right)
   L_H^2+\left(-\frac{184 \zeta_3}{3}+\frac{7850}{81}+\frac{20 \pi ^2}{27}\right) L_H+\frac{2356 \zeta_3}{27}+\frac{46 \pi ^4}{81}\\
   &-\frac{277 \pi ^2}{81}-\frac{85081}{486},\\
C_{\text{ns}}^{(3)}=&-\frac{32 L_H^6}{81}-\frac{80 L_H^5}{81}+\left(\frac{2486}{81}-\frac{128 \pi ^2}{81}\right)
   L_H^4+\left(\frac{736 \zeta_3}{9}-\frac{23284}{243}-\frac{416 \pi ^2}{243}\right) L_H^3+\left(\frac{11024 \zeta_3}{81} \right.\\
   &\left.-\frac{209686}{729}+\frac{7052 \pi ^2}{243}-\frac{4124
   \pi ^4}{1215}\right) L_H^2 +\left[-\frac{235168 \zeta_3}{81}+\pi ^2 \left(\frac{3776 \zeta_3}{81}+\frac{5356}{729}\right) \right.\\
   &\left.+\frac{15328 \zeta_5}{9}+\frac{4877080}{2187}-\frac{514 \pi ^4}{405}\right] L_H-\frac{87112 \zeta_5}{81}+\pi ^2 \left(\frac{928 \zeta_3}{243} -\frac{124987}{729}\right)-\frac{25664 \zeta_3^2}{27}  \\
   &  +\frac{4274126 \zeta_3}{729}-\frac{492512 \pi ^6}{229635}+\frac{326479 \pi ^4}{21870}-\frac{145304189}{39366},\\
\end{split}\label{eq:Cns per order}
\end{equation}
where $L_H=\ln[(-M_L^2 - i \epsilon)/\mu^2]$. 

Next, the expression for the vector singlet contribution $C_\text{s}^V$ starts at the third-loop level and can be given as 
\cite{Baikov:2009bg,Gehrmann:2010ue}
\begin{equation} 
  \begin{split}
    C_\text{s}^V
    =
      \left(\frac{\alpha_s}{4\pi}\right)^3 
      \left(
        \frac{280\zeta_3}{27}-\frac{1600\zeta_5}{27}+\frac{100\pi^2}{27}
        +\frac{80}{9}-\frac{2 \pi ^4}{81}
      \right)\;.
  \end{split}\label{eq:CVS per order}
\end{equation}
Note that as there are no $\epsilon$-poles confronted   by $C_\text{s}^V$at $\mathcal{O}(\alpha_s^3)$,  the expression in eq.\ \eqref{eq:CVS per order} contains no logarithmic terms.

At last, the axial vector function $C_{\text{s}}^A$ induced by the $\mathcal{O}_s$ operator will be determined.  Similarly, we also introduce the perturbative expansion for $C_{\text{s}}^A$ here, 
\begin{equation}
\begin{split}
C_{\text{s}}^A=\sum_{i=0}^{\infty}   \left(  \frac{\alpha_s}{4\pi} \right)^i  C_{\text{s}}^{A,(i)},
\end{split}
\end{equation}
where $C_{\text{s}}^{A,(i)}$ encodes the $\mathcal{O}_s$  contribution in each order. Topologically,  $\mathcal{O}_s$ can conduct both the singlet and non-singlet Feynman diagrams (see Fig.~\ref{fig:methods:feyndia_hard}).  Considering that the singlet part will start to work at $\mathcal{O}(\alpha_s^2)$,  $C_{\text{s}}^A$ should be identical to $C_{\text{ns}}$ up to NLO. 
Hence we have
\begin{equation}
\begin{split}
C_{\text{s}}^{A,(0)}=&\;1\;,  \\
C_{\text{s}}^{A,(1)}=&\; -\frac{4}{3}  L_H^2+4 L_H+\frac{2 \pi ^2}{9}-\frac{32}{3}\;.
\end{split}\label{eq:CAS per order:01}
\end{equation}
From \NNLO, $C_{\text{s}}^A$ however comprises both the singlet and non-singlet contributions. To discuss,  it is convenient to subtract the $C_{\text{ns}}$ 
from $C_\text{s}^A$, and define the pure singlet contribution as $C_\text{ps}^A\equiv (C_\text{s}^A-C_{\mathrm{ns}})/N_F$ (see Fig. \ref{fig:methods:feyndia_hard:sing_axial}). 
For now, the \NNLO\ calculation on $C_\text{ps}^A$ has been 
carried out in Ref.~\cite{Bernreuther:2005rw} based on the prescription in \cite{Larin:1993tq}, while the $C_{\text{ns}}$ expressions have been specified above.  After the combination, we get
\begin{equation}
\begin{split}
C_{\text{s}}^{A,(2)}=& \frac{8 L_H^4}{9}-\frac{52 L_H^3}{27}+\left(\frac{28 \pi ^2}{27}-\frac{418}{27}\right) L_H^2+\left(-\frac{184 \zeta_3}{3}+\frac{20 \pi ^2}{27}+\frac{11090}{81}\right) L_H+\frac{2356 \zeta_3}{27}\\
&+\frac{46 \pi ^4}{81}+\frac{83 \pi ^2}{81}-\frac{143401}{486}.\\
\end{split}\label{eq:CAS per order:2}
\end{equation}
In comparison to $C_{\text{ns}}^{(2)}$, it is observed that the terms proportional to $L_H^2$ and those with the higher power remain the same, while the participation of singlet contribution has modified the others. This phenomenon is in agreement with the expectation of eq.\ \eqref{eq:methods:res_rge_rage_C}, where the cusp anomalous dimension receives no changes but the non-cusp one endures an extra term, $\gamma_t$. We also have checked that up to $\mathcal{O}(\alpha_s^2)$, $C_{\text{s}}^A$ indeed satisfies the corresponding RGEs in eq.\ \eqref{eq:methods:res_rge_rage_C}.

In order to obtain the third order contribution, we resort to the perturbative solution of the RGE in eq.\ \eqref{eq:methods:res_rge_rage_C}, which gives, 
\begin{equation}
\begin{split}
C_{\text{s}}^{A,(3)} =& -\frac{32}{81}  L_H^6-\frac{80 }{81}L_H^5+\left(\frac{2486}{81}-\frac{128 \pi ^2}{81}\right) L_H^4+\left(\frac{736 \zeta_3}{9}-\frac{416 \pi ^2}{243}-\frac{36244}{243}\right) L_H^3+\left(\frac{11024 \zeta_3}{81} \right.\\
&\left.+\frac{5612 \pi ^2}{243}-\frac{199966}{729}-\frac{4124 \pi ^4}{1215}\right) L_H^2+\left[\pi ^2 \left(\frac{3776 \zeta_3}{81}-\frac{24884}{729}\right)+\frac{15328 \zeta_5}{9}-\frac{235168 \zeta_3}{81} \right.\\
& \left.-\frac{514 \pi ^4}{405}+\frac{8541520}{2187}\right] L_H +c_{\text{s}}^{A,(3)}.\\
\end{split}\label{eq:CAS per order:3}
\end{equation}
It is also seen that the coefficients in front of $L_H^4$, $L_H^5$ as well as $L_H^6$ all stay still with respect to $C_{\text{ns}}^{(3)}$, while    $\gamma_t$ make differences in the others. Also note that there is one constant term  $c_{\text{s}}^{A,(3)}$ which relies on the third order expressions of $C_\text{ps}^A$ .  In this work,  we take $c_{\text{s}}^{A,(3)}=C_{\text{ns}}^{(3)}|_{_{L_H=0}}$ to contain the non-singlet contributions\footnote{
  Very recently a three-loop result has been presented in 
  Ref.~\cite{Gehrmann:2021ahy}.
}.